\documentclass[a4paper,11pt]{article}
\usepackage{jheppub} 
\usepackage{lineno}
\usepackage[utf8]{inputenc}
\usepackage{amsmath, amssymb, amsthm}
\usepackage{framed}
\usepackage{physics}
\usepackage{graphicx}
\usepackage{mathtools}
\usepackage{hyperref}
\usepackage{import}
\usepackage{esint}
\usepackage{upgreek}
\usepackage{xifthen}
\usepackage{hyperref}
\usepackage{comment}
\usepackage{enumitem}
\usepackage[dvipsnames]{xcolor}
\usepackage{pifont} 
\usepackage{multirow}
\usepackage[export]{adjustbox}
\usepackage{tikz}
\usepackage{array}
\usepackage[normalem]{ulem}
\def\be{\begin{equation}}
\def\ee{\end{equation}}
\def\ba#1\ea{\begin{align}#1\end{align}}
\def\bg#1\eg{\begin{gather}#1\end{gather}}
\def\bm#1\em{\begin{multline}#1\end{multline}}
\def\bmd#1\emd{\begin{multlined}#1\end{multlined}}

\def\a{\alpha}
\def\b{\beta}

\def\D{\Delta}

\def\l{\lambda}

\def\r{\rho}

\def\W{\Omega}

\def\SBPS{S_\text{BPS}}

\def\({\left(}
\def\){\right)}
\def\[{\left[}
\def\]{\right]}
\def\<{\langle}
\def\>{\rangle}

\def\tr{\operatorname{tr}}
\def\Tr{\operatorname{Tr}}

\newcommand{\ff}[1]{\left(#1\right)!!}

\newcommand{\cmark}{\ding{51}}%
\newcommand{\xmark}{\ding{55}}%

\renewcommand{\(}{\left(}
\renewcommand{\)}{\right)}
\renewcommand{\[}{\left[}
\renewcommand{\]}{\right]}

\newcommand{\inlinefig}[2][10]{
{\includegraphics[height=#1\fontcharht\font`\B, valign=c]{#2}}
}

\title{

Living on the edge: a non-perturbative resolution to the negativity of bulk entropies  \\

}

\author{Stefano Antonini,}
\author{Luca V. Iliesiu,}
\author{Pratik Rath,}
\author{Patrick Tran}
\affiliation{Leinweber Institute for Theoretical Physics and Department of Physics, University of California,
 Berkeley, CA 94720 USA}

\emailAdd{santonini@berkeley.edu}
\emailAdd{liliesiu@berkeley.edu}
\emailAdd{pratik\_rath@berkeley.edu}
\emailAdd{patrick.tran@berkeley.edu}

\abstract{Lin, Maldacena, Rozenberg, and Shan (LMRS) presented a new information paradox in black hole physics by noticing that the entanglement and R\'enyi entropies in a two-sided black hole can become negative when the geometry contains a very large number of matter excitations behind the black hole horizon. While originally this puzzle was presented in the context of BPS two-sided black holes in two-dimensional supergravity, the negativity in fact persists for more general two-sided black holes in the presence of a large number of matter excitations. Since the entanglement and R\'enyi entropies in ordinary quantum systems cannot be negative, resolving this puzzle is a necessary step towards understanding the quantum mechanical description of black holes. In this paper, we explain how to address the entanglement negativity puzzle, both in the original setting discussed by LMRS and in more general non-supersymmetric settings, by summing over all non-perturbative contributions to the gravitational path integral.
We then interpret this result from the point of view of a dual matrix integral, which we use to extend our analysis beyond the regime of validity of the genus re-summation performed in the gravitational path integral. In this regime, positivity is rescued by new saddles of the matrix integral, a one-eigenvalue instanton and a two-eigenvalue instanton. Finally, we formulate a similar puzzle and its resolution using random tensor network techniques.

}

\begin{document}
\maketitle
\flushbottom

\section{Introduction}

In the past few years, the gravitational path integral has led to tremendous progress in understanding why black holes are described by conventional quantum systems. Examples include the correct recovery of the Page curve from replica wormhole contributions to the path integral \cite{Penington:2019npb, Almheiri:2019psf, Penington:2019kki, Almheiri:2019qdq}, detecting the chaotic statistics in the spectrum of black hole microstates \cite{Saad_Shenker_Stanford_2019,Saad_2019}, and exact state counts of BPS states by localizing the supergravity path integral \cite{Iliesiu:2022kny}. The common thread among all these developments is understanding the contributions to the path integral that are non-perturbatively small, but become important when probing extreme regimes.

Despite these developments, there are still several outstanding puzzles. One such puzzle \cite{Lin_Maldacena_Rozenberg_Shan_2023,Lin:2022rzw} arises when computing the entanglement or R\'enyi entropy between the two sides of a two-sided black hole when a large number of matter excitations are inserted behind the black hole horizon. Such two-sided states can be prepared using the Euclidean gravitational path integral. Consider inserting $k$ simple matter operators on the Euclidean asymptotic boundary, all projected to the same energy window, in the preparation of the thermofield-double state within a fixed energy window,\footnote{Similar states, often called partially-entangled thermal states (PETS), were also studied in e.g. \cite{Goel:2018ubv,Hsin:2020mfa,Sasieta:2022ksu,Balasubramanian:2022gmo,Balasubramanian:2022lnw,Antonini:2023hdh,deBoer:2023vsm,Antonini:2024mci,Magan:2025hce,Antonini:2025ioh,Engelhardt:2025vsp}.}
\be 
\ket{\psi_k} = \inlinefig[8]{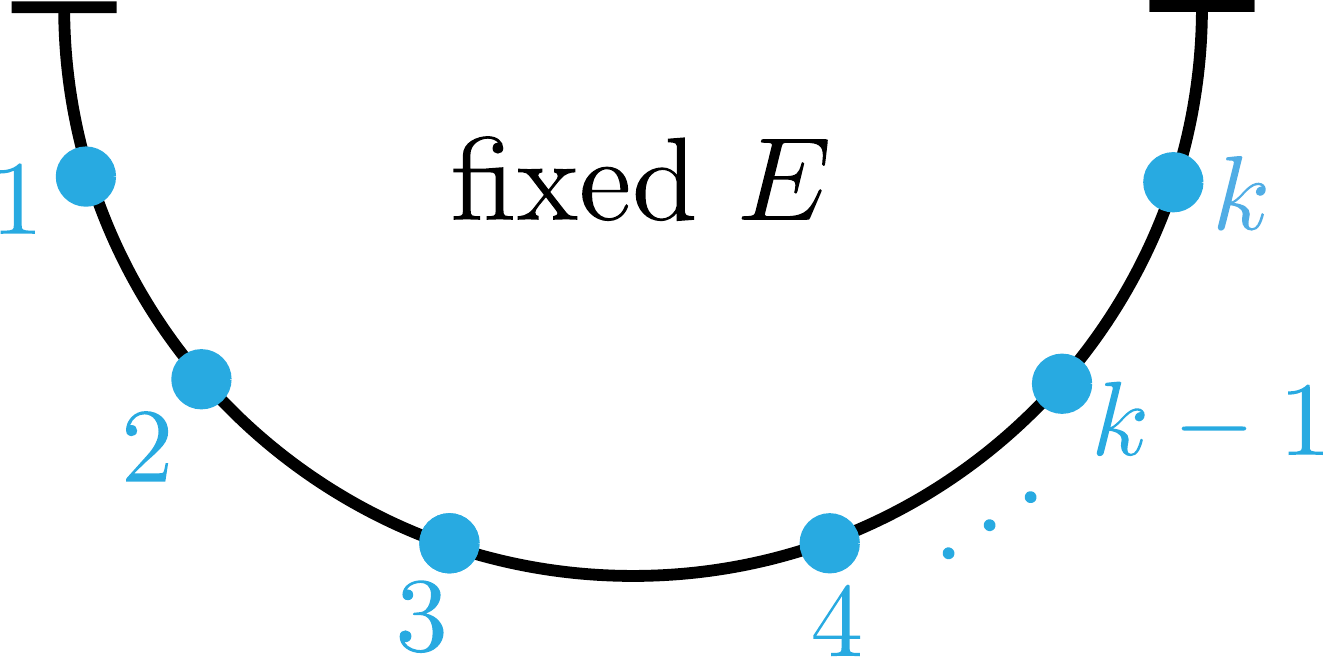} \,.
\label{eq:state-preparation-intro}
\ee 
As the number $k$ of matter insertions is increased, the entanglement entropy and R\'enyi entropies, 
 \be 
 \label{eq:EE-and-RE-def}
 S^{(n)} = \frac{1}{1-n} \log \left(\frac{\tr \rho^n}{(\tr \rho)^n}\right), \qquad S_{EE} = \lim_{n \to 1} S^{(n)} =  -\tr \left( \frac{\rho}{\tr \rho} \log \frac{\rho}{\tr \rho}\right)\,,
 \ee 
of the density matrix $\rho = \tr_L \ket{\psi_k}  \bra{\psi_k} $ of either one of the two sides (say the right side, $R$) eventually becomes negative when computed using the gravitational path integral at disk level. This is displayed by the green curve in Figure \ref{fig:moneyplot}, which shows the analytically computed second R\'enyi entropy as a function of $k$ in the specific example of two-sided BPS black holes with operator insertions projected to the BPS state. Such negativity should be impossible in a conventional quantum system whose density matrices only have eigenvalues between 0 and 1, consequently implying $S_{EE} \geq 0$ and $S^{(n)} \geq 0$. This negativity of the entanglement entropy and  R\'enyi entropies can be viewed as yet another version of the black hole information paradox---while in the Page curve formulation of the paradox the problem is that the semi-classical answer for the entanglement entropy of Hawking radiation exceeds its theoretically allowed maximum value, here the problem is that the entanglement entropy of the black hole exceeds its theoretically allowed minimum value of zero. The goal of this paper is to address the negativity puzzle by using the gravitational path integral for a variety of two-sided black holes, both in the original supersymmetric setup described by LMRS as well as in non-supersymmetric settings. 

 \subsection*{Summary of Results}
 
We start by formulating a precise version of the negativity puzzle in Section \ref{sec:2}. For that, we explain the extent to which the gravitational path integral is capable of precisely computing R\'enyi  and entanglement entropies. To exactly compute all these quantities, the path integral should be capable of computing all the elements of the density matrix $\rho$. Unfortunately, given the current capabilities, the gravitational path integral cannot exactly compute $\rho$; rather, it can only compute coarse-grained statistics of $\rho$. Thus, when computing the entropies in \eqref{eq:EE-and-RE-def}, there are various possible ways in which one can coarse-grain, take powers, take fractions, or compute logarithms. Luckily, with enough care, the gravitational path integral can distinguish between all these different orderings and, in principle, is capable of computing a zoo of entropies---annealed, quenched, semi-quenched, and quasi-quenched---that capture different properties of coarse-grained systems. By clarifying which entropies---the quenched, semi-quenched, and quasi-quenched---are required to be positive, we thus make the negativity puzzle sharper. For computational convenience, we focus on proving that the semi-quenched entropy remains positive for any value of $k$. Such an entropy can be shown to stay positive for all $k$ only if the quantum mechanical systems we coarse-grain over have isolated matter ``ground states''. Here, by matter ``ground state'' we mean the eigenstate associated with the eigenvalue of largest magnitude (the lowest or highest eigenvalue) of the projected operator used in \eqref{eq:state-preparation-intro}. This not only ensures that the quenched and quasi-quenched entropies also remain positive, but provides a direct probe for the discreteness of the black hole spectrum. 
  
Rephrasing the puzzle in terms of the eigenvalues of the unnormalized density matrix $\rho$ appearing in \eqref{eq:EE-and-RE-def} also helps clarify its origin. At leading order, the gravitational path integral only sees a continuous eigenvalue spectrum for $\rho$, which is in turn determined by the eigenvalue spectrum of the projected operators used in the state preparation \eqref{eq:state-preparation-intro}. For a large number of operator insertions, the eigenvalue spectrum of $\rho$ is supported almost exclusively close to the edge of the spectrum for the projected operator; hence, fewer states (those near the ground state or the maximum energy state) contribute to the entropies. Eventually, when the number of insertions is large enough that the effective number of states contributing to the entropy from this continuous spectrum becomes smaller than $1$, the entanglement and Rényi entropies become negative. This argument suggests that the negativity puzzle is generic and occurs for all two-sided black holes in AdS/CFT beyond the realm of super-JT gravity. In this generic setting, simple operators projected to a finite-dimensional subspace are also expected to have a spectrum with finite width. When these operators are projected onto a window where the spectrum consists almost entirely of black hole microstates, the path integral (at least at leading order) is also incapable of distinguishing the individual eigenvalues of the projected operator, just as it is incapable of distinguishing individual energy levels of black hole microstates. 

  \begin{figure}
    \centering
    \includegraphics[width=0.8\linewidth]{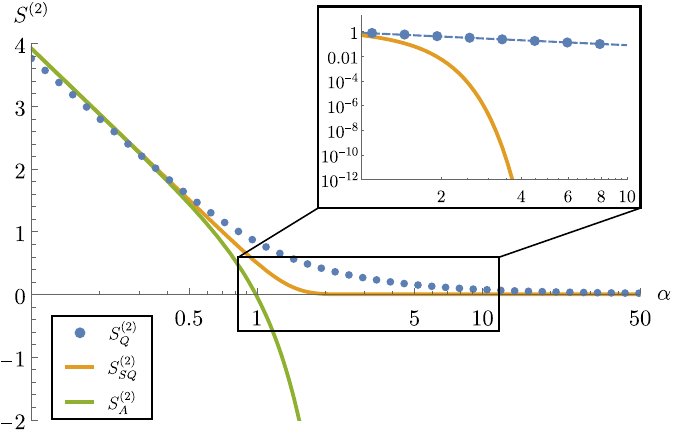}
    \caption{Comparing semi-quenched, annealed and quenched second R\'enyi-2 entropies, zoomed in near the Airy edge $k=\alpha e^{2S_{\text{BH}}/3}$.
    Semi-quenched and annealed R\'enyis are obtained analytically, whereas the quenched R\'enyi is obtained numerically.
    Inset: We zoom in on the quenched and semi-quenched entropies at large $k$ and plot them on a log-log scale. Notice the different behavior: the quenched entropy has an $\alpha^{-1}$ power-law decay (dashed blue line) which can be obtained analytically (see Appendix \ref{app:quenched}), whereas the semi-quenched entropy falls off exponentially. Nevertheless, as expected (and unlike the annealed entropy), our calculation shows that both entropies remain positive.}
    \label{fig:moneyplot}
\end{figure}

  In Section \ref{sec:bulk-resolution}, we then provide a bulk resolution to the negativity puzzle in the case where the operators used to prepare the state have a large scaling dimension. We first address the puzzle in the specific setup described by LMRS and then explain how to extend this result to the more general setting of non-supersymmetric two-sided black holes in JT gravity. We primarily focus on the computation of the R\'enyi-2 semi-quenched entropy, and explain how the positivity of higher R\'enyi entropies can also be proven, although we do not provide explicit formulas for them since computing them precisely is not necessary for proving the isolation of the ``matter ground state''. To compute entropies, we first explain how to compute matter correlation functions in super-JT and then JT gravity at arbitrary order in the genus expansion. The sum over all possible Wick contractions between the matter insertions can be reorganized as a finite sum over genus. In the supersymmetric LMRS setup, each given Wick contraction contributes precisely one diagram to this sum. The genus of such a diagram is the minimum possible genus in which the Wick contraction can be realized without any intersection between matter lines. We call this diagram the \textit{minimal embedding diagram} of the Wick contraction. In the non-supersymmetric case, higher-genus diagrams can also contribute for each Wick contraction. However, in the limit of our interest where the number of insertions is large (in particular, $k=O(e^{2\SBPS/3})$), the minimal embedding diagrams dominate the genus expansion and the analysis is identical to the supersymmetric case.
  
  When $k = O( e^{2\SBPS/3})$, namely in the regime in which the annealed entropy becomes negative, the correlators greatly simplify and the genus expansion can be explicitly re-summed (at least for one- and two-boundary correlators). In the same regime, a novel equality arises: in both supersymmetric and non-supersymmetric JT gravity, the contributions to matter correlation functions coming from surfaces of a given genus $g$ are equal to the contributions to the low-temperature thermal partition function in pure, non-supersymmetric JT gravity from spacetimes with the same genus $g$. For example, on a genus-1 surface, we find that the sum over all Wick contractions between the matter insertions can be rewritten as
  \begin{equation}
\begin{gathered}
     \inlinefig[10]{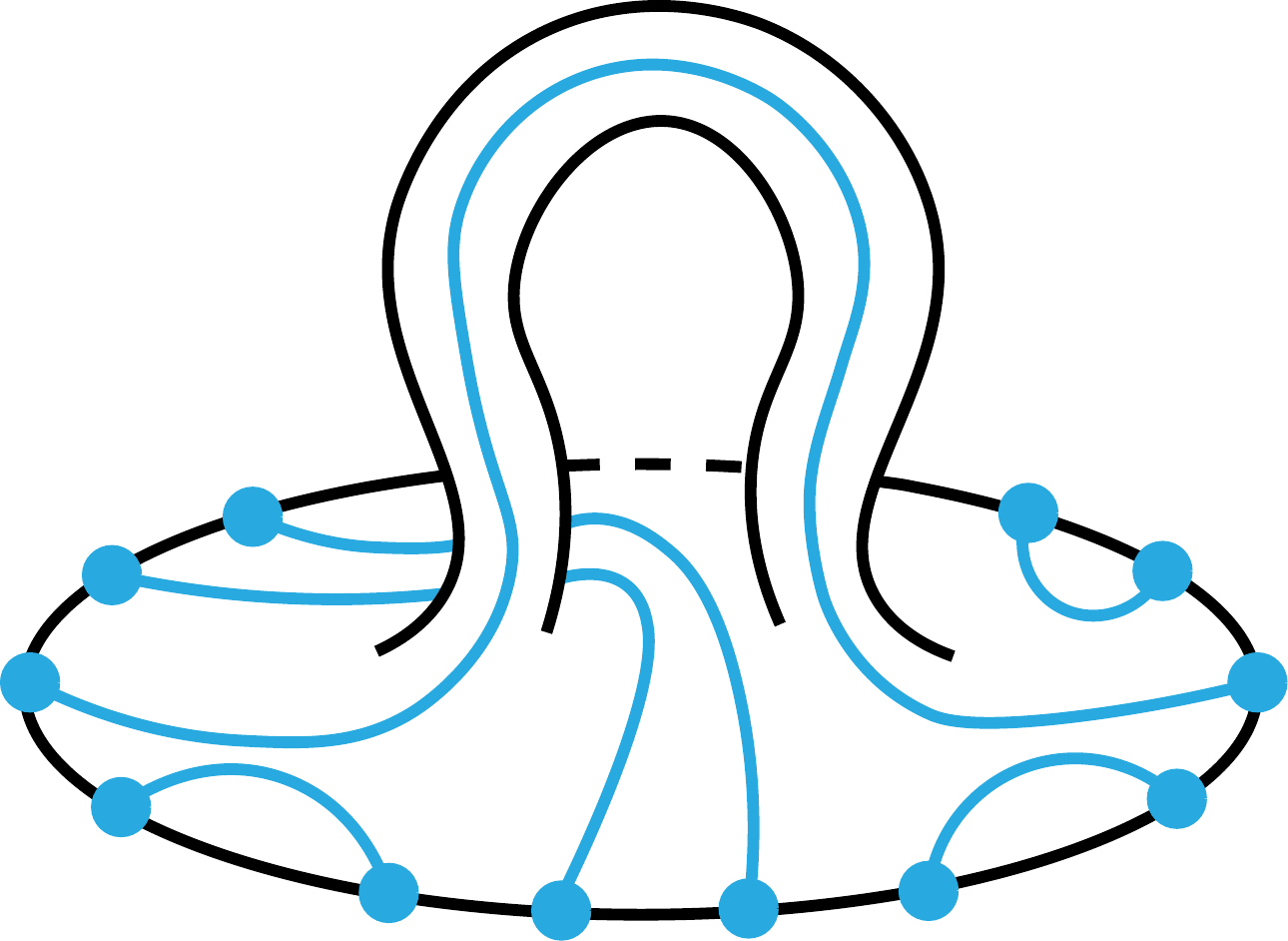}+\inlinefig[10]{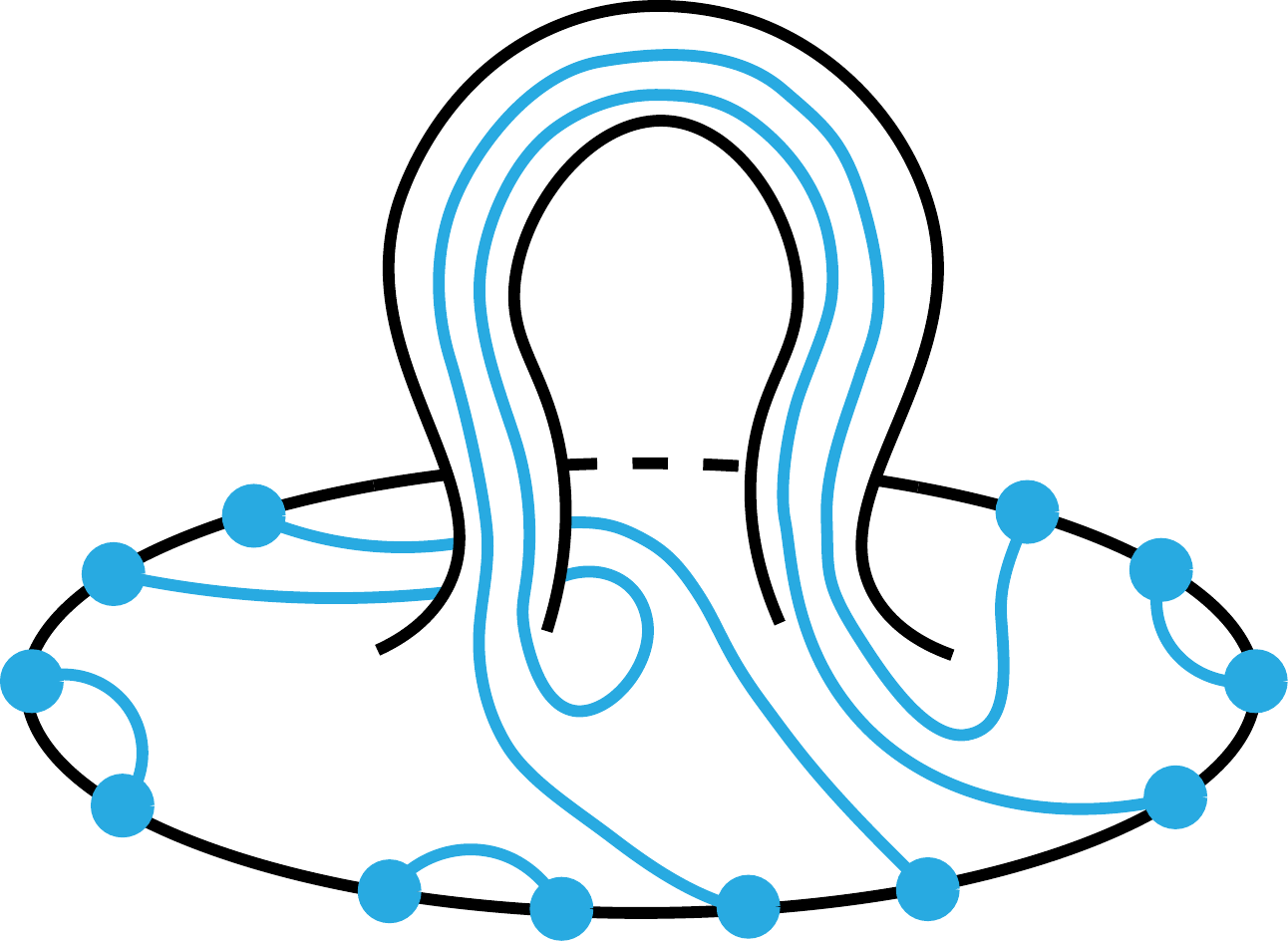} +\quad \dots\quad=\quad 2\times \inlinefig[10]{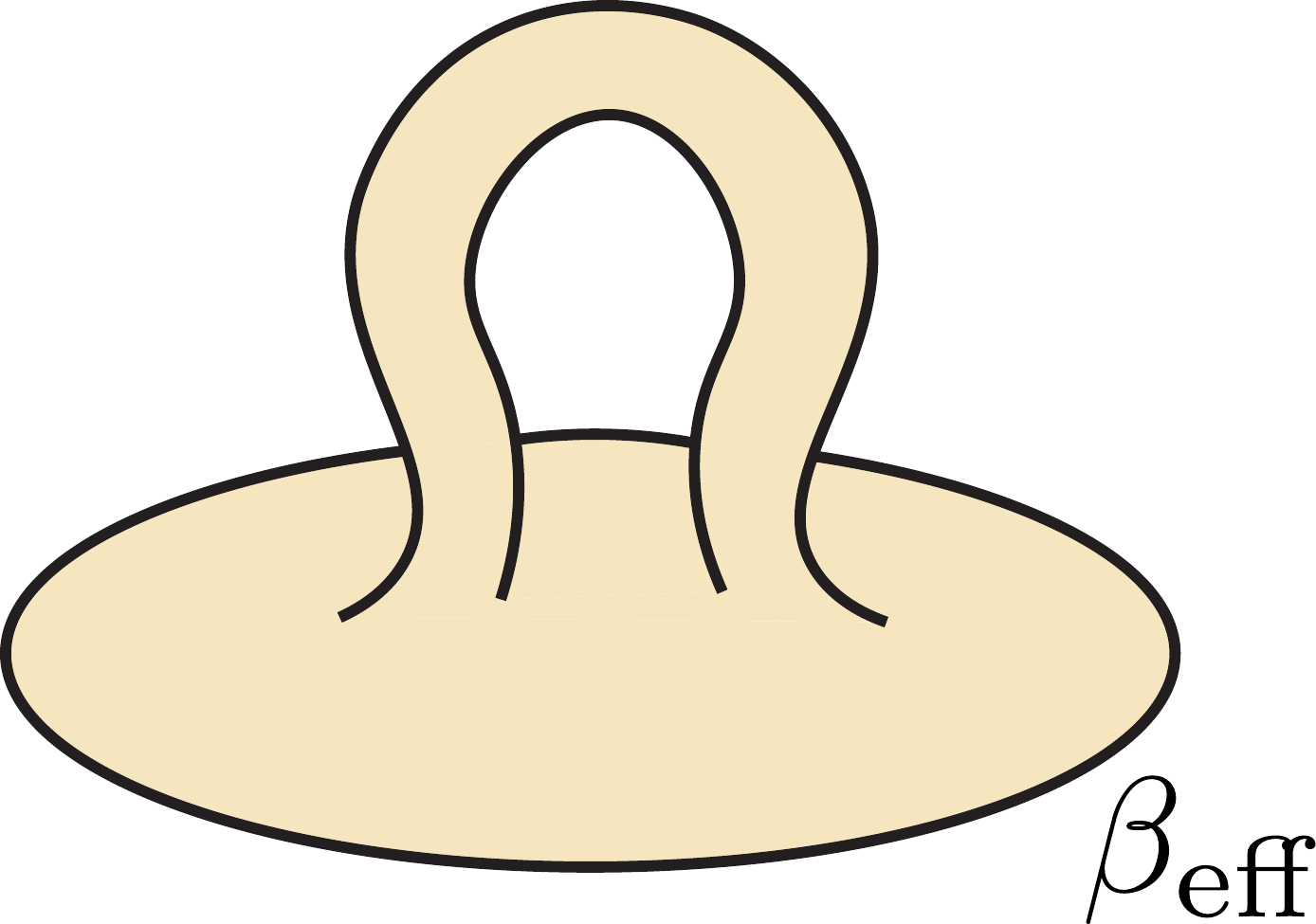}
    \end{gathered}
\label{eq:duality-with-effective-description}
\end{equation}
where the thermal partition function is evaluated at an inverse effective temperature  $\beta_\text{eff} = k$. This equality, which continues to hold for any genus and any number of boundaries, implies that the puzzle of the two-sided entanglement entropy negativity discussed above can be directly mapped to the puzzle of thermal entropy negativity encountered in JT gravity at very low temperatures \cite{Engelhardt_Fischetti_Maloney_2021,Hernandez-Cuenca_2024}, which was recently addressed in \cite{Antonini_Iliesiu_Rath_Tran_2025}.   
  
Using this novel equality, we then prove that the semi-quenched entropy, which became negative when only keeping the leading order geometry, stays positive after re-summing the entire genus expansion and taking into account multi-boundary wormhole contributions.

The above result is no longer valid when $k = \omega(e^{2\SBPS/3})$ because the re-summation over all topologies has to be done exactly, rather than by discarding terms subleading in $\beta$ in the regime where $\beta/e^{2\SBPS/3} \sim O(1)$. To solve the problem for such a large number of operator insertions, in Section \ref{sec:matrixmodels} we show that there is an exact duality between matter correlators with large scaling dimension in gravity and observables in a Gaussian GUE random matrix integral. Schematically, we show that at every order in the genus expansion 
\be 
\overline{\tr(\tilde O^{2k_1}) \dots \tr(\tilde O^{2k_n})}_g = \langle \tr(M^{2k_1}) \dots \tr(M^{2k_n}) \rangle_g\,,
\label{eq:matrix-integral-gravity-duality}
\ee
where the left-hand-side is given by the contribution on surfaces of genus $g$ of all correlators of the projected matter operator $\tilde O$, and the right-hand-side is given by the matrix integral non-planar diagrams at genus $g$ that contribute to the above correlator.\footnote{This equality was conjectured by \cite{Lin_Maldacena_Rozenberg_Shan_2023} and was checked at disk level. In this paper, we explain why this equality continues to hold for all orders in the genus expansion. To do that, we also build on the results of \cite{Boruch:2023trc} that computed the correlation functions in BPS states on surfaces with an arbitrary genus and an arbitrary number of boundaries. } Using this equality, we explain how to compute the entropy when $k = \omega(e^{2\SBPS/3})$ directly in the matrix integral using new saddles that dominate the matrix integral: a one-eigenvalue instanton and a two-eigenvalue instanton. In this regime, we show that the semi-quenched entropy still remains positive. The duality \eqref{eq:matrix-integral-gravity-duality} not only allows us to extend the regime of validity of the entropy positivity, but has several other benefits: it is useful in calculating the quenched entropy in addition to the semi-quenched entropy; it provides a clear explanation for the effective description introduced in \eqref{eq:duality-with-effective-description} in terms of the universal Airy behavior of matrix integrals when zooming onto their spectral edge; and it paves the way towards understanding the resolution of the puzzle for arbitrary values of $\Delta$ by making use of this universal regime.

In Figure \ref{fig:moneyplot}, we put all these results together and show our results for the annealed, semi-quenched, and quenched entropy in the regime $k=O(e^{2\SBPS/3})$.

In Section \ref{sec:TN}, we provide a different perspective on the negativity puzzle by using a random tensor network (RTN) model. This approach allows us to explore a different version of the puzzle: rather than considering two-sided black hole states constructed by many identical Hermitian operator insertions as in \eqref{eq:state-preparation-intro}, the same negativity puzzle arises when using sufficiently many non-identical non-Hermitian operator insertions in the state preparation. Using tensor network techniques, we once again show that the semi-quenched entropy remains positive when re-summing all-genus contributions. We also explain how different choices of configurations of permutations in the RTN are in one-to-one correspondence with possible Wick contractions of matter fields in the gravitational theory. The RTN should thus be considered a tool for formulating and solving this alternative version of the gravitational puzzle. The advantages of the RTN formulation are in its simplicity, its ability to explicitly compute R\'enyi-$n$ entropies with $n>2$, and its potential for a generalization to higher-dimensional setups.

In section \ref{sec:discussion}, we discuss the broader implications of our results.

In the appendices, we focus on technical details and generalizations of the above results. In Appendix \ref{6j}, we argue why for large scaling dimensions we can neglect the contribution of diagrams that contain intersecting Wick contractions in non-SUSY JT gravity, a result that we make use of when computing the R\'enyi entropies in Section \ref{sec:bulk-resolution}. We also discuss why, for a small scaling dimension, all diagrams are weighted equally, independent of the number of intersections; the same arguments were applied in the supersymmetric case in \cite{Lin_Maldacena_Rozenberg_Shan_2023}. In Appendix \ref{sec:derivation-2n-pt-func-non-SUSY-JT} we explain how, in the large $\Delta$ limit, we can compute the $2k$-point correlator for identical operators projected to a small energy window in non-SUSY JT gravity. This is used in Section \ref{sec:bulk-resolution} to generalize the entropy positivity results found in the BPS case to the non-SUSY case. 

In Appendices \ref{sec:ETH} and \ref{sec:genericdresolution}, we explain how our positivity results can also be generalized away from large $\Delta$. The small $\Delta$ limit proves to be particularly interesting. In that case, the naive computation of $2k$-correlation function reveals that the entropy decreases linearly with $k$ and goes negative when $k$ is linear in the entropy of the black hole; this should be contrasted with the limit of large scaling dimension where the entropy decreases logarithmically with $k$ and goes negative when $k$ is exponential in the entropy of the black hole. We explain that this is due to an order of limits issue: when $k \gg 1/\Delta$, the entropy changes behavior and starts decreasing logarithmically with $k$, rather than linearly, making the entropy once again only go negative for exponential values of $k$ in physically motivated scenarios (i.e, $\Delta \gg e^{-S_{\text{BH}}}$). In this regime, the finite edge of the projected operator becomes visible, and the techniques described above can still be used to resolve the entropy negativity. In Appendix \ref{sec:ETH}, we explain the difficulties in describing the puzzle and its resolution for generic values of $\Delta$, and introduce a toy model, the $q$-deformed Gaussian matrix model, which captures the relevant features of the problem. In particular, the $q$-deformed Gaussian is able to interpolate between the large $\Delta$ and small $\Delta$ limits, while retaining a square-root edge for any value of $\Delta$. In Appendix \ref{sec:genericdresolution}, we then use this toy model to solve the puzzle at generic $\Delta$ with techniques similar to those employed in Section \ref{sec:matrixmodels}.

In Appendix \ref{appendix:f}, we derive a generic, explicit expression for the coefficients of the genus expansion for two-boundary correlation functions in JT (super)gravity coupled to heavy matter (i.e., two-trace correlation functions in the GUE, as explained in Section \ref{sec:matrixmodels}), and study their simplification in the limit of many operator insertions. These results are used in Section \ref{sec:bulk-resolution} to compute the R\'enyi-2 semi-quenched entropy and solve the LMRS puzzle.

In Appendix \ref{app:quenched}, we use the matrix integral description to show that, in the large $\Delta$ limit, the quenched entropy also stays positive. We also show that it decays as $1/k$ for large enough $k$, and explain the origin of this behavior.

Finally, in Appendix \ref{app:RTNs}, we give explicit expressions for the R\'enyi-3 and R\'enyi-4 entropies in the random tensor network setup of Section \ref{sec:TN}. We also extend the setup of Section \ref{sec:TN} to Hermitian operators using RTNs constructed using random orthogonal (rather than unitary) tensors, explain how to compute entropies by mapping to an Ising-like model, and compute the semi-quenched R\'enyi-2 entropy, which once again is always positive.

\section{An Information Paradox}
\label{sec:2}

Let us begin by setting up the puzzle that is at the core of this paper. It can be thought of as a version of the information paradox, which was originally discussed in the context of supersymmetric black holes in \cite{Lin_Maldacena_Rozenberg_Shan_2023}. We will first review it in their setup and then discuss how it is a much more general problem, which we will address in the following sections.

\subsection{Review of the puzzle}
\label{sec:review-of-the-paradox}

Ref.~\cite{Lin_Maldacena_Rozenberg_Shan_2023} worked with $\mathcal{N}=2$ JT supergravity, a two-dimensional toy model of quantum gravity that is computationally tractable whilst exhibiting many of the puzzles encountered in higher-dimensional gravitational theories. Besides serving as a toy model, $\mathcal{N}=2$ JT supergravity also provides an effective description for higher-dimensional black holes in AdS, capturing the behavior of nearly-supersymmetric black holes close to extremality \cite{Boruch:2022tno}.\footnote{A similar puzzle arises for supersymmetric black holes in flat space where the effective theory is described by $\mathcal N=4$ super-JT gravity \cite{Heydeman:2020hhw}. To see this, one can use the more complicated matter correlators in $\mathcal N=4$ super-JT gravity discussed in \cite{Lin:2025wof}.  } To pose the entanglement entropy negativity puzzle, let us first review the basics of this theory.  The theory is defined in terms of a metric $g_{\mu\nu}$ and dilaton field $\phi$, as well as their super-partners, the gravitino and dilatino, and other bosonic components belonging to the same supermultiplet, a zero-form Lagrange multiplier $b$, and a $U(1)$ gauge field $A$ (with a field strength $F=dA$). The action is schematically given by
\be 
\label{eq:action-N=2-super-JT}
I_{\mathcal{N}=2\text{ JT}} = -S_0 \chi_{g,n} - \frac{1}2 \int_{M_{g,n}} \left[\phi \left(R+\Lambda\right) +  b\,F + \text{(fermions)}\right] + \text{(bdy.~terms)}\,,
\ee
where the boundary terms depend on the boundary condition that we choose for all the fields above. To compute observables in this theory, one performs a path integral with the action \eqref{eq:action-N=2-super-JT} summing over all spacetime surfaces $M_{g,n}$ with any genus $g$ and a given number of boundaries $n$. The parameter $S_0$ multiplies the Euler characteristic of the spacetime, $\chi_{g,n}$. When taking $S_0$ large, this term suppresses the contribution of higher genus topologies when summing over geometries, making surfaces with $g=0$ dominate the sum over geometries. 
Since we thus need to only be concerned with disconnected geometries at leading order, we can take $n=1$ for the remainder of this section. 

To compute observables, we can integrate out the dilaton, dilatino, and zero-form Lagrange multiplier in the path integral, making the boundary terms in \eqref{eq:action-N=2-super-JT} the only surviving terms in the action. If imposing Dirichlet boundary conditions,\footnote{E.g.,~ fixing the value of the dilaton and induced metric at the boundary $\partial M_{0,1}$ of $M_{0,1}$.} this term can be rewritten in terms of the $\mathcal N=2$ super-space parametrization of the boundary $\partial M_{0,1}$. The action of these super-reparametrizations is given by the $\mathcal N=2$ super-Schwarzian
\be 
I_{\mathcal N=2\text{ Schw}} = E_\text{brk.}^{-1} \int_0^\beta d\tau \left[\text{Sch}(f, \tau) + 2(\partial_\tau \sigma)^2 + (\text{fermions})\right]\,,
\ee
where $\tau$ parametrizes the boundary time, while the bosonic fields $f(\tau)$ and $e^{i r \sigma(\tau)} \in U(1)$ as well as their fermionic superpartners describe the shape of the boundary in super-space.\footnote{The action could also contain a topological term $I_\text{topological} = i \theta r \int d\tau (\partial_\tau \sigma)$, coming from a bulk $\theta$-angle for the gauge field $A$. Adding this term affects observables in the theory, in particular changing the number of BPS states \cite{Boruch:2022tno}. For simplicity, we will take $\theta = 0$.} The inverse temperature $\beta$ and the coupling $E_\text{brk.}$ are determined by the boundary conditions for the induced metric and dilaton. The parameter $r$ determines the periodic identification of $\sigma (\tau)$ with $\sigma \sim \sigma+ \frac{2\pi}r$.  From the bulk perspective, $r$ is determined by the smallest R-charge among all matter fields that could possibly be coupled to the U(1) R-symmetry gauge field $A$ in \eqref{eq:action-N=2-super-JT}. Consistency with $\mathcal N=2$ supersymmetry requires $1/r$ to be an integer. 

The partition function in this theory can be computed exactly \cite{Stanford:2017thb, Boruch:2022tno}: its spectrum exhibits a degeneracy of $O(e^{S_0})$ at zero energy, and a continuum of states cleanly separated by a non-zero energy gap (assuming $1/r$ is odd).\footnote{For $1/r$ even, the spectrum does not have a gap, but there is still a large degeneracy of BPS states. } The number of BPS states is given by
\be
Z(\beta\to \infty) = \sum_{\substack{Z\in r \cdot \mathbb Z \\ |Z| < \frac{1}2 }} e^{S_0} \cos\left(\pi Z\right) \equiv e^{\SBPS}\,,
\label{eq:BPS-degeneracy}
\ee 
where the sum is over the possible $U(1)$ R-symmetry charges that BPS states can have. For simplicity, in the remainder of the paper, we will simply focus on a theory with $r=1$, in which all BPS states have $Z=0$.\footnote{This is a well-motivated setup since the $\mathcal N=2$ super-Schwarzian with $r=1$ is actually the relevant effective theory to describe the near-BPS limit of black hole microstates in $\mathcal N=4$ super Yang-Mills.} In such a case, note from \eqref{eq:BPS-degeneracy} that $\SBPS = S_0$.

The fact that an exact degeneracy of BPS states is visible directly in the gravitational theory allows one to sharply define a projector $P_0$ onto the subspace of BPS states. The insight of LMRS \cite{Lin_Maldacena_Rozenberg_Shan_2023} was that one could then consider the preparation of an arbitrary BPS wormhole state by considering the insertion of simple operators followed by a projection onto the BPS subspace, which is implemented using an infinite amount of Euclidean evolution, i.e.,
\begin{equation}
    \tilde{O} = P_0 O P_0 = \lim_{\b\to\infty} e^{-\b H} O e^{-\b H},
\end{equation}
where $O$ is a simple operator and the operator $\tilde{O}$ will be referred to as an LMRS operator. For simplicity, we will only consider simple operators neutral under the $U(1)$ R-symmetry. 

Such LMRS operator insertions allow one to prepare wormhole states that correspond to BPS boundary states, but which break all supersymmetry in the bulk. To set up the paradox, let us now choose a particular $O_{\D}$ which is a primary with conformal dimension $\D$ in the 1D boundary CFT, which we take to be a generalized free field in the bulk. For most of this paper, except in Section \ref{sec:TN}, we will take this operator to be Hermitian. We can then consider a one-parameter family of BPS states given by
\begin{equation}
    \ket{\psi_k}= \(\widetilde{O}_\D\)^k \ket{\text{TFD}_{0}} = \inlinefig[8]{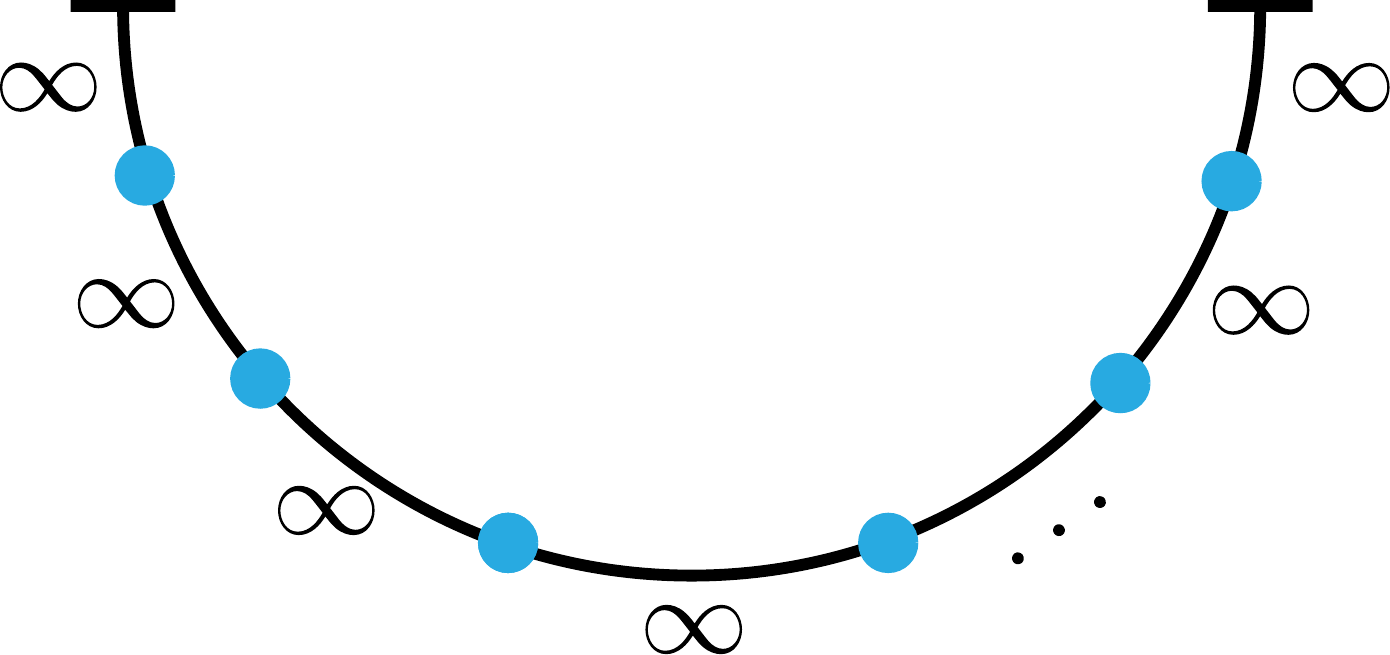} ,
\end{equation}
where $\ket{\text{TFD}_0}$ is the zero temperature thermofield double state. Such states have $k$ operator insertions in the Euclidean past that generate a long wormhole with many particles in the interior of the two-sided black hole.

The paradox then arises from computing the R\'enyi entropies $S_n$ of either asymptotic boundary (say the right boundary, $R$) given by
\begin{equation}
\label{eq:Renyi-entropy-definition}
    S_n\(\r\)=\frac{1}{1-n}\log \[\frac{\tr \(\r^n\)}{\(\tr \r\)^n}\],
\end{equation}
where we have allowed for an unnormalized density matrix $\r = \tr_L \(\ket{\psi_k} \bra{\psi_k}\)$. It is then straightforward to see from the Euclidean path integral that 
\begin{equation}
    \tr\(\r^n\) = \tr\[\(\widetilde{O}_\D\)^{2kn}\],
\end{equation}
where we remind the reader that the trace can be restricted to just the BPS subspace, since either side vanishes outside this subspace. Thus, computing the $n$-th R\'enyi entropy requires us to compute the $2kn$-point correlation function of the LMRS operator $\widetilde{O}_\D$.

While such correlation functions could potentially be computed for general $\D$ using the methods of \cite{Penington:2024sum}, it is generally complicated. Instead, we can discuss two simplifying limits following \cite{Lin_Maldacena_Rozenberg_Shan_2023}. 

\paragraph{The $\Delta \to \infty$ limit. }The first limit of interest is when $\D\to\infty$. In this limit, any Wick contraction that involves crossing diagrams is exponentially suppressed in $\D$.\footnote{Specifically, compared to a diagram where two given Wick contractions are non-intersecting, the intersecting diagram is suppressed as \cite{Lin_Maldacena_Rozenberg_Shan_2023}, 
\be 
\frac{\text{Intersecting}}{\text{Non-intersecting}}  \sim e^{-4\Delta \log\left(1+\sqrt{2}\right)}\,, \qquad \text{ for } \Delta \to \infty.
\ee 
} 
Thus, one finds that the only contributing diagrams are planar, resulting in\footnote{To simplify the calculation, below we will assume that $\Delta$ is the largest parameter in the problem. However, following the discussion in Appendix \ref{sec:ETH} the results below should be valid even when taking $\Delta\gg 1$ up to an overall small shift in the location of the edge of the eigenvalue spectrum for the operator $\tilde O_\Delta$.}
\begin{equation}
\begin{split}
    \tr(\r^n)&=
    \inlinefig[10]{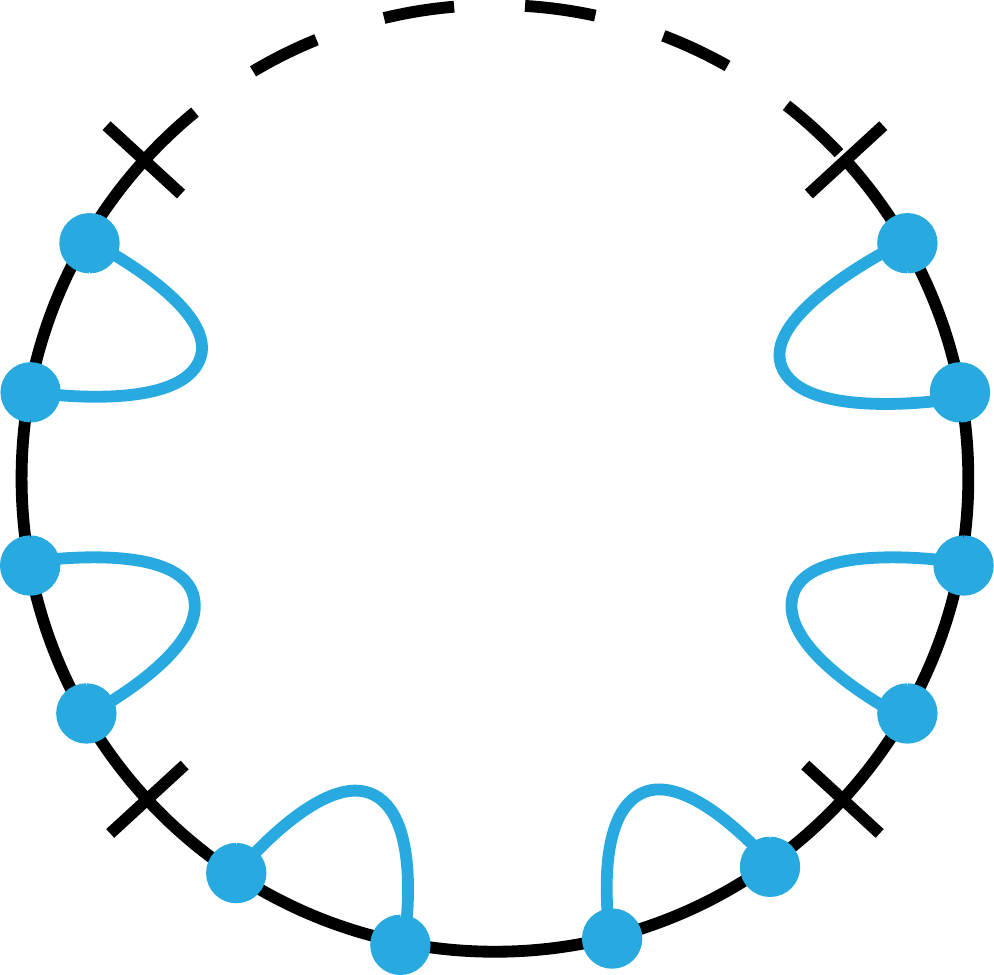}+ \inlinefig[10]{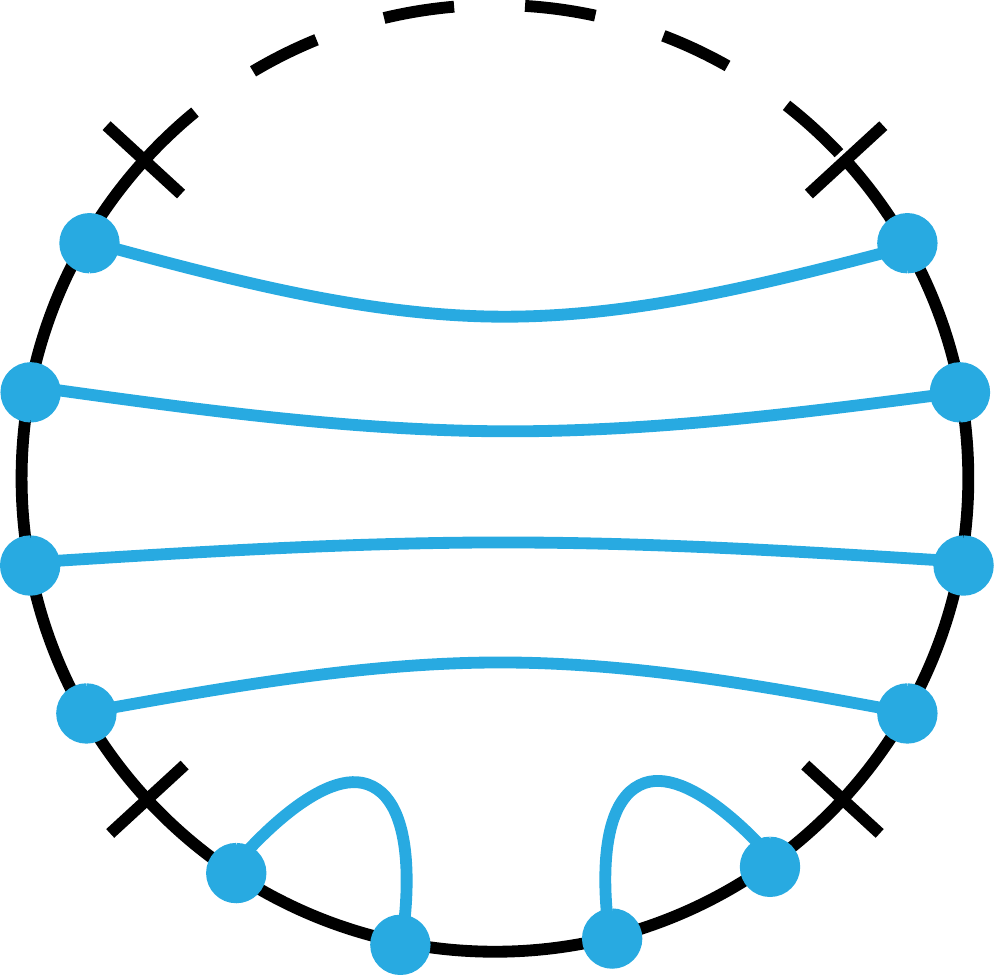}+\dots
    \\& = e^{\SBPS} C_{kn} (P_{\Delta \to \infty})^{kn}.
\end{split}
\label{eq:moments-of-rho-Delta-to-infinity}
\end{equation}
The black dashes indicate the separation between any two copies of the density matrix $\rho$; in total, each contribution thus has $n$ dashes. 
Above, $C_n=\frac{(2n)!}{(n+1)!n!}$ is the Catalan number that counts the number of non-crossing diagrams, and shows up frequently in the context of random matrix theory. $P_{\Delta \to \infty}$ is the contribution from the propagator associated with a Wick contraction, and in this example, is independent of the positions of the operator insertions since all operators are separated by infinite Euclidean time.\footnote{\label{foot:PBPS} Specifically, in $\mathcal N=2$ super-JT gravity $P_{\Delta} = \frac{1}{2\pi}\frac{\Delta(\Gamma(\Delta))^2}{\Gamma(2\Delta)}(\Gamma(\Delta+\frac{1}{2}))^2$ \cite{Boruch:2023trc}. Note that as $\Delta \to \infty$, $P_\Delta \to \infty$. This appears counter-intuitive since in the path integral we are inserting $e^{-\Delta \ell}$. However, the growth in $\Delta$ is due to the strong backreaction of the insertion of the heavy operators on the shape of the AdS$_2$ boundary, with operator insertions coming very close together in the large $\Delta$ limit.
} The normalization factor for the density matrix $\rho$ in \eqref{eq:Renyi-entropy-definition} is similarly given by
\begin{equation}
    (\tr\r)^n = e^{n \SBPS} (C_{k})^n (P_{\Delta\to\infty})^{kn}.
\end{equation}
Finally, computing the R\'enyi entropies, we find that at large $k$
\begin{equation}
\label{eq:Sn-Delta-equals-infinity}
    S_n(\r)=\SBPS - \frac{3}2 \log(k) + O(1)\,,
\end{equation}
which is independent of the propagator $P_{\Delta\to\infty}$ whose value consequently proved unimportant.
Thus, when $k$ becomes sufficiently large, $k > e^{2\SBPS/3} + O(1)$, we find that all R\'enyi entropies and the entanglement entropy become negative, something that should be impossible in an ordinary quantum system.

\paragraph{The $\Delta \to 0$ limit. } The second limit that is under analytic control is the $\Delta \to 0$ limit. In this limit, individual crossing diagrams and non-crossing diagrams contribute exactly the same amount to the $n$-point function of BPS projected simple operators. Thus, to compute $\tr\[\(\widetilde{O}_{\D \to 0}\)^{2kn}\]$, we simply need to count the number of Wick contractions between $2kn$ operators. In total, one finds:
\be 
\label{eq:moments-of-rho-Delta-to-0}
\begin{split}
\tr(\r^n)&=\inlinefig[10]{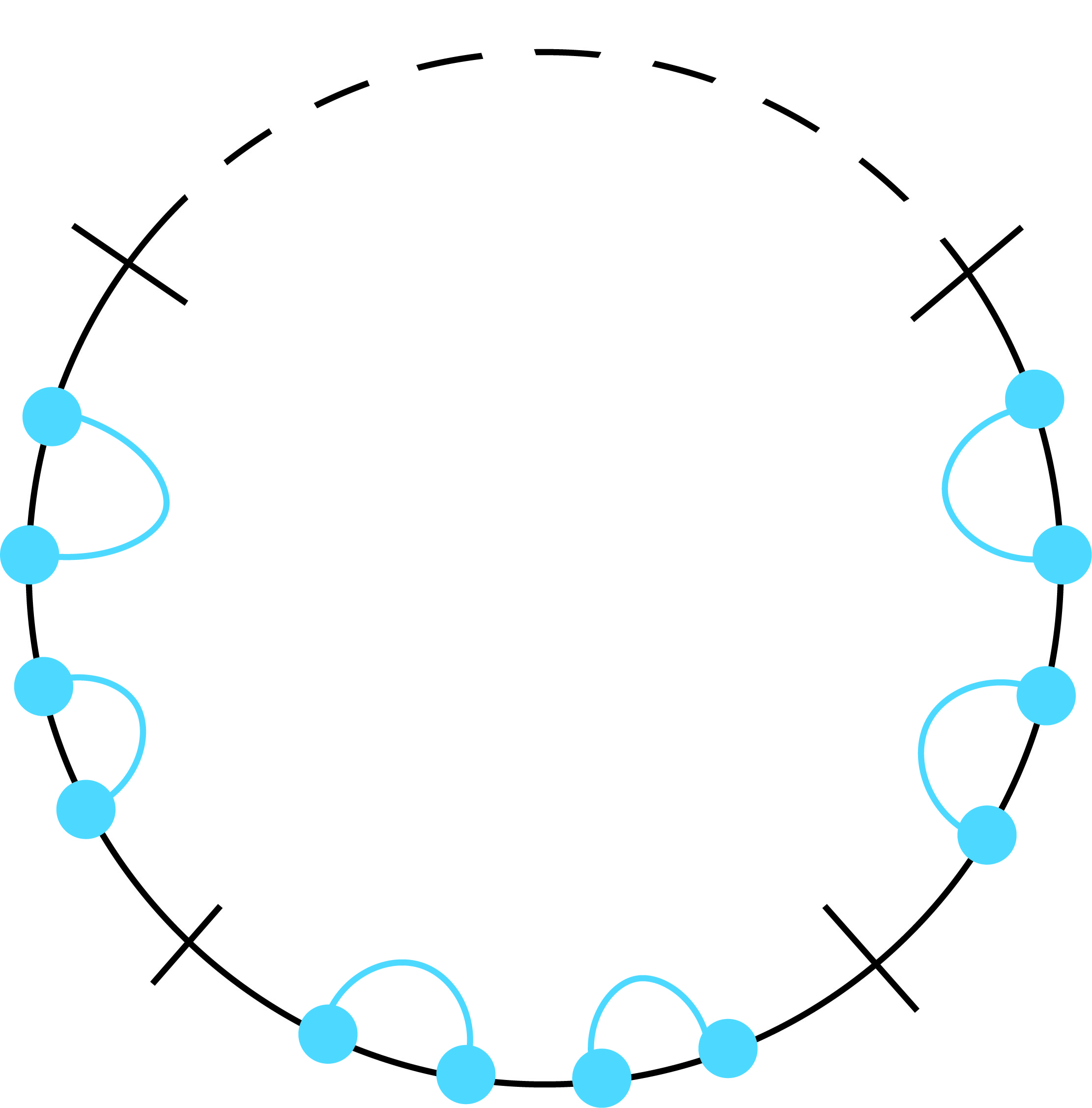}+ \inlinefig[10]{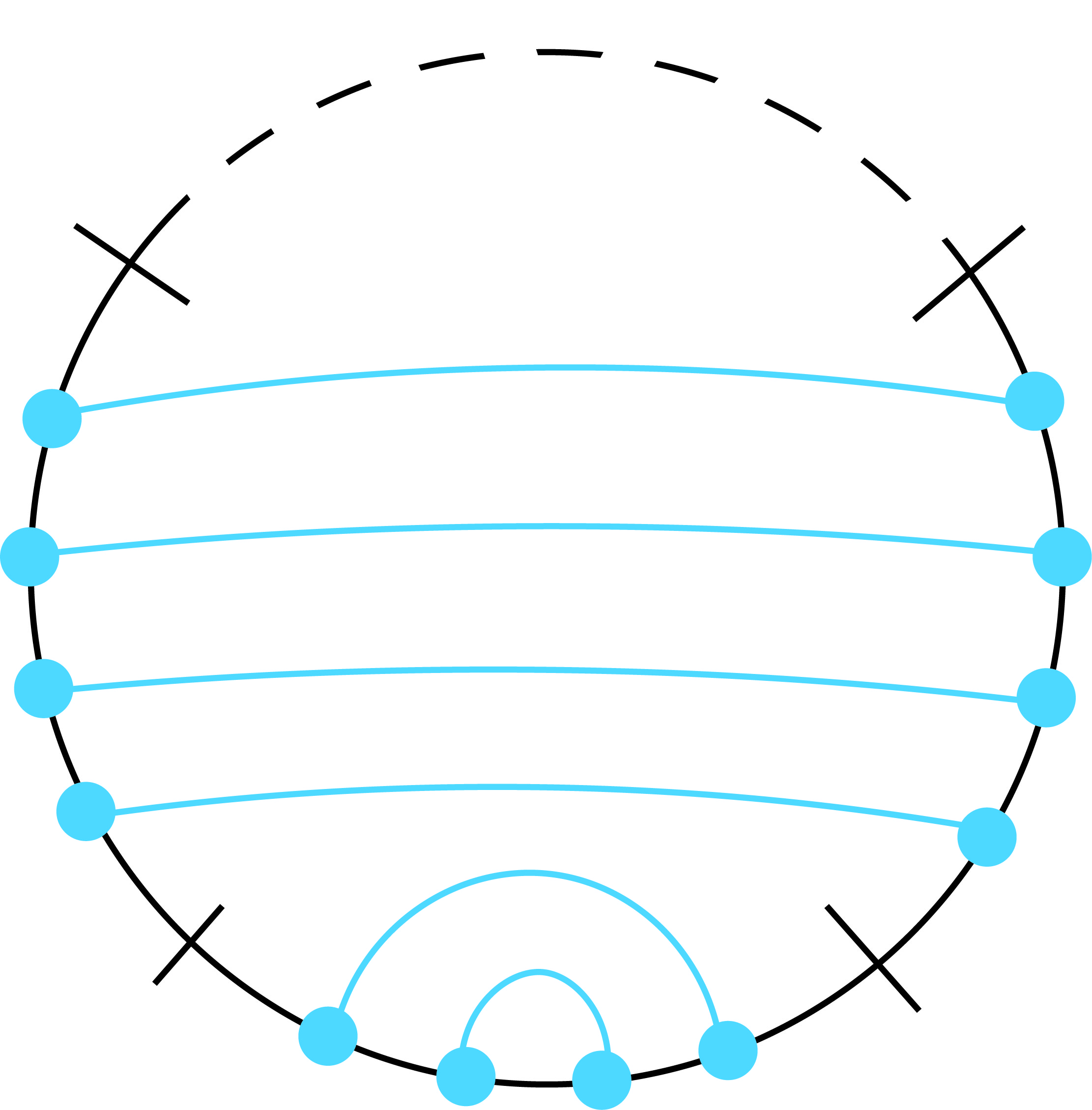}+\inlinefig[10]{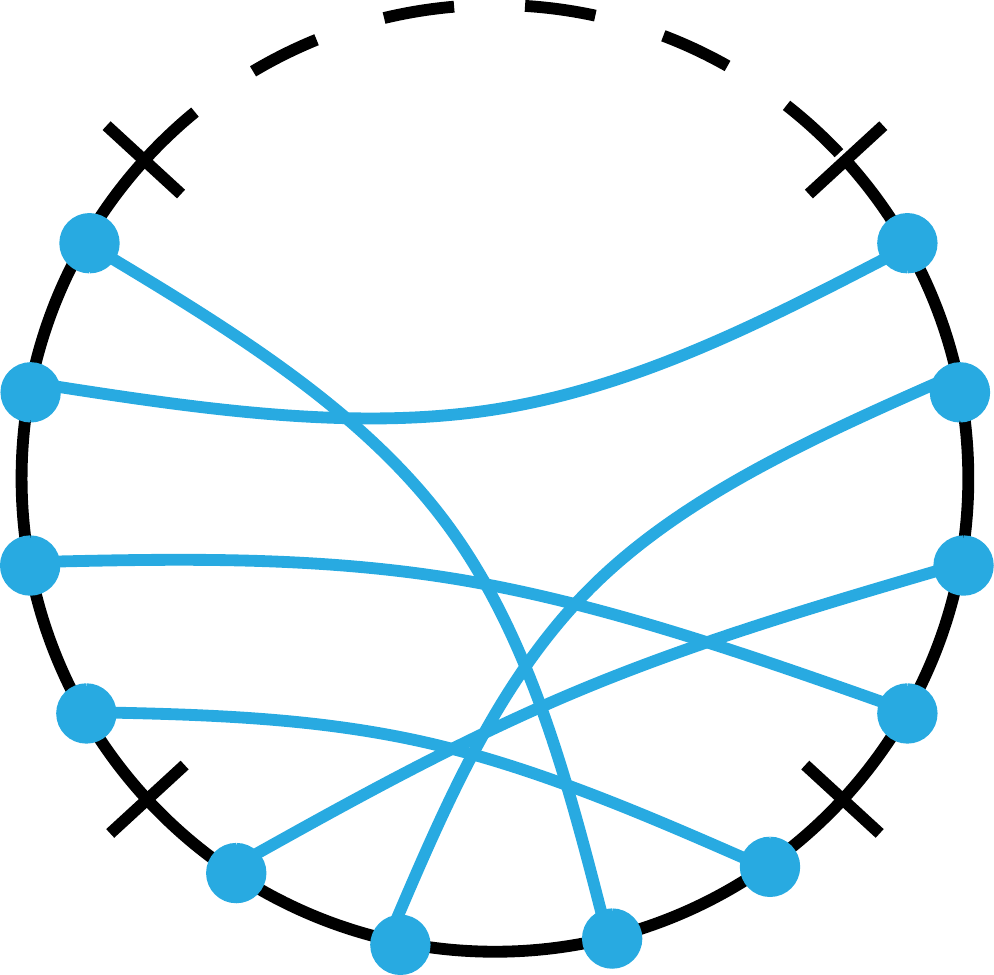}+ \inlinefig[10]{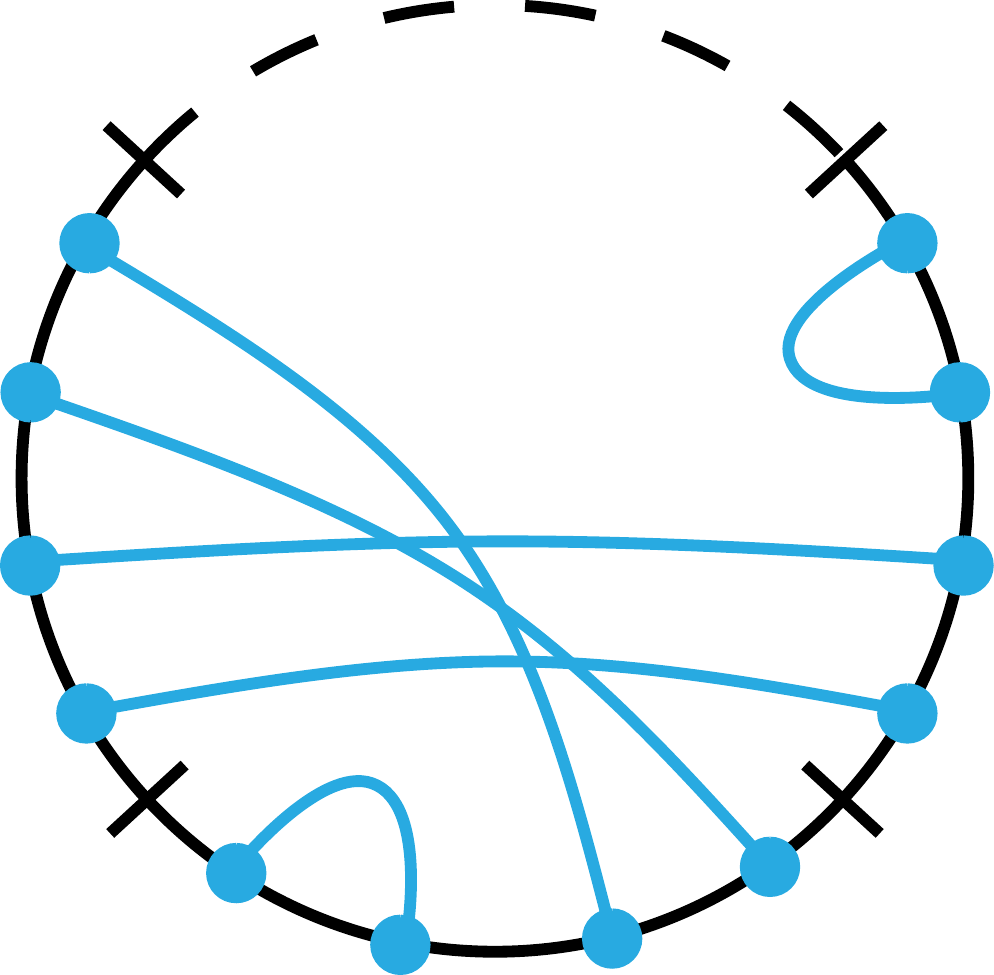}+\dots\\
& = e^{\SBPS} (P_{\Delta \to 0})^{kn} (2k n-1)!!
\end{split}
\ee
where $P_{\Delta \to 0}$ is the propagator between two BPS projected simple operators. The R\'enyi entropies at large $k$ are thus given by 
\be
\label{eq:Sn-Delta-equals-zero}
S_n(\rho) = \SBPS - k \frac{n \log n}{n-1} + O(1)\,.
\ee
This now becomes negative for $k > \frac{\SBPS(n-1)}{n \log n} + O(1)$. Similarly, the entanglement entropy 
\be 
S_{\text{EE}}(\rho) = \SBPS - k + O(1)\,,
\ee
goes negative for $k> S_0$. Thus, as $\Delta \to 0$, the puzzle of entropy negativity is encountered for a much smaller number of operator insertions compared to the large $\Delta$ limit, where the puzzle arose for a number of operator insertions that is exponential in $S_0$.

\paragraph{The root of the problem: a continuous eigenvalue distribution.} Why does the gravitational path integral seemingly find negative entropies? Instead of computing the entropy from the moments $\tr\(\r^n\)$, we can take a different perspective on this puzzle and use these moments to compute the density of eigenvalues for the operator $\tilde{O}_\Delta$. The resulting spectrum is continuous.\footnote{This should perhaps not be surprising. Only a very special set of moments that satisfy trace relations between lower and higher moments originates from discrete spectra.  } For example, as $\Delta \to \infty$, we obtain a density of eigenvalues $\sigma_\Delta(x)$ \cite{Lin_Maldacena_Rozenberg_Shan_2023}\footnote{Notice that this is the density of eigenvalues for the operator $\tilde{O}_\Delta$, and not the density of states of $\mathcal{N}=2$ JT supergravity, where the BPS sector is entirely degenerate and has constant energy.}
\be
\label{eq:eval-density-Delta-to-infty}
\sigma_{\Delta\to \infty} (x) = e^{\SBPS}\frac{1}{2\pi P_{\Delta \to \infty}}\sqrt{4P_{\Delta\to\infty}-x^2}
\ee
which has a square-root edge at $x_*^{\pm}  = \pm 2 \sqrt{P_{\Delta \to \infty}}$. Above, the eigenvalue density is normalized such that 
\be 
\int_{x_*^-}^{x_*^+} dx\,\sigma_{\Delta\to \infty} (x)  = e^{\SBPS},
\ee
i.e., it recovers the correct number of BPS states. The density of eigenvalues \eqref{eq:eval-density-Delta-to-infty} can indeed be seen to satisfy 
\be 
\label{eq:moments-Delta-to-infty-from-eval-density}
\tr\[\(\widetilde{O}_{\D \to \infty}\)^{2kn}\] = \int dx \,\sigma_{\Delta\to \infty} (x) x^{2kn } = e^{\SBPS} C_{kn} \left(P_{\Delta \to \infty}\right)^{kn}\,.
\ee
When $\Delta \to 0$, the density of states becomes,\footnote{In the $\Delta \to 0$ limit, we see that the edge of the spectrum has moved to $x_{*}^\pm \to \pm \infty$. First, $\Delta \geq 0$ is the unitarity bound for CFT${}_1$, see e.g. \cite{Qiao:2017xif}. Taking $\Delta=0$ exactly corresponds to considering an operator whose two-point function is either constant or scales logarithmically. In the former case, this simply means that the operator is the identity, and our analysis does not apply. In the latter case, the operator is the alternative quantization of a massless scalar and its two-point function can become negative at small distances, hence violating reflection positivity. The strict $\Delta\to 0$ limit is therefore sick. Thus, when using $\Delta\to 0$ we mean $\Delta \neq 0$ with $\Delta \ll 1$, without scaling $\Delta$ with $S_0$. We will discuss this limit further in Appendices \ref{sec:ETH} and \ref{sec:genericdresolution}.\label{foot:delta0}}
\be 
\label{eq:eval-density-Delta-to-0}
\sigma_{\Delta\to 0} (x) = \frac{e^{\SBPS}}{\sqrt{2\pi P_{\Delta \to 0}}}e^{-\frac{x^2}{2P_{\Delta \to 0}}}, 
\ee  
also normalized to recover the correct number of BPS states upon integration. The moments of this density of eigenvalues agree with \eqref{eq:moments-of-rho-Delta-to-0}, 
\be 
\label{eq:moments-Delta-to-0-from-eval-density}
\tr\[\(\widetilde{O}_{\D\to 0}\)^{2kn}\] = \int dx \,\sigma_{\Delta\to \infty} (x) x^{2kn } = e^{\SBPS} (P_{\Delta \to 0})^{kn} (2k n-1)!!\,.
\ee 
This is because the moments of a Gaussian distribution are given by the number of Wick contractions among $2kn$ points. 

Thus, for both large and small $\Delta$, we see that the moments \eqref{eq:moments-Delta-to-infty-from-eval-density} and \eqref{eq:moments-Delta-to-0-from-eval-density} come from a continuous density of eigenvalues for the operator $\tilde O_\Delta$. The fact that $\sigma_\Delta(x)$ is continuous is, in fact, the root of the entropic puzzle. When studying a very large number of operator insertions, the eigenvalue spectrum of the density matrix $\rho =  \frac{(\tilde O_{\Delta})^{2k}}{\Tr(\tilde O_{\Delta})^{2k}}$  is also continuous and has a density skewed either towards the edge of the spectrum (if $\tilde O_{\Delta}$ has an edge, as we have seen for $\Delta \to \infty$) or towards large values of $x$ (if $\tilde O_{\Delta}$ does has no edge, as we have seen for $\Delta \to 0$). In both cases, the average number of states that the density matrix $\rho$ has non-vanishingly small support on\footnote{I.e.,~the number of states $\int_a^bdx\,\sigma_{\Delta}(x)$ for which $\frac{\sigma_{\Delta}(x) x^{2k}}{\int dx\,\sigma_{\Delta}(x) x^{2k}}>\epsilon$ for some chosen small value $\epsilon$.} decreases as $k$ is increased. Eventually, for a continuous density of eigenvalues, the effective number of states can become much smaller than $1$. When this happens, the R\'enyi entropies
\be 
S_n(\rho) = \frac{1}{1-n} \log\left[\frac{\int dx \,\sigma_{\Delta}(x) x^{2kn}}{\left(\int dx \,\sigma_{\Delta}(x) x^{2k}\right)^n}\right]
\ee
become negative, thus resulting in the puzzle that we encountered above.

\paragraph{Boundary expectations: why $S\to 0$ with a large number of insertions.} In a genuine quantum system, we expect the spectrum of eigenvalues in a finite-dimensional subspace to be discrete rather than continuous. Since it is useful to understand what to expect when correcting the density of eigenvalues found above, we can analyze the behavior of the R\'enyi entropies assuming that the projected operator $\widetilde{O}_\D$ has a discrete set of eigenvalues $\lambda_i$. In the large $k$ limit, $
\tr \rho^n$ is dominated by the eigenvalue of the operator $\tilde O_\Delta$ whose absolute value is largest. Thus, at leading order, the R\'enyi entropy can be written as
\begin{align}
S_{n}(\rho) &= \frac{1}{1-n} \log\left[\frac{\tr(\rho^n)}{(\tr \rho)^n}\right] = \frac{1}{1-n} \log \left[\frac{1+\sum_{i\neq \text{max}} \left(\frac{\lambda_i}{ \lambda_\text{max} } \right)^{2kn}}{\left(1+\sum_{i\neq \text{max}} \left(\frac{\lambda_i}{ \lambda_\text{max} } \right)^{2k}\right)^n}\right] \nonumber\\ &\approx \frac{n}{n-1}   \left(\frac{\lambda_{\text{max}-1}}{ \lambda_\text{max} } \right)^{2k} + O\left[\left(\frac{\lambda_{\text{max}-2}}{ \lambda_\text{max} } \right)^{2k}\right]
\label{eq:Renyi-entropy-approaching-zero}
\end{align}
where $\lambda_\text{max}$ is the eigenvalue of $\widetilde{O}_\D$ with largest absolute value and $\lambda_{\text{max}-1}$ is the eigenvalue with the second largest absolute value. At large $k$, we thus expect the R\'enyi entropies to always remain positive and to vanish in the $k\to\infty$ limit. 
Furthermore, Eq.~\eqref{eq:Renyi-entropy-approaching-zero} also indicates that the way in which the  R\'enyi entropies approach zero from above sensitively depends on the spacing between the two largest eigenvalues of  $\tilde O_\Delta$.

\subsection{Sharpening the paradox: which entropies need to be positive }
\label{sec:entropies}

\bgroup
\def\arraystretch{1}
\begin{table*}[t]
    \centering
    \begin{tabular}{|c|c|c|c|}
        \hline
        \textbf{Types} & \textbf{Rényi Entropies}  & 
        \begin{tabular}{@{}c@{}}\textbf{Positivity}\\\textbf{requirement}\end{tabular} &
        \textbf{Entropies} ($n \to 1$) \\
        \hline\hline 
        & & &\\[-15pt]
        Quenched 
        & $ S_Q^{(n)} \equiv \frac{1}{1-n} \overline{\left(\log \frac{\tr(\rho^n)}{(\tr\rho)^n}\right)}$ 
        & \cmark 
        & \multirow{2}{*}[-1.4ex]{$\overline{\log \tr \rho } -\overline{ \left( \dfrac{\tr(\rho \log \rho)}{\tr \rho} \right)}$} \\
        & & &\\[-15pt]
        \cline{1-3}
        & & &\\[-15pt]
        Quasi-Quenched 
        & $S_{QQ}^{(n)} = \frac{1}{1-n} \log \overline{\left( \frac{\tr \rho^n}{(\tr \rho)^n} \right)}$ 
        & \cmark 
        & \\
        & & &\\[-15pt]
        \hline 
        & & &\\[-15pt]
        Semi-Quenched 
        & $S_{SQ}^{(n)} \equiv \frac{1}{1-n} \log \frac{\overline{\tr\, \rho^n}}{\overline{ (\tr \,\rho)^n }}$ 
        & \cmark 
        & $\dfrac{\overline{ \tr \rho \log \tr \rho }- \overline{ \tr(\rho \log \rho) }}{\overline{ \tr \rho }} $ \\
        & & &\\[-15pt]
        \hline
        & & &\\[-15pt]
        Annealed 
        & $S^{(n)}_{A} \equiv \frac{1}{1-n} \log \frac{\overline{\tr \, \rho^n }}{\(\overline{ \tr \, \rho }\)^n}$ 
        & \xmark 
        & $\log\overline{ \tr \rho } - \dfrac{\overline{ \tr(\rho \log \rho) }}{\overline{ \tr \rho }}$ \\[9pt]
        \hline
    \end{tabular}
    \caption{A summary of the different types of entropies that can be defined for ensembles of theories and a review of their properties. The difficulty of computation in gravity decreases as we go down the table.}
    \label{tab:entropy}
\end{table*}
\egroup

 The gravitational path integral fails to correctly capture the positivity of the R\'enyi or entanglement entropies because, at leading order in $G_N$, it only captures the coarse-grained spectrum of eigenvalues for the operator $\tilde O_\Delta$ instead of its exact spectrum. Since the coarse-grained spectrum can be continuous, the resulting R\'enyi entropies can become negative, as we have explained above. As we shall review in detail in Section \ref{sec:matrixmodels}, by accounting for subleading contributions in $G_N$ the path integral can instead recover more detailed statistics for the density of eigenvalues $\sigma_\Delta(x)$. The resulting statistics of $\sigma_\Delta(x)$ can oftentimes be seen to come from an ensemble of theories rather than an individual theory, explaining why the density of eigenvalues $\tilde O_\Delta$ can be continuous rather than discrete. 
Using such detailed statistics, one can hope that the subleading corrections in $G_N$ can resolve the negativity of the R\'enyi entropies seen in \eqref{eq:Sn-Delta-equals-infinity} and \eqref{eq:Sn-Delta-equals-zero}. We will see that this is indeed the case. 

To understand the statistical properties of the R\'enyi entropies, we first need to discuss the different generalizations of  \eqref{eq:Renyi-entropy-definition} that differ in their ordering of operations: taking logarithms, ratios, powers, and ensemble averages. Since not all generalizations turn out to be positive, the negativity puzzle discussed above will only apply to the versions of the R\'enyi entropies that are required to be positive. Conventionally, the two quantities discussed in the literature are the quenched entropies, which have a positivity requirement, and annealed entropies, which are not required to be positive.  In \cite{Antonini_Iliesiu_Rath_Tran_2025}, we discussed a larger class of R\'enyi and entanglement entropies which are more easily computable than the quenched entropy and retain its positivity requirements. These entropies are
compared in Table~\ref{tab:entropy} and we shall review their properties below.

\vspace{1em}
\paragraph{Annealed entropy.} 
We begin with the simplest quantities, the \emph{annealed Rényi entropies}, defined by
\begin{equation}\label{eq:ann}
    S^{(n)}_{A} \equiv \frac{1}{1-n} \log \frac{\overline{\tr \, \rho^n }}{\(\overline{ \tr \, \rho }\)^n}.
\end{equation}
These entropies treat the unnormalized density matrix $\rho$ in a fully averaged manner. They are the simplest entropies to compute using the gravitational path integral since, at integer $n$, they do not require any analytic continuation. As mentioned above, $S_A^{(n)}$ are not necessarily positive like entropies in an ordinary quantum system. This is because these quantities compute average moments of the matrix $\tilde\rho = \frac{\rho}{\overline{\tr\rho}}$. Because $\tilde\rho$ is only normalized on average, $\tilde\rho$ can have support on eigenvalues larger than 1, making it possible for the R\'enyi entropies to become negative. 

Taking the $n\to1$ limit, one obtains the annealed entanglement entropy
\begin{subequations}
\begin{align}
    S_A &= \log \overline{\tr \rho } - \frac{\overline{ \tr(\rho \log \rho) }}{\overline{ \tr \rho }}\,,
\end{align}
\end{subequations}
which similarly does not have any positivity requirements. 
\vspace{1em}

\paragraph{Quenched entropy.}
One can instead compute the R\'enyi entropies first and coarse-grain after. This results in the  \emph{quenched Rényi entropies}, defined as
\begin{equation}
    S_Q^{(n)} \equiv \frac{1}{1-n} \overline{\left(\log \frac{\tr(\rho^n)}{(\tr\rho)^n}\right)}.
    \label{eq:quenched-entropy}
\end{equation}
Since the R\'enyi entropies in each member of the ensemble are positive before coarse-graining, Eq.~\eqref{eq:quenched-entropy} retains the positivity of the standard R\'enyi entropies.

Compared to their annealed version, the quenched R\'enyi entropies are much harder to compute since they require the following replica trick:
\begin{equation}
    S_Q^{(n)}  =\frac{1}{1-n} \lim_{m\to 0} \left[\frac{\overline{(\tr \,\rho^n)^m}-\overline{(\tr\, \rho)^{nm}}}{m}\right],
\end{equation}
which can be computed at positive integer $n$ and $m$ using the gravitational path integral and require an analytic continuation for $m$. In the von Neumann limit ($n \to 1$), this gives the quenched entanglement entropy 
\begin{subequations}
\begin{align}
    S_Q &= \overline{\log \tr \rho } - \overline{\( \frac{\tr(\rho \log \rho)}{\tr \rho}\) },
\end{align}
\end{subequations}
which again requires a complicated replica limit.

\vspace{1em}
\paragraph{Semi-Quenched entropy.}
Since the annealed entropies go negative and the quenched entropies are hard to compute, we can instead define intermediate quantities that are forced to remain positive but are easier to compute. As we shall see shortly, the positivity of these quantities, as well as the positivity of the quenched entropies, will probe the existence of an isolated largest or smallest eigenvalue for the operator $\tilde O_\Delta$. These quantities are the \emph{semi-quenched R\'enyi entropies} defined in \cite{Antonini_Iliesiu_Rath_Tran_2025}\footnote{We remark that, although these quantities were formally introduced and studied in a similar context in \cite{Antonini_Iliesiu_Rath_Tran_2025}, they have been already studied in the literature in various settings as approximations to the quenched entropies, see e.g. \cite{Hayden:2016cfa,Antonini:2022sfm,Antonini:2023hdh}.} as
\begin{equation}
    S_{SQ}^{(n)} \equiv \frac{1}{1-n} \log \frac{\overline{\tr\, \rho^n}}{\overline{ (\tr \,\rho)^n }}.
\end{equation}
These quantities do not require any analytic continuation to be computed at integer $n>1$ and are thus easier to compute using the gravitational path integral. 

As explained in \cite{Antonini_Iliesiu_Rath_Tran_2025}, it is easy to show that the semi-quenched entropies are always positive, because $(\tr \rho^n)\leq (\tr \rho)^n$ for $n\geq 1$ for any member of the ensemble. Taking an ensemble average of this inequality results in the positivity of $S_{SQ}^{(n)}$. For $n<1$, the argument is similar except that the inequalities are reversed. Finally, for $n \to 1$, we define the semi-quenched version of the entanglement entropy,
\begin{equation}
    S_{SQ} = \frac{\overline{\tr \rho \log \tr \rho }- \overline{ \tr(\rho \log \rho) }}{\overline{ \tr \rho }},
\end{equation}
which is also always positive. 

\paragraph{Quasi-quenched entropy.} Another
possibility is to define the \emph{quasi-quenched Rényi entropies},
\begin{align*}
    S_{QQ}^{(n)} &= \frac{1}{1-n} \log \overline{\left( \frac{\tr \rho^n}{(\tr \rho)^n} \right)} = \frac{1}{1-n} \log \lim_{k\to-n}\overline{ (\tr \rho^n) (\tr \rho)^k }.
\end{align*}
These entropies are also positive by a similar argument as above, but, just like the quenched entropies, a replica limit is necessary to compute them. Since they are as difficult to compute as the quenched entropy, we will not focus on them in this paper. We remark that, in the $n\to1$ limit, the quasi-quenched and quenched entanglement entropies are identical.

\paragraph{Inequalities between the different entropies.} Since we have found that some of the entropies introduced above are required to be positive, one may wonder whether the different entropies bound each other. Applying Jensen's inequality, we have found the following bounds,  
\be 
S_{SQ}^{(n)}\geq S_A^{(n)} \quad\text{ for all }n, \,\qquad\, S_{Q}^{(n)}\geq S_{QQ}^{(n)}\quad \text{ for } n\geq 1\,, \qquad S_{Q}^{(n)}\leq S_{QQ}^{(n)}\quad \text{ for } n\leq 1\,,
\ee
and have found counterexamples for all other potential inequalities among the four classes of R\'enyi entropies.

\paragraph{Positivity of semi-quenched, positivity of quenched, and the isolation of the matter ``ground state'' are all equivalent. } While the semi-quenched entropy does not bound the quenched entropy, proving the positivity of the former for an arbitrarily large number $k$ of insertions actually implies the positivity of the latter. Specifically, we will argue that proving that either the semi-quenched or quenched R\'enyi entropies are positive for all $k$ and a given value of $n$ implies that the lowest (or highest) eigenvalue in the spectrum of $\tilde O_\Delta$ is isolated, in which case all other semi-quenched and quenched R\'enyi entropies are also positive. We will refer to the eigenstate associated with the eigenvalue of largest magnitude (lowest or highest eigenvalue) as the matter ``ground state''. This serves as a strong physical motivation for computing the simplest quantity, the semi-quenched entropy, to confirm that all other entropies remain positive. The semi-quenched entropies can only go negative if 
\begin{equation}
\label{eq:inequality-that-leads-to-neg}
    \overline{\tr \rho^n}>    \overline{(\tr \rho)^n} ,\quad\quad\quad n\geq 1\,,
\end{equation} 
or the reversed inequality if $n<1$.
We shall prove that if the eigenvalue spectrum of $\tilde O_\Delta$ is continuous all the way to the matter ``ground state'', then the inequality \eqref{eq:inequality-that-leads-to-neg} is indeed satisfied.  Let us assume there exists at least one member of the ensemble that has a continuous eigenvalue spectrum all the way to the matter ground state. Since we have to deal with the eigenvalue of largest magnitude in the spectrum of $\tilde O_\Delta$, it will be more convenient to work with the density of squared eigenvalues $\lambda = x^2$.  We can then write a Laurent expansion for this density around the maximum $\lambda$, denoted by $\lambda_* = x_*^2$, as
\begin{equation}
   \sigma_\Delta(\sqrt{\lambda}) = e^{S_\text{edge}}(\lambda-\lambda_*)^\alpha + o((\lambda-\lambda_*)^\alpha) 
  \label{eq:DOS-expansions}
\end{equation} 
where $e^{S_\text{edge}}$ keeps track of the overall scaling of the density of eigenvalues close to the edge of the spectrum, and $\alpha$ determines the behavior of the squared eigenvalue density near the edge. Since the total number of states ($\int_0^{\l_*} \frac{d\l}{\l^{1/2}}  \sigma_\Delta(\sqrt{\lambda})$) in the window that we project to is finite, we should have $\alpha>-1$. For such an ensemble member, Eq.~\eqref{eq:inequality-that-leads-to-neg} can be rewritten at large $k$ as 
\begin{equation}
    k>e^{\frac{S_\text{edge}}{1+\alpha}} n^{\frac{1}{n-1}} \Gamma(1+\alpha)^{\frac{1}{1+\alpha}} \lambda_*^{\frac{1}2 \frac{1+2\alpha}{1+\alpha}}\,.
\end{equation} 
Thus, given an edge for the spectrum, there always exists some sufficiently large value of $k$ for which the inequality \eqref{eq:inequality-that-leads-to-neg} is satisfied for this ensemble member. Therefore, the contribution to the semi-quenched entropy of ensemble members with continuous eigenvalue spectrum all the way to the matter ``ground state'' is strictly negative at large $k$, and its magnitude increases as $k$ is increased. For ensemble members with isolated matter ``ground states'', Eq.~\eqref{eq:DOS-expansions} is invalid and one can show that ${\tr \rho^n} \to  {(\tr \rho)^n}$  as $k \to \infty$, with  $\tr \rho^n\leq (\tr \rho)^n $ for all $k$ and $n\geq 1$ (and reversed for all $n<1$).\footnote{This can be easily checked. For example for $n=2$ we can rewrite the inequality ${\tr \rho^2} \leq  {(\tr \rho)^2}$ in terms of the continuous part of the spectrum $\sigma(\lambda)$ and the value of the discrete isolated ground state $\lambda_*$,
\be 
\left(\int_{-\lambda_e}^{\lambda_e} d\lambda\,\lambda ^{2 k} \sigma_{\Delta} (\lambda )\right) \left(2 \lambda_*^{2 k}+\int_{-\lambda_e}^{\lambda_e} d\lambda\, \lambda ^{2 k} \sigma_{\Delta} (\lambda ) \right)>\int_{-\lambda_e}^{\lambda_e} d\lambda\,\lambda ^{4 k} \sigma_{\Delta} (\lambda ),
\ee 
which is true since, by definition, $|\lambda_*| > |\lambda_e|$ and because $\sigma_\Delta (\lambda)>0$.
}  Thus, such ensemble members will have a vanishingly small positive contribution at large $k$.  Thus, averaging over an ensemble where even a single member has a continuum all the way down to the matter ``ground state'' leads to the inequality \eqref{eq:inequality-that-leads-to-neg}, making the semi-quenched entropy negative. Consequently, the positivity of the semi-quenched entropy for all $k$ (even for a single fixed $n$) indicates that all members of the ensemble have an isolated matter ``ground state''. Since the presence of an isolated matter ``ground state'' guarantees that all entropies remain positive in each member of the ensemble, the positivity of the quenched entropies also follows.\footnote{This follows from the fact that if there is an isolated matter ``ground state'' then ${\tr \rho^n}\leq{(\tr \rho)^n} $ in each member of the ensemble, making the entropy in each member positive.  } 

\paragraph{Summary.}At leading order in $G_N$, the calculations of the R\'enyi entropies discussed in Section \ref{sec:review-of-the-paradox} apply to all the entropies defined above, and all entropies become negative when the number $k$ of operator insertions becomes sufficiently large. However, this is only a problem for the semi-quenched, quasi-quenched, and quenched R\'enyi entropies, which are required to be positive, and is not a problem for the annealed entropies, which can be negative. Thus, by including subleading corrections in $G_N$, we will show that the former three entropies are positive in the large $k$ limit; we will also find that the annealed entropies remain negative even when including subleading corrections in $G_N$. Since semi-quenched entropies are easier to compute, the primary focus of this paper will be to resolve the negativity puzzle for this class of entropies (see Section \ref{sec:bulk-resolution}). Nevertheless, we shall also explicitly compute the quenched entropy for very large $k$ using the matrix model description of our setup (see Section \ref{sec:matrixmodels}). 

\subsection{Extending the negativity puzzle: non-SUSY JT and higher dimensions}

While \cite{Lin_Maldacena_Rozenberg_Shan_2023} introduced the entropy negativity puzzle solely for two-sided BPS black holes in $\mathcal N=2$ super-JT gravity, a fairly specific setup, the same puzzle arises in more general settings. 

Specifically, consider a two-sided black hole state in any holographic theory. We can restrict the left and right energies (associated to the ADM masses on the left and right boundaries) to be within some energy band $\mathfrak E$ with average energy $\bar{E}$ and width $\delta E$. In the preparation of the state analogous to $\ket{\psi_k}$ (which we call $\ket{\psi_{k, \mathfrak E}}$), we can once again consider the insertion of $k$ simple operators with scaling dimension $\Delta$ that are projected onto the energy band $\mathfrak E$,\footnote{Such a state can be concretely constructed by first considering two-sided states in which there is a period of Euclidean evolution $\beta_1$, $\beta_2$, \dots, $\beta_{k+1}$ between subsequent operator insertions of $O_\Delta$, 
\be 
\ket{\psi_{\beta_1, \beta_2, \dots, \beta_{k+1}}} = \inlinefig[10]{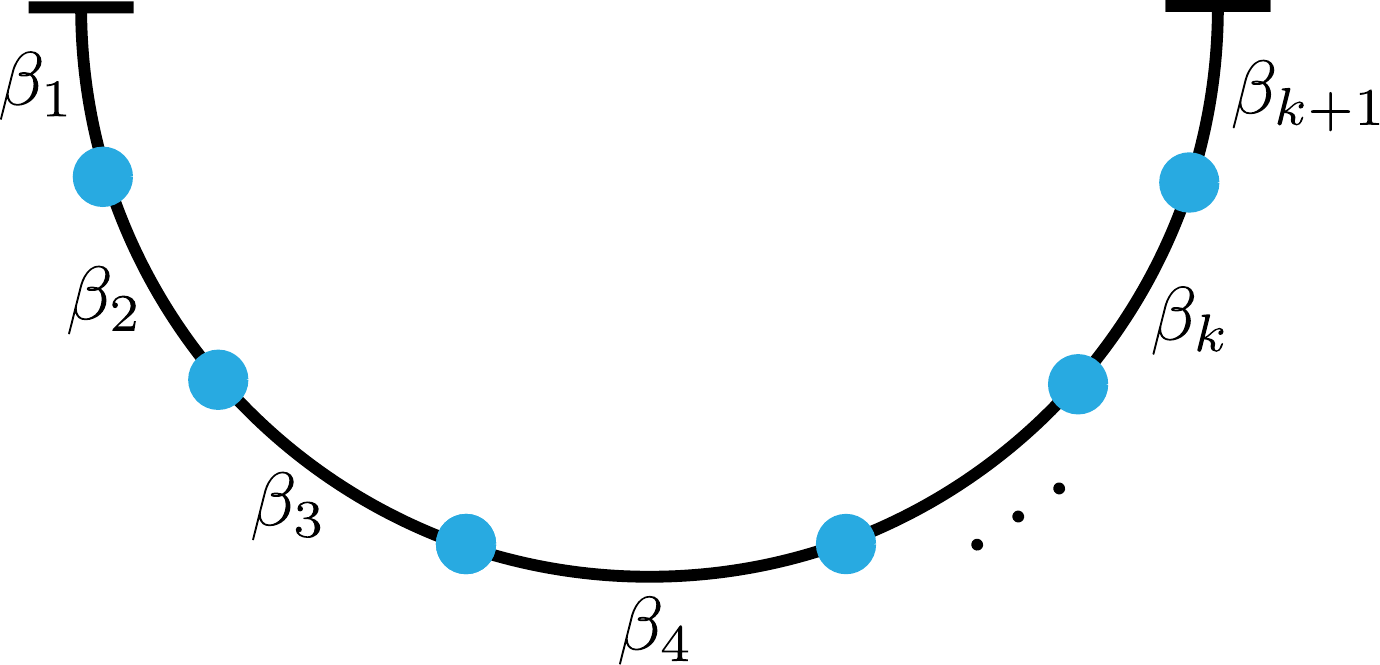}.
\ee
Then, we can consider the linear combination of such states that implements the appropriate projection onto the energy band  $\mathfrak E$,
\be 
\ket{\psi_{k, \mathfrak E}} = \int d\beta_1 \dots d\beta_{k+1} \frac{\left(e^{\beta_1 E_+} - e^{\beta_1 E_-}\right) \dots \left(e^{\beta_{k+1} E_+} - e^{\beta_{k+1} E_-}\right)}{\beta_1 \dots \beta_{k+1}} \ket{\psi_{\beta_1, \beta_2, \dots, \beta_{k+1}}}\,.
\ee
where $E_+$ and $E_-$ are the upper and lower bounds in the energy band $\mathfrak E$ and the contours of integration run parallel to the imaginary axis. } 
\be 
\tilde O_\Delta = P_{\mathfrak E} O_\Delta P_{\mathfrak E}\,.
\label{eq:microperator}
\ee
Consider the density matrix $\rho = \tr_L \ket{\psi_{k, \mathfrak E}} \bra{\psi_{k, \mathfrak E}} $ after tracing out one of the two sides (say, the left side). The moments of $\rho$ are still determined by the eigenvalues of the projected operator $\tilde O_\Delta$, $\tr(\rho^n) = \tr_{\mathfrak{E}}\[(\tilde O_\Delta)^{2kn}\]$.
Eventually, when $k$ becomes large, the R\'enyi entropies and entanglement entropy of $\rho$ will still probe the statistics of the smallest/largest eigenvalues of $\tilde O_\Delta$. From a bulk point of view, in the $G_N\to 0$ limit, we expect on generic grounds that the moments of $ \tr_{\mathfrak{E}}\[(\tilde O_\Delta)^{2kn}\]$ arise from a continous density of eigenvalues for the operator $(\tilde O_\Delta)$.\footnote{If this were not the case, then once $k$ or $n$ become sufficiently large, the moments of $\tilde O_\Delta$ 
would become subject to trace relations; such trace relations should not be visible from the bulk in the $G_N \to 0$ limit. } Because of this, generically, the R\'enyi and entanglement entropies will become negative when $k$ is sufficiently large.

To further sharpen the puzzle in this non-supersymmetric setup, we can once again revert to JT gravity, this time without any supersymmetry, coupled to a free field with scaling dimension $\Delta$. Once again, consider the cases of large and small $\Delta$, for a tiny energy band $\mathfrak E$ of size $\delta E$, centered at $\bar{E}$.\footnote{Specifically, take $\delta E \ll E_\text{brk}$, where $E_\text{brk}$ is the JT coupling determined by the boundary value of the dilaton.} As we argue in Appendix \ref{6j} and further discuss in Section \ref{sec:matrixmodels}, similar to the BPS case,  in the $\Delta \to \infty$ limit all contributions to $\tr_{\mathfrak{E}}\[(\tilde O_\Delta)^{2kn}\]$ that have intersecting Wick contractions between the various operator insertions are exponentially suppressed. Thus, the only surviving contributions are the same as in \eqref{eq:moments-of-rho-Delta-to-infinity}. Consequently,  
\be 
\tr_{\mathfrak E}(\rho^n) = e^{S_0}  C_{kn} (P_{\Delta \to \infty}(\bar{E}))^{kn}(\rho_0(\bar{E}) \delta E)^{kn+1}
\ee
where $\rho_0(E) = \frac{1}{2\pi^2 E_\text{brk}} \sinh(2\pi \sqrt{2 \frac{E}{E_{\text{brk}}}})$ is the density of states of JT gravity rescaled to be independent of $S_0$, and $P_{\Delta \to \infty}(E)$ is now the propagator associated to the operator $\tilde O_{\Delta}$ in a non-supersymmetric theory.\footnote{\label{foot:non-susy-prop}Fixing the energies to the two sides of the propagator to be the same (namely the average energy in the microcanonical window), this is given by
\be 
P_{\Delta \to \infty}(E) = \frac{\left|\Gamma(\Delta) \Gamma\left(\Delta + i 2 \sqrt{2\frac{E}{E_\text{brk}}}\,\right) \right|^2}{(2E_\text{brk}^{-1})^{2\Delta} \Gamma(2\Delta)}\,.
\ee
} The associated R\'enyi entropies are therefore analogous to Eq.~\eqref{eq:Sn-Delta-equals-infinity}, 
\be 
S^{(n)}(\rho) = \log(e^{S_0}\delta E \rho_0(E)) - \frac{3}2 \log k + O(1) = S(\mathfrak{E}) - \frac{3}2 \log k + O(1)\,,
\ee 
which again become negative at exponentially large values of $k$.  The similarity to the BPS case continues to hold in the $\Delta \to 0$ limit, in which all diagrams contribute the same amount regardless of their number of crossings (see Appendix \ref{6j} and \ref{sec:smalld}). Thus, the R\'enyi entropies are still given by Eq.~\eqref{eq:Sn-Delta-equals-zero} (with $\SBPS$ replaced by $S(\mathfrak{E})$), which similarly become negative at values of $k$ scaling linearly with $S(\mathfrak{E})$.

\section{A bulk resolution to the puzzle in the large $\Delta$ limit}
\label{sec:bulk-resolution}

In this section, we will solve the puzzle in the large $\Delta$ limit using a purely bulk argument in JT (super)gravity. As we have discussed in Section \ref{sec:2}, the problem arises because the calculation of the (annealed) entropy we reviewed is insensitive to the discrete nature of the eigenvalue spectrum. We can therefore expect that non-perturbative quantum gravity effects, such as contributions from higher genus and connected geometries, which should contain information about the discreteness of the spectrum of operators, play a role in the resolution of the puzzle. Our first step will then be to understand which higher-genus geometries contribute to the calculation of $\overline{\[\tr(\rho^n)\]^m}$, where the overline indicates that we are computing the quantity using the gravitational path integral. In most of this section, we will drop the overline notation, but all powers are taken before evaluating the quantity with the gravitational path integral (i.e., before averaging over the dual matrix ensemble).

Let us start from the BPS sector of JT supergravity. We focus on the computation of $\tr (\rho^n)$, for which we have a single connected asymptotic boundary with $2kn$ operator insertions. The boundary conditions are simply given by
\be
\tr(\r^n)=\inlinefig[10]{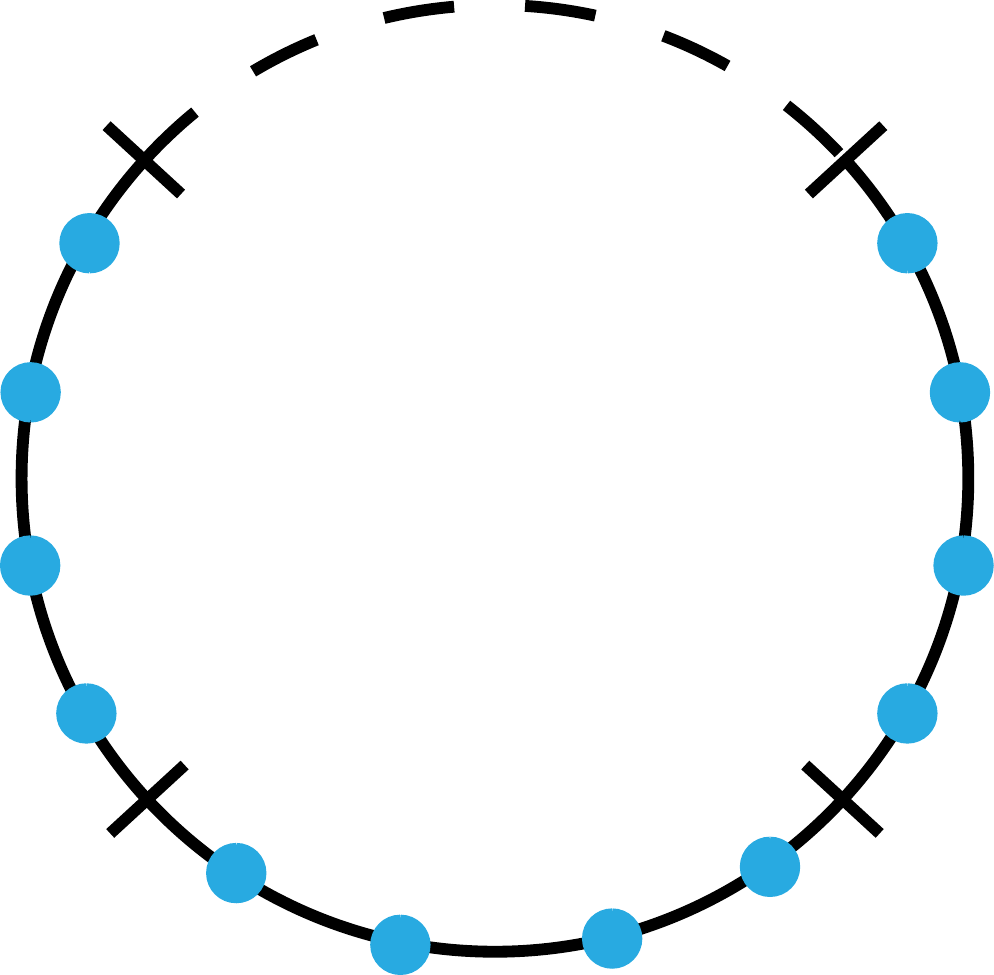}\,.
\ee
Recall that in the large $\Delta$ limit of interest here, the only bulk diagrams contributing to $\tr(\r^n)$ are pairwise Wick contractions of the operator insertions for which matter worldlines connecting the insertions do not intersect. At the disk (genus-0) level, the relevant diagrams were depicted in Eq.~\eqref{eq:moments-of-rho-Delta-to-infinity}. But which non-planar (higher genus) diagrams should we take into account?\footnote{One could worry that neglecting intersecting diagrams, which are exponentially suppressed in $\Delta$, while keeping higher genus diagrams, which are exponentially suppressed in $\SBPS$, is an inconsistent approximation. For simplicity, we can assume here that $\Delta$ is the largest parameter in the theory. However, we expect our results to be more general. In fact, in the $k= O(e^{2\SBPS/3})$ regime of interest, higher genus diagrams contribute at the same order as disk diagrams, and they guarantee positivity of the entropies. On the other hand, as we discuss in Appendix \ref{sec:qgaussian}, crossing diagrams are only responsible for shifting the location of the edge of the spectral density for the operator $\tilde{O}_{\Delta}$.} As we will now explain in detail, the only non-vanishing diagrams are those for which all possible closed geodesics intersect at least one matter worldline. Intuitively, these are higher genus diagrams in which every handle is ``threaded'' by a matter worldline. To understand why these are the only contributing diagrams, we first need to recall how to compute matter correlation functions in JT (super)gravity. As mentioned in Section \ref{sec:2}, for the BPS case we will be focusing on a theory with $r=1$, such that all BPS states have R-charge $Z=0$. The generalization of our results to $r>1$ and multiple R-charge sectors is straightforward. We will also explain how to generalize our results to the non-SUSY case.

\subsection{Computing matter correlation functions in JT (super)gravity}
\label{sec:3.1}

Let us review the computation of diagrams (arising from Wick contractions) contributing to a matter correlation function in $\mathcal{N}=2$ JT supergravity \cite{Lin_Maldacena_Rozenberg_Shan_2023,Boruch:2023trc}. A very similar technique, which we review in Appendix \ref{sec:derivation-2n-pt-func-non-SUSY-JT}, can also be used in non-SUSY JT gravity \cite{Mertens_Turiaci_Verlinde_2017,Blommaert:2018oro,Iliesiu:2019xuh,Yang_2019}. For simplicity, let us start with a two-point function in the BPS sector, for which only one planar diagram contributes:
\begin{equation}
    \tr\(\widetilde{O}_\D^{2}\)=\inlinefig[10]{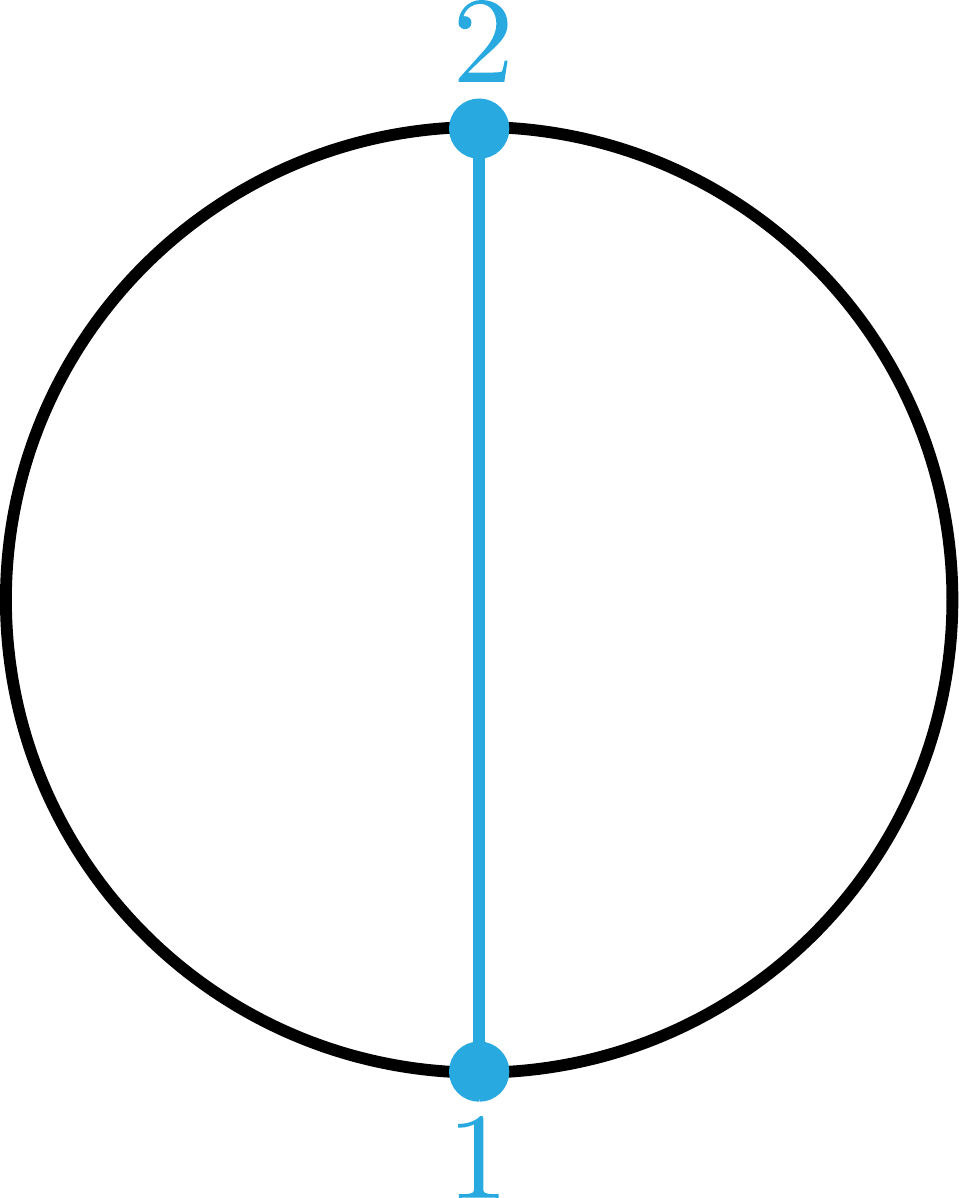}
\end{equation}
To evaluate this diagram, we can first cut it open along the matter worldline, whose bulk length we label by $\ell_{12}$. We obtain two half-disks whose boundaries are the union of the matter worldline of length $\ell_{12}$ and a portion of the asymptotic boundary, which is infinitely long in the BPS sector. We can think of each one of these geometries as preparing the zero-temperature Hartle-Hawking state on the bulk slice of length $\ell_{12}$. The corresponding Hartle-Hawking wavefunction is known and given by \cite{Lin_Maldacena_Rozenberg_Shan_2023,Boruch:2023trc}
\begin{equation}
    \Psi_{12}(\ell_{12},a_{12})=\frac{2}{\pi}e^{-\frac{\ell_{12}}{2}}\left[\xi_{12}e^{-\frac{ia_{12}}{2}}K_{\frac{1}{2}}(2e^{-\frac{\ell_{12}}{2}})+\eta_{12}e^{\frac{ia_{12}}{2}}K_{\frac{1}{2}}(2e^{-\frac{\ell_{12}}{2}})\right]
    \label{eq:HH}
\end{equation}
where $e^{ia_{12}}$ is the $U(1)_R$ Wilson line between the two boundaries, $K_n(x)$ is the modified Bessel function of the second kind, and $\xi_{12}=-\bar{\psi}_l/\sqrt{2}$, $\eta_{12}=-i\bar{\psi}_r/\sqrt{2}$, where $\bar{\psi}_l$, $\bar{\psi}_r$ are the superpartners of the length variable with respect to the right and left supercharges. Notice that the Hartle-Hawking wavefunction depends on the orientation of the boundary, since it is not invariant under exchange $1\leftrightarrow 2$. The wavefunction with opposite orientation is given by
\begin{equation}
    \Psi_{21}(\ell_{12},a_{12})=\frac{2}{\pi}e^{-\frac{\ell_{12}}{2}}\left[\eta_{12}e^{\frac{ia_{12}}{2}}K_{\frac{1}{2}}(2e^{-\frac{\ell_{12}}{2}})-\xi_{12}e^{-\frac{ia_{12}}{2}}K_{\frac{1}{2}}(2e^{-\frac{\ell_{12}}{2}})\right].
\end{equation}
The Hartle-Hawking wavefunctions satisfy the normalization condition 
\begin{equation}
    \int d\mu_{ij}\Psi_{ij}\Psi_{ji}=1\,,
\end{equation}
where we defined the measure
\begin{equation}
    \int d\mu_{ij}=\frac{1}{2}\int_{-\infty}^\infty d\ell_{ij}\int_0^{2\pi}da_{ij}\int d\eta\int d\xi\, .
    \label{eq:measure}
\end{equation}
We can now compute the two-point function by gluing the two half-disks along the matter worldline, inserting a factor of $e^{-\Delta \ell_{12}}$ and integrating over $\ell_{12}$. More precisely,
\begin{equation}
\begin{aligned}
    \tr\(\widetilde{O}_\D^{2}\)&=\int d\mu_{12}e^{-\Delta \ell_{12}}\inlinefig[10]{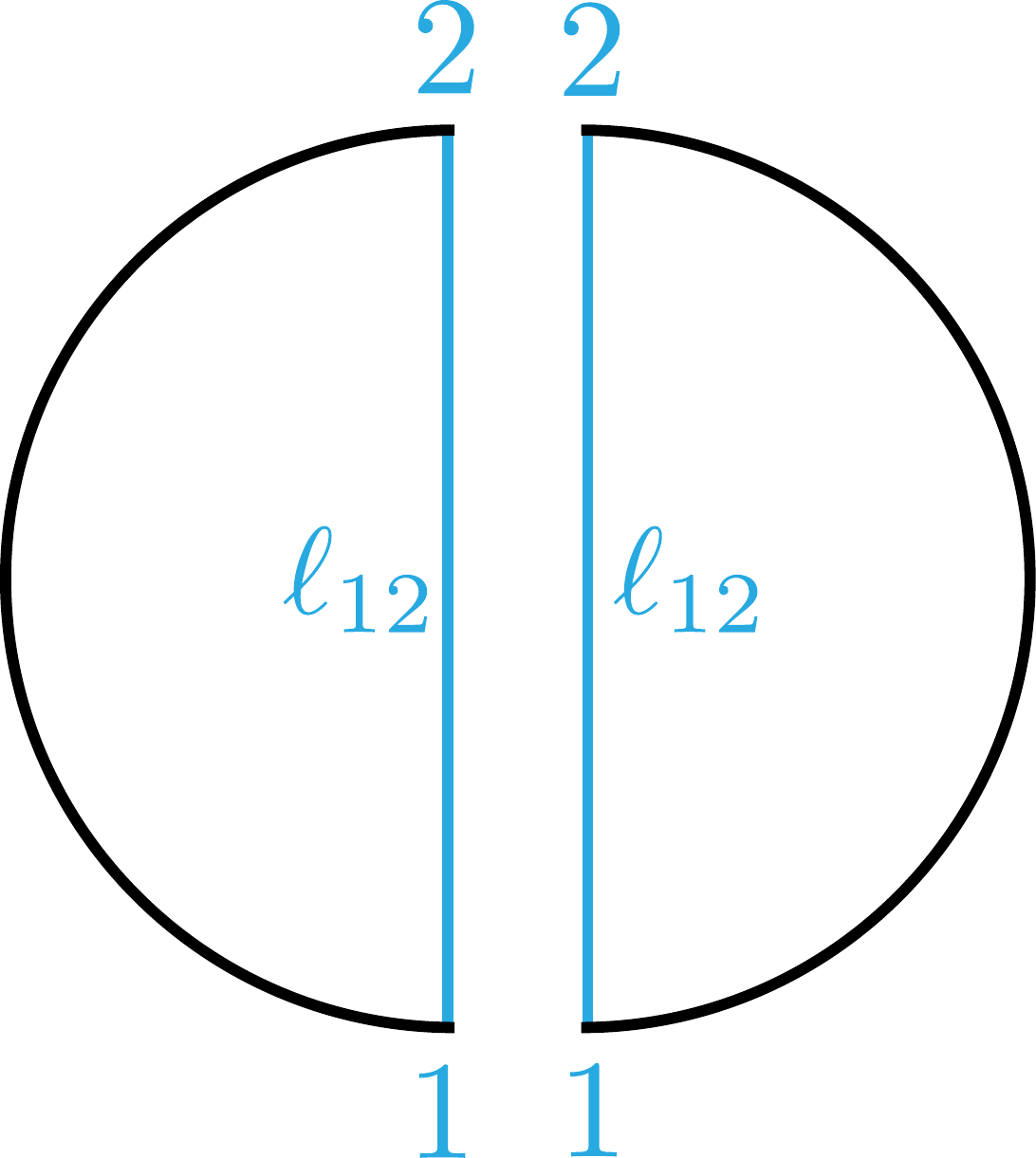}\\  &=e^{\SBPS}\int d\mu_{12}e^{-\Delta \ell_{12}}\Psi_{12}(\ell_{12},a_{12})\Psi_{21}(\ell_{12},a_{12})=e^{\SBPS}P_{\Delta\to\infty},
    \end{aligned}
\end{equation}
where we introduced the appropriate topological factor $e^{\SBPS}$ for the disk
and the explicit expression of $P_{\Delta\to\infty}$ is given in Footnote \ref{foot:PBPS}.

To compute more generic correlation functions, we need to introduce a generalization of the Hartle-Hawking wavefunction, which we will call the \textit{asymptotic polygon}.\footnote{The difference between the asymptotic polygon and the polygon defined in \cite{Boruch:2023trc} is that the (connected) boundary of the polygon in \cite{Boruch:2023trc} is made of only geodesic segments and no asymptotic boundary segments.} This is a connected, genus-0 bulk surface bounded by a single connected boundary made up of $n$ geodesic segments and $n$ asymptotic boundary segments, with geodesics and asymptotic segments alternating. We can think of the geodesic segments as arising from cutting a given diagram along a matter worldline. To such an asymptotic polygon, we can assign a product of Hartle-Hawking wavefunctions, each one associated with one geodesic segment:\footnote{More generically, for $r>1$, the asymptotic polygon is given by $I(i_1,i_2;...;i_{2n-1},i_{2n})=\sum_{Z'}\cos(\pi Z')\tilde{\Psi}^{Z'}_{i_1i_2}\tilde{\Psi}^{Z'}_{i_3i_4}...\tilde{\Psi}^{Z'}_{i_{2n-1}i_{2n}}$, where $\tilde{\Psi}^{Z}_{i_1i_2}=\Psi^Z_{i_1i_2}/\cos(\pi Z)$.}
\begin{equation}
    I(i_1,i_2;...;i_{2n-1},i_{2n})=\Psi_{i_1i_2}\Psi_{i_3i_4}...\Psi_{i_{2n-1}i_{2n}}.
    \label{eq:polygon}
\end{equation}
Notice that the Hartle-Hawking wavefunction is simply an asymptotic polygon with $n=1$.

By gluing together asymptotic polygons along geodesic boundaries, we can construct any diagram giving a non-vanishing contribution to a generic correlation function. Since each geodesic segment is associated with a matter worldline, we must also insert a factor of $e^{-\Delta \ell_{ij}}$ when gluing. As a simple example, let us compute a planar diagram contributing to the 4-point function:\footnote{Recall that the ordering of indices is important in the second line because the Hartle-Hawking wavefunctions are not invariant under exchange of indices.}
\begin{equation}
\begin{aligned}
    &\inlinefig[10]{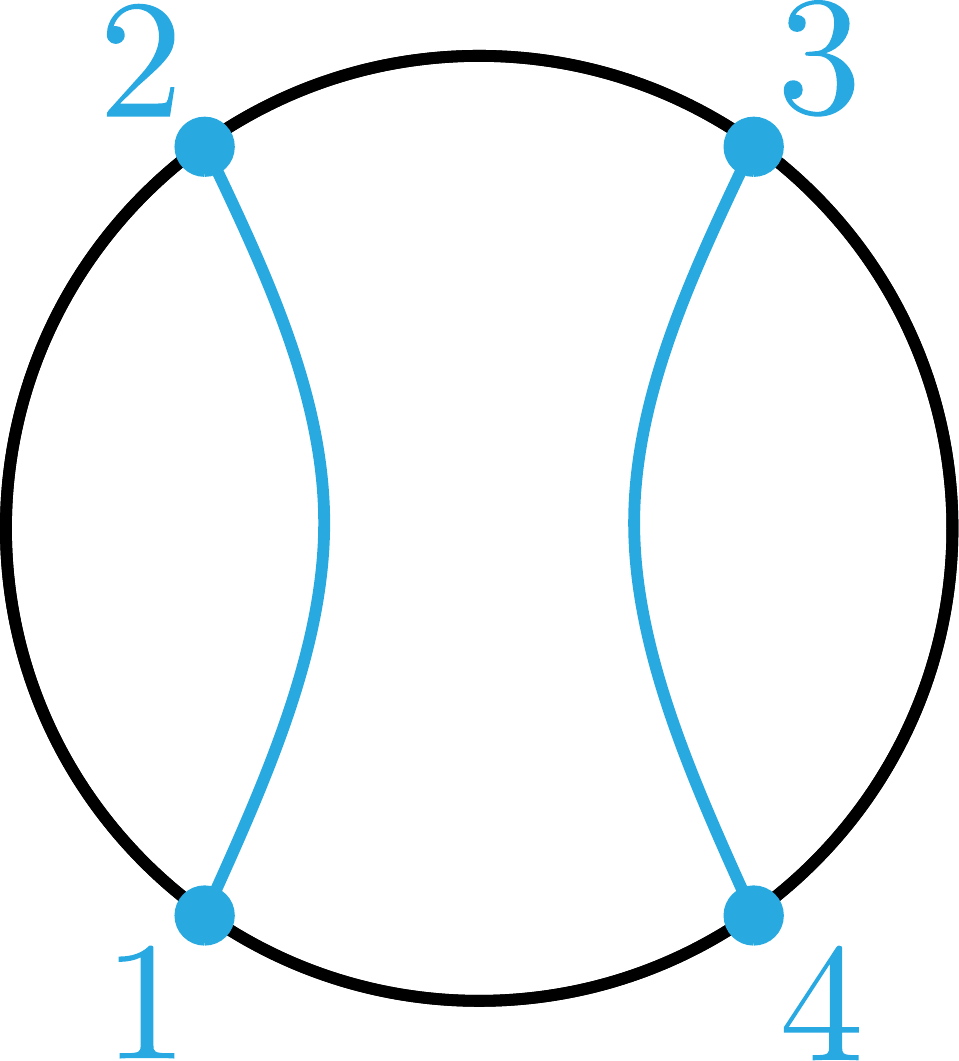}=\int d\mu_{12}d\mu_{34}e^{-\Delta \ell_{12}}e^{-\Delta \ell_{34}}\inlinefig[10]{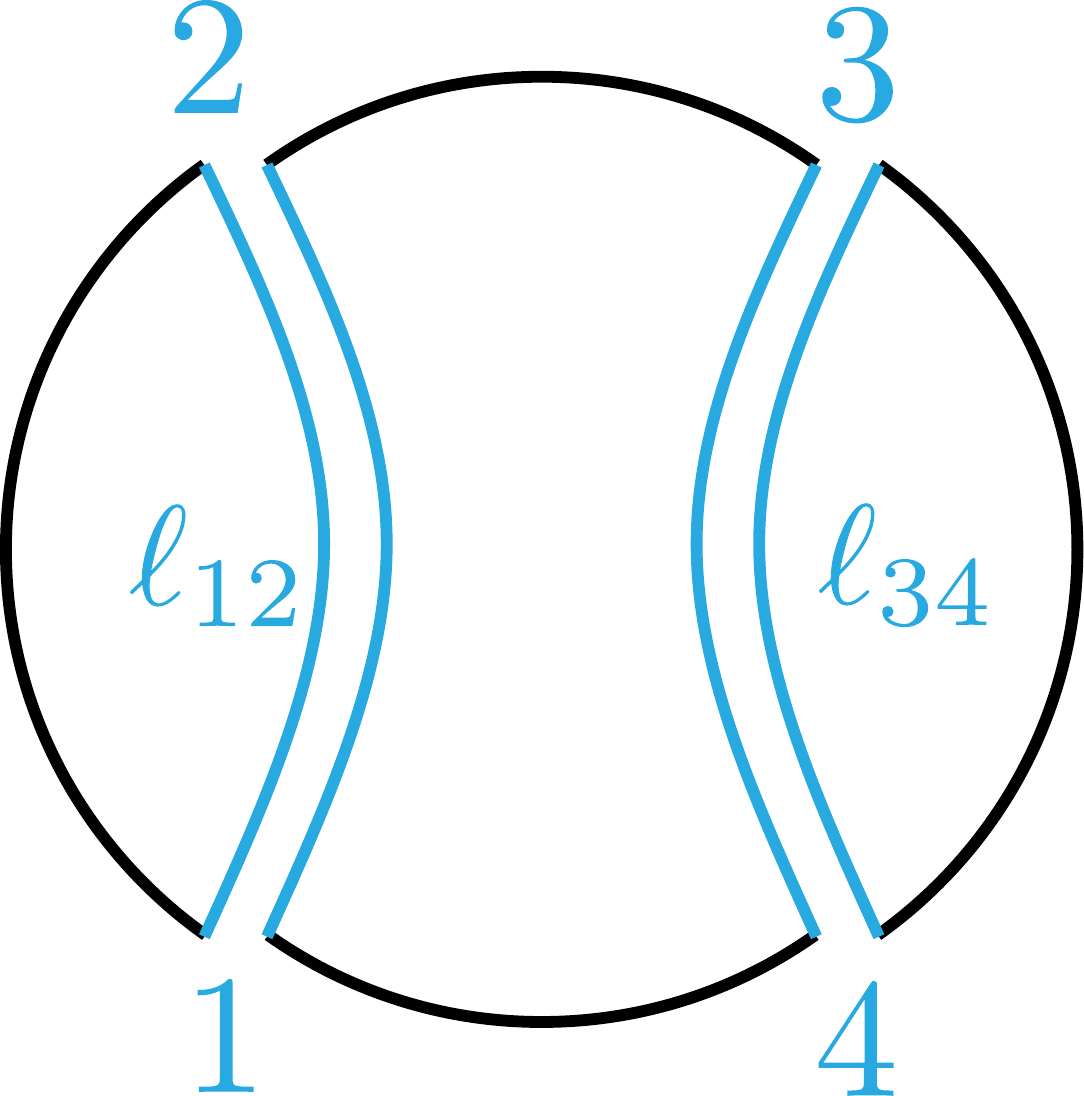}\\
    &=e^{\SBPS}\int d\mu_{12}d\mu_{34}e^{-\Delta \ell_{12}}e^{-\Delta \ell_{34}}I(2,1;4,3)\Psi_{12}\Psi_{34}=e^{\SBPS}\(P_{\Delta\to\infty}\)^2.
    \end{aligned}
\end{equation}
The only diagrams we cannot compute by gluing asymptotic polygons are those which, after cutting along all matter worldlines, have connected components with more than one connected boundary, or connected components with a single boundary but genus $g>0$, e.g. 
\begin{equation}
    \inlinefig[8]{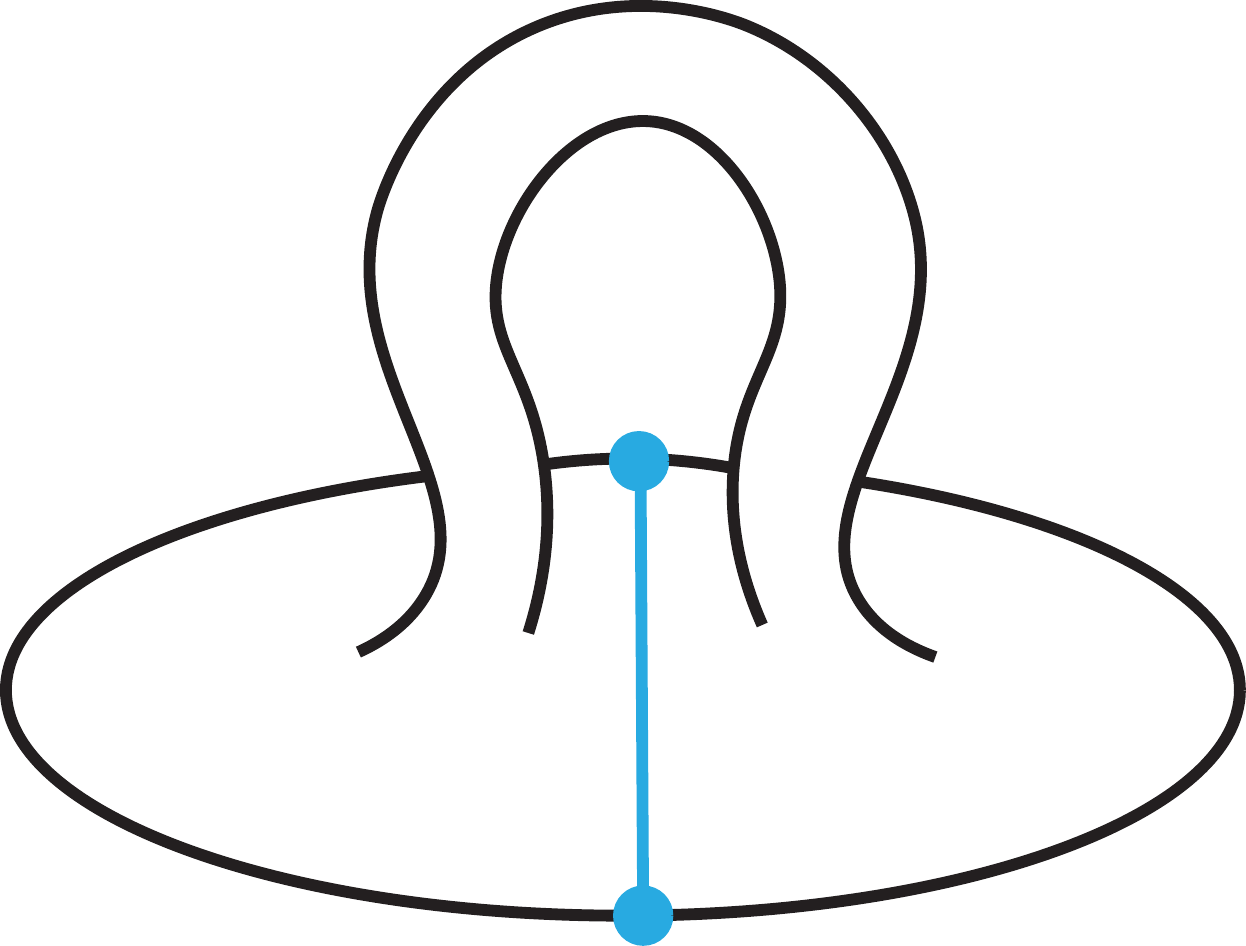}\longrightarrow \inlinefig[8]{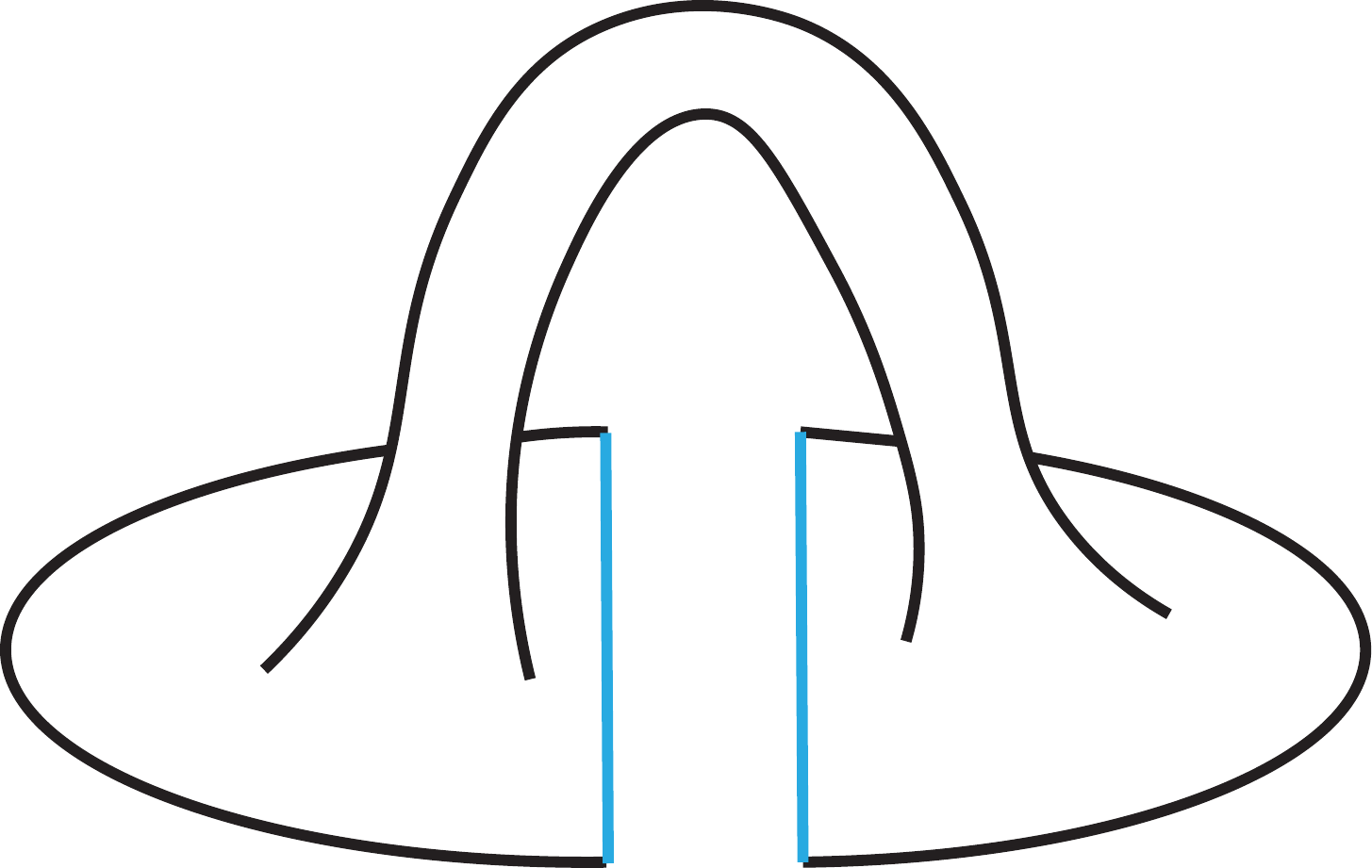}\quad\quad \inlinefig[8]{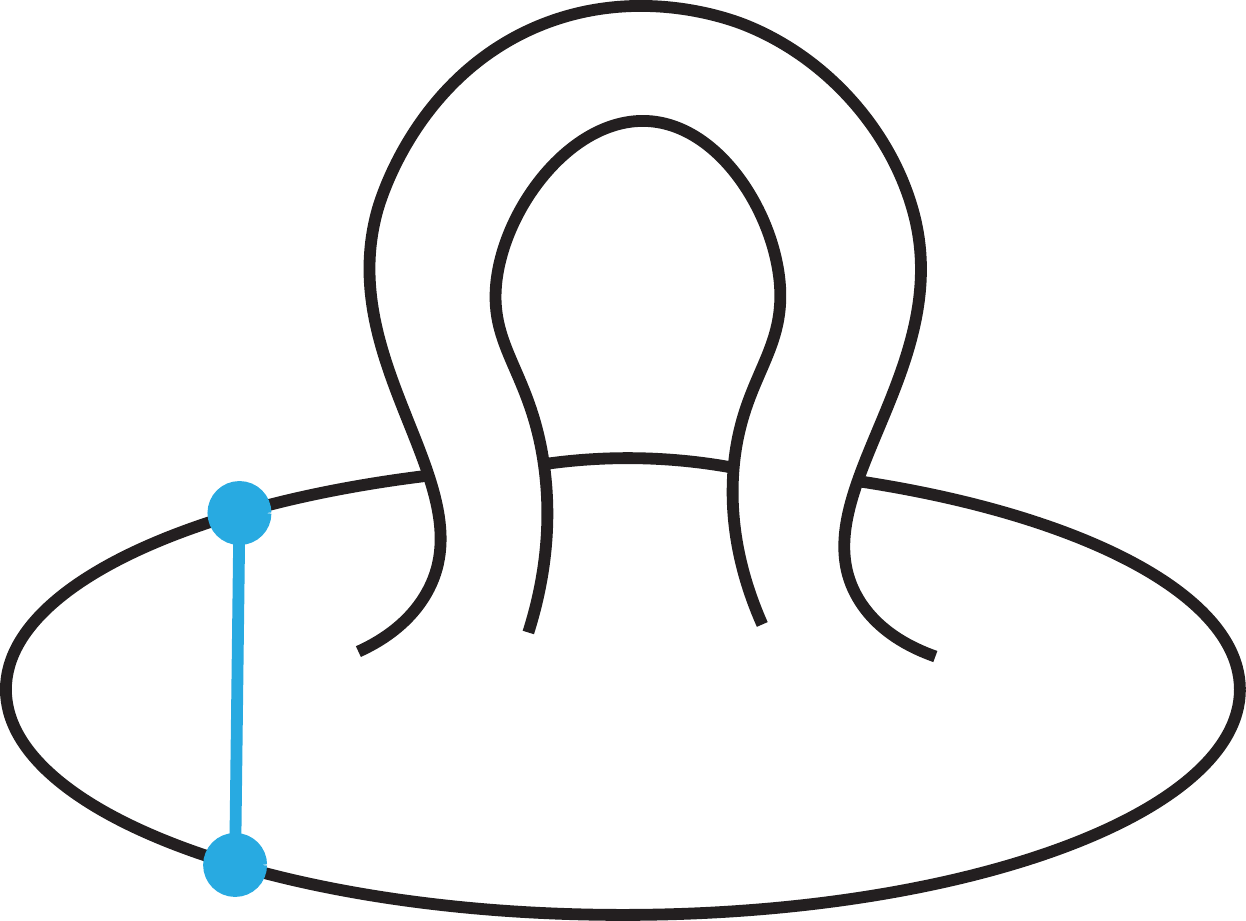}\longrightarrow \inlinefig[8]{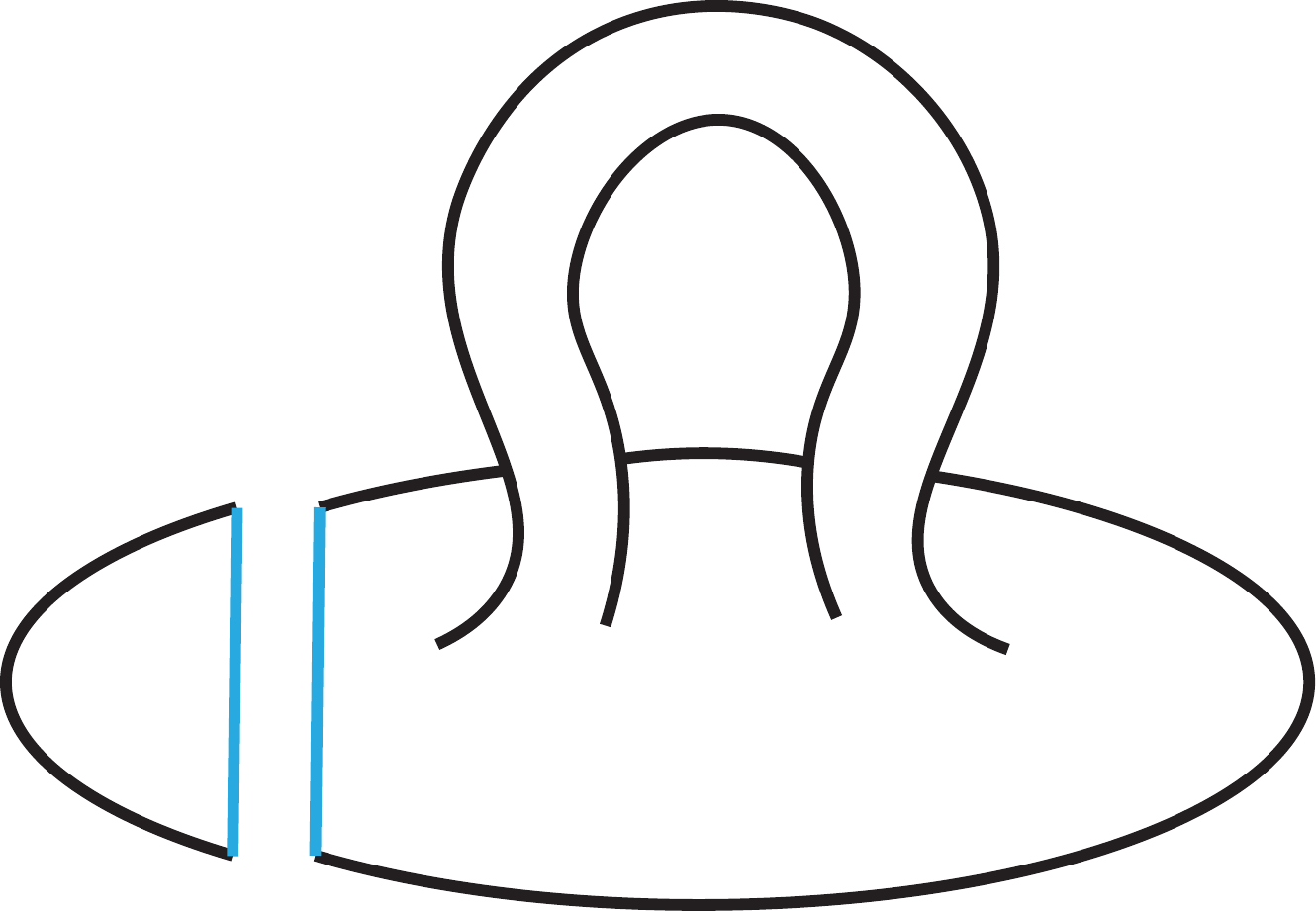}
    \label{eq:empty}
\end{equation}
We will refer to these diagrams as diagrams with ``empty handles''.

However, both these classes of diagrams vanish in the BPS sector, i.e., when taking the length of all asymptotic boundary segments between any operator insertions to be infinite. To see why, recall that in JT supergravity the trumpet with an asymptotic boundary of length $\beta$ and a closed geodesic boundary of length $b$ vanishes for $\beta\to\infty$ \cite{Iliesiu:2021are,Boruch:2023trc}. For the first class of diagrams with empty handles, each boundary has infinite length because it contains at least one segment of asymptotic boundary, which is infinitely long in our BPS case. These diagrams can then be obtained by gluing together multiple trumpets along closed geodesics, with each trumpet vanishing. For the second class of diagrams with empty handles, the connected component with genus $g>0$ can again be written as a gluing (along a closed geodesic) of one trumpet with infinitely long boundary to a bordered Riemann surface, and the result vanishes. For the example on the right of Eq.~\eqref{eq:empty}:
\begin{equation}
    \inlinefig[10]{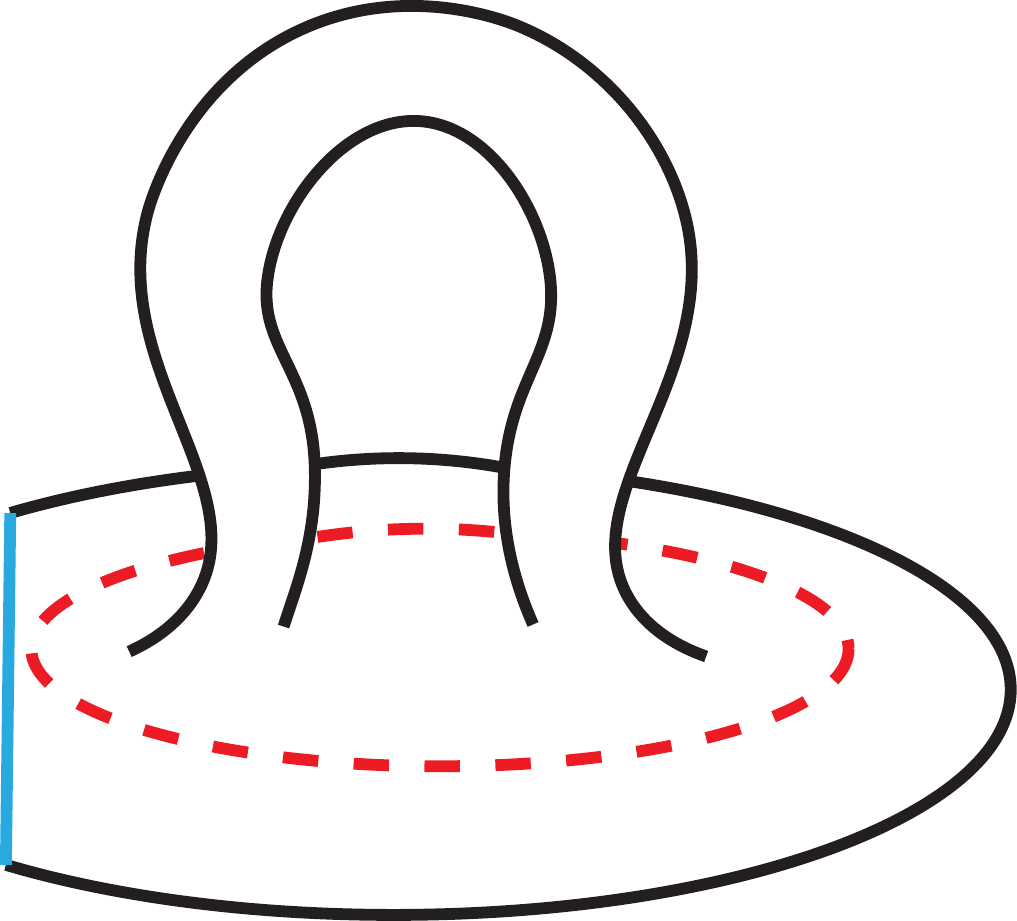}\quad \quad\longrightarrow \quad \quad\inlinefig[14]{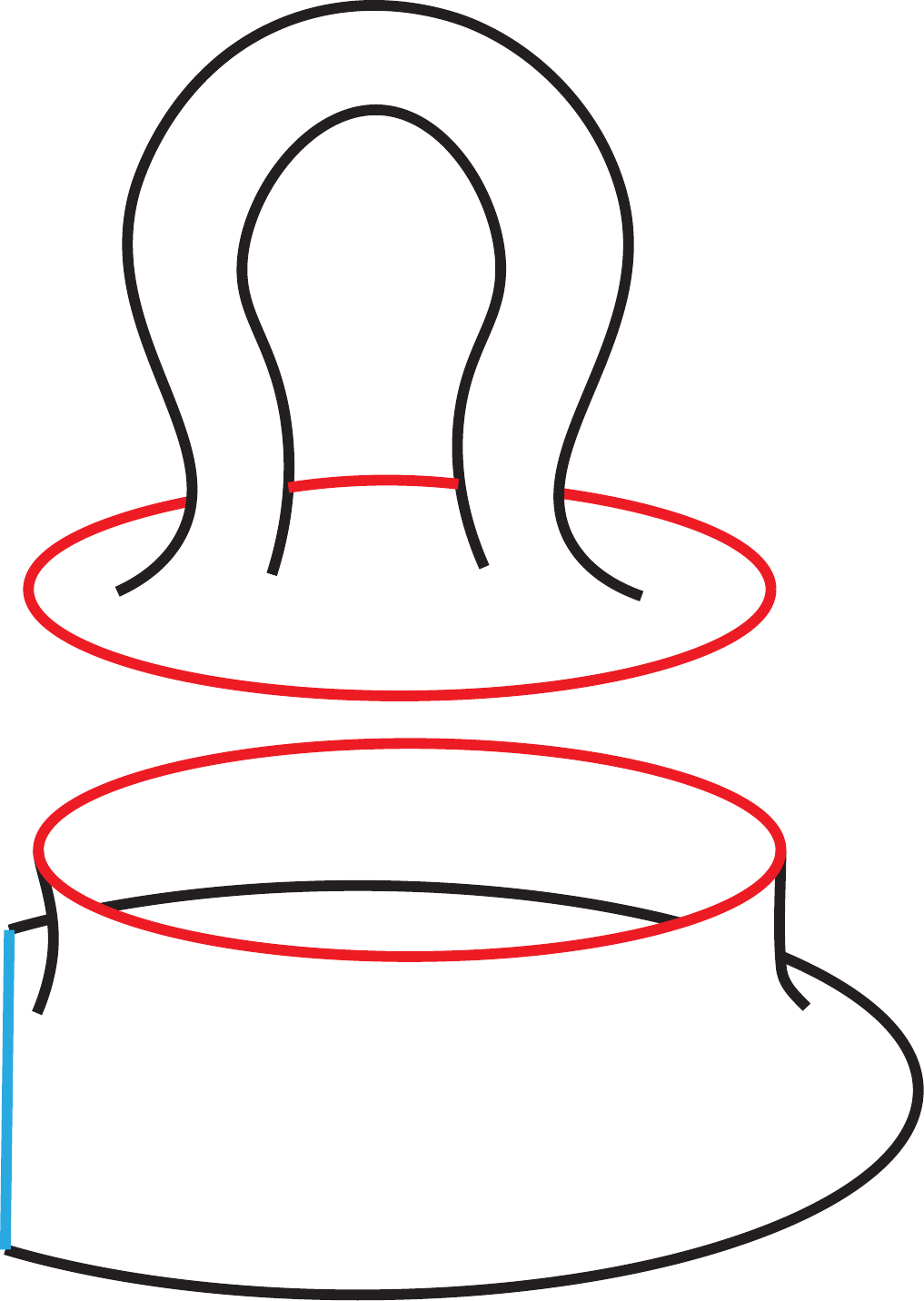}
\end{equation}
where the cut and gluing is along the closed red geodesic. The trumpet on the bottom of the right figure vanishes.
We then conclude that all non-vanishing diagrams can be obtained by gluing Hartle-Hawking wavefunctions and asymptotic polygons.

We now have a full recipe to compute diagrams contributing to a generic matter correlation function, which we summarize here:
\begin{enumerate}
    \item Cut the diagram along all matter worldlines;
    \item If the resulting diagrams have empty handles---i.e., have connected components with disconnected boundaries or genus $g>0$---the diagram vanishes. If they do not, proceed;
    \item Assign a function \eqref{eq:polygon} to each asymptotic polygon bounded by $n$ geodesic boundaries and $n$ asymptotic boundaries obtained after cutting;
    \item Compute the diagram by gluing the various components along bulk geodesics. This corresponds to assigning a factor $e^{-\Delta \ell}$ to each geodesic along which we are gluing and integrating with the appropriate measure \eqref{eq:measure};
    \item Assign the appropriate topological factor $e^{(2-2g-m)\SBPS}$, with $g$ genus and $m$ number of disconnected boundaries, to the fully-glued diagram.
\end{enumerate}
This prescription applies not only to planar diagrams with a single boundary, but also to higher genus diagrams and to diagrams contributing to multi-boundary correlation functions \cite{Boruch:2023trc}. Given a pairing of operators on a higher-genus surface, we have to sum over all contributions where the worldlines in a given operator pairing can wind in different ways around some of the non-contractible cycles of the spacetime; on any higher-genus surface, there are an infinite number of such contributions. For example, on a genus one surface with four $\tilde O_\Delta$ insertions, the following two configurations contribute to the same pairing between the four operators:
\be 
    \inlinefig[10]{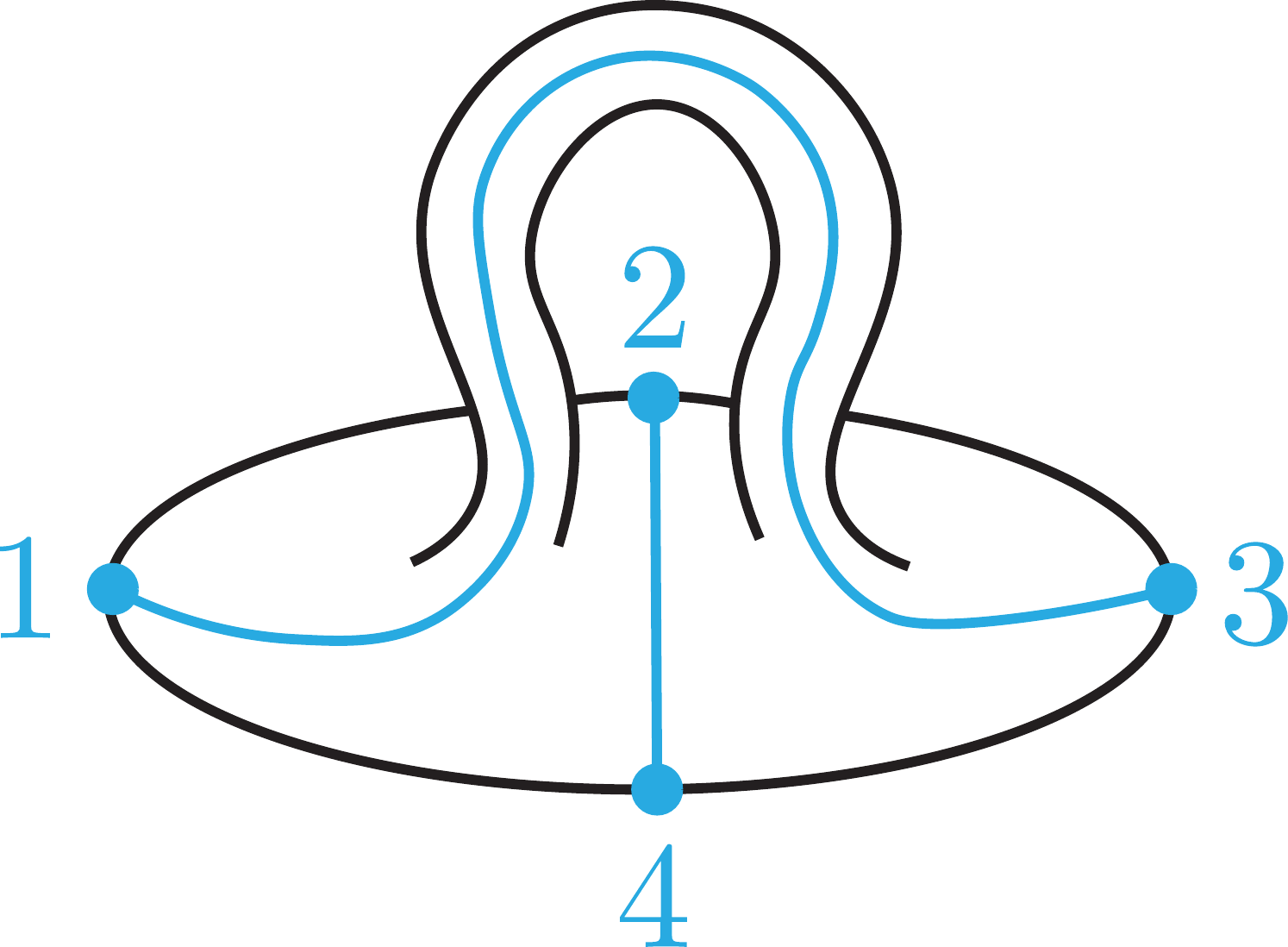}\quad\quad\quad\quad \inlinefig[10]{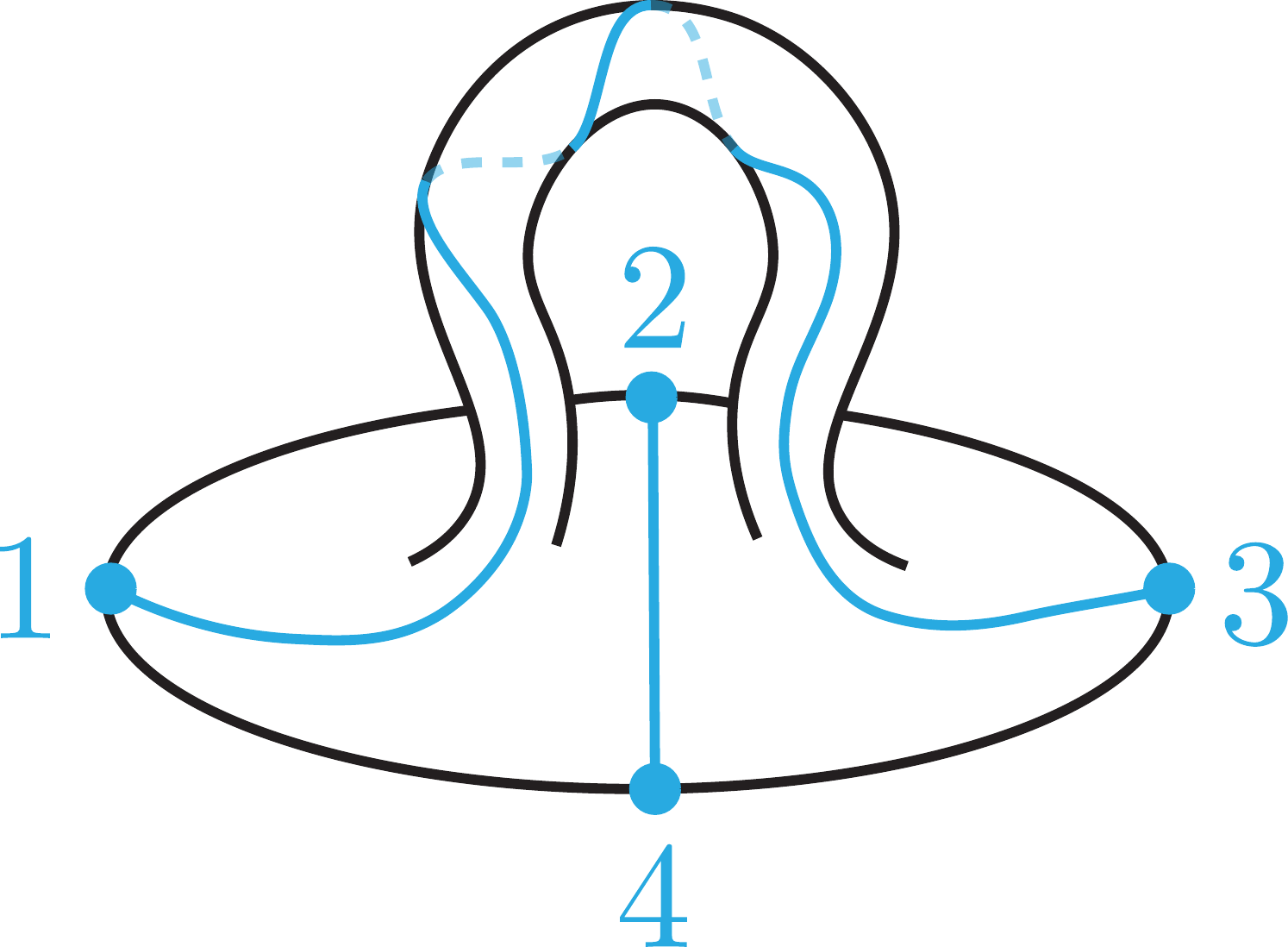}
\ee
The recipe described above correctly accounts for such windings and, at the same time, does not overcount surfaces that are related through a large super-diffeomorphism. More specifically, it does not overcount worldline configurations related through the action of the mapping class group, which represents the quotient group between large and small diffeomorphisms.  Such large transformations relate geodesics that have different windings around the various non-contractible cycles of the spacetime; therefore, when computing correlators in the gravitational path integral, which involve sums over all possible worldlines connecting a given pairing of points, one should be careful not to overcount worldline configurations that are, in fact, related by large super-diffeomorphisms. As explained in \cite{Boruch:2023trc}, in JT gravity and JT supergravity, the sum over all possible (non-intersecting) worldlines connecting a given pairing of boundary points precisely cancels out the effect of the large diffeomorphisms that
are generated by Dehn twists along the closed geodesic cycles that intersect the boundary-to-boundary worldlines. Thus, for a given pairing of boundary operators, instead of summing over all connecting geodesics, one can instead sum over a smaller set of representatives that are not related through the action of the mapping class group. Such representatives can be classified in terms of the topology of the surfaces obtained after cutting the spacetime along the worldlines connecting the different boundary points. As mentioned above in the special case of $\mathcal N=2$ JT supergravity, when focusing on the BPS sector, the only representatives that contribute are those that result in disk topologies after cutting along the worldlines connecting the different boundary points. Since disks have a trivial mapping class group, one no longer needs to implement any quotient by large diffeomorphisms in the remaining path integral. Thus, one can implement steps 3 and 4 in the algorithm above in precisely the same way as when starting with an initial surface that has disk topology in step 1.

To concretely explain how to use the above recipe, let us compute a genus-1 diagram contributing to the 4-point function on a single boundary. The first step is to cut along matter worldlines:
\begin{equation}
    \inlinefig[10]{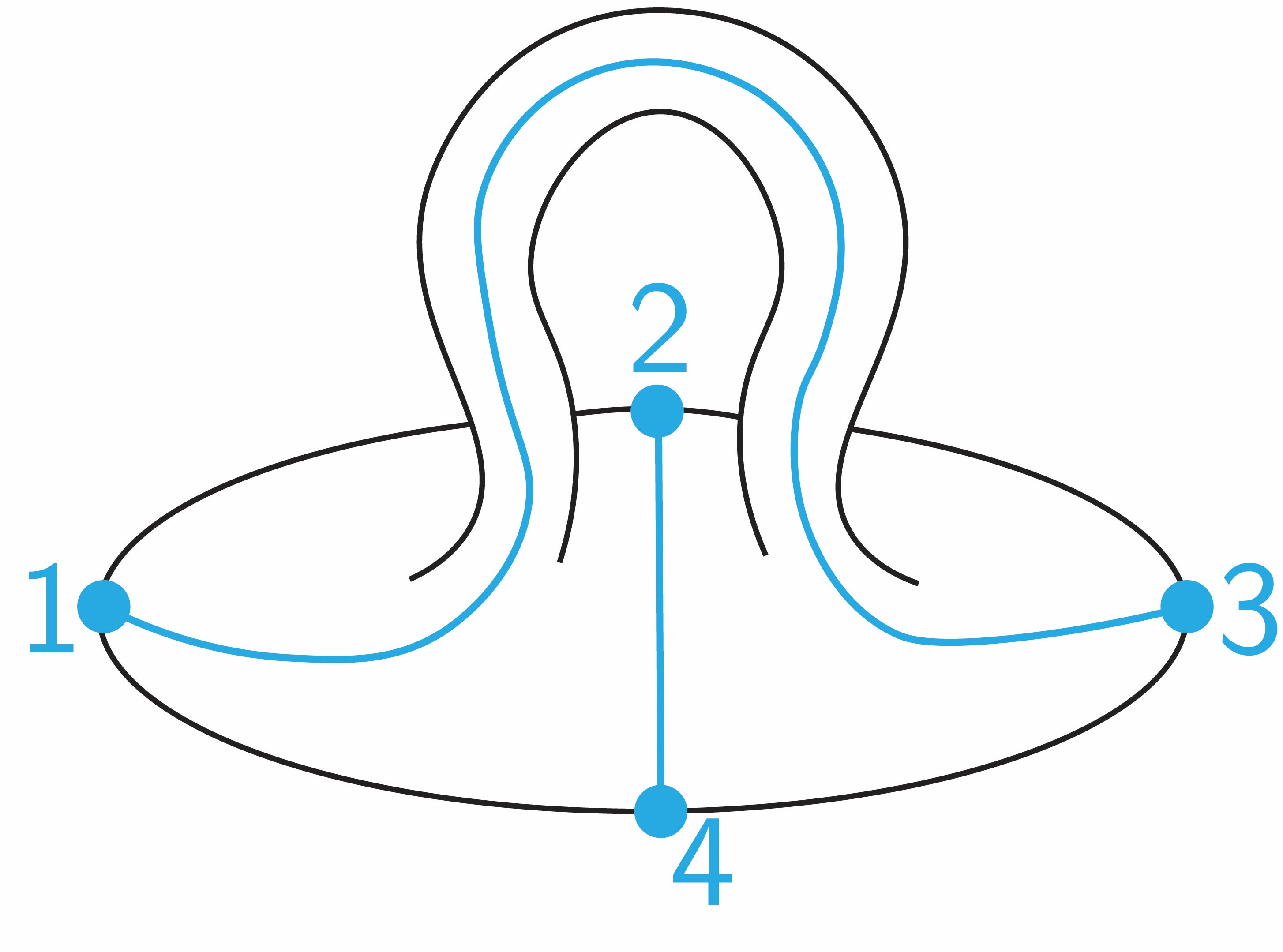}\quad\quad \xrightarrow{\textrm{Step 1}} \quad\quad \inlinefig[11]{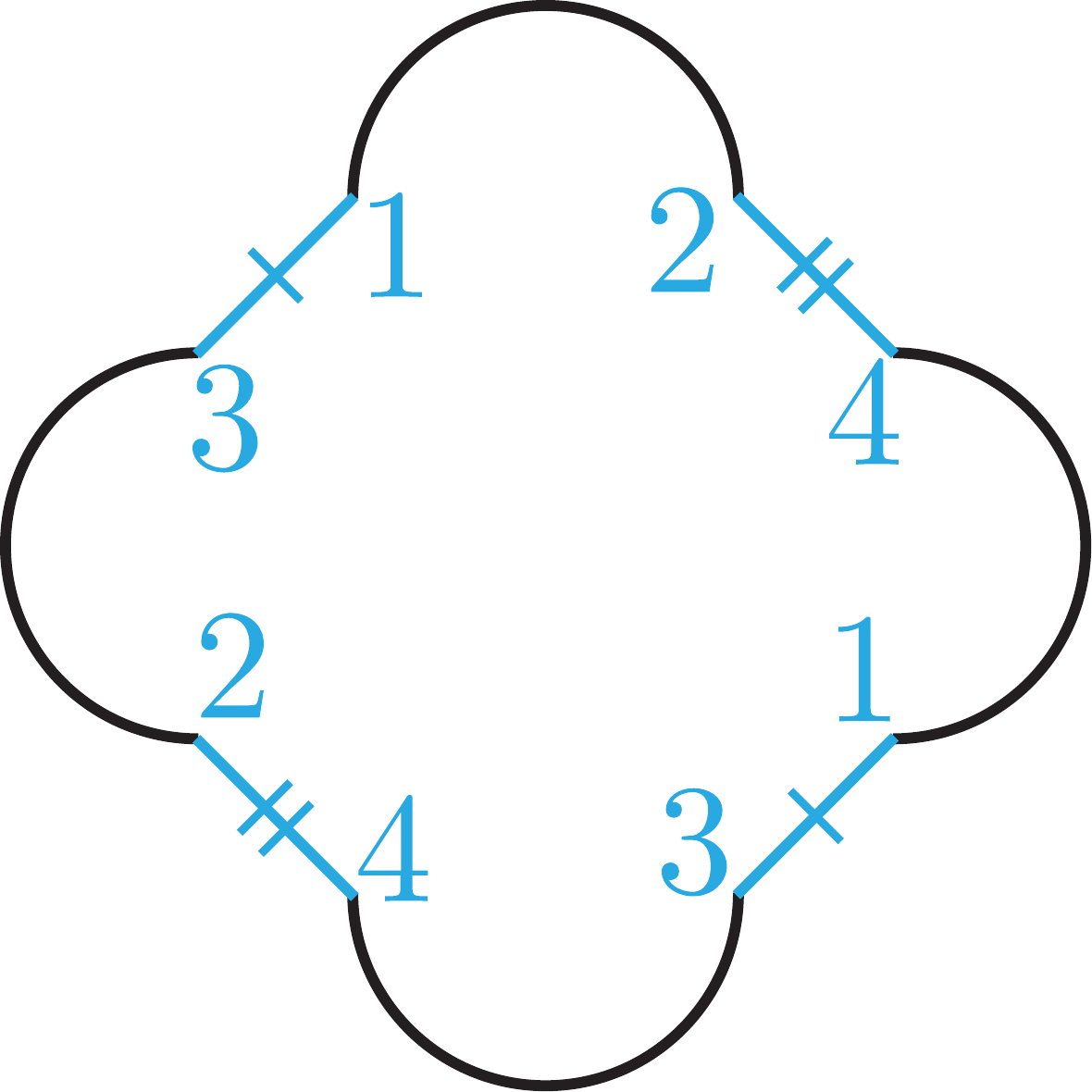}
\end{equation}
Since the resulting diagram has a single connected boundary and genus zero, it is non-vanishing. Following steps 3-5, we can then explicitly compute the diagram:
\begin{equation}
    \begin{aligned}
    &\inlinefig[10]{Figures/genus14pf1.pdf}=\int d\mu_{13}\,d\mu_{24}\,e^{-\Delta\ell_{13}}e^{-\Delta\ell_{24}}\inlinefig[11]{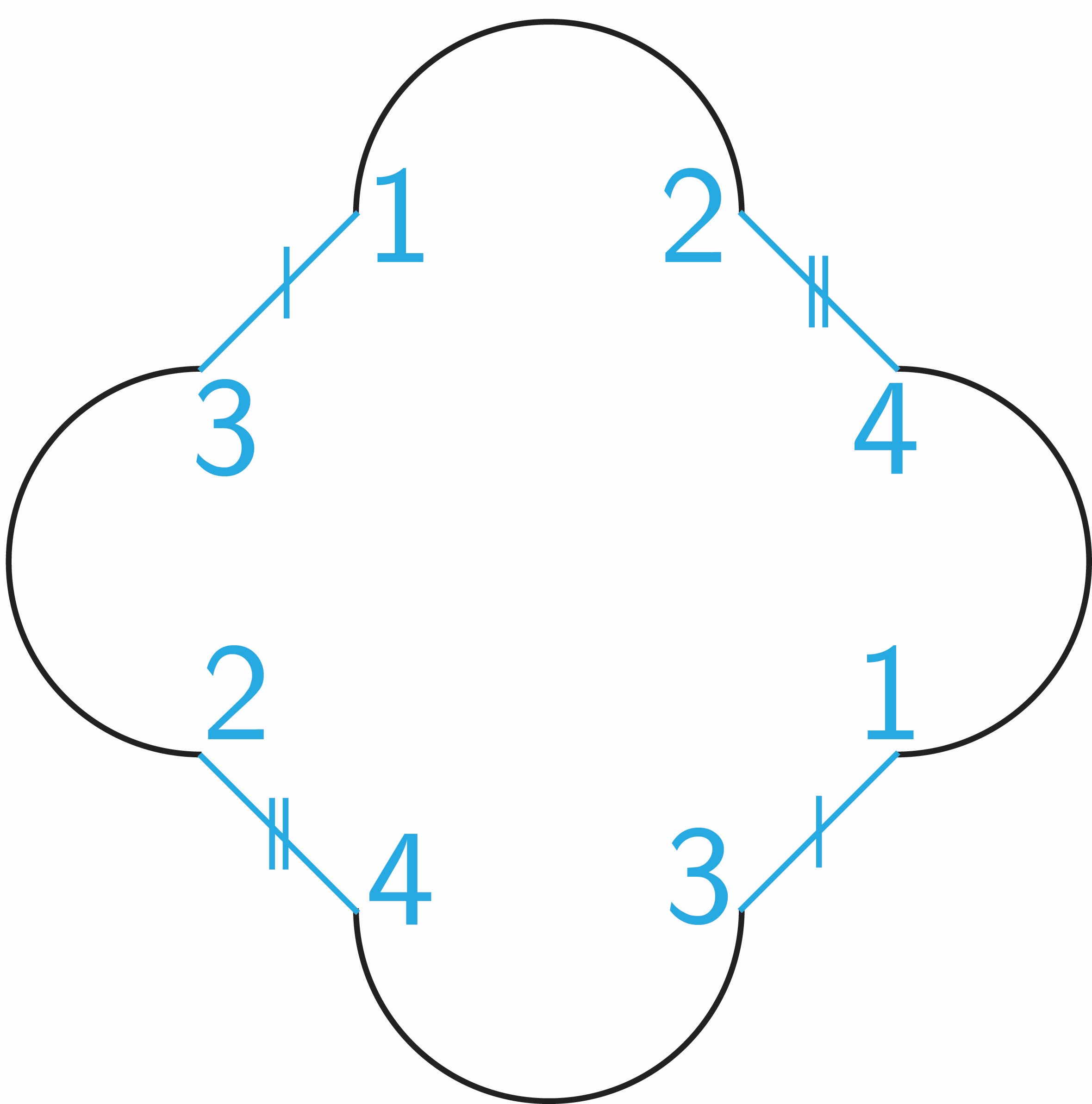}\\
    &=e^{-\SBPS}\int d\mu_{13}\,d\mu_{24}\,e^{-\Delta\ell_{13}}e^{-\Delta\ell_{24}}I(1,3;2,4;3,1;4,2)\\
    &=e^{-\SBPS}(P_{\Delta\to\infty})^2\,.
    \end{aligned}
\end{equation}
Given this prescription, it is easy to convince oneself that the calculation of all non-vanishing diagrams with fixed genus---arising from different Wick contractions and contributing to a given correlation function---involves the exact same integrals and therefore such diagrams contribute exactly the same amount. Specifically, consider a $2knm$ point function $[\tr(\rho^n)]^m$ on $m$ disconnected boundaries with $2kn$ insertions per boundary. The contribution of each non-vanishing, fully-connected genus-$g$ diagram to this correlation function is simply given by
\begin{equation}
    f_{2kn,m}(g) = e^{(2-2g-m)\SBPS}(P_{\Delta\to\infty})^{knm}.
    \label{eq:BPSdiagram}
\end{equation}
In order to determine the total contribution of fully-connected genus-$g$ diagrams to the correlation function, one then needs to simply multiply $f_{2kn,m}(g)$ by the total number of possible Wick contractions leading to non-vanishing diagrams and non-intersecting matter worldlines on the fully-connected genus-$g$ surface with $m$ boundaries. We will explain how to count such Wick contractions in Section \ref{sec:whichdiagrams}.

Before moving on, we would like to comment on the extension of these results to the case of non-SUSY JT gravity when inserting operators $\tilde{O}_{\Delta}$ involving projectors on a fixed microcanonical window, see Eq.~\eqref{eq:microperator}. The cutting and gluing prescription to compute diagrams remains largely unchanged, with an appropriate modification of the Hartle-Hawking wavefunction and integration measure, see e.g. \cite{Mertens_Turiaci_Verlinde_2017,Blommaert:2018oro,Iliesiu:2019xuh,Yang_2019, Boruch:2024kvv}. We review the details in Appendix \ref{sec:derivation-2n-pt-func-non-SUSY-JT}. There are two main differences. 

The first one is that, for diagrams without empty handles (i.e., those that do not vanish in the BPS case), each connected bulk patch comes with an integral over energies in the microcanonical window weighted by the JT gravity density of states. For a fully-connected diagram with $m$ boundaries, $2kn$ insertions per boundary, and genus $g$, the number of such connected patches is $p=knm+2-2g-m$ (see Appendix \ref{sec:derivation-2n-pt-func-non-SUSY-JT}). Thus, for a sufficiently small microcanonical window in which $\rho(E)\approx\rho(\bar{E})$, this results in a contribution
\begin{equation}
\begin{aligned}
    f_{2kn,m}^E(g)&=e^{(2-2g-m)S_0}(P_{\Delta\to\infty}(\bar{E}))^{knm}(\rho_0(\bar{E})\delta E)^{knm+2-2g-m}\\
    &=e^{(2-2g-m)S(\mathfrak{E})}(P_{\Delta\to\infty}(\bar{E})\rho_0(\bar{E})\delta E)^{knm}
    \label{eq:micro-factorized}
    \end{aligned}
\end{equation}
for each fully-connected diagram of genus $g$, where $P_{\Delta\to\infty}$ is now given by the propagator of non-SUSY JT gravity (see Footnote \ref{foot:non-susy-prop}) and we defined the microcanonical entropy $e^{S(\mathfrak{E})}=e^{S_0}\int_{\mathfrak E} dE\,\rho_0(E)\approx e^{S_0}\rho(\bar{E})\delta E$, where $S_0/(2\pi)$ is the background value of the dilaton.

The second difference is that, unlike the BPS case, diagrams with empty handles (e.g., those depicted in Eq.~\eqref{eq:empty}) do not vanish. As we explain in detail in Appendix \ref{sec:derivation-2n-pt-func-non-SUSY-JT}, these diagrams account for fluctuations of the number of states within the microcanonical window of interest. All diagrams with empty handles can be obtained by starting from diagrams without empty handles and genus $g$, and adding an arbitrary number of handles. The effect of summing over all diagrams obtained this way is to replace the product of JT gravity densities of states by the corresponding moment of the density of states $\rho_0(E_1)...\rho_0(E_n)\to \overline{\rho_0(E_1)...\rho_0(E_n)}$, which takes into account the correlation between different replicas in the matrix ensemble dual to JT gravity \cite{Saad_Shenker_Stanford_2019, Saad_2019,Blommaert:2020seb,Iliesiu:2021ari}. As a result, the contribution to a $2knm$-point function on $m$ boundaries (with $2kn$ insertions per boundary) of all diagrams obtained by starting with a given fully-connected diagram of genus $g$ without empty handles and adding all possible empty handles is given by
\begin{equation}
    f_{2kn,m}^{E,tot}(g)=e^{(2-2g-m)S_0}(P_{\Delta\to\infty}(\bar{E}))^{knm}\int_{\mathfrak E} dE_1...dE_{knm+2-2g-m}\overline{\rho_0(E_1)...\rho_0(E_{knm+2-2g-m})}\,.
    \label{eq:micro-nonfactorized}
\end{equation}
We remark that, at leading order in $e^{-S_0}$, $\overline{\rho_0(E_1)...\rho_0(E_{knm+2-2g-m})}$ factorizes and we recover Eq.~\eqref{eq:micro-factorized}. The corrections due to empty handles are therefore subleading for every fixed genus $g$ of the initial diagram without empty handles, and they are independent of the number of operator insertions $k$. As we will discuss in Section \ref{sec:mapping}, in the limit $k=O( e^{2S(\mathfrak{E})/3})$ of our interest, higher genus diagrams without empty handles all contribute at leading order because they are combinatorially enhanced in $k$. On the other hand, corrections due to diagrams with empty handles are not enhanced in $k$ and thus they always give subleading corrections to our results. As a result, in the regime of our interest, we can simply approximate Eq.~\eqref{eq:micro-nonfactorized} by Eq.~\eqref{eq:micro-factorized}, which has the same structure as Eq.~\eqref{eq:BPSdiagram} for the BPS case. We refer the reader to Appendix \ref{sec:derivation-2n-pt-func-non-SUSY-JT} for a more thorough discussion of these issues. 

\subsection{Full diagrammatic expansion of correlation functions}
\label{sec:whichdiagrams}

Now that we have a prescription for computing any matter diagram in JT (super)gravity, we can ask: how can we identify all possible diagrams contributing to a given $2knm$-point function on $m$ boundaries with $2kn$ insertions per boundary? Let us start by focusing on a single boundary, $m=1$.

To answer this question, recall that, in the large $\Delta$ limit, correlation functions can be computed by summing over all possible Wick contractions. For a given Wick contraction, we must consider only diagrams without intersections between matter worldlines. 

To gain some intuition, let us return to the simple case of the four-point function on a single boundary. There are three possible Wick contractions. Two of them can be realized on geometries with any genus $g\geq 0$. To avoid intersections, the third one can only be realized on geometries with genus $g\geq 1$. Let us define the \textit{minimal embedding diagram} for a given Wick contraction to be the geometry with the lowest possible genus on which the Wick contraction can be realized without any intersections. For each Wick contraction, only the minimal embedding diagram contributes a non-vanishing amount in the BPS case. This is because all higher topologies can be obtained by adding empty handles to the minimal embedding diagram. After cutting along matter worldlines, this leads to the vanishing diagrams discussed in Section \ref{sec:3.1}.\footnote{As we have discussed, in non-SUSY JT gravity in a microcanonical band, these diagrams do not vanish, but can be taken into account by simply replacing a product of densities of states by the corresponding moment of the density of states, and they are subleading in the limit $k=O(e^{2S(\mathfrak{E})/3})$ regime of our interest.} The full four-point function on a single boundary is then given by
\begin{equation}
\begin{aligned}
    \tr\left(\tilde{O}_{\Delta}^4\right) &= \inlinefig[9]{Figures/LMRS_4pt.pdf}\quad+\quad\inlinefig[9]{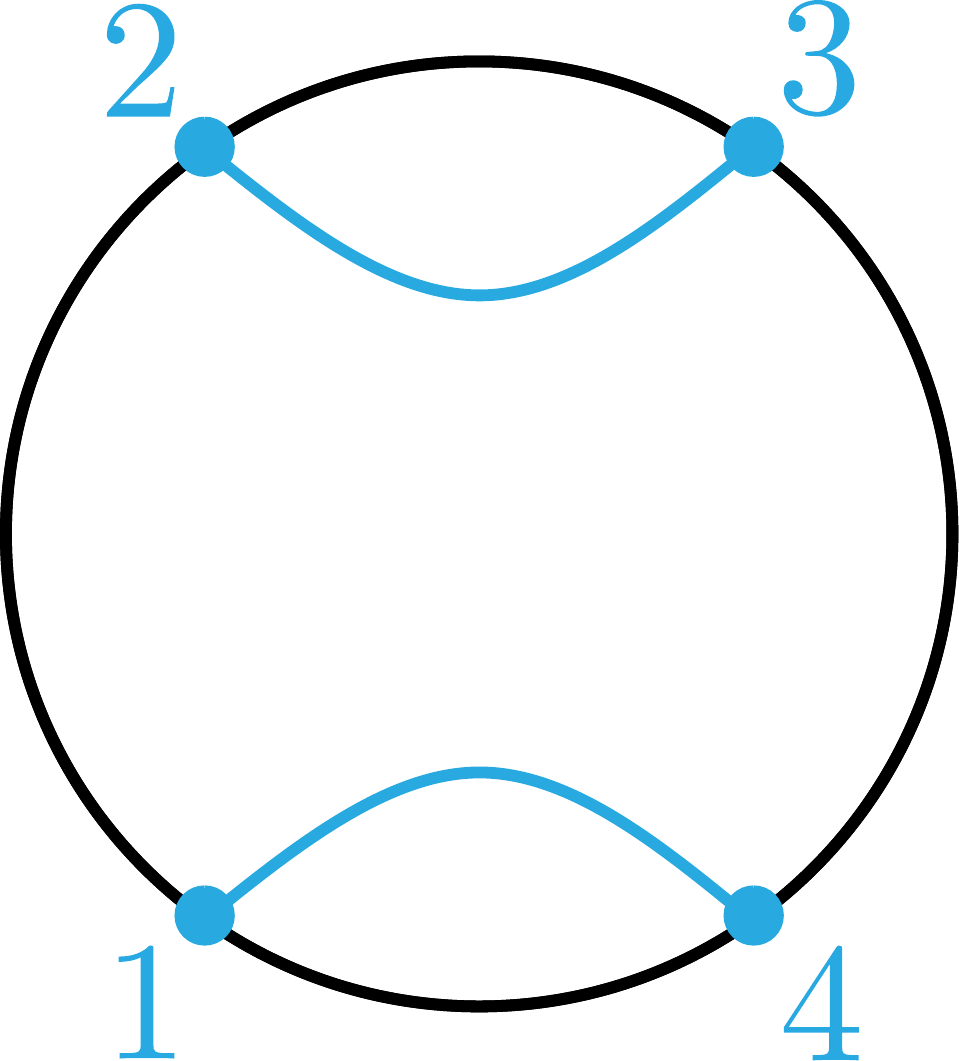}\quad+\quad\inlinefig[9]{Figures/genus14pf1.pdf}\\
    &=(2e^{\SBPS}+e^{-\SBPS})(P_{\Delta\to\infty})^2\,.
    \end{aligned}
\end{equation}
Thus, notice that the genus expansion truncates in this case at $g=1$ with all other diagrams vanishing per the arguments above.

The exact same reasoning applies to an arbitrary $2kn$-point function. The contribution of each Wick contraction to the $2kn$-point function is given by the minimal embedding diagram.\footnote{Notice that each Wick contraction contributes with one and only one minimal embedding diagram with genus $g$. All apparently different diagrams with the same genus $g$ are equivalent to each other under the mapping class group and are already taken into account by the cutting and gluing procedure, as explained in Section \ref{sec:3.1}.} Summing over Wick contractions, one obtains the $2kn$-point function. Note that the minimal embedding diagrams contributing to the $2kn$-point function are those for which all closed geodesics intersect at least one matter worldline, as stated at the beginning of Section \ref{sec:bulk-resolution}. In fact, if a closed geodesic does not intersect any matter worldline, after cutting along matter worldlines, the resulting diagram still has a non-intersecting closed geodesic after cutting (i.e., it has an empty handle), and it vanishes.

The sum over Wick contractions can be reorganized as a sum over the genus of the minimal embedding diagrams. What is the minimal embedding diagram with the highest possible genus for a $2kn$-point function on a single boundary? Let us start from the case in which $kn$ is even. The highest genus diagram is one that, after cutting along matter worldlines, gives rise to a single connected asymptotic polygon with $2kn$ geodesic boundaries:
\begin{equation}
    \inlinefig[12]{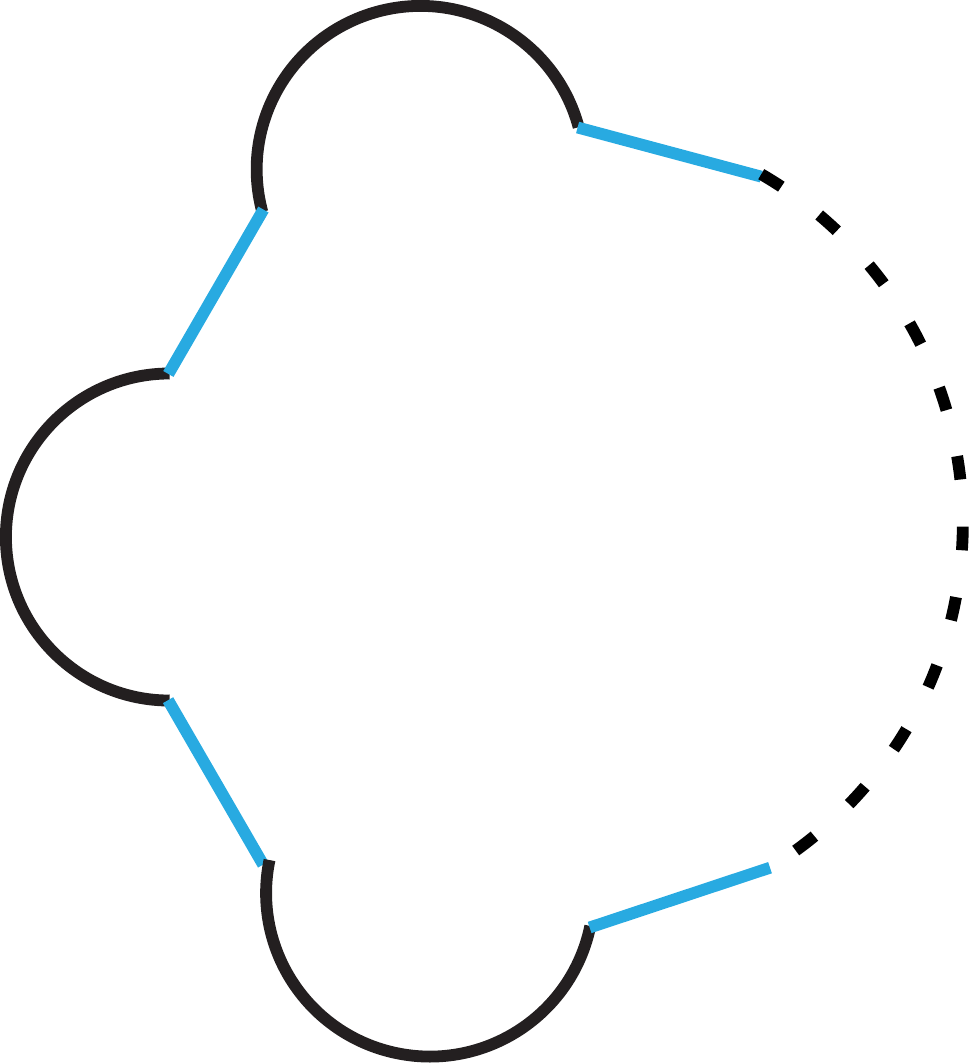}
    \label{eq:maxgenus}
\end{equation}
The $2kn$ geodesic boundaries are identified in pairs. Before gluing, the diagram in \eqref{eq:maxgenus} is a disk and its Euler characteristic is $\chi=2-1=1$. For each pair of geodesic boundaries we glue, we decrease the Euler characteristic by 1. Therefore, the Euler characteristic of the diagram of interest is $\chi^{\text{even}}_{\min}=1-kn\equiv 1-2g^{\text{even}}_{\max}$, i.e., $g^{\text{even}}_{\max}=kn/2$. To find $g_{\max}$ when $kn$ is odd, consider again the maximum genus diagram \eqref{eq:maxgenus} for $kn-1$ even, and add two insertions (recall that we are computing a $2kn$-point function) on any two of the asymptotic boundary segments of the asymptotic polygon. Connect the insertions by a worldline, and cut along the worldline. We then obtain two asymptotic polygons, each with (odd) $kn$ geodesic boundaries. The initial Euler characteristic is the sum of the two disks $\chi=1+1=2$, and after gluing one obtains $\chi^{\text{odd}}_{\min}=2-kn=1-(kn-1)\equiv 1-2g^{\text{odd}}_{\max}$, yielding $g^{\text{odd}}_{\max}=(kn-1)/2$. Therefore, in general we have $g_{\max}=\lfloor kn/2\rfloor$, where $\lfloor \cdot \rfloor$ indicates the integer part.

The $2kn$-point function on a single boundary can then finally be written as
\begin{equation}
    \tr(\rho^n)\equiv \tr\(\tilde{O}_{\Delta}^{2kn}\)=(P_{\Delta\to\infty})^{kn}\sum_{g=0}^{\lfloor kn/2\rfloor}e^{(1-2g)\SBPS}C^{(g)}_{2kn,[2^{kn}]}\,,
    \label{eq:1bdycorrelation}
\end{equation}
where $C^{(g)}_{2kn,[2^{kn}]}$ are combinatorial factors counting the number of Wick contractions with minimal embedding diagrams of genus $g$.\footnote{$[2^{kn}]$ indicates the partition type among the $2kn$ points, namely the partition in pairs.} These are a special case of genus-dependent Fa\`a Di Bruno coefficients and they take the form \cite{walsh1,walsh2,Zagier1986,Jackson1987CountingCI} (see \cite{Coquereaux} for a recent review)
\begin{equation}
    C^{(g)}_{2kn,[2^{kn}]}=\frac{(2kn)!}{(kn+1)!(kn-2g)!}\left[\(\frac{u/2}{\tanh(u/2)}\)^{kn+1}\right]_{u^{2g}}
    \label{eq:faadibruno}
\end{equation}
where $\left[X\right]_{u^\alpha}$ is the coefficient of $u^\alpha$ in the series expansion of $X$. For $g=0$, Eq.~\eqref{eq:faadibruno} correctly reduces to the Catalan numbers $C_{kn}$, which in fact count the number of Wick contractions for which the minimal embedding diagram is the disk.

Using our discussion at the end of Section \ref{sec:3.1}, we can give an analogous formula for the single-boundary $2kn$-point function in a microcanonical band in non-SUSY JT gravity, including all non-minimal embedding diagrams:
\begin{equation}
\begin{aligned}
    \tr_{\mathfrak{E}}(\rho^n)&=(P_{\Delta\to\infty}(\bar{E}))^{kn}\sum_{g=0}^{\lfloor kn/2\rfloor}e^{(1-2g)S_0}C^{(g)}_{2kn,[2^{kn}]}\int_{\mathfrak E}\overline{\prod_{i=1}^{kn+1-2g}dE_i\,\rho_0(E_i)}\\
    &\approx (P_{\Delta\to\infty}(\bar{E})\rho(\bar{E})\delta E)^{kn}\sum_{g=0}^{\lfloor kn/2\rfloor}e^{(1-2g)S(\mathfrak{E})}C^{(g)}_{2kn,[2^{kn}]}\,,
    \end{aligned}
    \label{eq:1bdycorrelationJT}
\end{equation}
where in the second line we kept only the disconnected part of the moment of the density of states, i.e., only minimal embedding diagrams.\footnote{We remark that, for generic values of $k$, this approximation is incorrect, because we are keeping higher genus minimal embedding diagrams while neglecting non-minimal embedding diagrams that in general can have the same genus. However, as we will see in Section \ref{sec:mapping}, for the $k=O( e^{2S(\mathfrak{E})/3})$ regime of interest, this approximation is well-justified, because higher genus minimal embedding diagrams are combinatorially enhanced in $k$, whereas non-minimal embedding diagrams are not.}

Finally, a completely analogous reasoning can be carried out for multi-boundary $2knm$-point functions on $m$ boundaries and $2kn$ insertions per boundary, namely $\[\tr\(\rho^n\)\]^m$. With a similar reasoning as above, one finds that the minimal embedding diagram with the highest possible genus has genus $g_{max}=\lfloor (knm+1-m)/2\rfloor$. We can then write down a formula for the fully-connected correlation function between $m$ boundaries\footnote{Once the generic fully-connected component with $m$ boundaries is known, one can immediately write the full correlation function by appropriately summing over products of fully-connected components.} in the BPS case (an analogous formula can be written for the non-SUSY JT gravity case)
\begin{equation}
    \[\tr\(\rho^n\)\]^m_{\text{conn.}}=(P_{\Delta\to\infty})^{knm}\sum_{g=0}^{\lfloor (knm+1-m)/2\rfloor}e^{(2-2g-m)\SBPS}C^{(m)}_g(2kn,...,2kn)\,,
    \label{eq:mbdycorrelation}
\end{equation}
where the combinatorial factors $C^{(m)}_g(2kn,...,2kn)$ count the number of Wick contractions giving rise to a minimal embedding diagram which is fully-connected between the $m$ boundaries and has genus $g$  for partitions of type $[2^{mkn}]$ among the $2mkn$ points (namely, partitions in pairs). Clearly, for $m=1$ we recover Eq.~\eqref{eq:1bdycorrelation} and the coefficients $C^{(m)}_g(2kn,...,2kn)$ are simply the genus-dependent Fa\`a Di Bruno coefficients \eqref{eq:faadibruno}. While for $m>2$ the coefficients $C^{(m)}_g(2kn,...,2kn)$ are not known in a closed form expression, it turns out that for $m=2$ they can be found explicitly in closed form.\footnote{We give the exact form of the coefficients in Appendix \ref{appendix:f}. To derive the coefficients, we used the results of \cite{dubrovin}, which give exact finite $N$ expressions for two-point functions of single trace operators in a Gaussian matrix integral. To relate them to $C^{(m)}_g(2kn,...,2kn)$ for $m=2$, we use that the combinatorics of diagrams in the BPS case exactly matches the combinatorics for correlation functions in a Gaussian Unitary Matrix Ensemble (GUE). We will discuss this in detail in Section \ref{sec:GUE}.} In the remainder of this section, we will instead introduce an exact map between our setup in the $k=O( e^{2\SBPS/3})$ regime and an effective pure non-SUSY JT gravity theory without matter, and use this effective theory to solve the LMRS puzzle for semi-quenched R\'enyi entropies.

\subsection{Effective non-SUSY JT gravity theory and puzzle resolution}
\label{sec:mapping}

Now that we know how arbitrary genus diagrams contribute to correlation functions, let us go back to the LMRS puzzle and focus on the semi-quenched entropies introduced in Section \ref{sec:entropies}. We will discuss here the BPS case, but our analysis also applies to the non-SUSY JT gravity case, as we will point out at the end of this subsection.

In general, we would like to compute the R\'enyi-$n$ semi-quenched entropy
\begin{equation}
    S^{(n)}_{SQ}=\frac{1}{1-n}\log\frac{\overline{\tr\(\rho^n\)}}{\overline{\(\tr\rho\)^n}}\,.
    \label{eq:semi-n}
\end{equation}
First, recall that the parameter $\SBPS$ is large. Therefore, for $k=O(1)$ not scaling with $\SBPS$, the genus expansions \eqref{eq:1bdycorrelation}, \eqref{eq:mbdycorrelation} are dominated by the $g=0$ (disk) term. In this regime, the semi-quenched entropy agrees with the annealed entropy. In fact, $n$-boundary connected contributions captured by Eq.~\eqref{eq:mbdycorrelation}, which in principle contribute to the denominator of the semi-quenched entropy \eqref{eq:semi-n} but not to the annealed entropy, are topologically suppressed with respect to the product of $n$ disks. At leading order in $e^{\SBPS}$, both the annealed and the semi-quenched entropy are then given by Eq.~\eqref{eq:Sn-Delta-equals-infinity}. 

The puzzle arises when the annealed entropy becomes negative, namely for $k=O\(e^{2\SBPS/3}\)$. In this regime, we expect some non-perturbative effect to rescue positivity of the semi-quenched entropy \eqref{eq:semi-n}. Let us thus focus on this regime and compute the R\'enyi-$n$ semi-quenched entropy in the double-scaling limit $e^{\SBPS}\to\infty$, $k\to\infty$, $ke^{-2\SBPS/3}\equiv \alpha$, where $\alpha$ is a fixed, order 1 parameter. 

As a first step, let us compute $\overline{\tr(\rho^n)}$, given by Eq.~\eqref{eq:1bdycorrelation}. In the large $k$ limit, the Fa\`a Di Bruno coefficients \eqref{eq:faadibruno} take a simplified form. Using the Taylor expansion for $\coth(x)$, we can write
\begin{equation}
    \(\frac{u/2}{\tanh(u/2)}\)^{kn+1}=\(1+\sum_{p=1}^\infty \frac{B_{2p}}{(2p)!}u^{2p}\)^{kn+1}=\sum_{s=0}^{kn+1}\binom{kn+1}{s}\(\sum_{p=1}^\infty \frac{B_{2p}}{(2p)!}u^{2p}\)^s
    \label{eq:bern}
\end{equation}
where $B_p$ are the Bernoulli numbers, and in the second equality we used the binomial theorem. We are interested in extracting the coefficient of the $u^{2g}$ term. Therefore, we are only interested in terms with $s\leq g$. In the $k\to\infty$ limit, the right-hand side of Eq.~\eqref{eq:bern} is then dominated by the $s=g$ term,\footnote{Notice that, for $k$ large but finite, $g
\leq kn/2$, as we have seen in Section \ref{sec:whichdiagrams}. Therefore, the largest binomial coefficient is always that of the $s=g$ term. Here we are taking the $k\to\infty$ limit first, so that $g/(kn)\to 0$ for any fixed order of $e^{-2g\SBPS}$ in the expansion. This guarantees that, for terms with $s<g$, combinatorial factors coming from the sum over $p$ (which only depend on $g$) do not offset the $1/k$-suppression of the binomial coefficient with $s<g$ with respect to the one with $s=g$.} leading to
\begin{equation}
    \[\(\frac{u/2}{\tanh(u/2)}\)^{kn+1}\]_{u^{2g}}\approx \binom{kn+1}{g}\(\frac{B_{2}}{2}\)^g\approx \frac{1}{g!}\(\frac{kn}{12}\)^g\,,
\end{equation}
where here and in the following we use the $\approx$ symbol to indicate that the expression is valid in the double-scaling limit we are considering. Plugging into the Fa\`a Di Bruno coefficients \eqref{eq:faadibruno} and further expanding at large $k$, one obtains
\begin{equation}
    C^{(g)}_{2kn,[2^{kn}]}\approx \frac{2^{2kn}}{\sqrt{\pi}g!(kn)^{3/2}}\(\frac{(kn)^3}{12}\)^{g}.
    \label{eq:faaexpansion}
\end{equation}
The single boundary $2kn$-point function \eqref{eq:1bdycorrelation} can then be evaluated explicitly for $k=\alpha e^{2\SBPS/3}$:
\begin{equation}
    \overline{\tr\(\rho^n\)}\approx \frac{2^{2kn}}{\sqrt{\pi}(\alpha n)^{3/2}}\(P_{\Delta\to\infty}\)^{kn}\sum_{g=0}^\infty \frac{1}{g!}\(\frac{(\alpha n)^3}{12}\)^g=\frac{2^{2kn}}{\sqrt{\pi}}\(P_{\Delta\to\infty}\)^{kn}\frac{e^{(\alpha n)^3/12}}{(\alpha n)^{3/2}}\,.
    \label{eq:resummed-disk-matter}
\end{equation}
Notice that, in the double-scaling limit of interest, diagrams of higher genera are enhanced in $k$. As a consequence, all genera contribute at the same order and the genus expansion had to be re-summed in order to correctly obtain the $2kn$-point function.\footnote{\label{footnote:suppression}In the non-SUSY JT gravity case, the same is true for minimal embedding diagrams. On the other hand, diagrams obtained by adding empty handles to a given minimal embedding diagram of genus $g$ are suppressed by factors of $e^{-S_0}$ with respect to the minimal embedding diagram, with no $k$ enhancement. Hence, they are subleading, justifying the approximation in the second line of Eq.~\eqref{eq:1bdycorrelationJT} in the double-scaling limit.}

Incidentally, using Eq.~\eqref{eq:resummed-disk-matter}, one can compute the all-genus correction to the annealed R\'enyi-$n$ entropy in the double-scaling limit:
\begin{equation}
    S^{(n)}_{A}=\frac{1}{1-n}\log \frac{\overline{\tr(\rho^n)}}{\(\overline{\tr(\rho)}\)^n}\approx \frac{1}{n-1}\log \left[\frac{n^{3/2}}{\pi^{(n-1)/2}}\frac{e^{-\alpha^3 n (n^2-1)/12}}{\alpha^{3(n-1)/2}}\right]\,.
    \label{eq:annealed-resummed}
\end{equation}
The annealed entropy clearly still becomes negative as $\alpha$ increases. As we have explained in Section \ref{sec:2}, this is not an issue because the annealed entropy does not need to remain positive. Nonetheless, this result shows that, although higher-genus corrections become important when $k=O( e^{2\SBPS/3})$, they are not enough to rescue positivity of the entropies. 

To compute the semi-quenched entropy \eqref{eq:semi-n}, which we know from the argument in Section \ref{sec:entropies} must stay positive, we need to evaluate the denominator $\overline{\(\tr\rho\)^n}$. This requires us to compute the fully-connected multi-boundary correlation function \eqref{eq:mbdycorrelation}. For $m>2$, this is beyond our reach because the $C^{(m)}_g(kn,...,kn)$ combinatorial factors are not known. For $m=2$, however, they are given explicitly in Appendix \ref{appendix:f}, and so one could explicitly compute $\overline{\(\tr\rho\)^2}$. Here we will employ a different but elegant approach, which is also generalizable for arbitrary $m$. We will map our problem for $k=O( e^{2\SBPS/3})$ to an analogous puzzle in which the (annealed) thermal entropy in pure, non-SUSY JT gravity becomes negative at very low temperatures $\beta=O( e^{2S_0^{\text{eff}}/3})$ \cite{Engelhardt_Fischetti_Maloney_2021,Hernandez-Cuenca_2024,Antonini_Iliesiu_Rath_Tran_2025}.\footnote{Here we are labeling the (large) background value of the dilaton in the non-SUSY JT gravity theory by $S_0^{\text{eff}}$ to distinguish it from $S_0$ in the non-SUSY case coupled to matter.} Using this effective theory, we will be able to explicitly compute $\overline{\(\tr\rho\)^2}$ in the $k=O( e^{2\SBPS/3})$ regime and resolve the negativity of the semi-quenched R\'enyi-2 entropy.

Let us introduce the effective theory, which surprisingly will be given by pure JT gravity without any matter insertions. In pure JT gravity, the single-boundary thermal partition function at inverse temperature $\beta$ can be written in terms of an asymptotic genus expansion \cite{Saad_Shenker_Stanford_2019}
\begin{equation}
    Z(\beta)=\tr \(e^{-\beta H}\)= \sum_{g=0}^\infty e^{(1-2g)S_0^{\text{eff}}} Z_{g,1}\(\beta\),
\end{equation}
where, for $g>1$, $Z_{g,1}(\beta)$ are given in terms of an integral of the Weil-Petersson volume of the moduli space of the hyperbolic Riemannian surface with genus $g$ and one geodesic boundary against the trumpet partition function \cite{Saad_Shenker_Stanford_2019}. At leading order at large $\beta$, $Z_{g,1}(\beta)\sim \beta^{3g-3/2}$ \cite{Engelhardt_Fischetti_Maloney_2021}. Therefore, for $\beta=\alpha e^{2S_0^{\text{eff}}/3}$ with $\alpha=O(1)$---which, for a reason that will become clear in Section \ref{sec:matrixmodels}, is called the Airy limit of JT gravity---all genera contribute at the same order and the genus expansion is not perturbatively under control. Nonetheless, each $Z_{g,1}(\beta)$ can be expanded in powers of $\beta$ and after swapping the two sums we obtain
\begin{equation}\label{eq:2sum}
    Z\(\b\) =e^{-\beta E_0} \sum_{l=0}^{\infty} \b^{-l} \sum_{g=0}^\infty \a^{3g-3/2} Z_{g,1,l}\,,
\end{equation}
where $E_0$ is the ground state energy of JT gravity. Typically, this is chosen to vanish through an appropriate choice of boundary counterterms, but in our case we will leave it arbitrary for a reason that will become clear shortly.
For $l=0$---which is the dominant term in the $\beta,e^{S_0^{\text{eff}}}\to\infty$ double-scaling limit---the $Z_{g,1,0}$ are known \cite{Engelhardt_Fischetti_Maloney_2021,Antonini_Iliesiu_Rath_Tran_2025}, the sum over genus is convergent, and it can be performed explicitly. In particular, we obtain\footnote{To match this result to the one reported in \cite{Antonini_Iliesiu_Rath_Tran_2025}, simply choose $\gamma=2^{-1/3}$ there.}
\begin{equation}
    Z(\beta)\approx \frac{e^{-\beta E_0}}{2\sqrt{\pi}\alpha^{3/2}}\sum_{g=0}^\infty\frac{1}{g!}\(\frac{\alpha^3}{12}\)^g= \frac{e^{-\beta E_0}}{2\sqrt{\pi}}\frac{e^{\alpha^3/12}}{\alpha^{3/2}}\,.
    \label{eq:airy-thermal-part}
\end{equation}
Notice that computing the annealed thermal entropy $S^{(n)}_A=\log\[\overline{ Z(n\beta)}/\overline{ Z(\beta)^n}\]/(1-n)$, one finds a result similar to Eq.~\eqref{eq:annealed-resummed}, with the annealed entropy becoming negative as $\alpha$ increases.

By comparing the result \eqref{eq:airy-thermal-part} with Eq.~\eqref{eq:resummed-disk-matter}, we immediately see that, after identifying $\beta_{\text{eff}}=k$, $S_0^{\text{eff}}=\SBPS$, and $E_0=-\log\(4P_{\Delta\to\infty}\)$, we obtain (we reintroduce here the overline notation to indicate the quantity is computed using the gravitational path integral)
\begin{equation}
    \overline{\tr\(\rho^n\)}=2\overline{ Z(n\beta_{\text{eff}})}\,.
\end{equation}
This equality is true only when on the left-hand side $k=\alpha e^{2\SBPS/3}$ and on the right-hand side the partition function is computed in the Airy limit of JT gravity $\beta_{\text{eff}}=\alpha e^{2S_0^{\text{eff}}/3}$. 

This match, in the double-scaling limit, of correlation functions in JT supergravity in the BPS sector to thermal partition functions in non-SUSY pure JT gravity is powerful. First, notice that diagrams at each genus can be mapped between the two theories. In particular, for $2kn$ operator insertions, the sum over all possible Wick contractions giving rise to minimal embedding diagrams of genus $g$ is captured by a single pure JT gravity diagram with genus $g$ and boundary length $n\beta_{\text{eff}}=nk$, e.g.
\begin{equation}
\begin{gathered}
    \inlinefig[10]{Figures/LMRS2.pdf}\quad+\quad\inlinefig[10]{Figures/LMRS3.pdf}\quad +\quad \dots\quad=\quad 2\times\inlinefig[10]{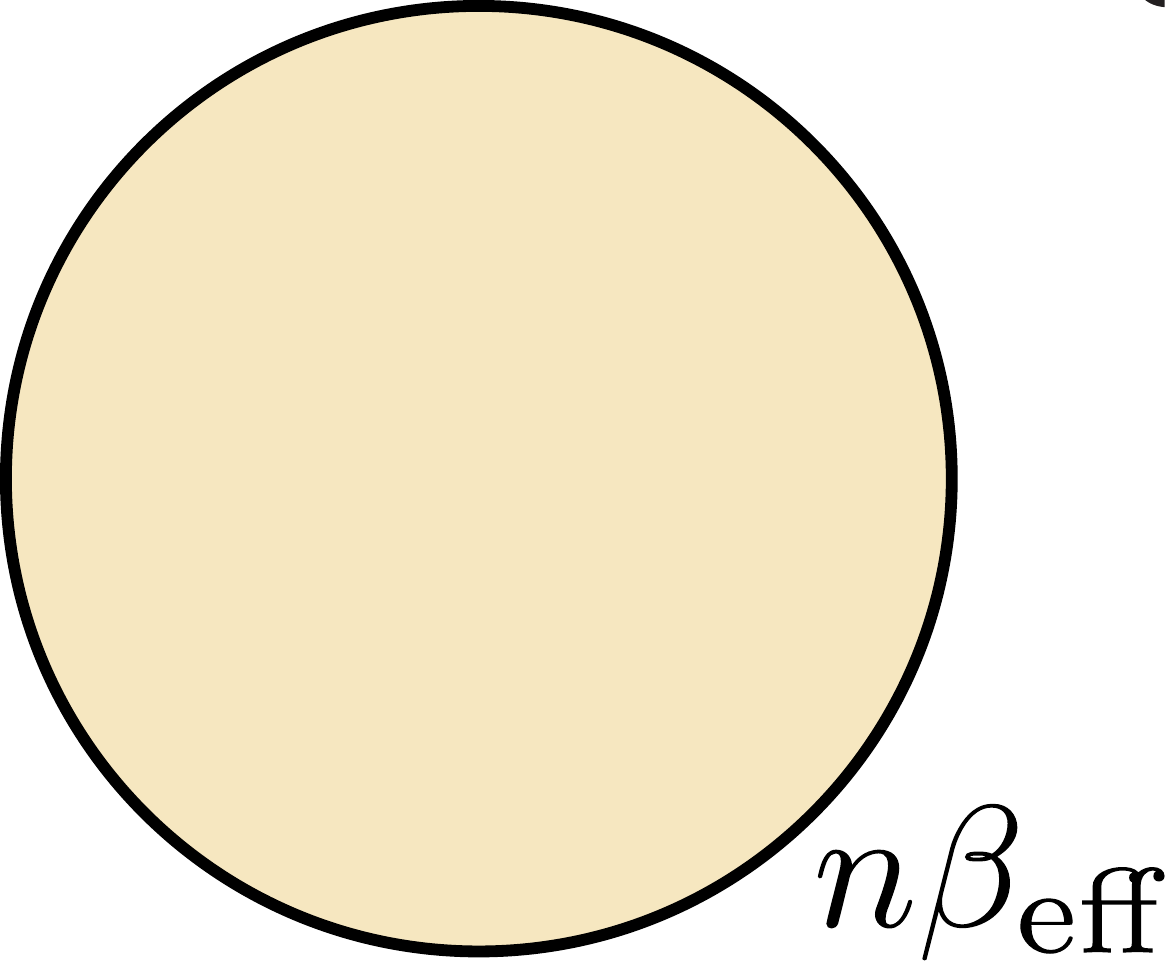}\\[10pt]
     \inlinefig[10]{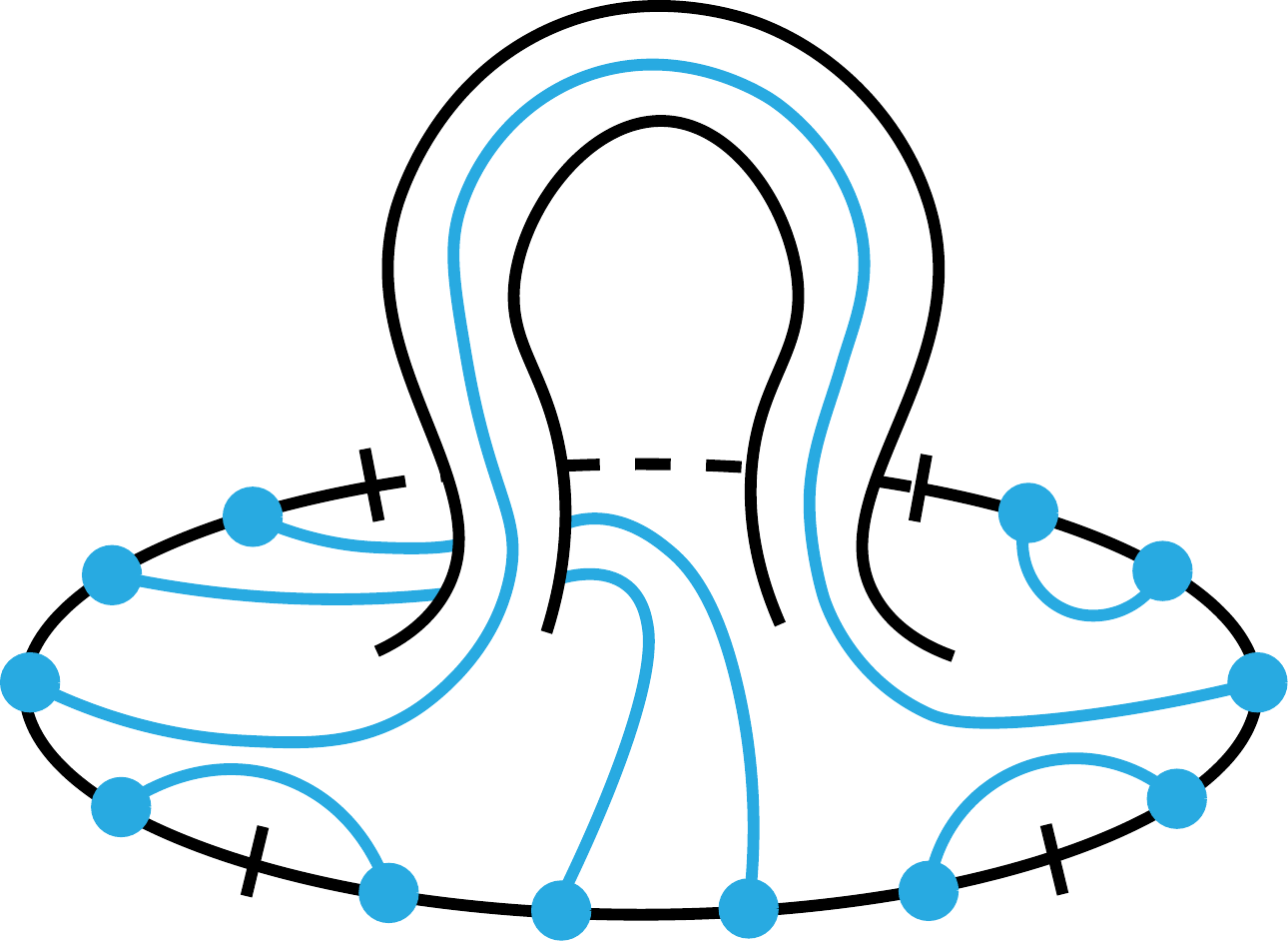}+\inlinefig[10]{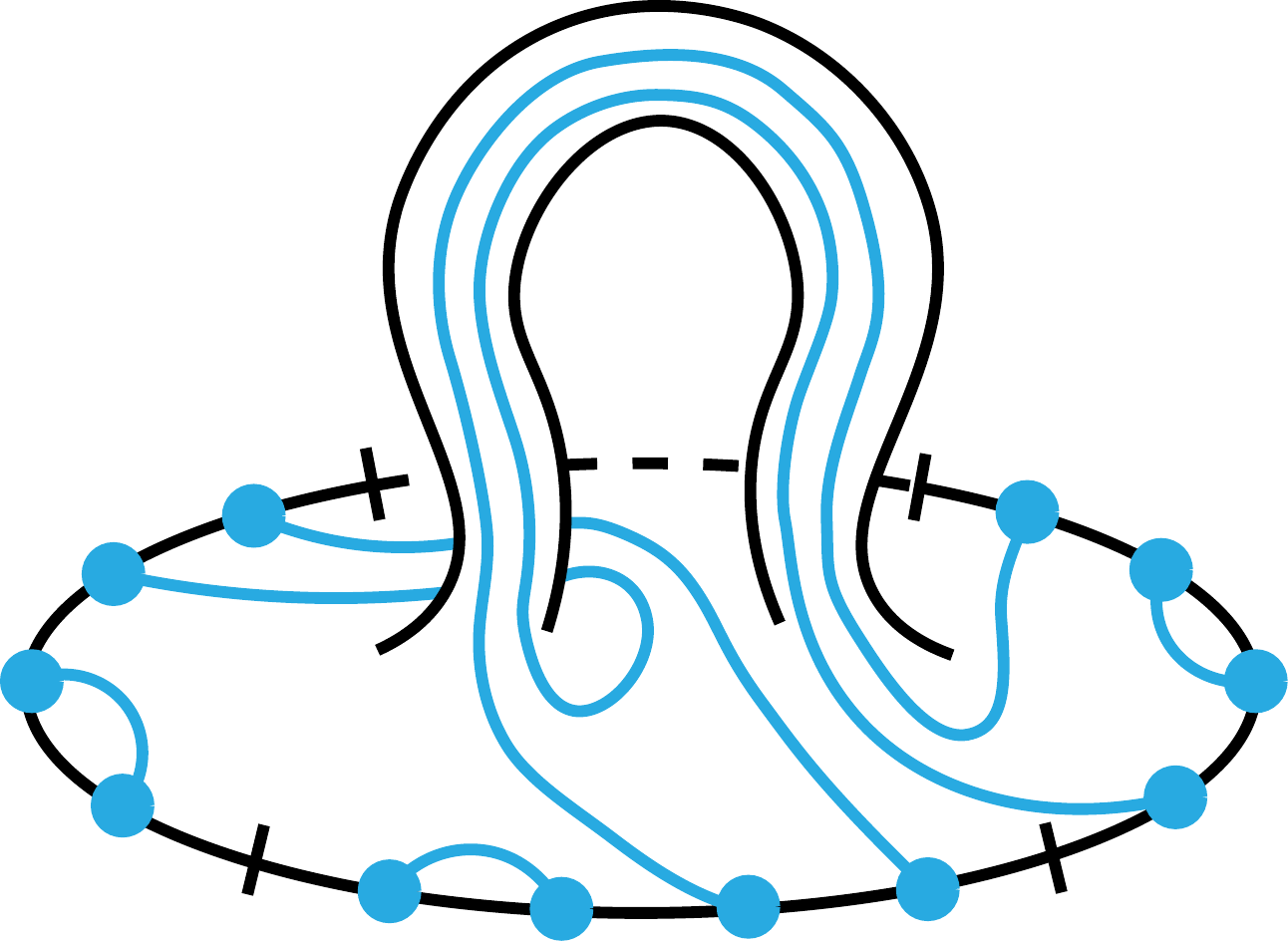} +\quad \dots\quad=\quad 2\times \inlinefig[10]{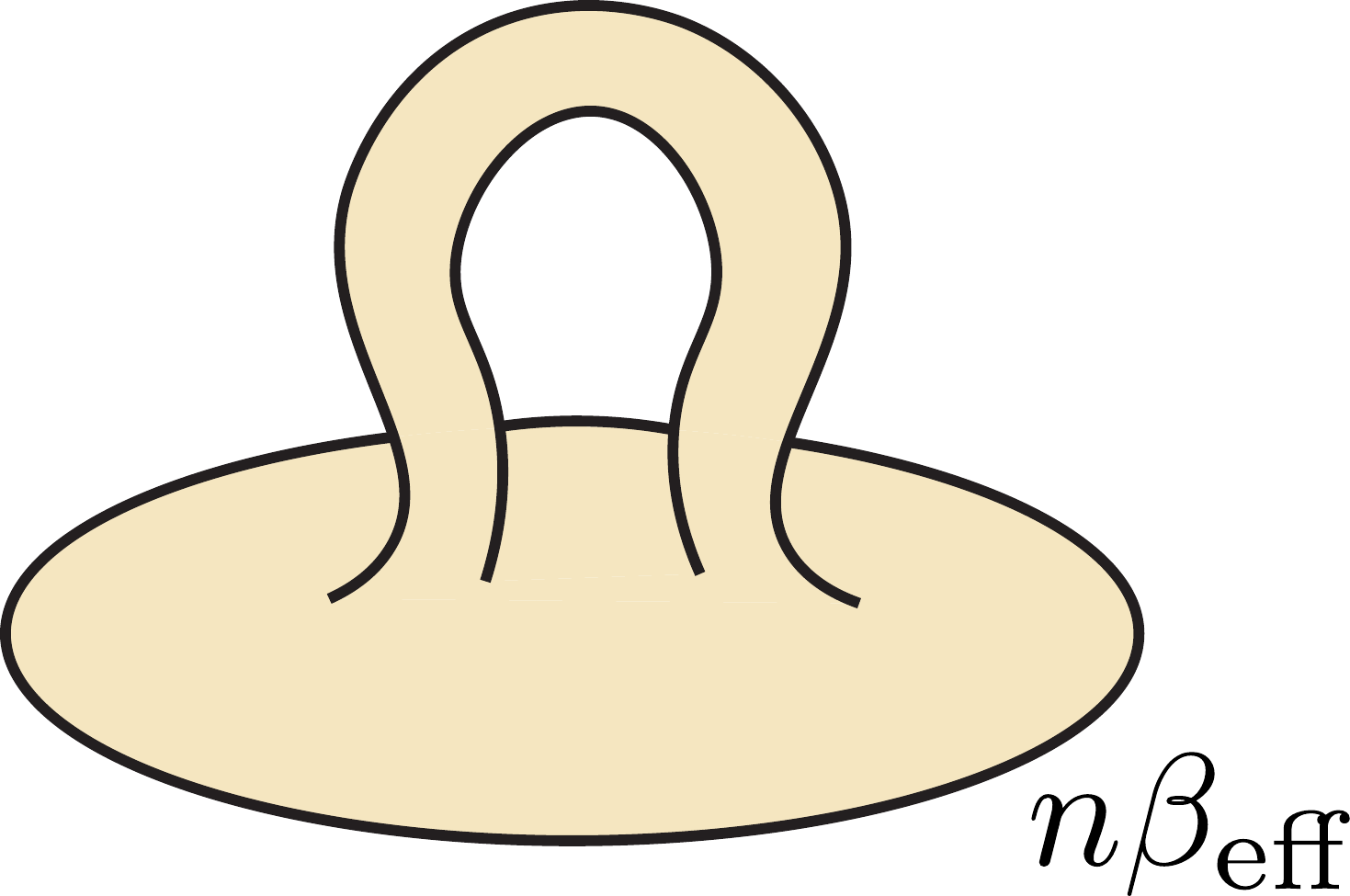}
    \end{gathered}
\end{equation}
Second, the match continues to hold for multi-boundary correlation functions and thermal partition functions. Generically,\footnote{Although plausible, this result is not fully justified at this stage. In this section, we will simply assume that Eq.~\eqref{eq:fullmatch} holds. In Section \ref{sec:GUE}, we will rigorously prove this assumption more generally. We report here a schematic outline of the proof in Section \ref{sec:GUE}. First, we realize that the relevant combinatorics for correlation functions in the BPS sector exactly match the combinatorics for correlation functions in a Gaussian Unitary Matrix Ensemble (GUE). Second, we show that, in the double-scaling limit of interest $k=O( e^{2\SBPS/3})$, GUE correlation functions reduce to appropriate thermal partition functions with effective inverse temperature $\beta_{\text{eff}}=k$ in the Airy matrix model up to the factor of 2 in Eq.~\eqref{eq:fullmatch}, which appears because our operators have two edges in their eigenvalue spectrum \eqref{eq:eval-density-Delta-to-infty}. Third, we connect the Airy model partition functions to the $\beta\to\infty$, $e^{S_0}\to\infty$, $\beta/e^{2S_0/3}=\alpha$ limit of partition functions in the SSS matrix model dual to JT gravity \cite{Saad_Shenker_Stanford_2019}, completing the proof of Eq.~\eqref{eq:fullmatch}. We have also checked that, for $m=2$, the large $k$ expansion of the coefficients $C_g^{(2)}(2kn,2kn)$ agrees with the prediction from the effective theory for $m=2$ (see Appendix \ref{appendix:f}). }
\begin{equation}
    \overline{\[\tr\(\rho^n\)\]^m}_{\text{conn.}}=2\overline{ \[Z(n\beta)\]^m}_{\text{conn.}}\,.
    \label{eq:fullmatch}
\end{equation}
 Once again, this match holds for each diagram contributing to the correlation function. For instance, for $m=2$ we have e.g.
\begin{equation}
\begin{gathered}
    \inlinefig[6]{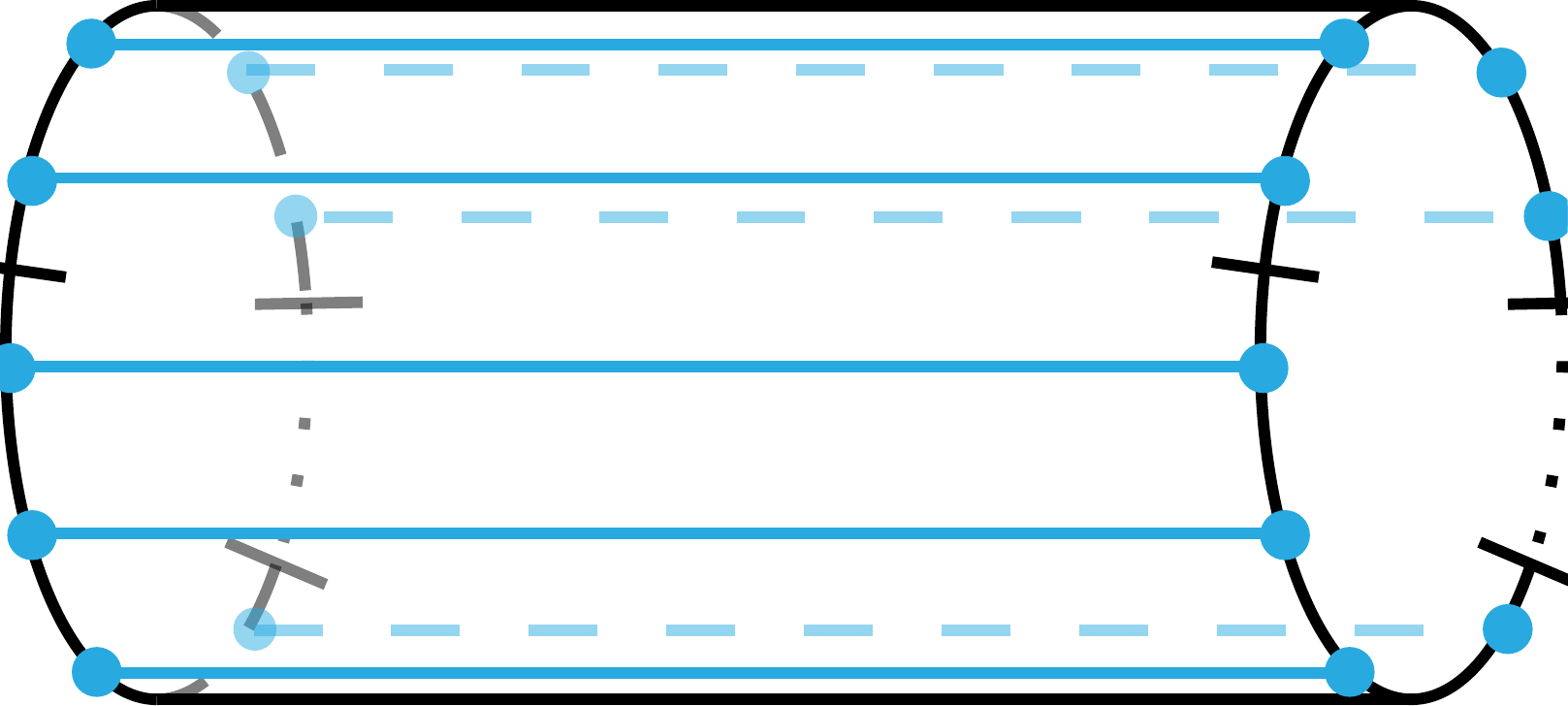}+\inlinefig[6]{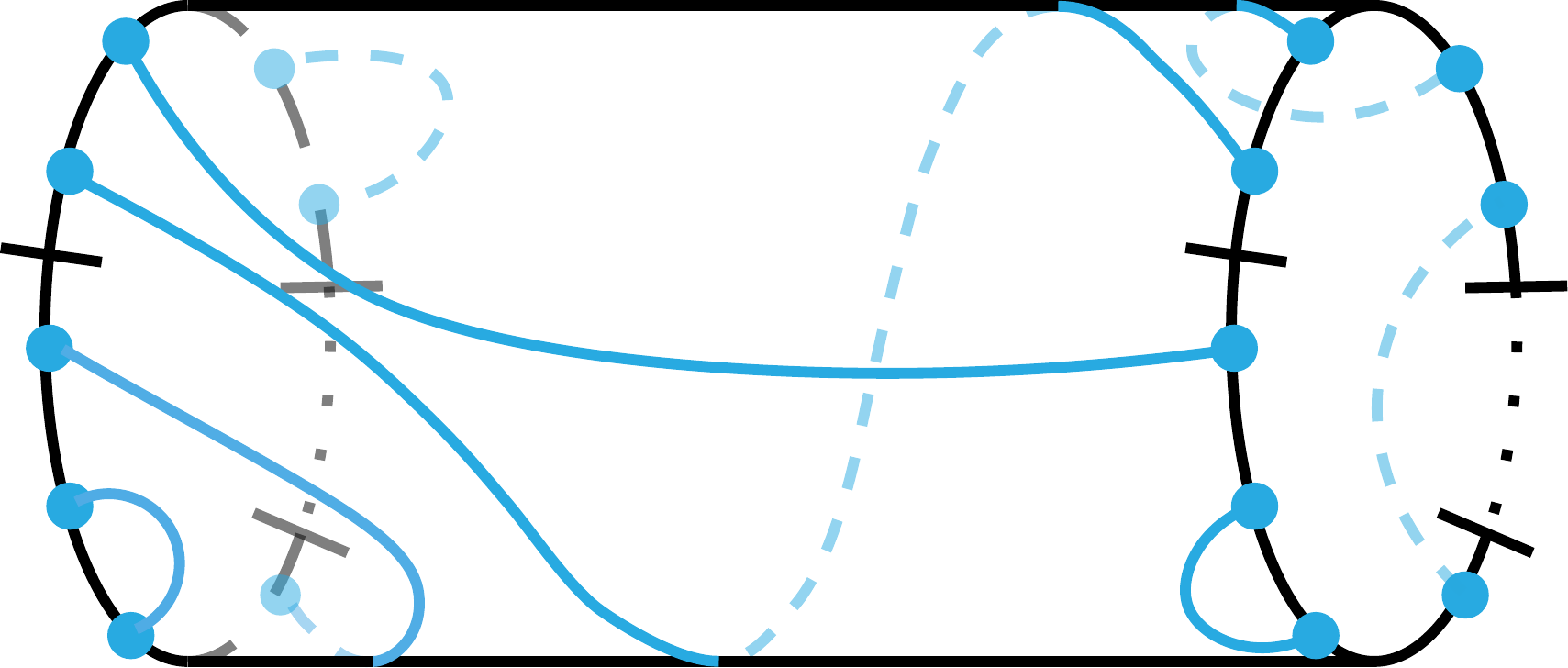}+ \dots= 2\times \inlinefig[6]{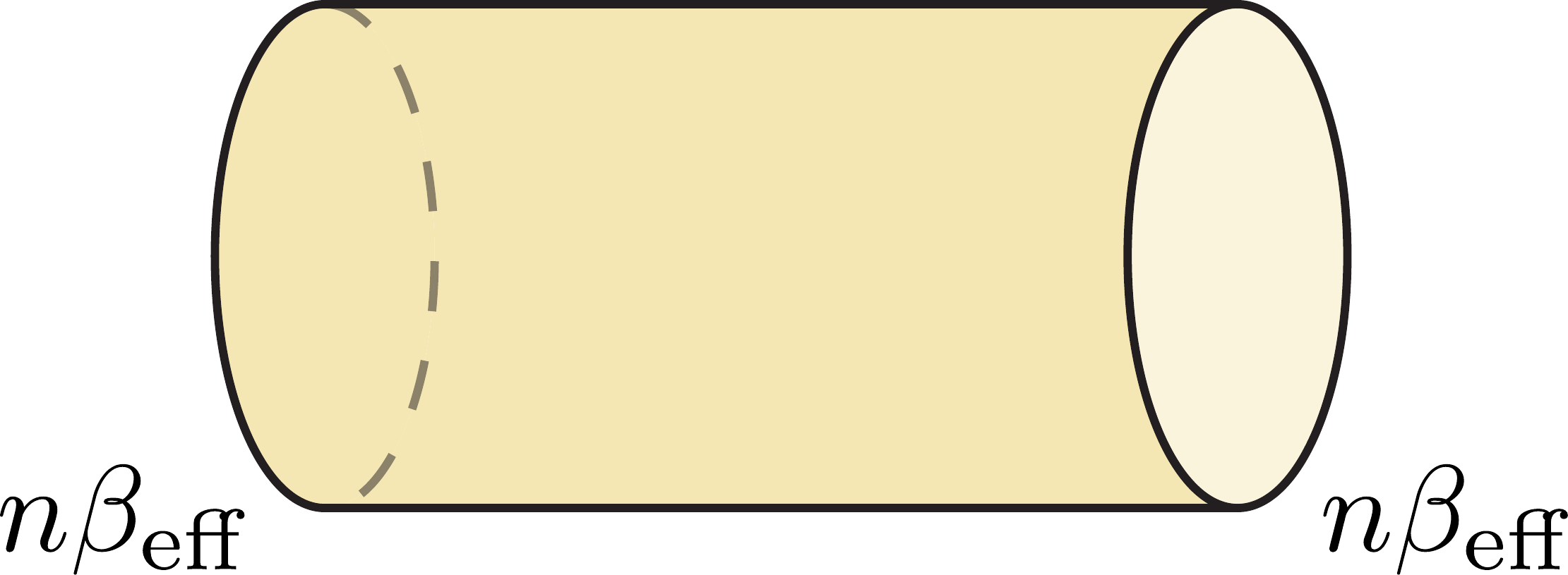}\\[10pt]
        \inlinefig[6]{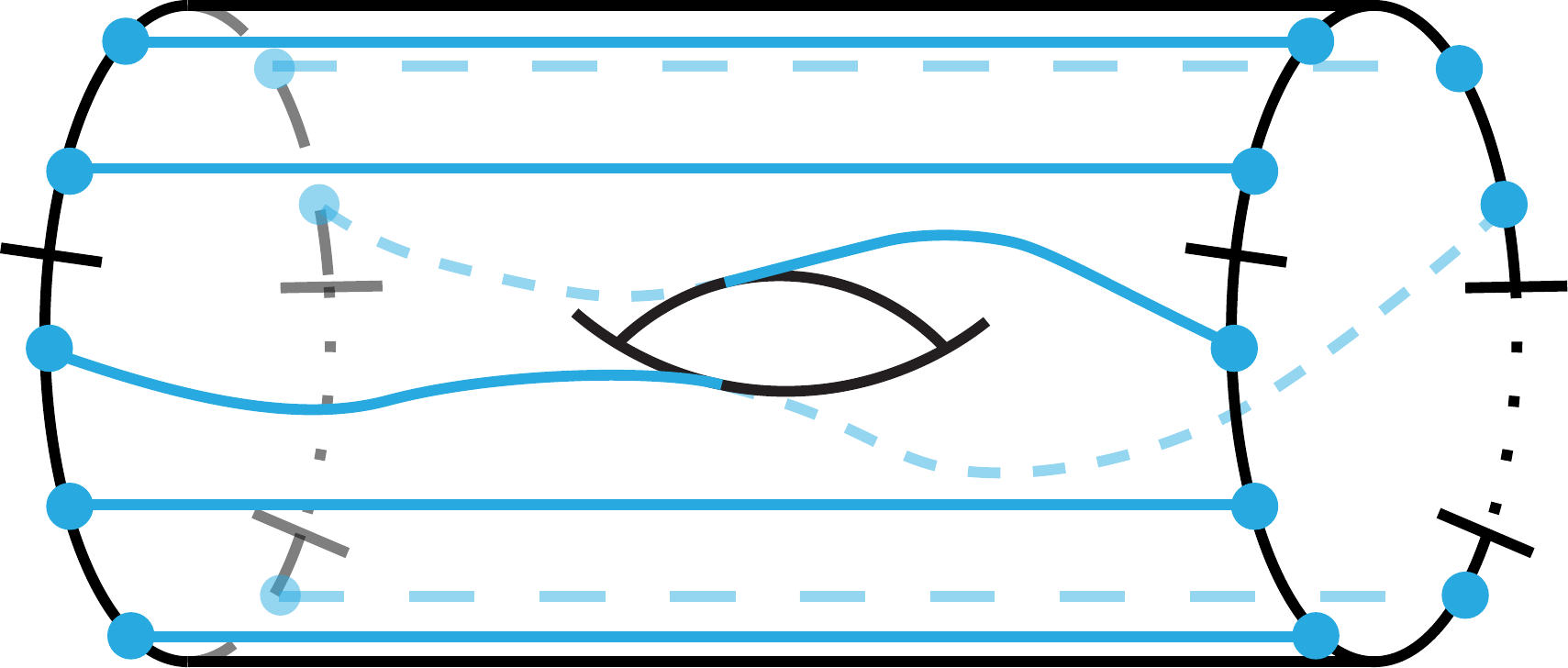}+\inlinefig[6]{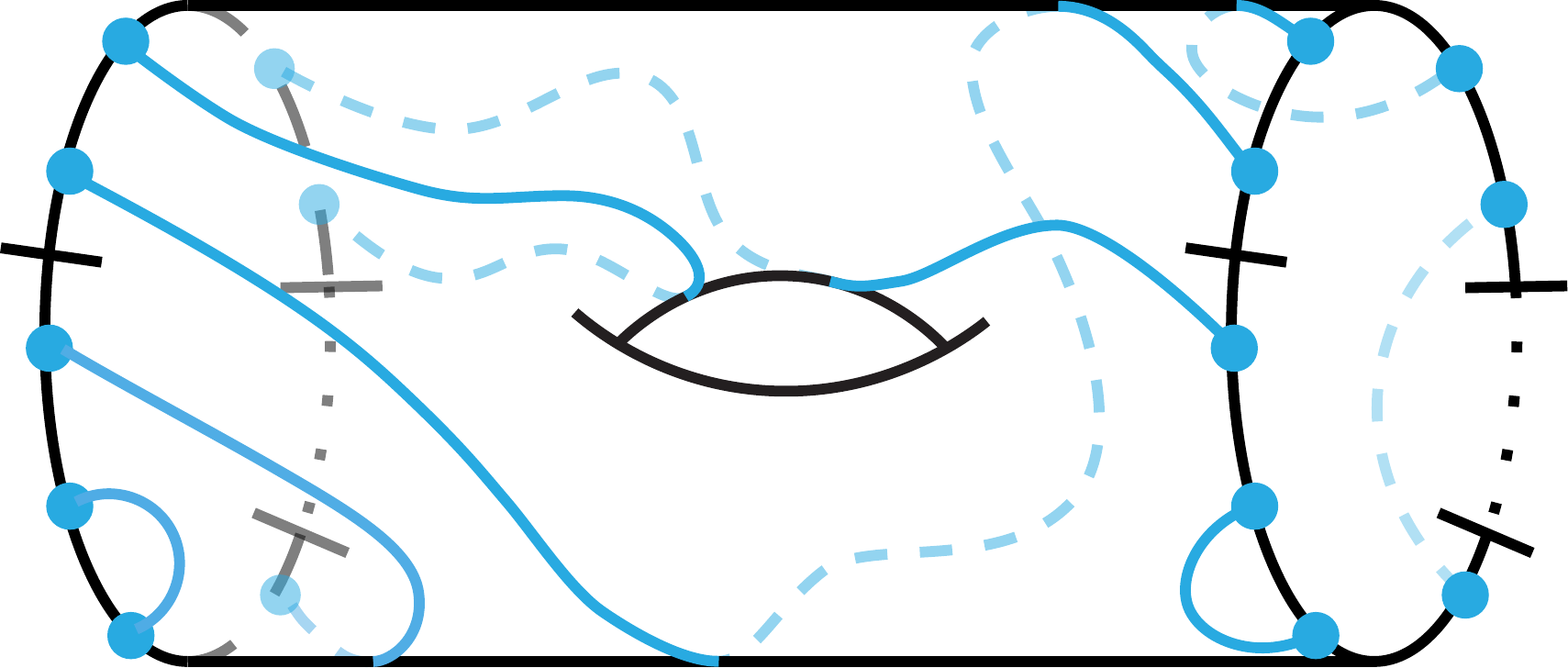} + \dots= 2\times \inlinefig[6]{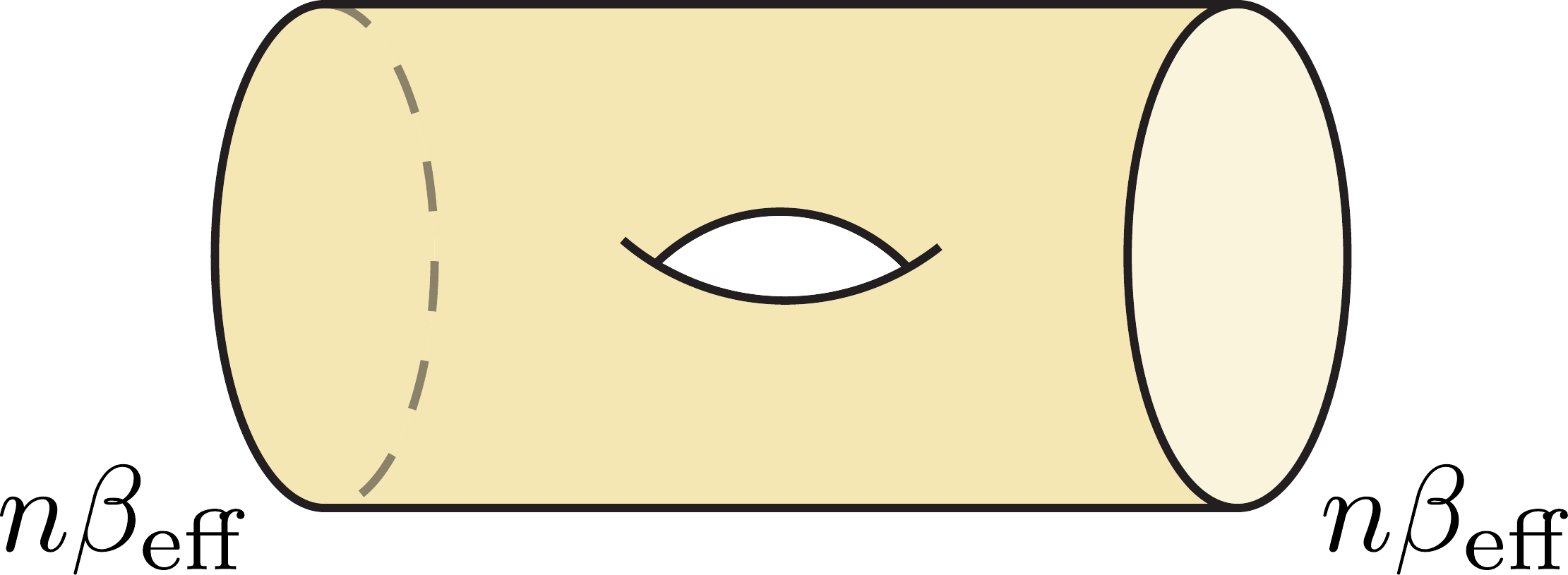}\\[10pt]
    \end{gathered}
\end{equation}

Let us give a physical intuition for this match, which we will also derive quantitatively in Section \ref{sec:resolution} using the dual matrix model descriptions of the two sides of the equivalence. Consider a boundary with $2kn$ operator insertions, and two points on this boundary. For each given Wick contraction between the operator insertions, we can give a discrete notion of ``distance'' between the two points, given by the minimum possible number of matter lines intersected by a line connecting the two points.
\begin{equation*}
    \inlinefig[12]{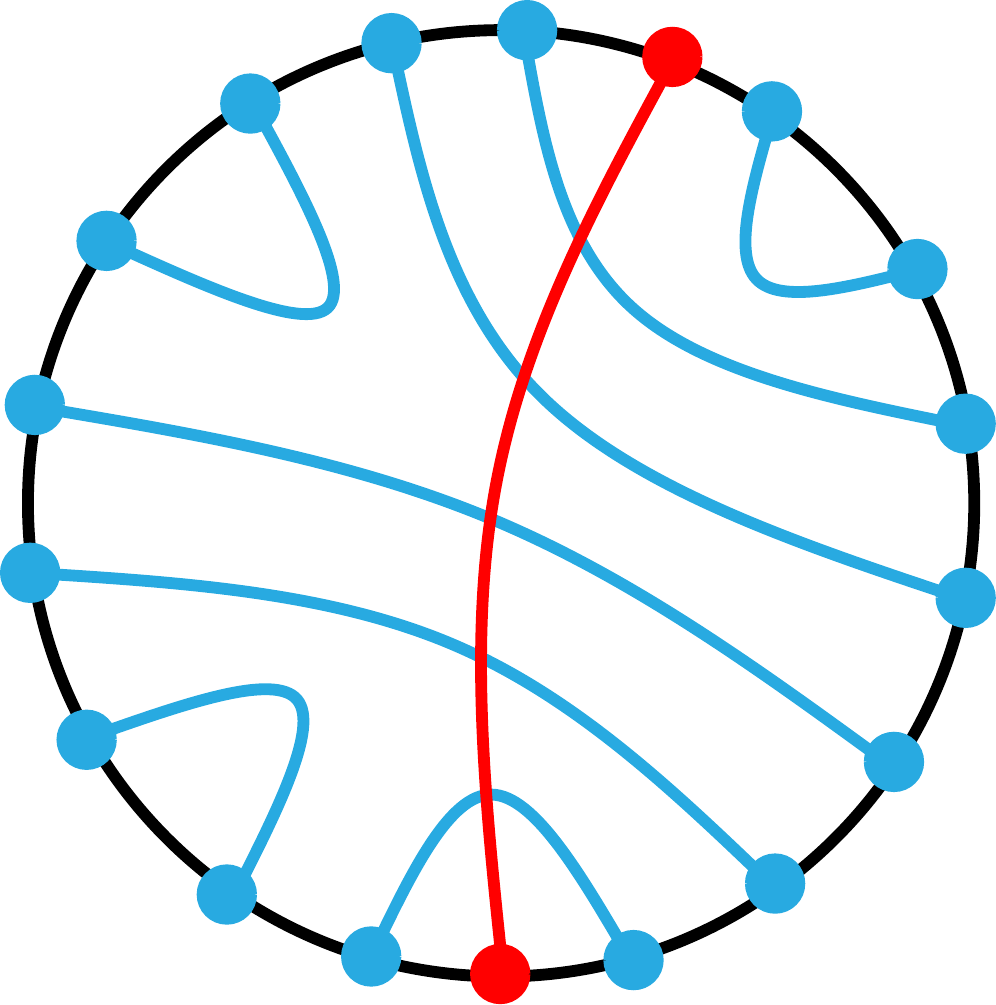}
\end{equation*}
We can define this distance for any two points on the boundary, and extend the same definition to multi-boundary correlators. 
In the large $k$ limit, this discrete notion of distance approaches a continuum notion of distance. 
Notice that the distance between two given points is different for different Wick contractions. We can then think of each given Wick contraction as a given geometry. The sum over Wick contractions then corresponds to a sum over geometries, which is precisely what a gravitational partition function is, hence the match between diagrams of genus $g$ contributing to the JT (super)gravity matter correlators and the genus-$g$ contribution to pure JT partition functions. We emphasize that this notion of distance and emergence of a geometry is very similar to that encountered in the double-scaled SYK model \cite{Berkooz:2018jqr, Lin:2022rbf}. In Appendix \ref{app:matter}, we examine this correspondence with the insertion of probe matter with arbitrary scaling dimension $\Delta_m$.

\begin{figure}
    \centering
    \includegraphics[width=0.8\linewidth]{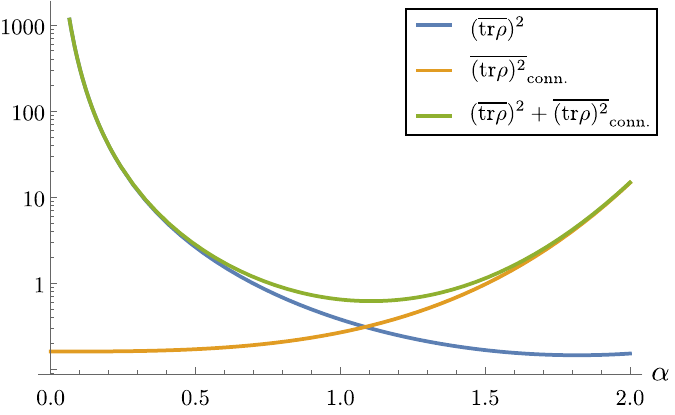}
    \caption{Comparing $(\overline{\tr\rho})^2$ (first term in the denominator of Eq.~\eqref{eq:renyi-2}) and $\overline{\(\tr\rho\)^2}_{\text{conn.}}$ (second term in the denominator of Eq.~\eqref{eq:renyi-2}). Here $\alpha\equiv ke^{-2\SBPS/3}$ and we set $E_0=0$ for simplicity. For small $\alpha$, the disconnected geometry $(\overline{\tr\rho})^2$ dominates. At large  $\alpha$, the connected geometry $\overline{\(\tr\rho\)^2}_{\text{conn.}}$ dominates.}
    \label{fig:cylvsdisk}
\end{figure}

We now have all the necessary ingredients to solve the LMRS puzzle for the semi-quenched R\'enyi entropy. In fact, we mapped the LMRS puzzle to a puzzle for the thermal entropy in pure JT gravity, which was recently addressed in \cite{Antonini_Iliesiu_Rath_Tran_2025}. Let us focus for simplicity on the R\'enyi-$2$ entropy. Schematically, we need to compute
\begin{equation}
\begin{aligned}
    &S^{(2)}_{SQ}=-\log\frac{\overline{\tr\(\rho^2\)}}{\overline{\(\tr\rho\)^2}}\\
    &=-\log \frac{\inlinefig[6]{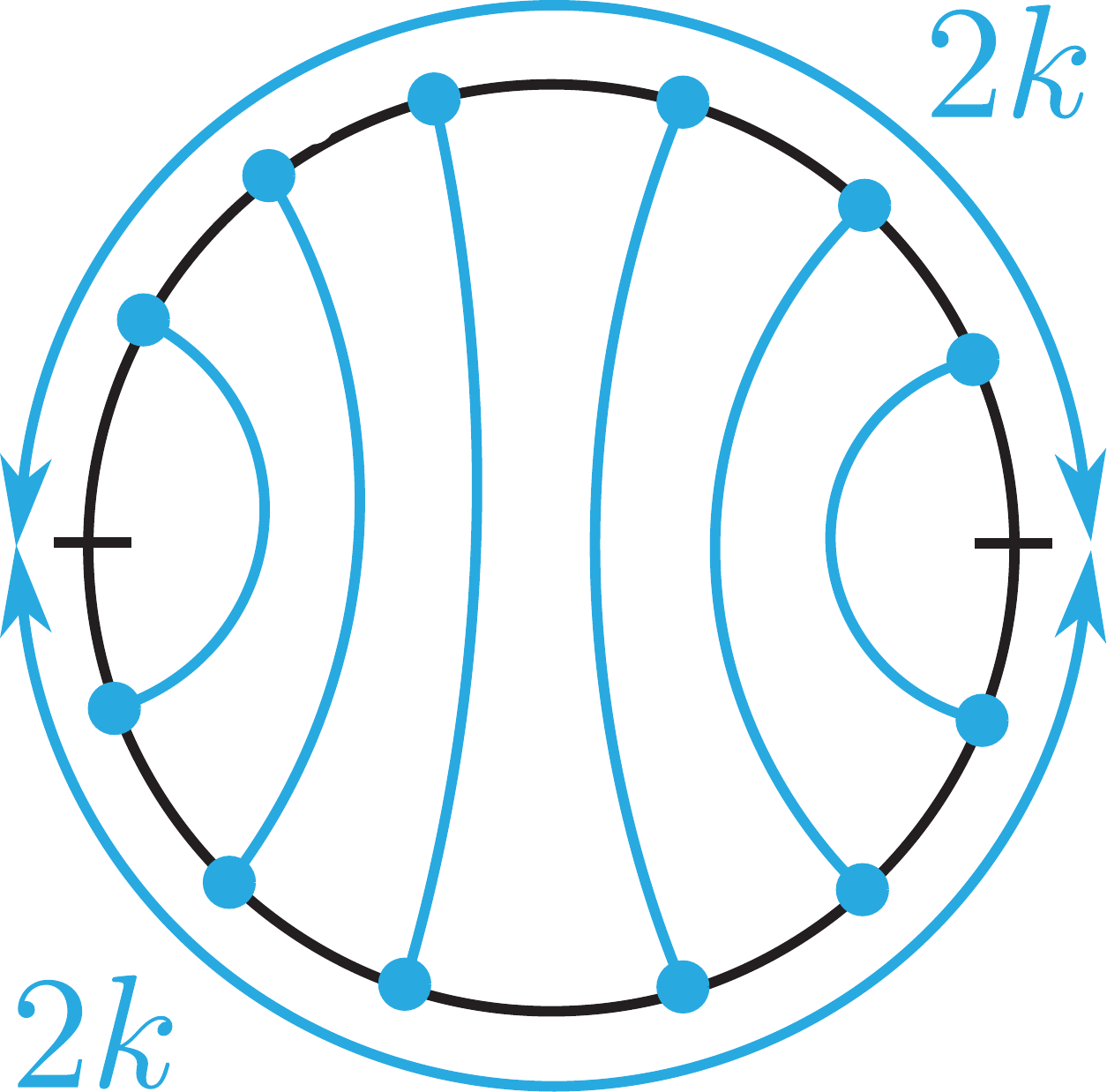}\quad +\quad \dots\quad +\quad \inlinefig[6]{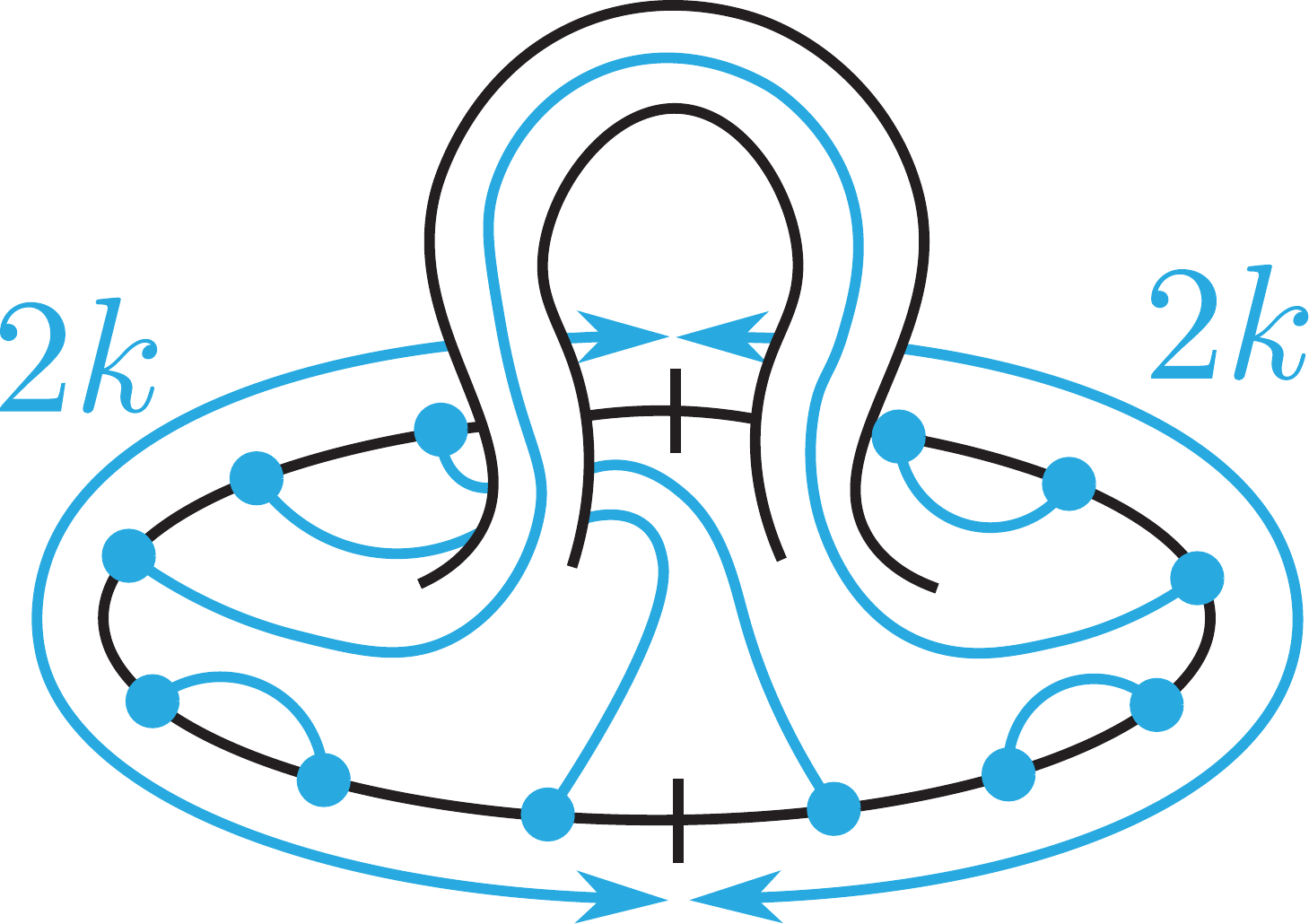}\quad +\quad \dots}{\(\inlinefig[5]{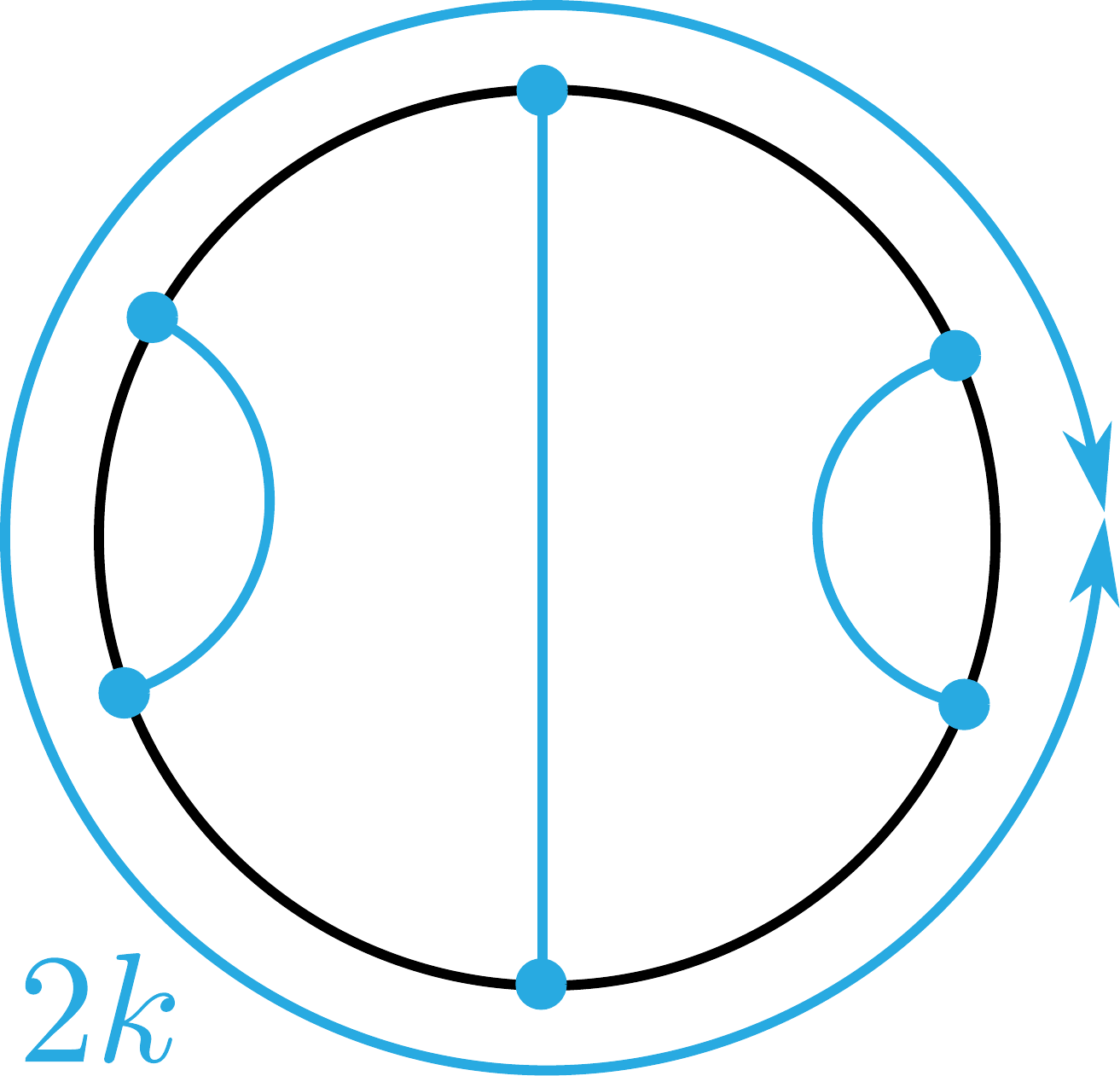}+ \dots + \inlinefig[5]{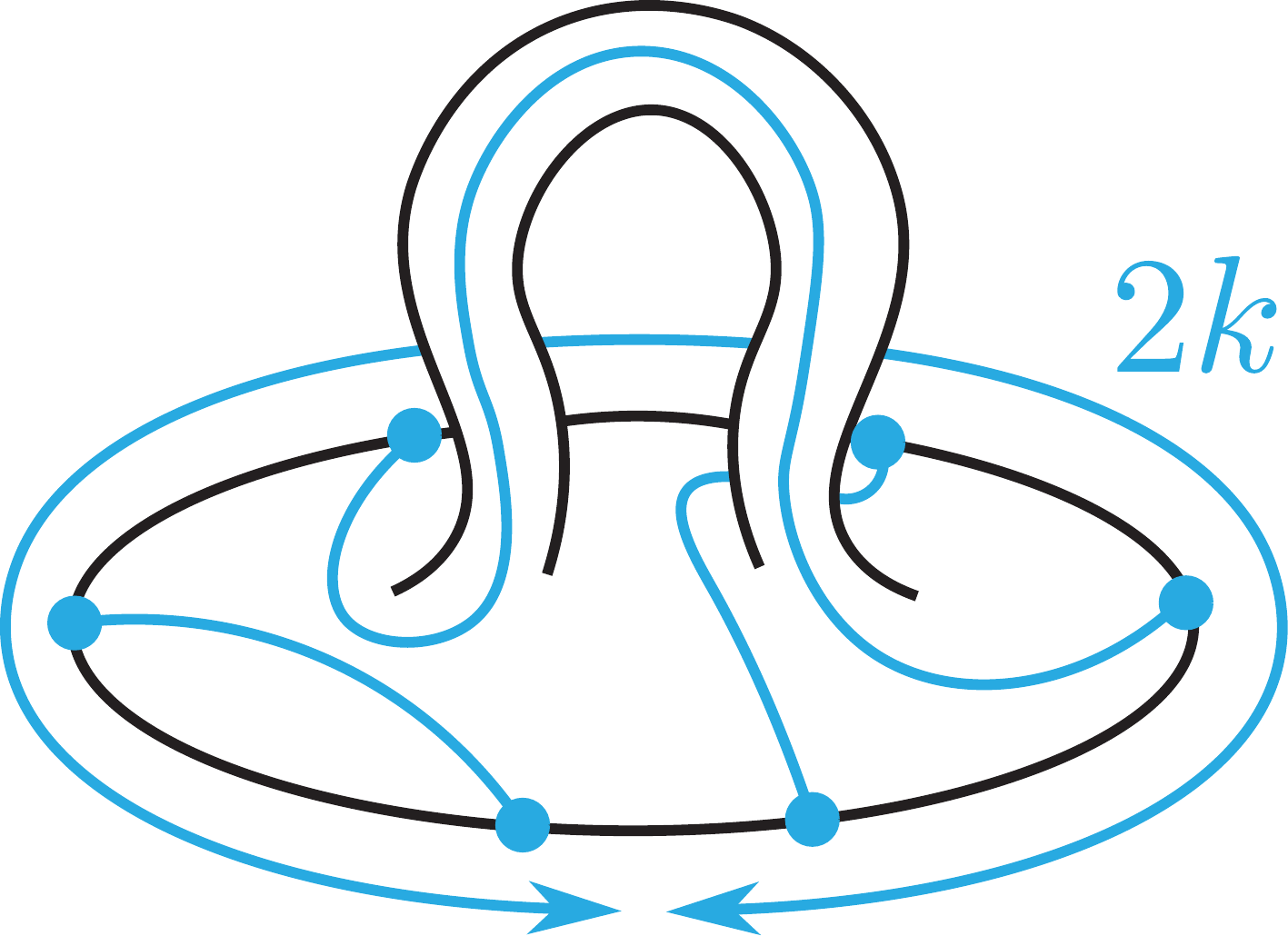}+ \dots\)^2+\(\inlinefig[5]{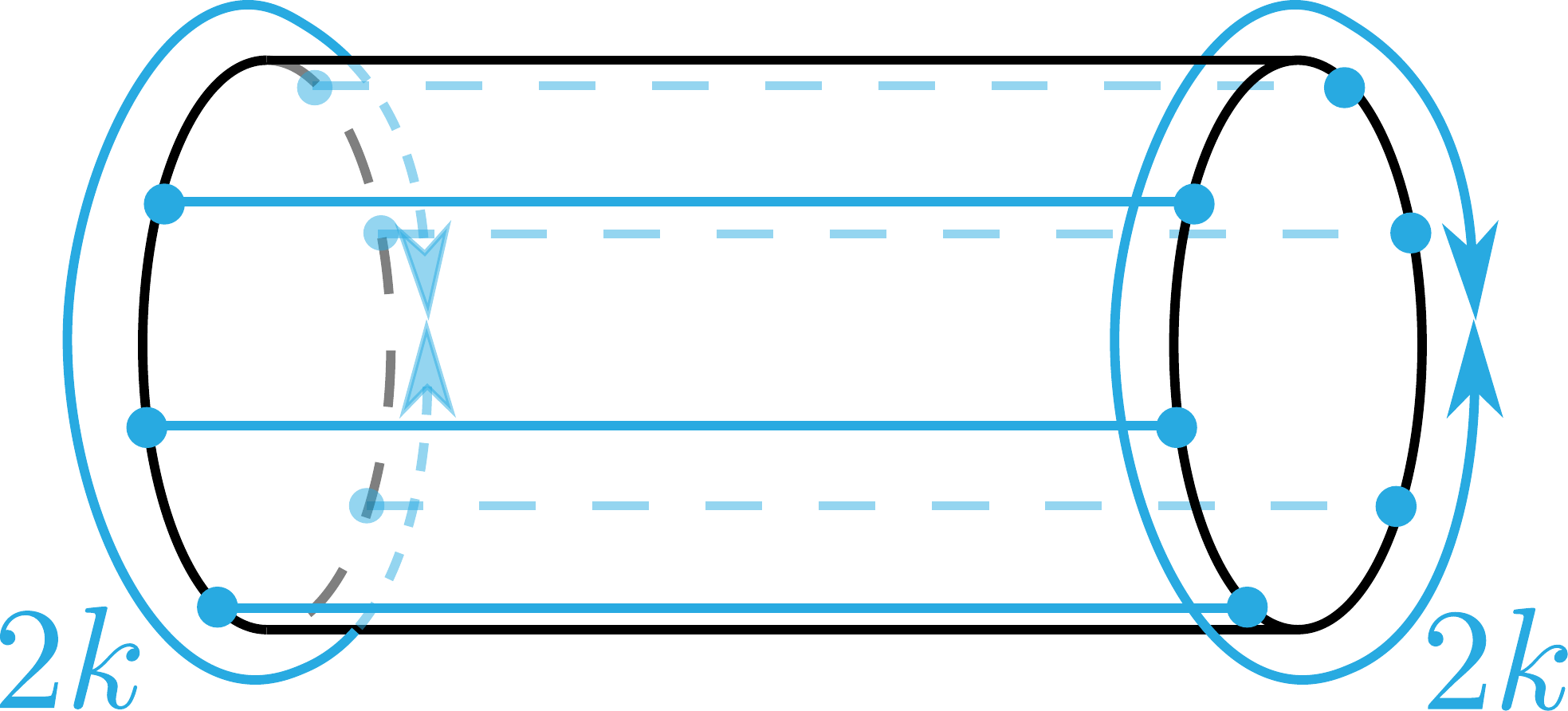}+ \dots + \inlinefig[5]{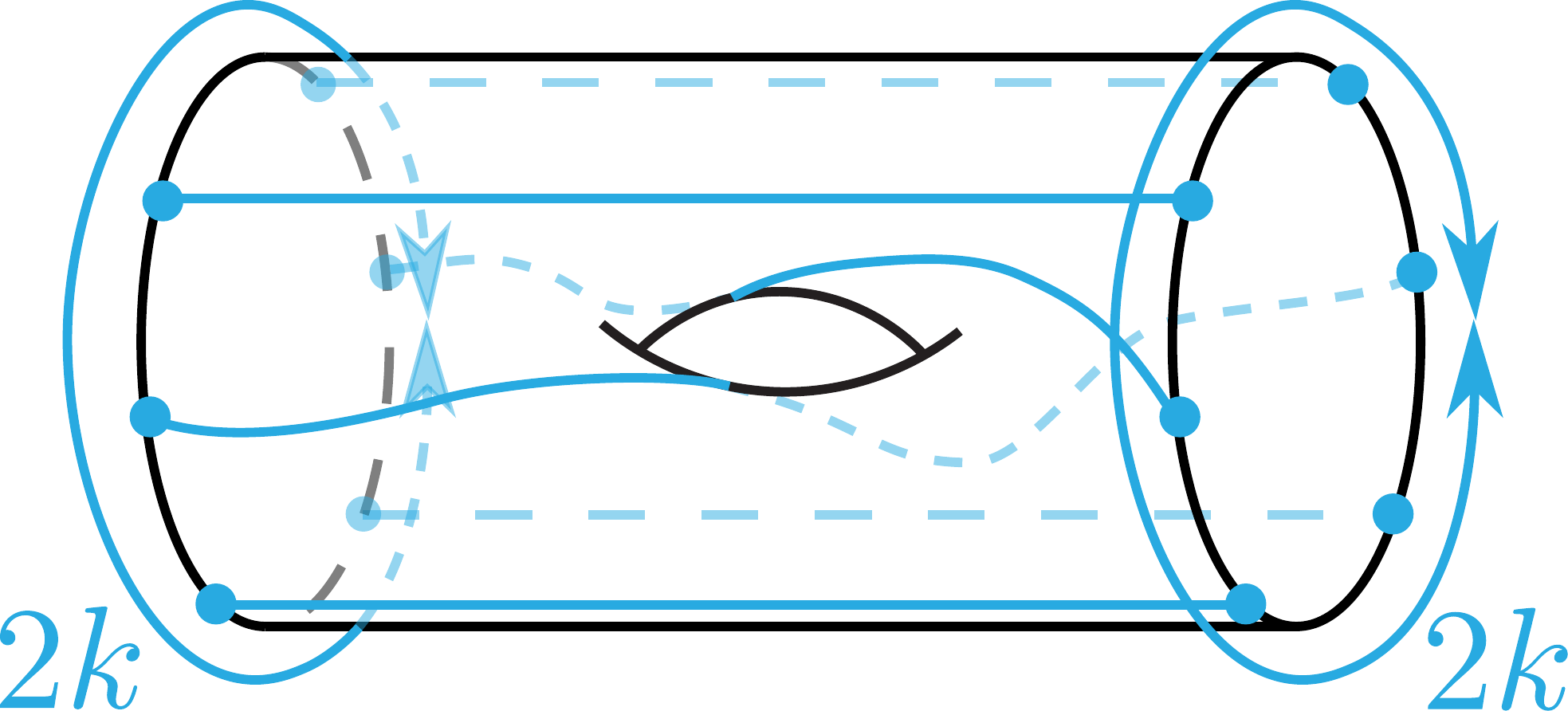}+ \dots\)}\\[10pt]
    &=-\log \frac{2\times \(\inlinefig[6]{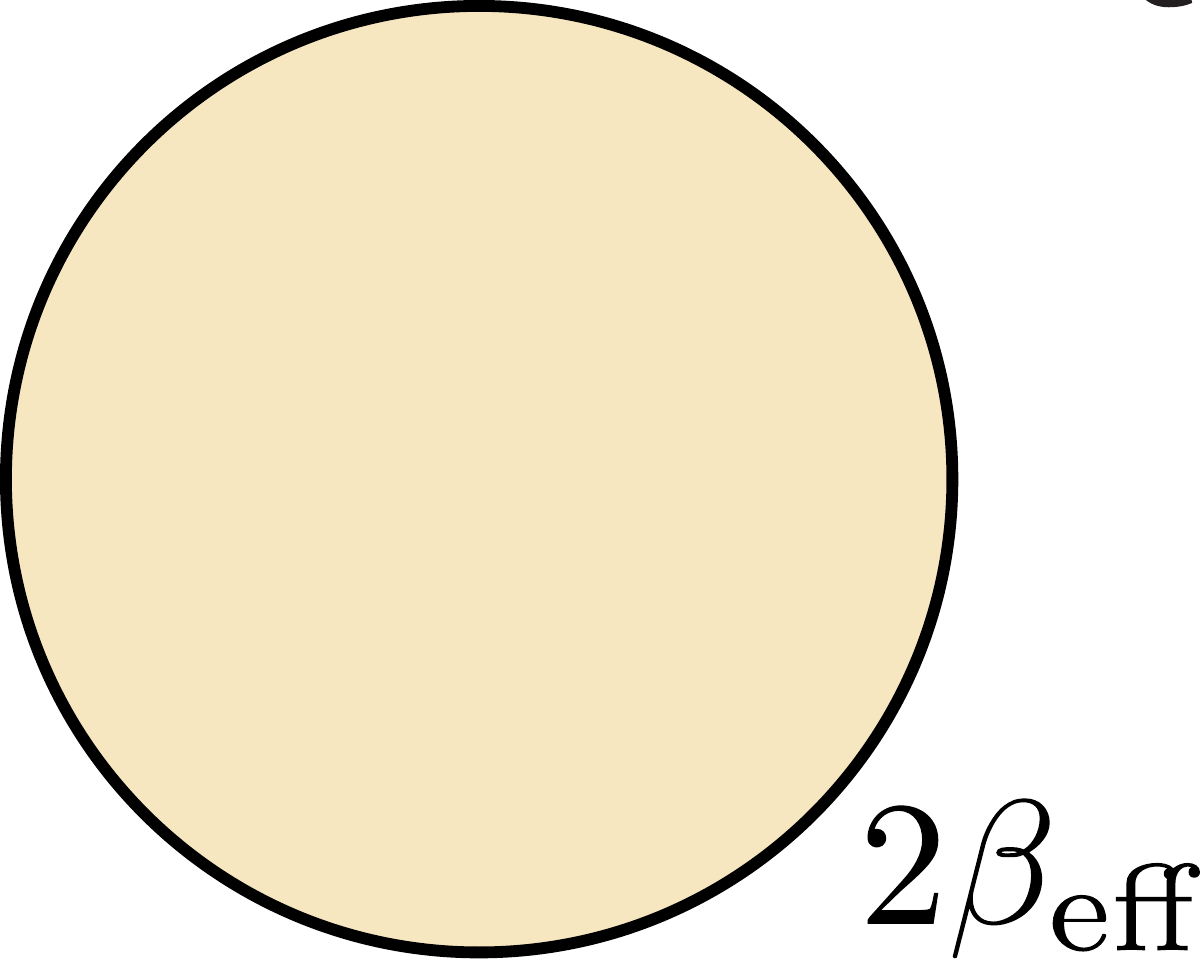}+ \inlinefig[6]{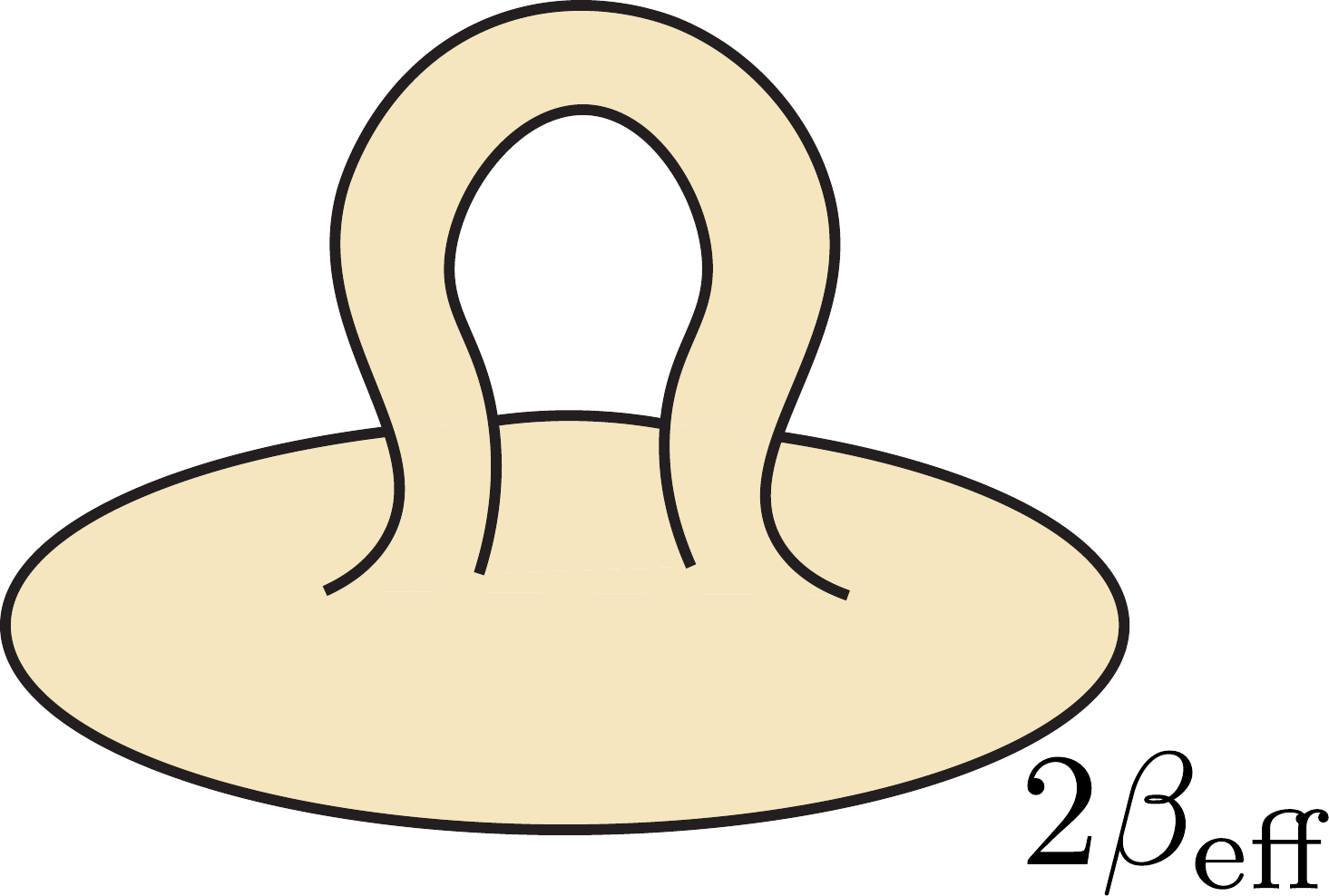}+\dots\)}{4\times\(\inlinefig[5]{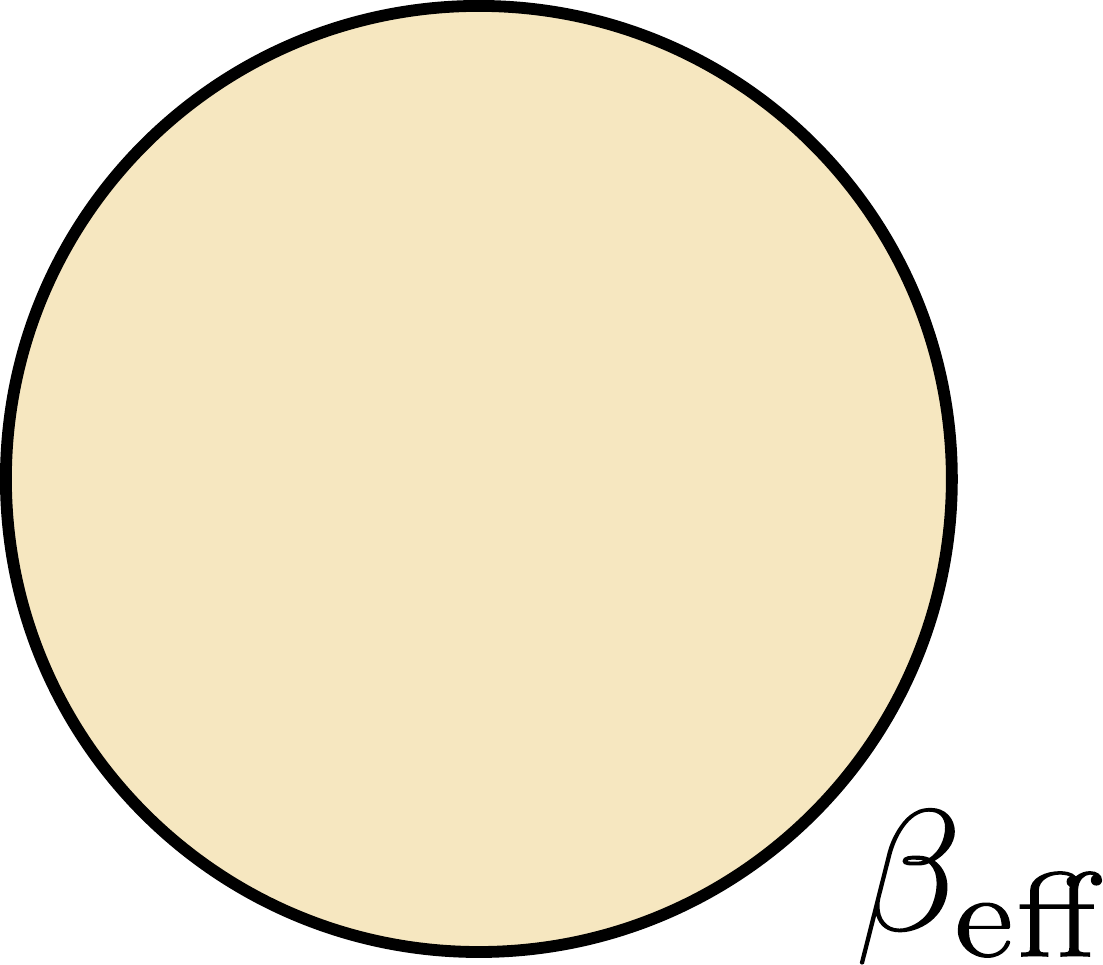}+\inlinefig[5]{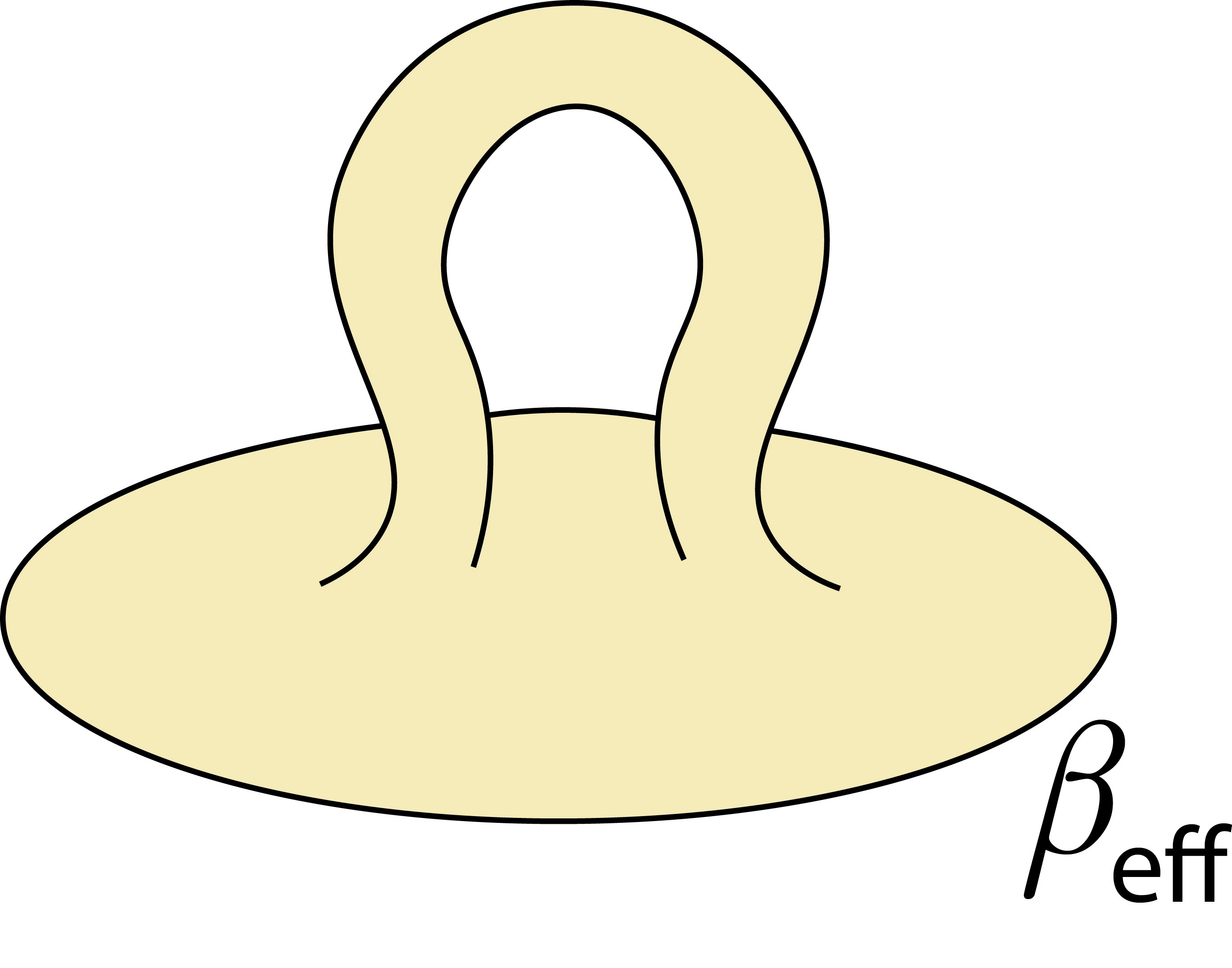}+\cdots\)^2+2\times \(\inlinefig[5]{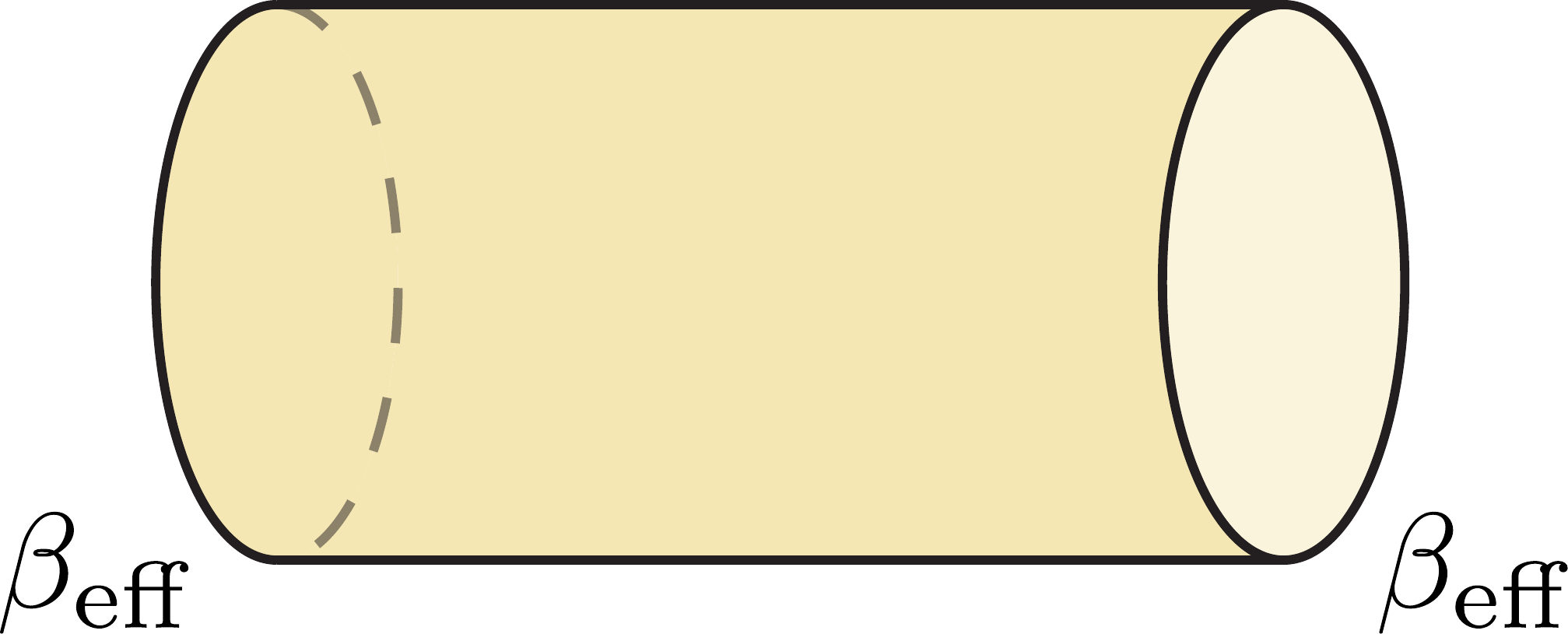}+\inlinefig[5]{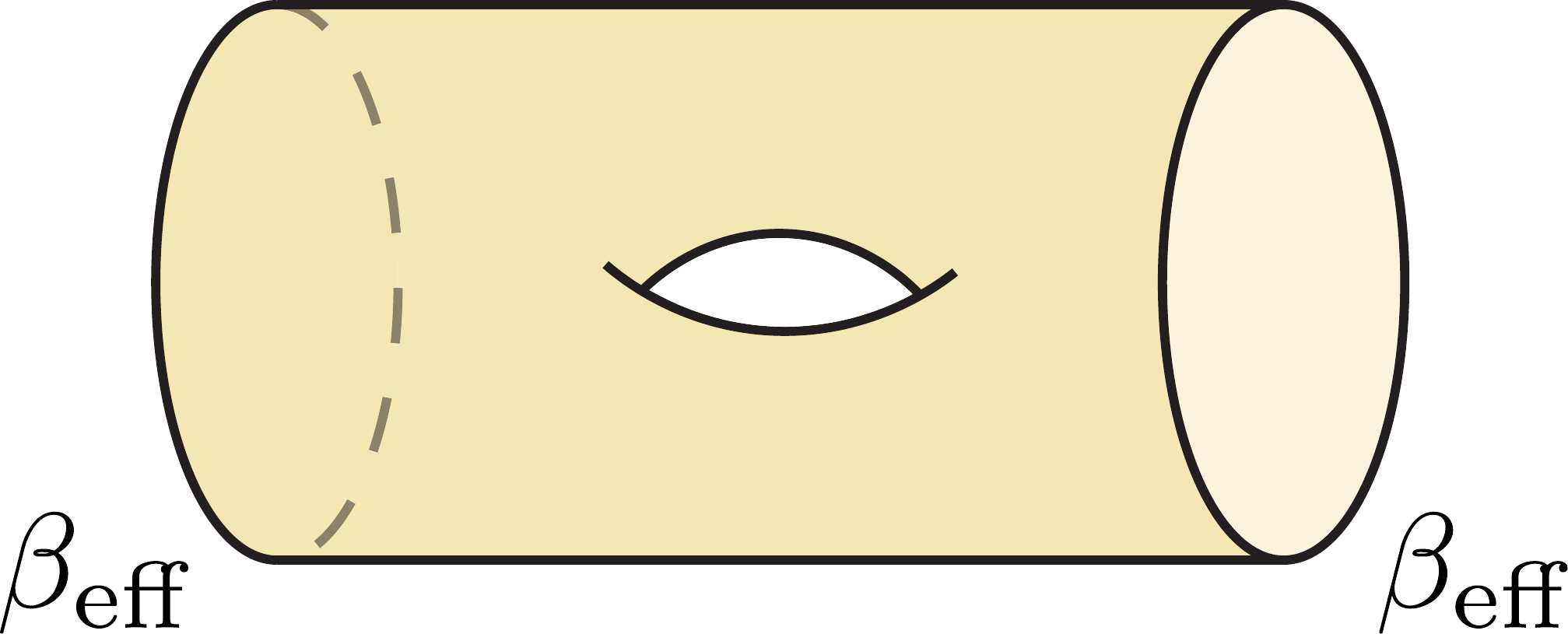}+\cdots \)}
    \end{aligned}
    \label{eq:renyi-2}
\end{equation}
We already explicitly computed in Eq.~\eqref{eq:resummed-disk-matter} (or, equivalently, \eqref{eq:airy-thermal-part}) the genus re-summation of single-boundary contributions. In the effective JT gravity theory in the Airy limit, the genus re-summation of the connected cylinder contribution (second term in the denominator) is also explicitly known and given by \cite{Antonini_Iliesiu_Rath_Tran_2025}
\begin{equation}
\begin{aligned}
    \frac{1}{2}\overline{\(\tr\rho\)^2}_{\text{conn.}}=\overline{ Z^2(\beta_{\text{eff}})}_{\text{conn.}}&=\inlinefig[5]{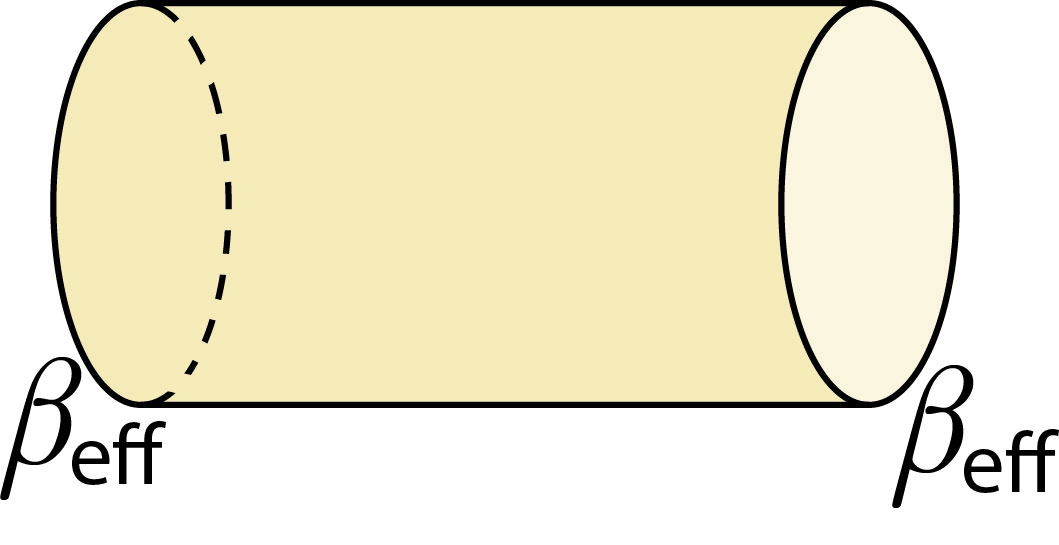}+\inlinefig[5]{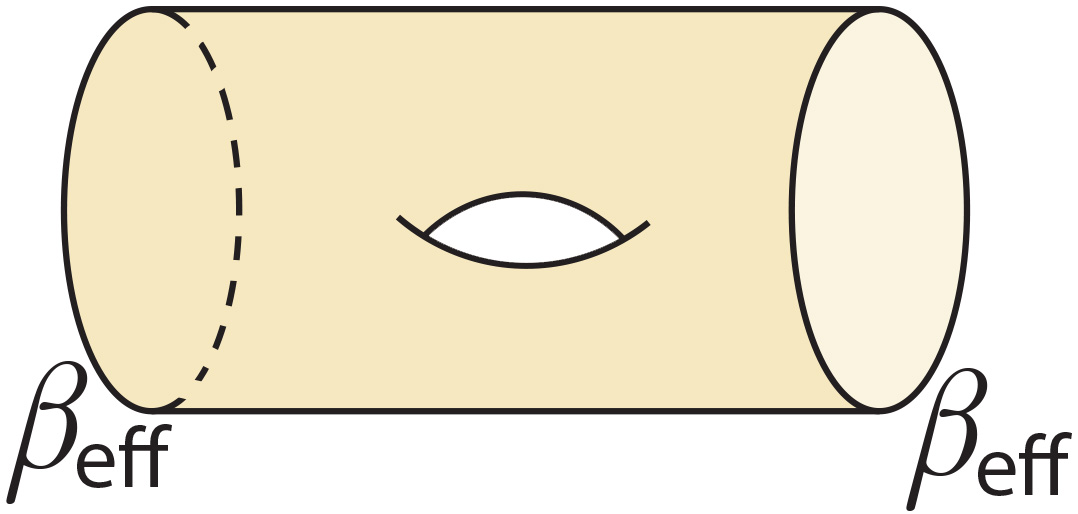}+\cdots\\[10pt]
    &= \frac{e^{-2\beta_{\text{eff}} E_0}e^{2\alpha^3/3}}{4\sqrt{2\pi}\alpha^{3/2}}\text{erf}\(\sqrt{\frac{\alpha^3}{2}}\)\,.
    \end{aligned}
    \label{eq:cyl-resummed}
\end{equation}
Notice that, at large $\alpha$, the re-summed cylinder contribution \eqref{eq:cyl-resummed} dominates in the denominator of Eq.~\eqref{eq:renyi-2} over the product of two re-summed disks \eqref{eq:airy-thermal-part}, signaling that connected wormhole effects become important, see Figure \ref{fig:cylvsdisk}.
Putting everything together, we obtain
\begin{equation}
    S^{(2)}_{SQ}=\log\left[2\sqrt{\frac{2}{\pi}}\frac{e^{-\alpha^3/2}}{ \alpha^{3/2}}+\text{erf}\(\sqrt{\frac{\alpha^{3}}{2}}\)\right]\xrightarrow[]{\alpha\gg 1}\sqrt{\frac{2}{\pi}}\frac{e^{-\alpha^3/2}}{\alpha^{3/2}}\,,
    \label{eq:final-semiquenched}
\end{equation}
which is always positive and decaying as $\alpha$ increases, see Figure \ref{fig:moneyplot}. We remark that the term guaranteeing positivity is the second one in the log in Eq.~\eqref{eq:final-semiquenched}, which arises from the connected re-summed cylinder contribution. This term is clearly absent for the annealed entropy, because the denominator of the log in the R\'enyi-2 entropy is simply given by $\(\overline{ Z(\beta)}\)^2$, which is a product of two re-summed disks. Therefore, we conclude that not only higher genus effects, but also connected wormhole effects are necessary to rescue positivity of the entropy.

How about generic semi-quenched R\'enyi-$n$ entropies with $n>2$? In that case, we do not have closed-form expressions for the connected contributions appearing in the denominator of $S^{(n)}_{SQ}$. However, a general formula for $\overline{ Z^n(\beta)}$ in the Airy limit of JT gravity is known \cite{Engelhardt_Fischetti_Maloney_2021}. Although an explicit expression for $S^{(n)}_{SQ}$ cannot be obtained, using this formula, it is easy to show that $S^{(n)}_{SQ}\xrightarrow[]{\alpha\to\infty} 0$. We refer the reader to Appendix A of \cite{Antonini_Iliesiu_Rath_Tran_2025} for the details of this calculation.

The solution of the puzzle for the semi-quenched entropy we just discussed is almost identical in the case of matter insertions in non-SUSY JT gravity in a microcanonical band.\footnote{This should not to be confused with the effective JT description that we just discussed.} In fact, as we have discussed, in the double-scaling limit of interest we can neglect diagrams with empty handles, and $\tr_{\mathfrak{E}}(\rho^n)$ is simply given by the second line of Eq.~\eqref{eq:1bdycorrelationJT}. We can then again map the problem to the effective pure JT gravity theory without any operator insertions. The only difference is in the identification of the parameters, for which we have $S_0^\text{eff}=S(\mathfrak{E})$, $E_0=-\log(4P_{\Delta\to\infty}(\bar{E})\rho(\bar{E})\delta E)$, and again $\beta_{\text{eff}}=k$. The rest of the solution is completely analogous.

In this section, we were able to solve the puzzle entirely from a bulk perspective, without making use of the matrix integral dual to JT gravity, and we showed that the semi-quenched entropy stays positive and approaches zero. However, this approach only works in the regime $k=O( e^{2\SBPS/3})$. What happens when $k=\omega\(e^{2\SBPS/3}\)$, or equivalently $\beta_{\text{eff}}=\omega\(e^{2S_0^{\text{eff}}/3}\)$ in the effective pure JT gravity theory? In that case, we lose control over the genus expansion and we do not know how to answer the question from a bulk perspective. In order to make progress, we need to use the dual random matrix integral. This will be the subject of Section \ref{sec:resolution}, where we will show that the matrix integral develops new saddle points, the one-eigenvalue and two-eigenvalue instantons \cite{Hernandez-Cuenca_2024,Antonini_Iliesiu_Rath_Tran_2025}, due to the many operator insertions. 

The matrix integral perspective, which we discuss in Section \ref{sec:matrixmodels}, also carries other advantages. First, it allows us to (numerically) compute the quenched entropy, whose bulk calculation is notoriously complicated and an interesting open problem \cite{Engelhardt_Fischetti_Maloney_2021,Hernandez-Cuenca_2024}. Second, it will allow us to go beyond the simple $\Delta\to\infty$ limit of the present section and solve the puzzle, at least at the level of a simple toy-model, for generic values of the scaling dimension $\Delta$ of the matter operator, see Appendices \ref{sec:ETH} and \ref{sec:genericdresolution}.

Finally, we remark that, as we have explained in Section \ref{sec:entropies}, our results for the positivity of the semi-quenched entropy for any value of $k$ imply the existence of an isolated matter ``ground state'' for every member in the ensemble the gravitational path integral is averaging over (which, as we will see, is a GUE matrix ensemble). This, in turn, implies that the quenched entropies, although they cannot be explicitly computed from the bulk side, are guaranteed to also stay positive.

\section{Matrix model description and solution to the puzzle at arbitrary $k$}
\label{sec:matrixmodels}

In recent years, Random Matrix Theory (RMT) has played a central role in our understanding of non-perturbative quantum gravity. Its power is exemplified by the duality between pure JT gravity and a matrix integral \cite{Saad_Shenker_Stanford_2019}. The JT gravity topological expansion can be exactly mapped to the diagrammatic expansion of the dual double-scaled matrix integral, which provides a non-perturbative completion of JT gravity. This duality allows one to explicitly compute observables in the bulk theory that require knowledge of non-perturbative quantum gravity effects, such as the plateau region of the spectral form factor \cite{Saad_Shenker_Stanford_2019}. It also establishes a duality between JT gravity and an \textit{ensemble} of dual theories, in which each matrix drawn from the double-scaled matrix integral is identified with the Hamiltonian of a microscopic theory.

The idea that observables computed using the gravitational path integral (GPI) are averaged over some ensemble has been widely discussed, see e.g. \cite{Saad_Shenker_Stanford_2019,Marolf:2020xie,Belin:2020hea,Penington:2019kki,Almheiri:2019qdq,Saad:2021uzi,Chandra:2022bqq,Chandra:2022fwi,Sasieta:2022ksu,Balasubramanian:2022gmo,deBoer:2023vsm}, and is strongly motivated by the factorization puzzle \cite{Maldacena:2004rf} arising in the context of the AdS/CFT correspondence. Consider two copies of a microscopic, holographic system, which we will denote by CFT${}_L$ and CFT${}_R$. The partition function of these two theories at finite temperature must necessarily factorize into the left and right partition functions: $Z(\beta_L,\beta_R)=Z_L(\beta_L)Z_R(\beta_R)$. However, when computing $Z(\beta_L,\beta_R)$ using the GPI, spacetime wormholes exist that connect the two boundaries where CFT${}_L$ and CFT${}_R$ live, leading to a non-factorizing answer. More generally, when computing inner products using the GPI, spacetime wormhole contributions imply that\footnote{Here the overline again means that the quantity is computed using the gravitational path integral.} $\overline{\langle\psi_i|\psi_j\rangle^n}\neq (\overline{\langle\psi_i|\psi_j\rangle})^n$ \cite{Penington:2019kki,Almheiri:2019qdq,Saad_Shenker_Stanford_2019,Chandra:2022fwi,Sasieta:2022ksu,Balasubramanian:2022gmo}, which is clearly incompatible with ordinary quantum mechanics of a single theory. Notice that both these issues arise not only in two-dimensional gravity, but in any dimension. If the GPI computes averages over an ensemble, both issues are resolved: an average over an ensemble of a factorizing quantity need not factorize; similarly, moments of overlaps capture the statistical properties of the ensemble, and therefore can be seen as providing the size of fluctuations of the inner product in the ensemble.

One potential issue with the ensemble interpretation is that---motivated by well-known explicit AdS/CFT examples in string-theoretic setups, e.g. \cite{Maldacena:1997re,Aharony:1999ti,Aharony:2008ug}---in higher dimensions we expect a single holographic dual theory to exist rather than an ensemble of theories. Why then does the GPI seem to be averaging over an ensemble even in higher dimensions? A plausible answer is that the GPI we typically employ is only able to capture coarse grained quantities in the dual theory \cite{Chandra:2022fwi,Sasieta:2022ksu,Balasubramanian:2022gmo,deBoer:2023vsm,Antonini:2024mci,Antonini:2025ioh}. In fact, the GPI is defined only in terms of low-energy gravitational EFT objects and does not take UV objects (such as strings, branes, etc.) into account. Thus, it should not be able to capture microscopic features of the full quantum gravity theory or, equivalently, of the dual theory. A full quantum gravity path integral including these UV objects should instead be able to compute all microscopic quantities in the single dual theory. Interestingly, this perspective can also be adopted for JT gravity. For example, \cite{Mukhametzhanov:2021hdi,Blommaert:2021fob,Saad:2021rcu, Blommaert:2022ucs} showed that a modified GPI for JT gravity with the inclusion of non-perturbative objects computes quantities consistent with the existence of an underlying single theory, e.g., a factorized two-boundary partition function. In other words, the path integral of \cite{Mukhametzhanov:2021hdi,Blommaert:2021fob,Saad:2021rcu} captures the dynamics of a single realization of the JT ensemble discussed above, which can therefore also be interpreted as an ``ignorance ensemble''. 

This interpretation is validated by the fact that large-$N$ holographic theories are believed to typically be chaotic. In particular, their high-energy spectrum is dense and satisfies some generic features, such as eigenvalue repulsion, which are well captured by random matrix theory. Therefore, computing observables after coarse-graining over microscopic data (e.g., over some small energy window) is equivalent to considering observables averaged over a random matrix ensemble \cite{Chandra:2022bqq,Chandra:2022fwi,Sasieta:2022ksu,Balasubramanian:2022gmo,deBoer:2023vsm}. In other words, in this case, the ensemble the GPI averages over arises from our ignorance of microscopic data of a single theory rather than a true ensemble of theories. Consequently, RMT is relevant for the description not only of JT gravity, but also of higher-dimensional gravity. In this setting, because the true microscopic theory can be seen as a typical draw from the ensemble, only self-averaging quantities give meaningful information about the dual theory, and therefore the full, non-perturbative quantum gravity theory.

So far, we have been discussing only the spectral properties of pure gravitational theories, but RMT can also account for the presence of matter coupled to gravity. In fact, for chaotic theories, the Eigenstate Thermalization Hypothesis (ETH) posits that, in the high-energy sector, matrix elements of light operators in the energy eigenbasis behave erratically, and can be modeled by random variables \cite{Srednicki:1994mfb, deutsch1991quantum}:
\begin{equation}
    \langle E_i|O|E_j\rangle = f_O(\bar{E})\delta_{ij}+g(\bar{E},\delta E)R_{ij}
\end{equation}
where $E_i,E_j=O(N^2)$ are large energy eigenvalues in a small energy band, $O$ is a light operator, $\bar{E}$ and $\delta E$ are the average energy and width of the band, $f_O(\bar{E})$ is the microcanonical expectation value of $O$, $g(\bar{E},\delta E)$ the square root of the microcanonical two-point function, and $R_{ij}$ is a random variable. In the original ETH formulation, $R_{ij}$ is simply taken to be Gaussian-distributed, but this ansatz can (and must) be refined to include non-Gaussianities \cite{Jafferis_Kolchmeyer_Mukhametzhanov_Sonner_2023,Jafferis:2022uhu}, as we will discuss in Appendix \ref{sec:ETH}. For CFTs, ETH is equivalent to the statement that light-heavy-heavy OPE coefficients behave erratically and can be approximated by random variables drawn from an appropriate ensemble \cite{Belin:2020hea}. ETH suggests that matter operators can also be modeled by random matrices \cite{Jafferis_Kolchmeyer_Mukhametzhanov_Sonner_2023}. One can then expect gravitational theories to be generically described (with the caveats explained above) by multi-matrix integrals, with one matrix modeling the Hamiltonian, and one matrix for each matter operator. The specific matrix potential will depend on the theory and can include interactions between the various sectors. 

Now that we have established a bird's-eye view of the general relationship between gravity and RMT, we can focus our attention on JT gravity, keeping in mind that it should be possible to generalize the lessons we learn here to higher dimensions and more generic gravitational theories. In particular, in Section \ref{sec:JTRMT} we will review the features of matrix integrals relevant to our work. In Section \ref{sec:GUE}, we will discuss how, for the BPS case and for the non-SUSY JT case within a microcanonical window, matter correlators in gravity for the operators $\tilde{O}_\Delta$ at large $\Delta$ correspond to correlators in the GUE matrix integral. In Section \ref{sec:resolution}, we use this correspondence to explain the equality, for $k=O(e^{2\SBPS/3})$, between matter correlators and pure JT partition functions encountered in Section \ref{sec:mapping}. We then extend our results to the $k=\omega(e^{2\SBPS/3})$ regime by finding new saddle points dominating the matrix integral, the one-eigenvalue instanton, and the two-eigenvalue instanton. We discuss the difficulties in the generalization of our results to generic values of $\Delta$ and propose a simplified toy model to describe matter operators at any $\Delta$ in Appendix \ref{sec:ETH}, and resolve the LMRS puzzle in this toy model in Appendix \ref{sec:genericdresolution}.

\subsection{Random matrix theory}
\label{sec:JTRMT}
In this subsection, we give a schematic review of the features of random matrix ensembles relevant for our discussion. We mostly adopt the notation of \cite{Hernandez-Cuenca_2024}, to which we refer the reader for a more complete review of random matrix theory (see also \cite{Anninos:2020ccj}).

In this paper, we will be interested in random matrix ensembles with a single-trace matrix potential. Their partition function is given by
\begin{equation}
    Z=\int dM\,e^{-N \tr [V(M)]},
\end{equation}
where $dM$ is a Lebesgue measure on each entry of the matrix $M$, $N$ is the size of the matrix, and $V(M)$ is a potential that defines the matrix model. It is useful to rewrite the matrix integrals in terms of an integral over eigenvalues of $M$
\begin{equation}
    Z=\int d\Lambda\,e^{-N\bar{I}(\Lambda)},
    \label{eq:matrixint}
\end{equation}
where $d\Lambda=\prod_id\lambda_i$, $\Lambda$ is the diagonal eigenvalue matrix, and we absorbed the Jacobian determinant (also known as Vandermonde determinant) into the definition of the matrix integral action $\bar{I}(\Lambda)$. In this paper we will focus on Wigner-Dyson ensembles \cite{Wigner1993,dyson1,dyson2}, for which the action reads
\begin{equation}
    \bar{I}(\Lambda)=\sum_{i=1}^N V(\lambda_i)-\frac{\upbeta}{2N}\sum_{\substack{i,j=1\\i\neq j}}^N\log|\lambda_i-\lambda_j|
    \label{eq:actiondiscrete}
\end{equation}
where $\upbeta=1,2,4$ correspond to three different choices of ensemble---respectively orthogonal, unitary, and symplectic---which are invariant under the action of the three corresponding Lie groups. The second term in Eq.~\eqref{eq:actiondiscrete}, comes from the Vandermonde determinant and is responsible for the eigenvalue repulsion mentioned above. We can also introduce the effective potential $\bar{V}_{\text{eff}}$ felt by a single eigenvalue $\lambda_k$ by only keeping the terms in the action that depend on $\lambda_k$:
\begin{equation}
    \bar{V}_{\text{eff}}(\lambda_k)=V(\lambda_k)-\frac{\upbeta}{N}\sum_{\substack{i=1\\ i\neq k}}^N \log|\lambda_i-\lambda_k|.
    \label{eq:effdiscrete}
\end{equation}

At large $N$, the matrix integral \eqref{eq:matrixint} can be evaluated in the saddle point approximation and is dominated by a set of eigenvalues that minimizes the effective action \eqref{eq:actiondiscrete}. The saddle point equations are thus a set of $N$ coupled equations, given by the minimization condition for the $N$ effective potentials. In order to study the saddle point equations at large $N$, it is typically valid to approximate the discrete set of eigenvalues by a continuum. To do this, we can replace the sum over eigenvalues by an integral over a continuum variable weighted by an eigenvalue density $\sigma$:
\begin{equation}
    \sum_{i=1}^N\longrightarrow N\int dx \,\sigma(x).
\end{equation}
The (real and positive) eigenvalue density is normalized as $\int dx \,\sigma(x)=1$. All integrals here are taken over the support of $\sigma(x)$. The integral over the $N$ discrete eigenvalues then turns into a functional integral over the density of eigenvalues. The partition function in the continuum limit takes the form
\begin{equation}
    Z=N\int \mathcal{D}\sigma\, e^{-N I[\sigma]},
\end{equation}
where $I[\sigma]$ is the continuum limit of the action \eqref{eq:actiondiscrete}:
\begin{equation}
    I[\sigma]=N\int dx\,\sigma(x) V(x) - \frac{\upbeta N}{2} \fint dx\,dy\,\sigma(x)\sigma(y)\log|x-y|.
\end{equation}
The principal value integral in the Vandermonde term captures the fact that the discrete sum is over eigenvalues that are different from each other. We can also define the continuum limit of the effective potential \eqref{eq:effdiscrete},
\begin{equation}
    V_{\text{eff}}[\sigma,x]=V(x)-\upbeta\fint dy\,\sigma(y)\log|x-y|.
    \label{eq:conteff}
\end{equation}
The saddle point of the matrix integral, which in the discrete case was given by a set of eigenvalues, is now given by a specific functional form $\bar{\sigma}(x)$ for the eigenvalue density. This can be found by extremizing the effective potential \eqref{eq:conteff} with respect to $x$ and solving the resulting saddle point equation
\begin{equation}
    V'(x)=\upbeta \fint dy\frac{\bar{\sigma}(y)}{|x-y|}, \quad  x\in \textrm{supp}(\bar{\sigma}).
    \label{eq:extremization}
\end{equation}
The general solution of this integral equation is known explicitly for Wigner-Dyson ensembles and can be found in \cite{Hernandez-Cuenca_2024}. In this paper, we focus our attention on some known solutions for specific forms of the matrix potential $V(x)$. We remark that, given a matrix potential, the form of the equilibrium eigenvalue density $\bar{\sigma}$ is uniquely determined by Eq.~\eqref{eq:extremization}. On the other hand, if the leading eigenvalue density is known, Eq.~\eqref{eq:extremization} determines the matrix potential $V(x)$ (up to a constant shift).

Finally, we introduce here one more quantity that will be useful in our analysis, and that is fully determined by knowledge of the effective potential and the leading eigenvalue density, namely the spectral curve
\begin{equation}
    y(z)\equiv -\frac{\partial_z V_{\text{eff}}[\bar{\sigma},z]}{\upbeta}=-i\pi \bar{\sigma}_c(z), \quad z\in \mathbb{C}\setminus \textrm{supp}(\bar{\sigma}).\label{eq:spectral}
\end{equation}
A proof of the second equality can be found in \cite{Hernandez-Cuenca_2024}. Notice that if we were to evaluate the spectral curve on the support of $\bar{\sigma}$, it would vanish by the definition of $\bar{\sigma}$. However, it is non-trivial off the support, and it will play a central role in the study of one-eigenvalue instanton and two-eigenvalue instanton saddles of the matrix integral in Sections \ref{sec:oneeigen} and \ref{sec:twoeigen}.

\subsubsection*{Scaling limits and edge statistics}

With our definition of the matrix potential $V(\lambda)$, where we removed a factor of $N$, eigenvalues are $O(1)$ and do not scale with $N$ as we take a large $N$ limit. In the continuum limit, this typically corresponds to the eigenvalue density being compactly supported over an interval of $O(1)$ width. A very simple example, which turns out to be important for our setup as we will explain in Section \ref{sec:GUE}, is that of the Gaussian Unitary Ensemble (GUE), in which the potential takes a simple form $V(\lambda)=\lambda^2/(2\xi^2)$, where $\xi$ is the standard deviation of the Gaussian potential. The corresponding leading eigenvalue density is given by the Wigner semicircle:
\begin{equation}
    \bar{\sigma}_{\text{GUE}}(x)=\frac{1}{2\pi\xi^2}\sqrt{4\xi^2-x^2}.
    \label{eq:semicircle}
\end{equation}
This example presents a number of properties that apply to generic Wigner-Dyson ensembles. First, it has what is known as a ``soft edge'', namely, the eigenvalue density vanishes at the edge. In particular, it vanishes as a square root. This feature turns out to be universal for Wigner-Dyson ensembles. Second, because $\bar{\sigma}$ is supported on the interval $[-2\xi,2\xi]$ independent of the value of $N$, this implies that the typical spacing between and fluctuation of eigenvalues is $O(1/N)$. One important fact is that this scaling is correct for eigenvalues in the bulk of the distribution, but not for eigenvalues near the square root edge, for which the correct scaling is given by $1/N^{\frac{2}{3}}$ \cite{Tracy_Widom_1993}. In both cases, the spacing and the fluctuations vanish as we take the large $N$ limit. 

If we are interested in studying fluctuations, we must therefore rescale the variable $x$ with an appropriate power of $N$ before taking the large $N$ limit. This is called a scaling limit of the matrix ensemble, and it has the effect of zooming into a specific part of the spectrum. In this paper, we will mostly be interested in the eigenvalue statistics near the square root edge. Therefore, we must first shift the variable $x$ to center on the edge of interest at $x=\pm 2\xi$, and then rescale $x\to x\cdot \xi/N^{\frac{2}{3}}$ before taking the large $N$ limit. 
For all Wigner-Dyson ensembles, this yields the Airy density of eigenvalues, which describes the eigenvalue density near the edge
\begin{equation}
    \sigma_{\textrm{Airy}}(x)=\textrm{Ai}'\left(-x\right)^2+x\textrm{Ai}\left(-x\right)^2.
    \label{eq:airy}
\end{equation}
The Airy distribution captures the square root behavior near the edge as well as fluctuations around it, with eigenvalues having a non-vanishing (but small) probability of being beyond the edge identified by the saddle point distribution.
The Airy limit introduced here will play a central role in our discussion in Section \ref{sec:resolution}.

\subsection{A GUE operator matrix integral for large $\Delta$}
\label{sec:GUE}

Given that we now have all the ingredients to compute matter correlators in JT (super)gravity for the matter operators $\tilde{O}_{\Delta}$ in the large $\Delta$ limit as well as correlation functions in matrix integrals, we can establish a precise dictionary between the two.\footnote{In Appendix \ref{sec:ETH} we comment about a more generic double-matrix model approach to JT gravity coupled to matter initially introduced for the non-SUSY case in \cite{Jafferis_Kolchmeyer_Mukhametzhanov_Sonner_2023}.} We will once again begin with the BPS case and comment at the end on the extension of our results to the non-SUSY case. In the regime of large $\Delta$, \cite{Lin_Maldacena_Rozenberg_Shan_2023} realized that at the disk level the $2k$-point functions of the projected operators $\tilde O_\Delta$ are computed by the leading order term in the large-$N$ expansion of $\langle \tr\, M^{2k} \rangle$ in a GUE Gaussian matrix ensemble. This is because each disk diagram corresponding to a given Wick contraction among the $2k$ operators in gravity has an associated diagram for the same Wick contraction that appears at leading order in the `t Hooft expansion of $\langle \tr\, M^{2k} \rangle$. For example, for the four-point function, one diagram contributing at leading order on either side is:
\be 
\inlinefig[10]{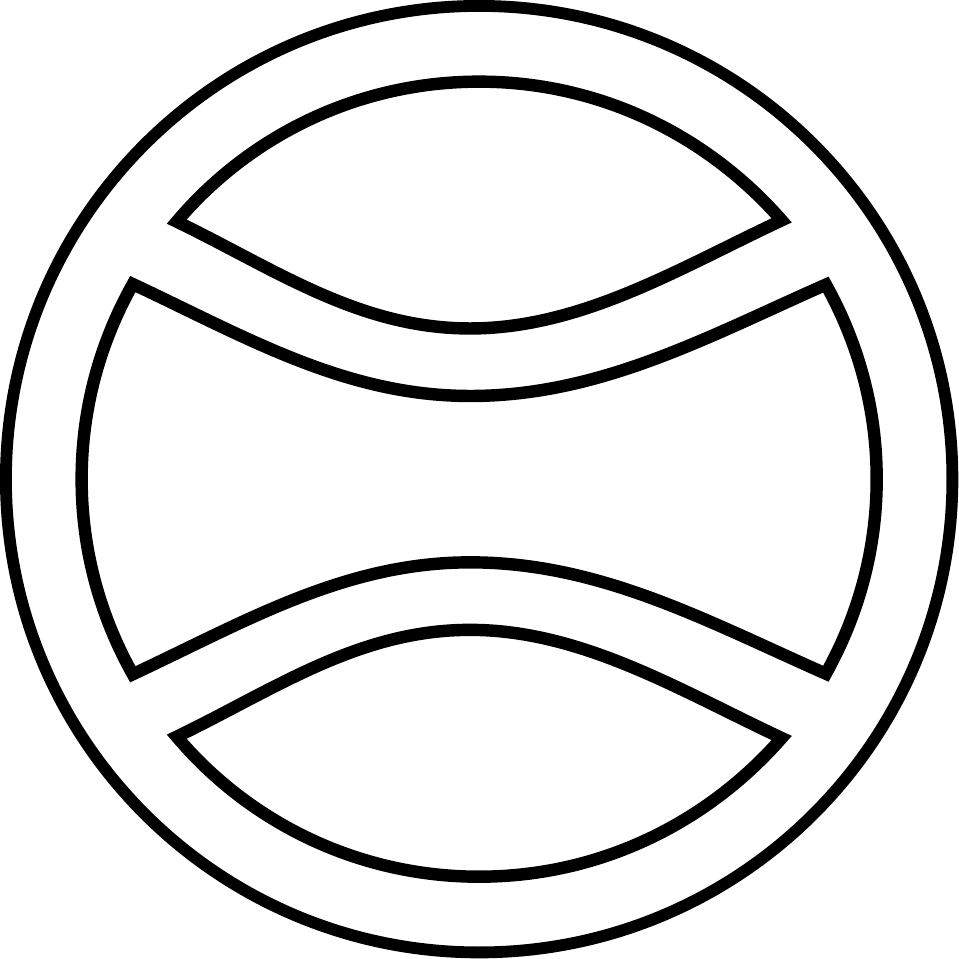}\quad\quad\longleftrightarrow\quad\quad\inlinefig[10]{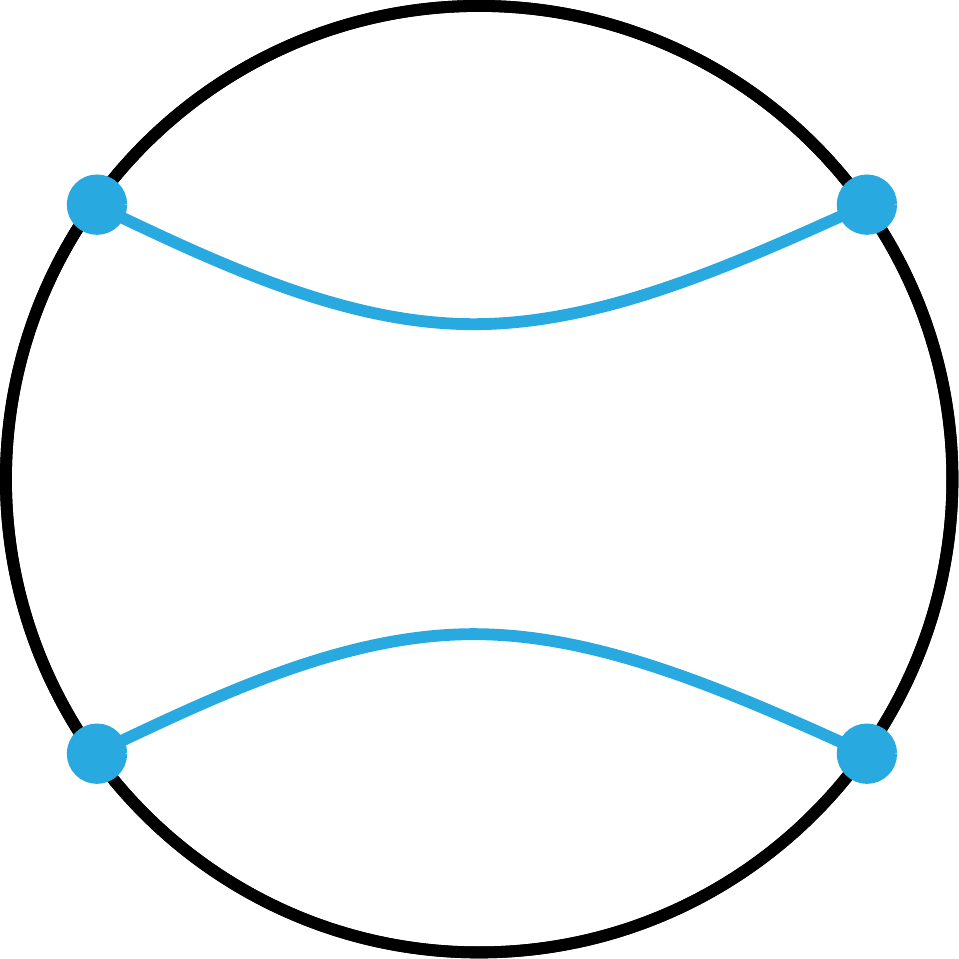}
\ee
As described in Section \ref{sec:bulk-resolution} in gravity, disk diagrams that include intersections are exponentially suppressed in $\Delta$. Correspondingly, in the matrix integral, intersections between the propagating legs are not allowed because of the Gaussian potential. Thus, when summing over all non-intersecting diagrams, we indeed find that at leading order 
\be 
\overline{ \tr \,\tilde O_{\Delta}^{2k}}_{g=0} =\<\tr M^{2k}\>_{g=0} = e^{\SBPS} C_{k} (P_{\Delta\to \infty})^{k}\,,
\label{eq:match-genus-0}
\ee
where $C_{k}$ are again the Catalan numbers, which simply count the overall number of diagrams that appear in the gravity or matrix integral descriptions. To achieve the match, we have chosen,
\be 
N = e^{\SBPS}\,, \qquad V(M) = \frac{1}{2 P_{\Delta\to \infty} }  \tr M^2\,.
\ee 
Notice that the standard deviation here is $\xi=\sqrt{P_{\Delta\to\infty}}$, and the edges of the eigenvalue distribution are at $\lambda=\pm 2\sqrt{P_{\Delta\to\infty}}\equiv\pm \lambda_{e}$. The density of eigenvalues is
\begin{equation}
    \bar{\sigma}_{\Delta\to\infty}(x)=\frac{1}{2\pi P_{\Delta\to\infty}}\sqrt{4P_{\Delta\to\infty}-x^2}\,.
    \label{eq:GUEedge}
\end{equation}
The $g=0$ subscript in Eq.~\eqref{eq:match-genus-0} indicates that on the gravity side, we have only accounted for the disk contribution, while in the matrix integral side, we have only accounted for `t Hooft diagrams that can be embedded on a genus-0 geometry (the disk).

We now explain why this equality continues to hold order-by-order, diagram-by-diagram, to all orders in the genus expansion. As explained in Section \ref{sec:bulk-resolution}, on the gravity side, the only configurations that contribute are minimal embedding diagrams for each Wick contraction, i.e., those for which the matter worldlines do not intersect and that result in surfaces with disk topology after cutting the spacetime along these matter worldlines. In the matrix integral, the non-planar `t Hooft diagrams that contribute to $\< \tr M^{2k}\>$ can be embedded on a genus $g$ surface such that the propagating legs in the `t Hooft diagrams do not intersect. Just as in gravity, after cutting the genus $g$ surface along these propagating legs, we again only obtain a set of disconnected surfaces, each with $g=0$. Thus, for each Wick contraction between the $2k$-operators, there is a direct map between the minimal embedding diagrams encountered in gravity and the diagrams encountered in the matrix Gaussian integral. For instance, at genus $1$, the only contribution to the 4-point function on the gravity and matrix integral sides is\footnote{As we have explained in Section \ref{sec:bulk-resolution}, in gravity there are no further higher genus contributions to the four-point function, because all diagrams except minimal embedding diagrams vanish; the same applies to the matrix integral.}
\begin{equation}
    \inlinefig[10]{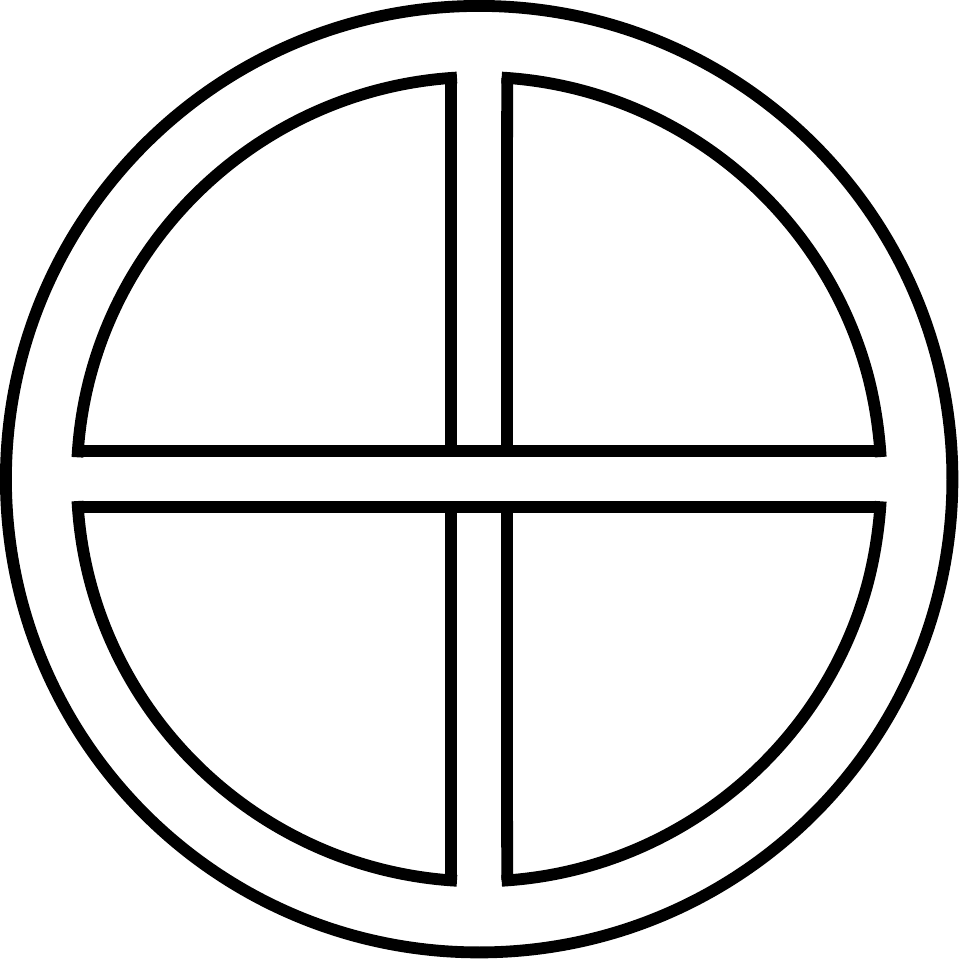}\quad\quad\longleftrightarrow\quad\quad\inlinefig[10]{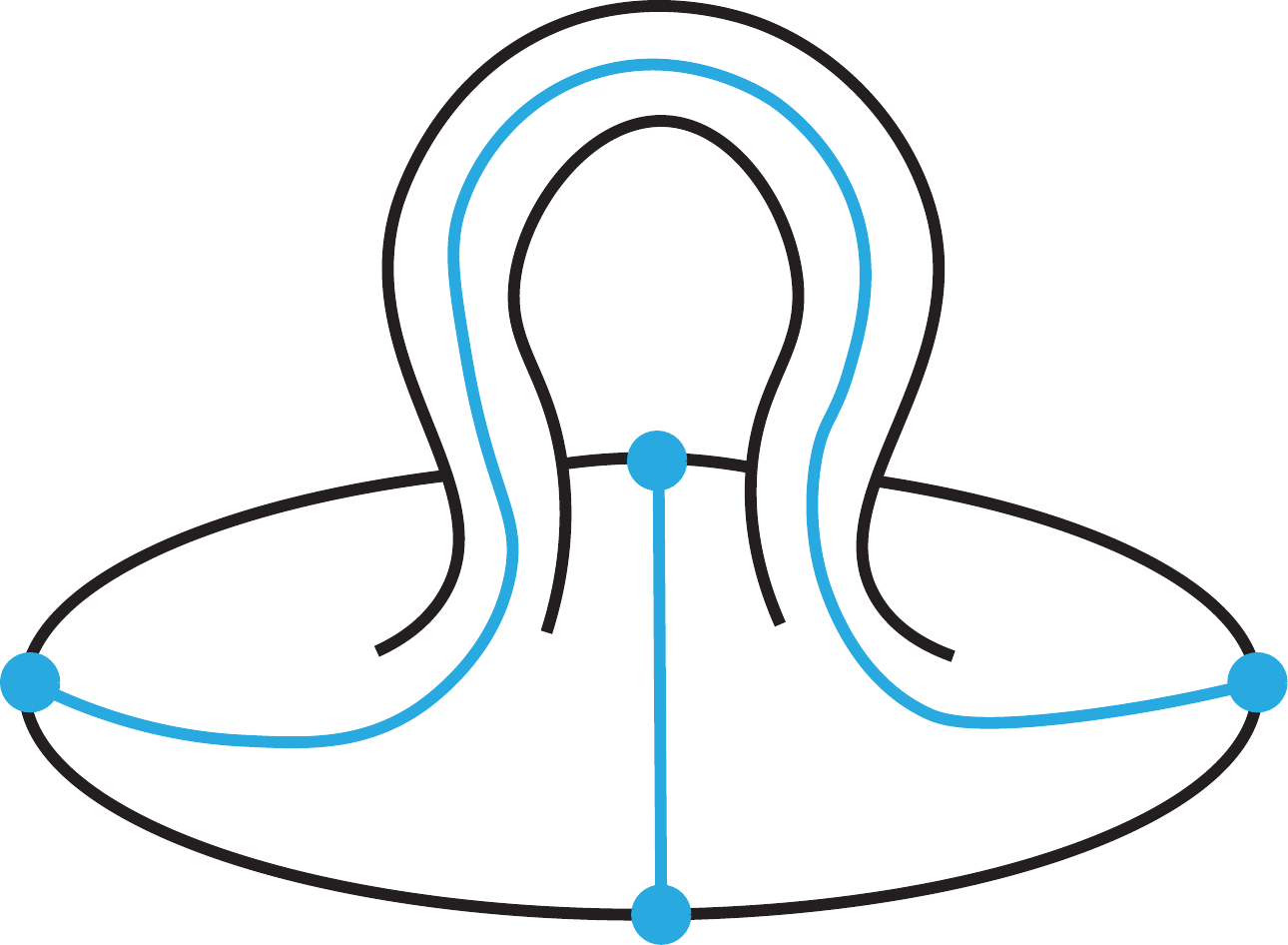}
\end{equation}

In gravity, all genus-$g$ diagrams  have the same contribution that is suppressed by $e^{\SBPS \chi_{g,1}} = N^{ \chi_{g,1}}$. In the matrix integral,  non-planar diagrams (that have overlapping propagating legs) are suppressed by $N^{\chi_{g,n}}$ where $\chi_{g,n}$ is the Euler characteristic of the surface with minimal genus $g$ and $n$ boundaries on which the diagram can be embedded. Thus, since $e^{\SBPS} = N$, the suppression on the two sides is the same. The count over all such diagrams is given by the same counting problem: in both cases, the genus-dependent Fa\'a di Bruno coefficients compute the number of pairings of $2k$-points that result in a graph that can be embedded on a genus $g$ surface when placing the $2k$-points on the boundary of the surface. We thus find 
\be 
\overline{\tr \,\tilde O_{\Delta}^{2k}}_{g} =\<\tr M^{2k}\>_{g} = e^{\SBPS(1-2g)} C^{(g)}_{2k, [2^k]} (P_{\Delta\to \infty})^{k}\,.
\label{eq:match-all-genus}
\ee
The appearance of the Fa\'a di Bruno coefficients \eqref{eq:faadibruno} can be explicitly checked since on the matrix integral side $\<\tr M^{2k}\>$ can be computed at finite $N$. Specifically, following \cite{Brezin_Hikami_2007,Brezin:2016eax} (see also \cite{Zagier1986}), 
 \be
 \frac{1}{N} \< \tr\,M^{2k}\> = (P_{\Delta\to \infty})^{k} \frac{(2k)!}{N^k}
\left[ \sum_{l=0}^k 
\frac{\Gamma(N)}{\Gamma(N - k + l)\,\Gamma(k - l + 1)\,\Gamma(k - l + 2)\,\Gamma(l + 1)\,2^l} \right] \,.
\label{eq:exact-finite-N-1-pt}
    \ee
The ratio of gamma functions in each term in the square parentheses yields polynomials in $N$ of degree $N^{k-l}$. Thus, in a large $N$ expansion Eq.~\eqref{eq:exact-finite-N-1-pt} truncates at order $1/N^k$. This precisely matches the truncation in the gravitational genus expansion described in Eq.~\eqref{eq:1bdycorrelation}. Grouping terms in a finite power series in $1/N$, we find
 \begin{align} 
     \frac{1}{N} \langle \mathrm{tr} M^{2k} \rangle &=
\frac{(2k)! (P_{\Delta\to \infty})^{k}}{k!\,(k+1)!} \bigg[
1 + \frac{k(k-1)(k+1)}{12\,N^2} \nonumber \\ &\hspace{2.5cm} +
\frac{k(k+1)(k-1)(k-2)(k-3)(5k-2)}{1440\,N^4}
+ \dots
\bigg] \,,
\label{eq:GUE1bdy}
      \end{align} 
where one can check that the coefficients in the expansion are precisely the Fa\'a di Bruno coefficients in Eq.~\eqref{eq:faadibruno}, with the leading order coefficient given by the Catalan numbers. 

Finally, the equality between gravity correlators and matrix integral observables continues to hold for surfaces with a higher number of boundaries.  Specifically, when computing $\overline{\tr(  O_{\Delta}^{2k_1} ) \dots \tr ( \tilde O_{\Delta}^{2k_n} )} $ in gravity, the contributing geometries have $n$ boundaries and once again result in surfaces with disk topology after cutting the spacetime along the matter worldlines connecting the various boundary operator insertions.  This matches the `t Hooft diagrams in the computation of  ${\langle\tr M^{2k_1} \dots \tr M^{2k_n}\rangle}$, which again satisfy the same properties. For example, when $n=2$, connected diagrams must have at least one worldline (in gravity) or one propagating leg (in the matrix integral) connecting the two boundaries. An example of the map between diagrams in the two descriptions is
\be 
\inlinefig[8]{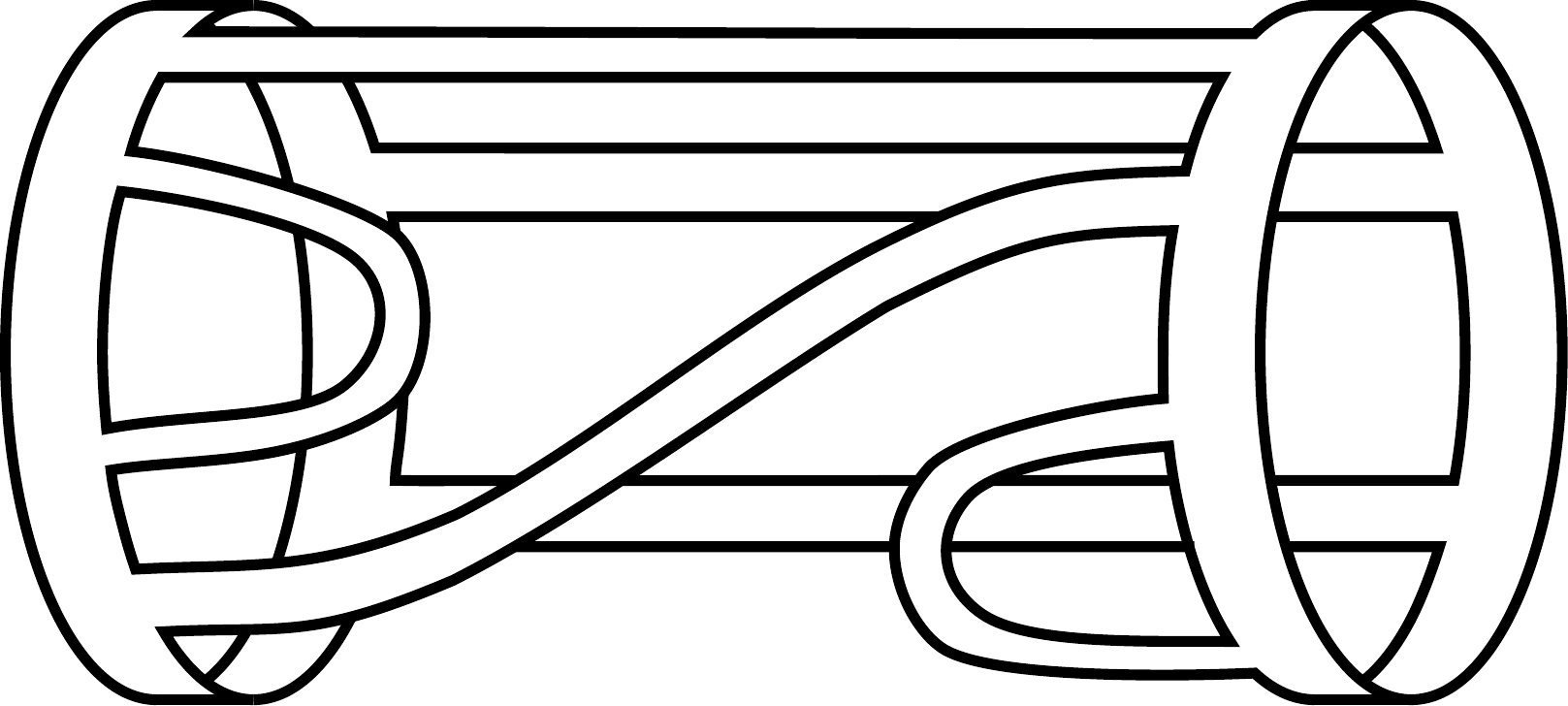}\quad\quad\longleftrightarrow\quad\quad\inlinefig[8]{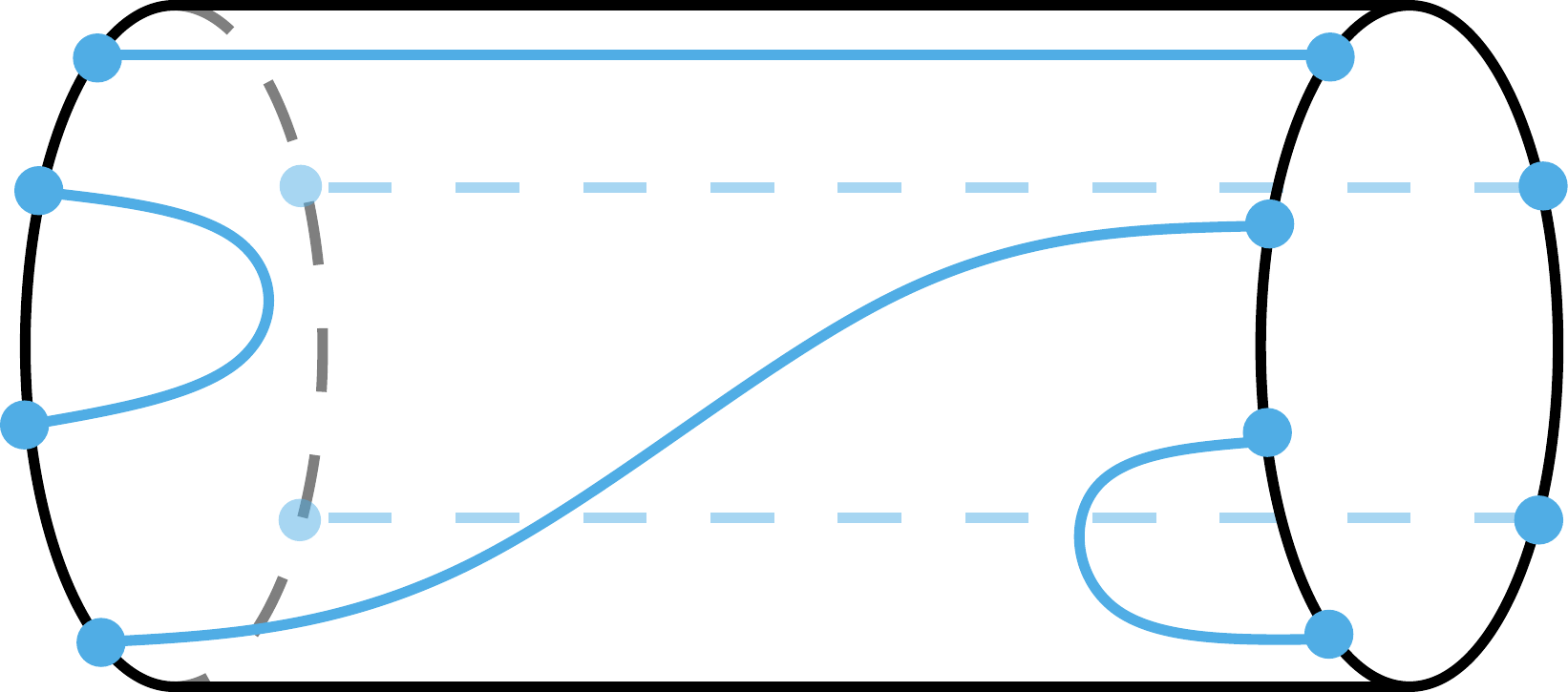}\,.
\ee

For $n>1$ boundaries, both on the gravity and the matrix integral side, the total number of diagrams that appear in the fully connected contribution is given by a generalization of the Fa\'a di Bruno coefficients, $C_g^{(m)}(2k_1,...,2k_m)$, for partition types $[2^{k_1+ \dots +k_m}]$. 
Since the combinatorial count for $n>1$ is quite complicated, exact results in the matrix integral help us compute the connected $m$-boundary correlators in gravity.  In fact, we take this approach in Appendix \ref{appendix:f} where we use the finite-$N$ expression of $\<\tr M^{2k_1}\, \tr\,M^{2k_2}\>$ to compute $C_g^{(2)}(2k_1,2k_2)$, i.e., the number of genus-$g$ diagrams contributing to $\overline{\tr( \tilde O_{\Delta}^{2k_1})\tr (\tilde O_{\Delta}^{2k_2} )}$. In the $k_1=k_2=\alpha N^{2/3}$ regime, these coefficients exactly reproduce the result \eqref{eq:cyl-resummed} for the connected two-boundary correlator.

The equality between matter correlators in gravity and matrix integral observables can also be generalized to the case of non-SUSY JT gravity in a microcanonical band. In this case, however, there is one additional subtlety related to the size $N$ of the matrix $M$, which is supposed to capture the number of energy levels within the small energy window $\mathfrak{E}$. In fact, $N$ itself fluctuates. This number comes from a matrix integral controlled by the matrix ensemble over Hamiltonians $H$, which thus provides a distribution for the size $N$ of the matrices. Considering such a distribution over $N$ recovers the first line in Eq.~\eqref{eq:1bdycorrelationJT} in the one-boundary case and yields a similar match between gravity and the Gaussian matrix model (when including the distribution over $N$) in the multi-boundary case. However, as we have already pointed out, in the large $k$ regime of our interest, these fluctuations---which are suppressed in $N$ and not enhanced in $k$---do not affect our results for the entropies, hence the approximation in the second line of Eq.~\eqref{eq:1bdycorrelationJT} (and its analog for multi-boundary correlators).

To summarize, the match between matter correlators in gravity for $\tilde{O}_\Delta$ at large $\Delta$ and GUE correlators shows that we can also use GUE techniques to compute correlators in gravity, both in the SUSY and non-SUSY cases. Below, we shall explain how the behavior of correlators for $k =O( N^{2/3})$ described in Section \ref{sec:bulk-resolution} is recovered from the Airy limit of the Gaussian matrix ensemble, and how this result can be extended beyond the Airy limit using matrix integral techniques. Since, as we have explained in Section \ref{sec:JTRMT}, the behavior of matrix models in the Airy limit is universal in the presence of a square root edge,\footnote{The square root edge itself is universal for a large class of matrix models.} regardless of their potential, the matrix model will serve as a useful tool to study the cases when $\Delta$ is not large (see Appendices \ref{sec:ETH} and \ref{sec:genericdresolution}) or when bulk matter interactions are introduced.

\subsection{Resolution to the paradox from the matrix integral point of view
}
\label{sec:resolution}

Now that we have established the relationship between our bulk theory and a GUE random matrix integral, we can use the latter to better understand the LMRS puzzle and its resolution. 

First, we will analyze the $k=O(e^{2\SBPS/3})\equiv O(N^{2/3})$ limit (studied from the bulk side in Section \ref{sec:bulk-resolution}) and clarify why the match between JT (super)gravity matter correlation functions and partition functions in the effective pure JT gravity theory arises. Intuitively, the reason is that, in this regime, we are probing the (universal) Airy edge of the matter GUE. Because the edge of the energy eigenvalue distribution of pure JT gravity is also controlled by the Airy distribution, we can expect a mapping between the two theories to arise in the appropriate limits. We will make this intuition quantitatively precise. Using this Airy limit, we will recover our bulk results for the semi-quenched entropy and compute the leading order behavior of the quenched entropy at large $k$.

Second, we will use the GUE matrix integral for matter to extend our results beyond the Airy limit to the $k=\omega(e^{2\SBPS/3})\equiv \omega(N^{2/3})$ regime. In this regime, it is unclear whether the genus re-summation carried out in Section \ref{sec:bulk-resolution} is still valid. On one hand, the semi-quenched entropy that results from such a re-summation is convergent for any value of $\alpha$, including when it depends non-trivially on $\SBPS$. On the other hand, it is unclear whether higher-order terms in the $1/k$ expansion, which we dropped in Eq.~\eqref{eq:faaexpansion}---or, equivalently, higher-order terms in the $1/\beta$ expansion we dropped in Eq.~\eqref{eq:airy-thermal-part} for the corresponding thermal partition function in the effective pure JT gravity theory---become important when $k=\omega(e^{2\SBPS/3})$. From the GUE point of view, we can analyze this regime. We will find a new saddle point, a one-eigenvalue instanton, in the matrix integral. Using this dominant saddle together with a subdominant two-eigenvalue instanton saddle, we will compute the semi-quenched entropy for $k=\omega(e^{2\SBPS/3})$. We find that for $k=\alpha N^{2/3}$ with $\alpha\gg 1$, these results agree with the Airy answer found in Section \ref{sec:bulk-resolution}. However, for $k=\omega(N^{2/3})$, the semi-quenched entropy behaves differently from the Airy limit, although it does remain positive for arbitrarily large values of $k$. This confirms that the genus re-summation in the Airy limit does not extend to the $k=\omega(e^{2\SBPS/3})\equiv\omega(N^{2/3})$ regime.

Once again, in this section, we will focus our attention on the BPS case, but the generalization to the non-SUSY case is immediate, as we have explained.

\subsubsection{$k=O(e^{2\SBPS/3})$: effective pure JT gravity theory from the Airy limit}
\label{sec:GUEairy}

First, let us focus on the $k=O( e^{2\SBPS/3})$ regime, which in the GUE corresponds to $k=O( N^{2/3})$. It is informative to first consider this regime for the exact finite $N$ GUE results presented in Section \ref{sec:GUE}. In this limit, the one-boundary correlation function \eqref{eq:GUE1bdy} becomes \cite{Brezin_Hikami_2007}
\be 
\langle \mathrm{tr}\, M^{2k} \rangle
\approx 2^{2k}\frac{N(P_{\Delta\to \infty})^{k}}{\sqrt{\pi}\,k^{3/2}}
\left[
1 + \frac{k^3}{12\,N^2}
+ \frac{k^6}{(12)^2\,2!\,N^4}
+ \dots
\right] =2^{2k} \frac{N (P_{\Delta\to \infty})^{k}}{\sqrt{\pi}k^{3/2}} e^{\frac{k^3}{12N^2}},
\label{eq:1pfairy}
\ee
where in the second term we have re-summed the series in the same way as we have discussed in Section \ref{sec:mapping}. A similar result can be obtained from the finite-$N$ GUE formulas in the two boundary case using the results in Appendix \ref{appendix:f},\footnote{\cite{Brezin_Hikami_2007} also derived this result from a matrix integral saddle-point calculation at large $k$. We will comment further on this result in Appendix \ref{appendix:f}.}
\be
     \begin{split}
     \< \tr M^{2k} \tr M^{2k} \>_\text{conn.}  &= \frac{2^{4k} (P_{\Delta \to \infty})^{2k}}{2\pi}\sum_{g=0}^\infty \frac{ k^{3g}}{N^{2g}}\sum_{p=0}^{g}\frac{(-1)^{g-p}}{(g-p)! (p!)(2g-2p+1)2^{g-3p}12^p }\\ 
&=\frac{2^{4k} (P_{\Delta \to \infty})^{2k}}{2\sqrt{2\pi}}\frac{N}{k^{3/2}}\exp\left(\frac{2k^3}{3N^2}\right)\erf\left(\sqrt{\frac{k^3}{2N^2}}\right)\,,
\end{split}
\label{eq:2pfairy}
\ee
which again agrees with the genus re-summation of cylinders seen in Section \ref{sec:mapping}. Recall that, as we have seen in Section \ref{sec:mapping}, these results are related to the $\beta_{\text{eff}}=O( e^{2S_0^{\text{eff}}/3})$ regime (the so-called Airy limit) of the thermal partition function in an effective pure JT gravity theory after the identifications $e^{S_0^{\text{eff}}}=N$, $\beta_{\text{eff}}=k$, $E_0=-\log(4P_{\Delta\to\infty})$:
\begin{equation}
    \< \tr M^{2k}\>\approx 2 \overline{Z(\beta_{\text{eff}})},\quad\quad
    \< \tr M^{2k} \tr M^{2k} \>_\text{conn.}\approx 2\overline{[Z(\beta_{\text{eff}})]^2}_{\text{conn.}}.\label{MMtoJT}
\end{equation}
Leveraging the connection between JT gravity coupled to matter and the GUE matrix model explained in Section \ref{sec:GUE}, we can now understand in detail why this surprising match with an effective pure JT gravity with no matter arises.

For sufficiently large $k$, the correlator $\langle\tr M^{2k}\rangle$ in the GUE is dominated by eigenvalues near the (square-root) spectral edges, whose statistics is governed by the Airy model \cite{Tracy_Widom_1993}. Labeling the eigenvalues of $M$ by $\lambda_i$ and the eigenvalue with largest magnitude at each edge by $\lambda_e$, we obtain
\begin{equation}
\begin{aligned}
    \langle\tr M^{2k}\rangle \approx 2 \int d\Lambda e^{-N\bar{I}_{\text{edge}}(\Lambda)}\sum_i \lambda_i^{2k}&=2\lambda_e^{2k}\int d\Lambda\,e^{-N\bar{I}_{\text{edge}}(\Lambda)}\sum_i \(1-\frac{\lambda_e-\lambda_i}{\lambda_e}\)^{2k}\\
    &\approx 2\lambda_e^{2k}\int d\Lambda \,e^{-N\bar{I}_{\text{edge}}(\Lambda)}\sum_i e^{-\frac{2k}{\lambda_e}\(\lambda_e-\lambda_i\)}\,,
\end{aligned}
\end{equation}
where the factor of $2$ comes from the (equal in the large $N$ limit) contribution of the two edges. Here, we focused on the right edge. In the second line we used the limit definition of the exponential for $k\to\infty$, which is valid in the double-scaling limit $k\to\infty$, $N\to\infty$, $k/N^{2/3}\equiv \alpha=O(1)$ because $\lambda_e-\lambda_i=O(N^{-2/3})$ for eigenvalues $\lambda_i$ near the edge (see Section \ref{sec:JTRMT}). 

Let us now take the continuum limit and replace $\lambda_i$ with the continuum variable $x$. Recall that for our GUE, the edges of the eigenvalue distribution are at $x=\pm 2\sqrt{P_{\Delta\to\infty}}\equiv \pm \lambda_e$ because the standard deviation of the GUE is $\sqrt{P_{\Delta\to\infty}}$ (see Eq.~\eqref{eq:GUEedge}). To make contact with the usual Airy model and pure JT gravity, first rescale the eigenvalues $x=\sqrt{P_{\Delta\to\infty}}y$, so that the edges are at $y=\pm 2$ and the standard deviation of the resulting GUE is $1$.\footnote{This leads to an overall constant factor $\(P_{\Delta\to\infty}\)^{N/2}$ coming from the eigenvalue measure, which can be reabsorbed by a shift in the potential and anyway cancels out in all entropy calculations.} Then shift the location of the edge to $z=0$ by defining $z=2-y$ (here we are focusing again on the right edge, but the same analysis can be performed for the left edge). We then obtain
\begin{equation}
\begin{aligned}
    \langle\tr M^{2k}\rangle&\approx 2\(2\sqrt{P_{\Delta\to\infty}}\)^{2k}N\int D\sigma\,e^{-NI_{\text{Airy}}[\sigma]}\(N\int dz \sigma(z)e^{-k z}\)\\
    &\approx 2\(2\sqrt{P_{\Delta\to\infty}}\)^{2k}\int dz\,\(e^{2\SBPS/3}\bar{\sigma}_{\text{Airy}}(e^{2\SBPS/3}z)\)e^{-kz}\,,
    \end{aligned}
    \label{eq:matterAiry}
\end{equation}
where in the first line we used that near-edge statistics is controlled by the Airy model and in the second line we performed the saddle point approximation, leading to $\bar{\sigma}_{\text{Airy}}(x)$ defined in Eq.~\eqref{eq:airy}, and we used that, for our GUE, $N=e^{\SBPS}$ as we have explained in Section \ref{sec:GUE}.\footnote{The factors of $N^{2/3}=e^{2\SBPS/3}$ appear because, unlike Section \ref{sec:JTRMT}, we did not rescale the eigenvalues $z$ so that the typical eigenvalue spacing and fluctuation near the edge is order 1.}

This result shows that, in the double-scaling limit of interest, GUE correlators correspond to thermal partition functions computed in the Airy model. On the other hand, it is well known \cite{Engelhardt_Fischetti_Maloney_2021,Hernandez-Cuenca_2024,Antonini_Iliesiu_Rath_Tran_2025} that the thermal partition function of pure JT gravity without matter at inverse temperature $\beta=O(e^{2S_0/3})$ also matches the Airy model partition function. This is simply a consequence of the square root behavior of the edge of the JT density of states near the ground state. In particular, the partition function at inverse temperature $\beta_{\text{eff}}$ is given by \cite{Engelhardt_Fischetti_Maloney_2021}
\begin{equation}
    \overline{Z(\beta_{\text{eff}})} \approx \int dE\,\(e^{2S_0^{\text{eff}}/3}\bar{\sigma}_{\text{Airy}}(e^{2S_0^{\text{eff}}/3}E)\)e^{-\beta_{\text{eff}} (E-E_0)}\,.
    \label{eq:JTAiry}
\end{equation}
Comparing equations \eqref{eq:matterAiry} and \eqref{eq:JTAiry}, we thus obtain the match $\langle\tr M^{2k}\rangle=2\overline{ Z(\beta_{\text{eff}})}$ observed in Section \ref{sec:bulk-resolution} between matter correlation functions in the BPS sector and an effective pure JT gravity theory after identifying $\SBPS=S_0^{\text{eff}}$, $k=\beta_{\text{eff}}$, and $E_0=-\log\(4P_{\Delta\to\infty}\)$. This relationship---which from the bulk side was very unexpected---has its roots in the universal edge statistics of matrix models, which control the behavior of eigenvalues near both edges of the matter GUE as well as near the ground state of JT gravity. The match only holds in the $k=O(e^{2\SBPS/3})$ regime we studied ($\beta_{\text{eff}}=O(e^{2S_0^{\text{eff}}/3})$), because only in that regime are we probing the near-edge Airy statistics. For $k$ ($\beta_{\text{eff}}$) much smaller or much larger, the eigenvalue statistics is not universal and is sensitive to the specific matrix potential, which differs greatly between the GUE and the SSS matrix integral dual to pure JT gravity.

So far, we have discussed the match for one-boundary correlators. Let us now consider a two-boundary correlator $\langle\tr M^{2k}\tr M^{2k}\rangle$. The results of this analysis will also make the multi-boundary correlator $\langle(\tr M^{2k})^m\rangle$ answer clear. For two boundaries, in the GUE, we need to compute
\begin{equation}
    \langle\tr M^{2k}\tr M^{2k}\rangle=\int d\Lambda\, e^{-N\bar{I}_{\text{GUE}}(\Lambda)}\sum_{i,j}\lambda_i^{2k}\lambda_j^{2k}\,.
\end{equation}
At large $k$, this correlation function is again dominated by the largest and smallest eigenvalues, i.e., by eigenvalues near the two edges of the GUE. However, for this two-trace correlator, we have two sets of eigenvalues we are summing over. Thus, there are four contributions dominating the integral: for two of them, we keep eigenvalues near opposite edges (namely, we only keep eigenvalues $\lambda_i$ near $\mp\lambda_e$ and, correspondingly, $\lambda_j$ near $\pm\lambda_e$); for the other two, we keep eigenvalues near the same edge (i.e., both $\lambda_i$ and $\lambda_j$ near $-\lambda_e$ or near $\lambda_e$). 

We can then take the Airy limit as above to study eigenvalues near these edges. For the first class of contributions, in which we zoom near the left edge for the first sum and near the right edge for the second sum (or vice versa), the answer factorizes into a product of single-trace correlators in the Airy limit. This is a consequence of the fact that the two edges of the GUE are uncorrelated in the large-$N$ limit \cite{Bornemann_2010}. As a result, we obtain
\begin{equation}
\begin{aligned}
    \langle\tr M^{2k}\tr M^{2k}\rangle_{\text{opposite edge}}&\approx 2\[\(2\sqrt{P_{\Delta\to\infty}}\)^{2k}\int dz\, \(e^{2\SBPS/3}\bar{\sigma}_{\text{Airy}}(e^{2\SBPS/3}z)\)e^{-kz}\]^2\\
    &\approx 2\(\overline{Z(\beta_{\text{eff}})}\)^2\,,
    \end{aligned}
    \label{eq:match1}
\end{equation}
where the factor $2$ comes from considering both contributions (i.e., picking the left edge for the first sum and the right edge for the second sum or vice versa). For the second class of contributions, in which both eigenvalues are taken near the same edge, we are simply computing a two-trace correlator in the Airy model,
\begin{equation}
\begin{aligned}
    \langle\tr M^{2k}\tr M^{2k}\rangle_{\text{same edge}}\approx 2\(2\sqrt{P_{\Delta\to\infty}}\)^{4k}\int dz_1dz_2 e^{4\SBPS/3} \overline{\sigma_{\text{Airy}}\sigma_{\text{Airy}}}e^{-k(z_1+z_2)}
    \end{aligned}
\end{equation}
where, to simplify notation, we omitted the arguments of the Airy distributions. In this expression, $\overline{\sigma_{\text{Airy}}(z_1)\sigma_{\text{Airy}}(z_2)}=\overline{\sigma}_{\text{Airy}}(z_1)\overline{\sigma}_{\text{Airy}}(z_2)+\overline{\sigma_{\text{Airy}}(z_1)\sigma_{\text{Airy}}(z_2)}_{\text{conn.}}$ is the second moment of the Airy spectral density, which contains a disconnected and a connected piece. The factor 2 again arises from the choice of the left or right edge of the GUE when taking the Airy limit. Recall that the two-boundary partition function in pure JT gravity for $\beta=O(e^{2S_0/3})$ is also computed by precisely the same Airy matrix model, and the connected and disconnected contributions are identified with the re-summed cylinder \eqref{eq:cyl-resummed} and a product of two single-boundary partition functions, i.e., two re-summed disks \cite{Engelhardt_Fischetti_Maloney_2021,Antonini_Iliesiu_Rath_Tran_2025}. Therefore, we obtain
\begin{equation}
    \langle\tr M^{2k}\tr M^{2k}\rangle_{\text{same edge}}\approx 2\(\overline{Z(\beta_{\text{eff}})}\)^2+2\overline{\[Z(\beta_{\text{eff}})\]^2}_{\text{conn.}}
    \label{eq:match2}
\end{equation}
and finally, combining equations \eqref{eq:match1} and \eqref{eq:match2},
\begin{equation}
    \langle\tr M^{2k}\tr M^{2k}\rangle\approx 4\(\overline{Z(\beta_{\text{eff}})}\)^2+2\overline{\[Z(\beta_{\text{eff}})\]^2}_{\text{conn.}}\,.
\end{equation}
The connected piece of the two-trace matter correlator thus satisfies $\langle\tr M^{2k}\tr M^{2k}\rangle_{\text{conn.}}\approx 2\overline{\[Z(\beta_{\text{eff}})\]^2}_{\text{conn.}}$, namely Eq.~\eqref{eq:fullmatch} for $m=2$ and $n=1$.\footnote{For the disconnected piece, the match (including the factor of 4) obviously follows from the match between one-boundary correlators.} The generalization for $n>1$ is straightforwardly obtained by replacing $k\to kn$. The generalization to $m$-boundary correlation functions is also immediate: the fully connected contribution to the correlator is obtained by considering the contribution to an $m$-trace correlator in which we keep eigenvalues near the same edge for all the $m$ sums over eigenvalues, which simply yields Eq.~\eqref{eq:fullmatch} for the connected component of the correlator. We have thus fully justified the match between matter correlators in JT (super)gravity and thermal partition functions in pure JT gravity in the Airy limit by using universal edge statistics of the dual matrix models.

Finally, let us comment on the various entropies that we can extract from these correlators. The results obtained in Section \ref{sec:bulk-resolution} for $m$-boundary correlators (or, equivalently, $m$-boundary thermal partition functions in JT gravity for $\beta_{\text{eff}}=O(e^{2S_0^{\text{eff}}/3})$) are exactly reproduced by the corresponding moments of thermal partition functions in the Airy model \cite{Kontsevich:1992ti}. Therefore, the result \eqref{eq:final-semiquenched} for the semi-quenched R\'enyi-2 entropy (as well as the positivity results for R\'enyi-$n$ with $n>2$) can be obtained entirely from the boundary side using the Airy model. One advantage of working with the Airy model on the boundary side is that we can also study the quenched entropy in the Airy limit $k=O(e^{2\SBPS/3})$, for which a bulk evaluation is more difficult. We provide details of this analysis in Appendix \ref{app:quenched}, and report here the result for the leading-order behavior of the quenched R\'enyi-$n$ entropy:
\begin{equation}
    S_Q^{(n)}=\frac{\pi^2\chi}{6}\(1+\frac{1}{n}\)\frac{1}{\alpha}+O\left(\frac{1}{\a^3}\right)\,,
\end{equation}
where we reintroduced $\alpha=k/N^{2/3}$ and $\chi$ is a constant. Notice the polynomial decay of the quenched entropy as opposed to the exponential decay of the semi-quenched entropy \eqref{eq:final-semiquenched}: this difference also arose for the thermal entropy in JT gravity in \cite{Antonini_Iliesiu_Rath_Tran_2025}, for which the quenched entropy from the Airy model was computed in \cite{Janssen:2021mek}. However, in our case, the decay of the quenched entropy is linear in $1/\alpha$ as opposed to the cubic decay in \cite{Janssen:2021mek}. As we explain in Appendix \ref{app:quenched}, this is again a consequence of the existence of two uncorrelated edges in the GUE of our interest, as opposed to the single edge for pure JT gravity. This behavior for the quenched entropy was shown in Figure \ref{fig:moneyplot}, matching results obtained numerically.

\subsubsection{Beyond the Airy limit: the one-eigenvalue instanton}
\label{sec:oneeigen}

As we have discussed at the beginning of Section \ref{sec:resolution}, it is unclear whether the genus re-summation we carried out to obtain the semi-quenched entropy \eqref{eq:final-semiquenched} extends beyond the $k=O(e^{2\SBPS/3})$ regime. From the matrix integral point of view, this corresponds to asking whether the Airy limit controls correlators also for $k=\omega(N^{2/3})$. As we will now discuss, the answer is no. The reason is that, in this regime, the GUE matrix model for our matter operator is dominated by new saddle points, a one-eigenvalue instanton (dominant saddle) \cite{Saad_Shenker_Stanford_2019,Hernandez-Cuenca_2024} and a two-eigenvalue instanton (first subdominant saddle) \cite{Antonini_Iliesiu_Rath_Tran_2025}, which appear due to the insertion of a very large number of operators. Similar eigenvalue instantons, but dominating the JT gravity partition function at very low temperatures, were introduced in \cite{Hernandez-Cuenca_2024} and further studied in \cite{Antonini_Iliesiu_Rath_Tran_2025}. Although the details and the physical interpretation of the instantons are different (in particular, we emphasize that the instantons in our context are for the spectrum of the matter operator instead of the energy spectrum), the generic procedure to derive them is very similar to that introduced in \cite{Hernandez-Cuenca_2024}. For this reason, in this section, we will concisely derive the instanton and refer the reader to \cite{Hernandez-Cuenca_2024} for further details.

Let us consider a generic multi-trace correlator (i.e., multi-boundary correlator)
\begin{equation}
    \left\langle\(\tr M^{2k}\)^m\right\rangle = \int d\Lambda\,e^{-N\bar{I}_{\text{GUE}}(\Lambda)}\(\sum_i\lambda_i^{2k}\)^m=\int d\Lambda\,e^{-N\bar{I}^{\text{eff}}_{\text{GUE}}(\Lambda)}\,,
\end{equation}
where we defined an effective action 
\begin{equation}
    \bar{I}^{\text{eff}}_{\text{GUE}}(\Lambda)=\sum_{i=1}^N V_{\text{GUE}}(\lambda_i)-\frac{1}{N}\sum_{\substack{i,j=1\\i\neq j}}^N\log|\lambda_i-\lambda_j|-\frac{m}{N}\log \sum_{i=1}^N\lambda_i^{2k}\,.
\end{equation}
As we will see, the last term, coming from exponentiating the multi-trace operator insertion, can contribute a non-negligible amount to the effective action for a single eigenvalue when $k$ is sufficiently large. We can thus expect that a new saddle-point arises when $k$ is very large due to this modification of the effective action. This new saddle is called a one-eigenvalue instanton. For $N-1$ eigenvalues, the equilibrium density of states is unchanged up to $1/N$ corrections: it is still given by the Wigner semicircle. But one eigenvalue $\lambda_0$, the one with the largest magnitude (i.e., the smallest or largest one), sharply separates from this continuum. To find its location, let us make an ansatz in which $\lambda_0$ is well-separated from all the other eigenvalues. Up to corrections exponentially suppressed in $k$, the effective action can then be rewritten as
\begin{equation}
    \bar{I}^{\text{eff}}_{\text{GUE}}(\Lambda)=\sum_{i=1}^N V_{\text{GUE}}(\lambda_i)-\frac{1}{N}\sum_{\substack{i,j=1\\i\neq j}}^N\log|\lambda_i-\lambda_j|-\frac{2km}{N}\log |\lambda_0|\,.
    \label{eq:effinst}
\end{equation}
The saddle-point equations for all eigenvalues other than $\lambda_0$ are thus unchanged, leading to the semicircle distribution for $N-1$ eigenvalues. The saddle-point equation for $\lambda_0$ takes the form
\begin{equation}
    V_{\text{eff}}'(\lambda_0^*)\equiv V'_{\text{GUE}}(\lambda_0^*)-\frac{1}{N}\sum_{j\neq 0}\frac{1}{|\lambda_0^*-\lambda_j|}=\frac{2km}{N\lambda_0^*}\,.
    \label{eq:1eigensaddle}
\end{equation}
Notice that the last term---which can be interpreted as a force pushing the eigenvalue $\lambda_0$ to separate from the continuum---contributes at leading order if $k=O(N)$. However, this analysis remains valid as $k=\alpha N^{2/3}$ with $\alpha\gg 1$ and any $k=\omega(N^{2/3})$, as we will see shortly. Solving Eq.~\eqref{eq:1eigensaddle} by using Eq.~\eqref{eq:spectral} and the GUE equilibrium density of states, we obtain
\begin{equation}
    \lambda_0^*=\pm\sqrt{2P_{\Delta\to\infty}}\sqrt{1+\sqrt{1+\left(\frac{km}{N}\right)^2}}\,.
    \label{eq:1-eigen}
\end{equation}
Notice that the eigenvalue is separated from the (right for $+$, left for $-$) edge by an amount that grows with $k$ and becomes order 1 when $k=O(N)$. In the Airy limit $k=\alpha N^{2/3}$, we obtain $|\lambda_0^*|\approx 2\sqrt{P_{\Delta\to\infty}}+m^2\alpha^2\sqrt{P_{\Delta\to\infty}}/(4N^{2/3})$. Notice that, for $\alpha\gg 1$, this single eigenvalue is again well separated from the edge and from its fluctuations, which are of order $N^{-2/3}$ (see Section \ref{sec:JTRMT}). If $\alpha$ is small (say, order 1), then the eigenvalue is not really separated from the continuum (once we include fluctuations), and we can expect our analysis to break down (our initial ansatz is not justified). This expectation is confirmed by realizing that the saddle-point approximation we performed to find the one-eigenvalue instanton is controlled, in the $k=\alpha N^{2/3}$ regime, by the parameter $\eta=N^{2/3}/(km)=(m\alpha)^{-1}$. If $\alpha$ is order 1 but large, or if it scales with any power of $N$, $\eta$ is small, the saddle-point approximation is under control, and our analysis is reliable. If $\alpha$ is small, $\eta$ is large, signaling that fluctuations around the saddle point are large and we cannot trust our approximation. Therefore, we conclude that the one-eigenvalue instanton is reliable only for $k=\omega(N^{2/3})$ or for $k=\alpha N^{2/3}$ with $\alpha\gg 1$.\footnote{Incidentally, we remark that the parameter $\eta$ also explains why the one-eigenvalue instanton cannot be trusted in the computation of the quenched entropy (an attempt in this direction was made in \cite{Hernandez-Cuenca_2024} for the thermal entropy in JT gravity at low temperature). For the quenched entropy, one must perform a no-replica trick in which the number of replicas $m$ goes to zero \cite{Engelhardt_Fischetti_Maloney_2021}. This limit is strict, meaning that the parameter $\eta$ blows up and the saddle-point approximation is not under control.}

In this regime, the correlator can be computed by evaluating the effective action \eqref{eq:effinst} for the one-eigenvalue instanton on-shell, which yields\footnote{\label{foot:1-eigenpart}More conveniently, the effective potential $V_{\text{eff}}(\lambda)$ can be obtained from Eq. \eqref{eq:spectral} using the GUE density of states \eqref{eq:GUEedge}. The first factor in Eq. \eqref{eq:1-eigenpart} can then be evaluated explicitly and is given by $$e^{-NV_{\text{eff}}(\lambda_0^*)}\(\lambda_0^*\)^{2km}=\(2P_{\Delta\to\infty}\)^{km}e^{km\[-1+\log\(1+\sqrt{1+\frac{k^2m^2}{N^2}}\)\]+N\log\[\frac{km}{N}+\sqrt{1+\(\frac{km}{N}\)^2}\]}\,.$$}
\begin{equation}
    \left\langle\(\tr M^{2k}\)^m\right\rangle_{\text{1-eigenvalue}}\approx 2 e^{-NV_{\text{eff}}(\lambda_0^*)}\(\lambda_0^*\)^{2km} \times \(\text{1-loop}\)\,,
    \label{eq:1-eigenpart}
\end{equation}
where the factor of $2$ is due to the existence of two one-eigenvalue instanton saddles, one where the smallest eigenvalue separates from the left edge ($-$ sign in Eq.~\eqref{eq:1eigensaddle}) and one where the largest eigenvalue separates from the right edge ($+$ sign in Eq.~\eqref{eq:1eigensaddle}). Notice that the continuum of eigenvalues does not contribute, because its contribution cancels out (up to $1/N$-suppressed corrections) with the normalization by the matrix integral partition function without any matter insertion. Notice that we did not explicitly evaluate the 1-loop determinant. We will comment more below on the role of this determinant. For the R\'enyi-$n$ entropy, we thus obtain
\begin{equation}
    S_{SQ}^{(n)}\approx \frac{1}{1-n}\log\frac{\left\langle\tr M^{2kn}\right\rangle_{\text{1-eigenvalue}}}{\left\langle\(\tr M^{2k}\)^n\right\rangle_{\text{1-eigenvalue}}}=0\,.
\end{equation}
This result, which is valid at arbitrarily large values of $k$, shows that the semi-quenched entropy vanishes at large $k$ and never becomes negative. However, it also does not give us information about the behavior of the semi-quenched entropy as it vanishes. To improve on this result, we need to study a subleading saddle, the two-eigenvalue instanton, which will be the subject of the next subsection.

Before studying the two-eigenvalue instanton, it is useful to analyze our result in the $k=\alpha N^{2/3}$, $\alpha\gg 1$ regime. As we have seen, in this regime, the one-eigenvalue instanton analysis is reliable. On the other hand, the Airy analysis of Section \ref{sec:GUEairy} is also valid. We therefore expect the one-eigenvalue instanton results to agree with the Airy answer. In fact, this is the case. By expanding the result \eqref{eq:1-eigenpart} (see Footnote \ref{foot:1-eigenpart}) for $k\ll N$, one obtains
\begin{equation}
    \left\langle\(\tr M^{2k}\)^m\right\rangle_{\text{1-eigenvalue}}\approx 2\(4P_{\Delta\to\infty}\)^{km}e^{m^3\alpha^3/12}\times \(\text{1-loop}\)\,.
    \label{eq:1-eigenairy}
\end{equation}
By setting $m=1$ and $m=2$ and comparing with equations \eqref{eq:1pfairy} and \eqref{eq:2pfairy}, it is easy to show that this result agrees with the single-trace correlator as well as the leading-order behavior at $\alpha\gg 1$ of the two-trace correlator in the Airy limit. The 1-loop determinant that we did not evaluate should yield the correct prefactors and the $\alpha^{-3/2}$ factor present in \eqref{eq:1pfairy} and in the leading-order term of \eqref{eq:2pfairy}. In fact, using equations \eqref{eq:1pfairy} and \eqref{eq:2pfairy} to compute the R\'enyi-2 semi-quenched entropy in the Airy limit by only keeping the leading-order term at large $\alpha$ in the denominator yields exactly zero, just like in our one-eigenvalue instanton calculation. To study the decay of the semi-quenched entropy, we need to obtain the first subleading term in the denominator. In the Airy limit, this comes from the sum of the disconnected contribution (product of two single-trace correlators \eqref{eq:1pfairy}) and the first subleading term of the connected contribution \eqref{eq:2pfairy}. We will see that this is captured by a two-eigenvalue instanton.

\subsubsection{Beyond the Airy limit: the two-eigenvalue instanton}
\label{sec:twoeigen}

In order to study the decay behavior of the semi-quenched R\'enyi entropies, we must compute the subleading correction to the multi-trace correlator appearing in the denominator.\footnote{\label{foot:2eigen}The exact match between the single-trace correlator \eqref{eq:1pfairy} in the Airy limit and the corresponding answer for the one-eigenvalue instanton (Eq.~\eqref{eq:1-eigenairy} with $m=1$) suggests that the single-trace correlators appearing in the numerator of the semi-quenched R\'enyi entropies do not receive any subleading contributions from other saddles.} This is captured by a different saddle point, in which now the two eigenvalues of largest magnitude separate from the continuum: a two-eigenvalue instanton.\footnote{A similar saddle was recently studied in \cite{Antonini_Iliesiu_Rath_Tran_2025} in the study of thermal entropy in pure JT gravity at low temperatures.} To find the location of the two eigenvalues, we first expand the multi-trace operator insertion keeping only leading terms involving only the two largest eigenvalues $\lambda_0$ and $\lambda_1$:\footnote{This expression is correct only for $m>2$. For $m=2$, we simply obtain $\lambda_0^{4k}+\lambda_1^{4k}+2\lambda_0^{2k}\lambda_1^{2k}$. We will comment on the $m=2$ case further towards the end of this section.}
\begin{equation}
    \(\sum_{i=1}^N\lambda_i^{2k}\)^m\approx \lambda_0^{2km}+\lambda_1^{2km}+m\lambda_0^{2k(m-1)}\lambda_1^{2k}+m\lambda_0^{2k}\lambda_1^{2k(m-1)}+...
    \label{eq:expansion}
\end{equation}
We must then compute the GUE integral for each of these terms in the saddle-point approximation. The first two terms simply yield the one-eigenvalue instanton studied in the previous subsection. The third and fourth terms give the first subleading contribution. The terms we omitted are further suppressed.

We will be interested in the third and fourth terms in Eq.~\eqref{eq:expansion}. Because of the symmetry of the effective action under the exchange of $\lambda_0$ and $\lambda_1$, they contribute identically to the correlator. We can thus study only the third term and double the result. The effective action is given by
\begin{equation}
    \bar{I}^{\text{eff}}_{\text{GUE}}(\Lambda)=\sum_{i=1}^N V_{\text{GUE}}(\lambda_i)-\frac{1}{N}\sum_{\substack{i,j=1\\i\neq j}}^N\log|\lambda_i-\lambda_j|-\frac{2k(m-1)}{N}\log |\lambda_0|-\frac{2k}{N}\log |\lambda_1|\,,
    \label{eq:eff2inst}
\end{equation}
leading to the saddle-point equations for the two eigenvalues $\lambda_0$ and $\lambda_1$
\begin{equation}
    V_{\text{eff}}'(\bar{\lambda}_0)=\frac{2k(m-1)}{N\bar{\lambda}_0}\,, \quad\quad\quad V_{\text{eff}}'(\bar{\lambda}_1)=\frac{2k}{N\bar{\lambda}_1}\,.
\end{equation}
Notice that these are very similar to the saddle-point Eq.~\eqref{eq:1eigensaddle} for the one-eigenvalue instanton. The solutions are then immediately given by
\begin{equation}    \bar{\lambda}_0=\pm\sqrt{2P_{\Delta\to\infty}}\sqrt{1+\sqrt{1+\left(\frac{k(m-1)}{N}\right)^2}}\,,\quad\quad \bar{\lambda}_1=\pm\sqrt{2P_{\Delta\to\infty}}\sqrt{1+\sqrt{1+\left(\frac{k}{N}\right)^2}}\,.
    \label{eq:2-eigen}
\end{equation}
Notice that the eigenvalue $\bar{\lambda}_0$ is at the location of the one-eigenvalue instanton Eq. \eqref{eq:1-eigen} for $m\to m-1$ and $\bar{\lambda}_1$ is at the location for $m=1$; they are thus both well-separated from the spectral edge and from each other (the latter is not true for the special $m=2$ case, which we will discuss below). The remaining $N-2$ eigenvalues again sit in the usual Wigner semicircle (up to $1/N$-suppressed corrections). Notice that there are four possible saddles depending on whether the separated eigenvalues are on the same side or on opposite sides of the spectrum. Taking into account the exchange symmetry $\lambda_0\leftrightarrow\lambda_1$ (i.e., the fourth term in Eq.~\eqref{eq:expansion}), we obtain a total of eight saddles.
The on-shell action for each one of these saddles is identical, leading to
\begin{equation}
    \left\langle \(\tr M^{2k}\)^m\right\rangle_{\text{2-eigenvalue}}\approx e^{-NV_{\text{eff}}(\bar{\lambda}_0)-NV_{\text{eff}}(\bar{\lambda}_1)}\(\bar{\lambda}_0\)^{2k(m-1)}\(\bar{\lambda}_1\)^{2k}\times \left(\text{1-loop}\right)\,,
    \label{eq:2-eigenpart}
\end{equation}
where again the first factor can be evaluated explicitly (we omit here the explicit expression for the sake of brevity, but it can be obtained as explained in Footnote \ref{foot:1-eigenpart}).

Before analyzing the Airy limit of this result, let us compute the semi-quenched R\'enyi-$n$ entropy including the subleading 2-eigenvalue instanton correction in the denominator. Ignoring for now the 1-loop determinants, we obtain\footnote{We are not including a two-eigenvalue instanton contribution in the numerator for the reason explained in Footnote \ref{foot:2eigen}.}
\begin{equation}
\begin{aligned}
    S_{SQ}^{(n)}&\approx \frac{1}{1-n}\log\frac{\left\langle\tr M^{2kn}\right\rangle_{\text{1-eigenvalue}}}{\left\langle\(\tr M^{2k}\)^n\right\rangle_{\text{1-eigenvalue}}+\left\langle\(\tr M^{2k}\)^n\right\rangle_{\text{2-eigenvalue}}}\\
    &\approx \frac{1}{n-1}\frac{\left\langle\(\tr M^{2k}\)^n\right\rangle_{\text{2-eigenvalue}}}{\left\langle\(\tr M^{2k}\)^n\right\rangle_{\text{1-eigenvalue}}}\,,
    \end{aligned}
    \label{eq:2-eigenentropy}
\end{equation}
where in the second line we used that the two-eigenvalue instanton contribution is subleading with respect to the one-eigenvalue instanton contribution and that, for the one-eigenvalue instanton, $\left\langle\tr M^{2kn}\right\rangle_{\text{1-eigenvalue}}=\left\langle\(\tr M^{2k}\)^n\right\rangle_{\text{1-eigenvalue}}$.
The explicit expression for the semi-quenched entropy, which is not very enlightening, can be obtained explicitly by substituting equations \eqref{eq:1-eigenpart} and \eqref{eq:2-eigenpart} into \eqref{eq:2-eigenentropy}. 

Let us now focus on the $k=\alpha N^{2/3}$, $\alpha\gg 1$ regime. Once again, we expect our results to match the decay of the semi-quenched entropy obtained in the Airy limit, see Eq.~\eqref{eq:final-semiquenched}. First, let us analyze the exponential behavior of the two-eigenvalue instanton contribution to the multi-trace correlator. In the $k=\alpha N^{2/3}$, $\alpha\gg 1$ regime, we obtain
\begin{equation}
    \left\langle \(\tr M^{2k}\)^m\right\rangle_{\text{2-eigenvalue}}\approx \(4P_{\Delta\to\infty}\)^{km}e^{\frac{m[m(m-3)+3]\alpha^3}{12}}\times \(\text{1-loop}\)\,.
    \label{eq:2-eigenairy}
\end{equation}
Setting $m=2$, we obtain the exponential behavior $\sim e^{\alpha^3/6}$, which agrees with the exponential behavior of both the square of the single-trace correlator \eqref{eq:1pfairy} (disconnected contribution) and the first subleading term in the connected contribution \eqref{eq:2pfairy}. Plugging equations \eqref{eq:1-eigenairy} and \eqref{eq:2-eigenairy} into the expression \eqref{eq:2-eigenentropy} for the semi-quenched entropy, we obtain
\begin{equation}
    S_{SQ}^{(n)}\approx \frac{1}{n-1}e^{-3n(n-1)\alpha^3/12}\,,
    \label{eq:sqrenyin}
\end{equation}
where we neglected the multiplicative one-loop determinant contribution. Eq.~\eqref{eq:sqrenyin} gives a precise result for the exponential decay of the semi-quenched R\'enyi-$n$ entropy in the Airy limit. It would be interesting to compute this quantity from a bulk perspective using the techniques of Section \ref{sec:bulk-resolution}, which should reproduce the exponential decay \eqref{eq:sqrenyin} and would also yield the coefficients and power-law dependence encoded in the one-loop determinant. This would require an explicit expression for generic $n$ for the Airy limit of the coefficients $C_g^{(n)}(2k,...,2k)$ appearing in Eq.~\eqref{eq:mbdycorrelation}. To the best of our knowledge, this expression is unknown for generic $n$, and known only for $n=1$ (Fa\`a Di Bruno coefficients) and $n=2$ (derived in Appendix \ref{appendix:f}).
Setting $n=2$ in Eq.~\eqref{eq:sqrenyin}, we obtain the decay $S_{SQ}^{(2)}\approx e^{-\alpha^3/2}$, which matches the behavior \eqref{eq:final-semiquenched} obtained in the Airy limit. 

Finally, let us discuss (somewhat qualitatively) one-loop determinants. For $m\geq 3$, the two eigenvalues are well separated from each other in all of the eight two-eigenvalue instanton saddles. We can then expect the one-loop determinant to be the product of two one-loop determinants for a one-eigenvalue instanton. In fact, the two separated eigenvalues are nearly uncorrelated. 

For the special case $m=2$, we have two different classes of saddles and two different behaviors of the one-loop determinant. For the (four) saddles in which the eigenvalues are sitting on opposite sides of the spectrum (picking opposite signs in Eq.~\eqref{eq:2-eigen}), they are again well-separated from the edge and from each other. The one-loop determinant is then again the product of two one-loop determinants for a one-eigenvalue instanton. As we have explained in Section \ref{sec:oneeigen}, for $m=2$ the one-loop determinant for the one-eigenvalue instanton must contribute a factor of $\alpha^{-3/2}$ to the correlator. Thus, for this first class of saddles, we can expect the one-loop determinant to contribute a factor $\alpha^{-3}$ for $m=2$. Plugging this into Eq.~\eqref{eq:2-eigenentropy} with $m=2$ (as well as the one-loop determinant factor $\alpha^{-3/2}$ for the one-eigenvalue instanton), we obtain the correct $\alpha^{-3/2}$ power seen in Eq.~\eqref{eq:final-semiquenched} in the semi-quenched R\'enyi-2 entropy. 

On the other hand, for the (four) saddles for which the two eigenvalues are on the same side of the spectrum (picking identical signs in Eq.~\eqref{eq:2-eigen}), the two eigenvalues sit very close to each other at the location of an $m=1$ one-eigenvalue instanton. In this case, their mutual Vandermonde repulsion then affects the one-loop determinant non-trivially. We thus expect this one-loop determinant to yield some different power of $\alpha$ for this second class of saddles. This expectation is supported by the results of \cite{Antonini_Iliesiu_Rath_Tran_2025} for two-eigenvalue instantons in the computation of the thermal entropy in pure JT gravity at low temperatures. In fact, this second class of saddles is the only one present in \cite{Antonini_Iliesiu_Rath_Tran_2025} (because only one edge is present in the JT energy spectrum), and they must reproduce the power $\alpha^{-9/2}$ in the semi-quenched R\'enyi-2 entropy obtained from the Airy analysis in \cite{Antonini_Iliesiu_Rath_Tran_2025}. Their one-loop determinant should thus contribute a factor $\alpha^{-6}$ to the partition function. Because they are precisely the same saddles in our case, we expect the one-loop determinant for this second class of saddles to also contribute a power $\alpha^{-6}$ to the correlators. We then conclude that the first class of saddles, for which the power-law suppression in $\alpha$ was milder ($\alpha^{-3}$), dominates the matrix integral. We expect this result to continue to hold for $k=\omega(N^{2/3})$.

To summarize, using one-eigenvalue and two-eigenvalue instantons, we computed the R\'enyi-$n$ semi-quenched entropy for $k=\omega(N^{2/3})$ and found that our result smoothly interpolates with the Airy result obtained in Sections \ref{sec:bulk-resolution} and \ref{sec:GUEairy} in the $k=\alpha N^{2/3}$, $\alpha\gg 1$ regime in which both analyses are reliable. On the other hand, these results hold for arbitrarily large $k$. We remark that, for $k=\omega(N^{2/3})$ (i.e., going away from the Airy limit), the eigenvalue instanton analysis, as well as the results for the correlators and the semi-quenched entropy, is not universal and depends strongly on the specific matrix potential under examination (the GUE potential in this case). In particular, the eigenvalue instanton results for the correlators and the semi-quenched entropy differ from the Airy results when $k=\omega(N^{2/3})$. This can be checked by evaluating explicitly equations \eqref{eq:1-eigenpart}, \eqref{eq:2-eigenpart}, and \eqref{eq:2-eigenentropy}. This confirms that the genus re-summation carried out in Section \ref{sec:bulk-resolution} for $k=\alpha e^{2\SBPS/3}$ (recall that $N=e^{\SBPS}$) is not valid for $k=\omega(e^{2\SBPS/3})$: higher order terms in the $1/k$ expansion, which we dropped in our re-summation, must become important. In future work, we hope to be able to reproduce the non-universal result from the one and two-eigenvalue instanton analysis by keeping these subleading terms in the $1/k$ expansion.

\section{A similar puzzle and its resolution: a random tensor network approach} 
\label{sec:TN}

\begin{figure}
    \centering
    \includegraphics[width=\linewidth]{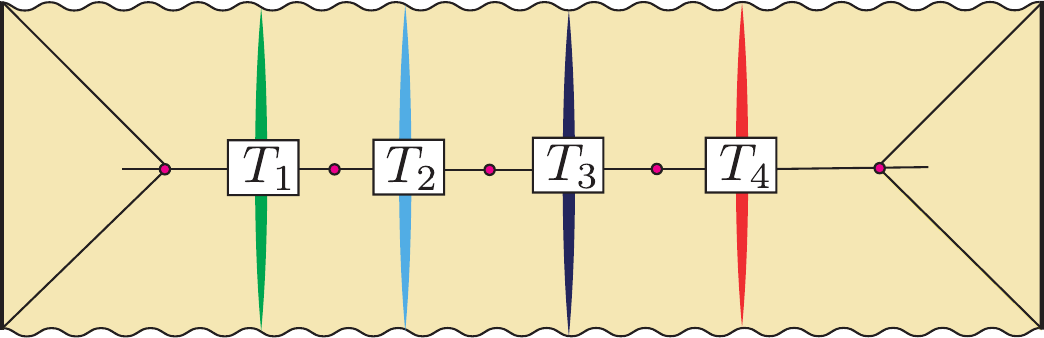}
    \caption{The random tensor network preparing the state \eqref{eq:TNstate} dual to a two-sided black hole with a long wormhole. The structure of the network provides a bulk interpretation of the state. The different tensors can be regarded as insertions of particles of different flavor (emphasized by the different colors) supporting the long wormhole. The red dots represent $k+1$ degenerate extremal surfaces. 
    }
    \label{fig:TN}
\end{figure}

In Section \ref{sec:2}, we discussed a negative entropy paradox that arises when considering a large number $k$ of insertions of a single Hermitian operator $\tilde{O}_{\Delta}$ in the preparation of two-sided black hole states, with the operators $\tilde{O}_{\Delta}$ including projectors on a fixed energy sector (the BPS sector in $\mathcal{N}=2$ JT supergravity or a microcanonical band $\mathfrak{E}$ in non-SUSY JT gravity). In this section, we study a similar paradox that arises when considering the insertion of $k$ different non-interacting\footnote{Here we mean that the only interactions are gravitational.} and non-Hermitian operators $\tilde{O}_\Delta^{(i)}$.\footnote{LMRS commented on the existence of a similar alternative version of the paradox (but involving Hermitian operators) without working out its details \cite{Lin_Maldacena_Rozenberg_Shan_2023}.} For example, we can consider each operator $\tilde{O}_{\Delta}^{(i)}$ to be charged under some internal $U(1)$ symmetry. We comment on the extension to non-interacting, Hermitian operators in Appendix \ref{app:RTNs}. We focus again on the large $\Delta$ limit and on the BPS case, although the same results can be generalized to non-SUSY JT gravity in a microcanonical band as explained in the previous sections. 

\subsection{Puzzle and resolution in RTN}

The index $i$ labeling operators can be thought of, for example, as an internal flavor or as being associated with slightly different scaling dimensions for the different operators. The corresponding two-point function takes the form
\begin{equation}
    \tr\((\tilde{O}_\Delta^{(i)})^\dagger\tilde{O}_\Delta^{(j)}\)=e^{\SBPS}P_{\Delta\to\infty}\delta_{ij}\,,
    \label{eq:TN2pf}
\end{equation}
where the Hermitian conjugate is needed because the operators are non-Hermitian. Notice that the two-point function vanishes if the two operator insertions have different flavors. We can then consider the state
\begin{equation}
    \ket{\psi_k} = \prod_{i=1}^k \tilde{O}^{(i)}_\Delta \ket{\text{TFD}_0},
    \label{eq:TNstate}
\end{equation}
where $\ket{\text{TFD}_0}$ is again the zero temperature thermofield double state (i.e., the maximally entangled state on two copies of the BPS subspace). Notice that the operators $\tilde{O}_{\Delta}^{(i)}$ do not commute. One can then consider many inequivalent states that differ by the ordering of the operator insertions. The same puzzle and the same resolution we will introduce arise for all these states, and we will thus focus on the specific state \eqref{eq:TNstate}.

Clearly, the presence of different, non-Hermitian operators with the property \eqref{eq:TN2pf} modifies the combinatorics of diagrams in JT (super)gravity coupled to matter with respect to those studied in Section \ref{sec:bulk-resolution}. Taking into account the different flavors of the operators $\tilde{O}_{\Delta}^{(i)}$ and their non-Hermiticity, we could pose the puzzle and find its resolution by an appropriate counting argument directly in the bulk as in Sections \ref{sec:2} and \ref{sec:bulk-resolution}, or use the connection between the large $\Delta$ limit of the operators and a simple matrix ensemble (in this case a complex Ginibre ensemble \cite{Ginibre:1965zz}, which is a generalization of the GUE to non-Hermitian matrices) as in Section \ref{sec:matrixmodels} to study the puzzle in the dual matrix model. However, in this section, we will take a different but equivalent route and map the problem to a random tensor network (RTN) setup \cite{Hayden:2016cfa}, which captures precisely the properties of such operators. This has the advantage of greatly simplifying an otherwise involved combinatorial problem, while offering a different perspective on the puzzle and its resolution, whose bulk interpretation is immediately clear due to the relationship between the structure of tensor networks and the emergence of spacetime in holography \cite{Swingle:2009bg,Almheiri:2014lwa,Pastawski:2015qua,Akers:2018fow,Dong:2018seb,Dong:2019piw}.\footnote{Other such uses of RTN technology to understand the gravitational path integral include \cite{Penington:2019kki,Akers:2020pmf,Dong:2021clv,Akers:2021pvd,Antonini:2022sfm,Penington:2022dhr,Wang:2022ots,Cheng:2022ori,Akers:2022zxr,Akers:2022qdl,Milekhin:2022zsy,Antonini:2023hdh,Dong:2023bfy,Dong:2024gud,Penington:2024jmt,Kaya:2025vof,Akers:2025ahe}.}

As we have explained in Section \ref{sec:GUE}, in the large $\Delta$ limit, a Hermitian operator $\tilde{O}_\Delta$ is a Gaussian random matrix. 
In the setup of this section, in which we have non-Hermitian operators $\tilde{O}^{(i)}_\Delta$, a completely analogous argument leads to the conclusion that each operator is an independent random matrix drawn from a complex Ginibre ensemble, which has potential $V(\tilde{O}^{(i)}_\Delta)=\frac{1}{2P_{\Delta\to\infty}}(\tilde{O}^{(i)}_\Delta)^\dagger \tilde{O}^{(i)}_\Delta$. The state \eqref{eq:TNstate} can be represented by a random tensor network (RTN) built out of a chain of $k$ (complex, non-Hermitian) Gaussian random tensors of dimension $D\times D$, where $D=e^{\SBPS}$, see Figure \ref{fig:TN}.\footnote{Earlier discussions of RTNs usually use Haar random tensors \cite{Hayden:2016cfa}, whereas here it is more natural to pick Gaussian random tensors. This does not change the statistics of normalized quantities we will compute and only affects the measure concentration properties, e.g., see the relation between the results in \cite{Akers:2022qdl} and \cite{Kar:2022qkf}.}
As it has been understood in AdS/CFT, such an RTN accurately represents both the boundary state as well as a coarse-grained description of the bulk geometry \cite{Swingle:2009bg,Almheiri:2014lwa,Pastawski:2015qua,Hayden:2016cfa}. On the boundary side, this RTN prepares a bipartite state on subregions $L$ and $R$, which are the two asymptotic boundaries at the two ends of the network. In the bulk picture, the random tensors correspond to heavy particles of different flavors that form a long wormhole, as shown in Figure \ref{fig:TN}. The RTN models a geometry where the Cauchy slice is a long wormhole that has $k+1$ extremal surfaces, each of which has roughly equal area (with the area having small fluctuations \cite{Dong:2018seb,Akers:2018fow,Dong:2019piw}). Each matter operator insertion sends a heavy particle into the bulk, which creates a bulge separating two extremal surfaces. In higher dimensions, the same model works in situations of spherical symmetry where we have spherical shells in the bulk \cite{Chandra:2022fwi,Sasieta:2022ksu,Balasubramanian:2022gmo,deBoer:2023vsm}. For instance, precisely this RTN model with pseudo-random tensors can be constructed in a 2D CFT to model the AdS$_3$ bulk \cite{Chandra:2023dgq}.

We are interested in computing the R\'enyi entropy of one of the two subregions, say $R$, of this system. To do so, we can import some technology from the RTN literature and discuss the corresponding bulk picture afterwards. We remark that our RTN calculation of the R\'enyi entropy will be exact provided that the operators $\tilde{O}_{\Delta}^{(i)}$ are Gaussian random matrices drawn from a complex Ginibre ensemble, and is therefore on the same page as the calculation of entropies in Sections \ref{sec:2}, \ref{sec:bulk-resolution}, and \ref{sec:matrixmodels}. In the following, we will schematically review and apply the rules to compute R\'enyi entropies in RTNs. For further details and a derivation of these rules, we refer the reader to the original literature \cite{Hayden:2016cfa}.

In analogy with the previous sections, let us define the density matrix of the right side to be $\rho_R\equiv \rho$. Notice that $\rho$ has exactly one insertion of $\tilde{O}_{\Delta}^{(i)}$ and one insertion of $(\tilde{O}_{\Delta}^{(i)})^\dagger$ for each index $i=1,..,k$. To compute $\overline{\tr\(\rho^n\)}$ in the RTN,\footnote{In this section we will use the overline to denote both averages over the random tensors in the network and quantities computed using the gravitational path integral.} one should consider $n$ copies of the network in Figure \ref{fig:TN} and $n$ copies of its complex conjugate, and appropriately glue them together at the endpoints according to the boundary conditions set by the replica trick. We then must sum over all possible contractions of the tensors in the network between the various replicas. One can reformulate this calculation in terms of a sum over permutations $g_x$ on $n$ elements applied at each tensor (labeled by $x$) in one copy of the network depicted in Figure \ref{fig:TN}, with boundary conditions at the endpoints of the network fixed by the replica trick. The computation of $\overline{\tr\(\rho^n\)}$ can then be mapped to the evaluation of the free energy in a model with an Ising-type action \cite{Hayden:2016cfa}. Each given set of permutations on the $k$ tensors is weighted by the corresponding on-shell action. Neighboring permutations $g_x$ and $g_y$ that do not align cost $d(g_x,g_y)\log D$ in action, where $d(g,h)$ is the Cayley distance given by the number of swaps required in going from $g$ to $h$. As we will discuss in more detail in Section \ref{sec:TNmapping}, from a bulk perspective, different sets of permutations $\{g_1,...,g_k\}$ are in one-to-one correspondence with bulk diagrams similar to those discussed in Section \ref{sec:bulk-resolution}. For simplicity, let us focus on the case of the R\'enyi-2 entropy, for which the combinatorics are simplest and most intuitive. In Appendix~\ref{app:RTNs}, we will discuss how to generalize to higher $n$. 

For the R\'enyi-2 entropy, because the calculation only involves two replicas, we can simply label the permutations as $\uparrow$ and $\downarrow$ to literally obtain an Ising model, with each tensor representing a spin. A permutation $\uparrow$ for a given tensor means that we are contracting it with the corresponding (Hermitian conjugate) tensor in the other replica. A permutation $\downarrow$ means that we are contracting it with the corresponding (Hermitian conjugate) tensor in the same replica. To formulate the puzzle, let us first compute the annealed R\'enyi-2 entropy 
\begin{equation}
     S_{A}^{(2)}(\rho) = -\log \frac{\overline{\tr\(\rho^2\)}}{\(\overline{\tr\(\rho\)}\)^2}
\end{equation}
in the limit $D\to\infty$ with $k$ finite. Let us start by computing the purity $\overline{\tr(\rho^2)}$. The boundary conditions set by the replica trick to compute the purity of the right side are $\uparrow$ at the right end of the network (which is glued to the other replica according to the replica trick) and $\downarrow$ at the left end (which is glued within the same replica):
\begin{equation}
    \inlinefig[5]{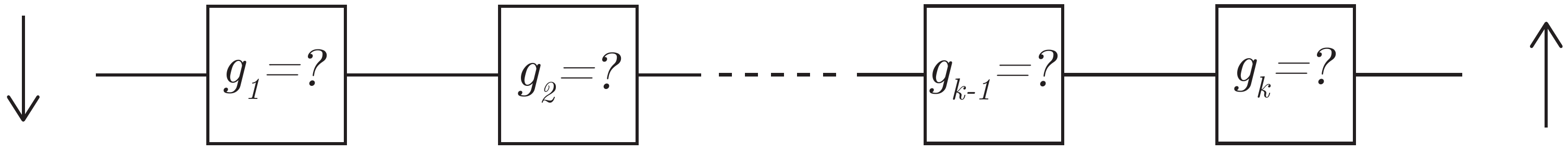}
\end{equation}
We must then sum over all possible configurations of permutations, in which each tensor can be labeled by $\uparrow$ or $\downarrow$. The purity is then given by
\begin{equation}
    \overline{\tr\(\rho^2\)} = \(P_{\Delta\to\infty}\)^{2k} D^2 \sum_{\{\vec{g}\}}D^{-d(\downarrow,g_1)-d(g_k,\uparrow)} \prod_{\<xy\>} D^{-d(g_x,g_y)},
\end{equation}
where the sum is over all possible configurations of permutations, the product is over all the $k-1$ internal edges $\<xy\>$ in the network between two tensors $x$ and $y$, $g_x$ take values $\uparrow$ or $\downarrow$, and $d(g_x,g_y)$=1 if $g_x\neq g_y$ and 0 otherwise. The factor $D^{-d(\downarrow,g_1)-d(g_k,\uparrow)}$ accounts for the boundary conditions (namely the two external edges at the left and right ends of the network), the factor $D^2$ is an overall normalization factor which matches our choice of normalization in the previous sections as well as previous RTN literature \cite{Hayden:2016cfa}, and we also included a factor $(P_{\Delta\to\infty})^n$, with $n=2$ number of replicas, for each tensor in the network to match the JT (super)gravity propagator.

There is a very simple, graphical rule to evaluate the contribution of each configuration of permutations to $\overline{\tr\(\rho^2\)}$. For a given configuration, draw domain walls separating regions within which all spins are aligned, e.g.
\begin{equation}
    \inlinefig[5]{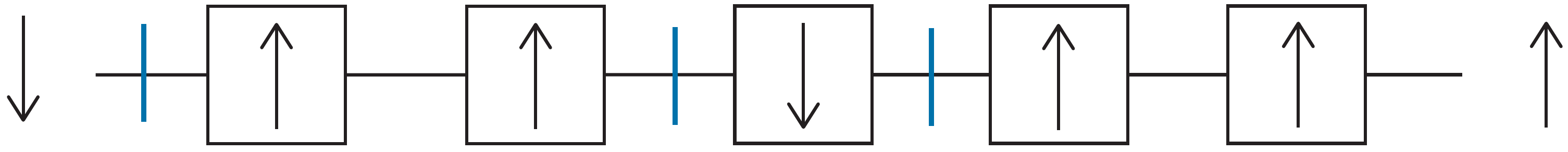}
\end{equation}
where we have drawn the domain walls in blue. Then count the number $W$ of domain walls. The contribution of the configuration to the purity is then simply given by $\(P_{\Delta\to\infty}\)^{2k}D^{2-W}$. Notice that for $\overline{\tr(\rho^2)}$, for which the boundary conditions at the two ends of the network are opposite, there must always be an odd number of domain walls. How many configurations with $W$ domain walls are there? Since we can place domain walls on any of the $k+1$ edges of the network, the answer is simply $\binom{k+1}{W}$. In the $D\to\infty$, $k$ finite limit we are considering for now, the only configurations contributing to the purity at leading order in $D$ are those with a single domain wall separating the two boundaries. There are $k+1$ such configurations, with the domain wall sitting on any of the $k+1$ edges. One such configuration is, for instance,
\begin{equation}
    \inlinefig[5]{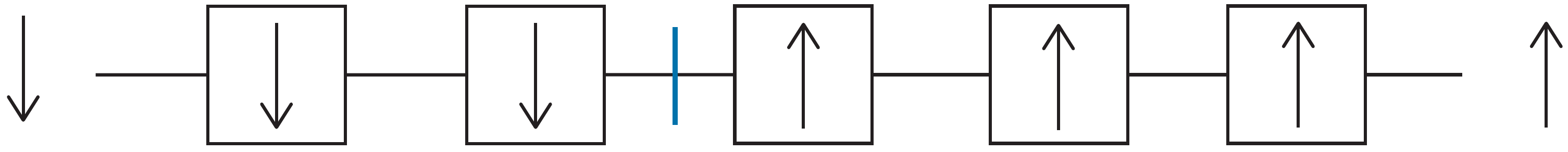}
\end{equation}
This has a natural bulk interpretation in terms of the $k+1$ degenerate RT surfaces in the geometry described by the tensor network, see Figure \ref{fig:TN}. At leading order in $D$, the purity is then simply given by
\begin{equation}
    \overline{\tr\(\rho^2\)}\approx\(P_{\Delta\to\infty}\)^{2k}(k+1)D\,,
\end{equation}
where the $\approx$ sign here means that we are taking the $D\to\infty$ limit while keeping $k$ finite.

The calculation of $\overline{\tr(\rho)}$ is very simple. From the bulk point of view, the answer is immediately clear: there is only one possible (and planar) configuration, leading to $\overline{\tr(\rho)}=\(P_{\Delta\to\infty}\)^{k}e^{\SBPS}$. In the RTN, the same result can be obtained by doing a one-replica calculation. Equivalently, we can compute directly $\(\overline{\tr(\rho)}\)^2$ by considering our two-replica Ising model, setting boundary conditions $\downarrow$ at both ends of the network, and only considering the trivial configuration of permutations in which all tensors are contracted within the same replica, namely $g(x,y)=\downarrow$ for any $x,y$. This configuration needs no domain walls, and the result is therefore
\begin{equation}
    \(\overline{\tr(\rho)}\)^2=\(P_{\Delta\to\infty}\)^{2k}D^2\,,
    \label{eq:TNannealednorm}
\end{equation}
which immediately implies $\overline{\tr(\rho)}=(P_{\Delta\to\infty})^{k}D$, matching the bulk result quoted above. Notice that this result is exact, independent of the value of $D$ or $k$. We can now finally compute the annealed entropy, obtaining
\begin{equation}
    S_A^{(2)}(\rho)=\log D-\log (k+1)\,,
    \label{eq:TNannealed}
\end{equation}
which is decreasing as a function of $k$. Thus, we arrived at a puzzle very similar to that studied in the previous sections: the (annealed) entropy becomes negative if we extrapolate our result to $k=O\(D\)$. We remark that, as we will see, the combinatorics of permutations for the RTN matches exactly that of bulk matter diagrams in JT (super)gravity. Thus, we could have arrived at the same puzzle by simply carefully studying bulk matter diagrams as in Section \ref{sec:2} and only keeping the leading order (disk) diagrams for every quantity.

Notice that the puzzle arises for $k=O(D)$ instead of $k=O(D^{2/3})$. To understand this difference, we can gain some insight about why this negative entropy is occurring from the boundary perspective. Unlike the case of identical operators, where the two-sided state eventually becomes arbitrarily close to a tensor product of two matter ``ground states'' for a very large number of operator insertions, it is more complicated to understand in this example. The density matrix, which computes the annealed entropy, is not given by a standard product of a single random matrix, but instead by
\begin{equation}
    \rho= \(\tilde{O}_\Delta^{(k)}\)^\dagger...\(\tilde{O}_\Delta^{(1)}\)^\dagger\tilde{O}_\Delta^{(1)}...\tilde{O}_\Delta^{(k)}\,,
\end{equation}
since we multiply together $k$ freely independent random matrices chosen with the same statistics. The spectrum of the density matrix $\rho$ is given by a free multiplicative convolution over the spectra of the matrices $\tilde{O}_\Delta^{(i)}$. The corresponding distributions of eigenvalues of $\rho$ have been worked out in Refs.~\cite{collins2010random,penson2011product} and are called the Fuss-Catalan distributions $\sigma_k(x)$.\footnote{These were labeled $\pi^{(k)}(x)$ and $P_k(x)$ respectively in \cite{collins2010random} and \cite{penson2011product}.} Here, we use a slightly different normalization, and the spectrum of the density matrix is instead given by $D \sigma_k(D x)$. This spectrum is supported in the range $\[0,\frac{(k+1)^{k+1}}{D k^k}\]$. For large $k$, the upper limit behaves like $e k/D$. Thus, we see that when $k=O(D)$, there is a range of eigenvalues that goes beyond 1. $D \sigma_k(D x)$ has sufficient support on eigenvalues larger than 1 to make the annealed entropy (as well as any other version of the entropy computed at leading order) negative for sufficiently large $k$.

Relatedly, some more intuition can be gained from understanding the action of random Gaussian matrices on an arbitrary vector. When we act with the same random matrix, it lengthens a vector exponentially along the direction of the eigenvector with the largest eigenvalue, with subleading corrections dictated by the spacing between the second largest and largest eigenvalue. When we instead act with independent random matrices, there is no simple characterization for the resulting limiting vector, but it is nevertheless true that there is an exponential growth of the size of the vector with Lyapunov exponents that have been computed previously \cite{newman1986distribution}. One way to see this is to focus on the factor $\mu_i$ by which the vector's length grows at each step $i$. Then, $\mu_i$ are iid variables and we are interested in the distribution of $\sum_i \log \mu_i$, which can be obtained using the central limit theorem. Computing the relevant Lyapunov exponents, one finds that they have a spacing of $O\(\frac{1}{D}\)$, unlike the spectrum of a given random matrix which has a spacing of $O\(\frac{1}{D^{2/3}}\)$ near the edge. This further helps to understand why we see the puzzle arise at $k=O(D)$ in this case.

Similar to the previous sections, to solve the puzzle, we can take a double-scaling limit $D\to\infty$, $k\to\infty$ with $\alpha\equiv k/D=O(1)$, and study the semi-quenched entropy
\begin{equation}
    S_{SQ}^{(2)}(\rho)=-\log\frac{\overline{\tr\(\rho^2\)}}{\overline{\(\tr \rho\)^2}}\,.
\end{equation}
Let us start again by computing $\overline{\tr\(\rho^2\)}$. In this limit, all terms in the sum over configurations of permutations are of the same order, and we must resum them all. Using the previous results, we obtain
\begin{equation}
   \begin{aligned}
        \overline{\tr\(\rho^2\)}_{\text{res.}}&=\(P_{\Delta\to\infty}\)^{2k}\sum_{\substack{W=1\\W \text{odd}}}^{k+1}\binom{k+1}{W}D^{2-W}=\frac{D^2}{2}\left[\(1+\frac{1}{D}\)^{k+1}-\(1-\frac{1}{D}\)^{k+1}\right]\\
        &\approx \(P_{\Delta\to\infty}\)^{2k}D^2\sinh(\alpha)\,,
 \end{aligned}
   \label{eq:TNtrrho2}
\end{equation}
where, in the first line, we used the binomial theorem; in the second line, we took the double-scaling limit (signaled by the $\approx$ sign) and used the limit definition of the exponential.

\begin{figure}
    \centering
    \includegraphics[width=0.8\linewidth]{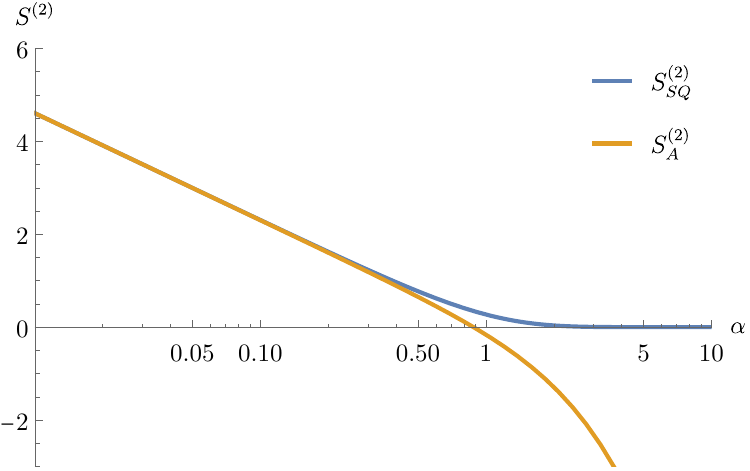}
    \caption{The annealed (yellow, Eq.~\eqref{eq:TNannealedres}) and semi-quenched (blue, Eq.~\eqref{eq:TNSQres}) R\'enyi-2 entropies obtained from our RTN calculation as a function of $\alpha=k/D$. For a large number of insertions, the annealed entropy becomes negative, even when including all-genus contributions. The semi-quenched entropy stays positive and vanishes for $k\to\infty$ thanks to connected wormhole contributions.}
    \label{fig:TNmoneyplot}
\end{figure}

Similarly, we can compute $\overline{\(\tr \rho\)^2}$. The only difference is in the boundary conditions at the two ends of the network, which are now $\downarrow$ at both ends. This implies that only configurations with an even number of domain walls contribute. We thus obtain
\begin{equation}
    \begin{aligned}
        \overline{\(\tr \rho\)^2}_{\text{res.}}&=\(P_{\Delta\to\infty}\)^{2k}\sum_{\substack{W=0\\W \text{even}}}^{k+1}\binom{k+1}{W}D^{2-W}=\frac{D^2}{2}\left[\(1+\frac{1}{D}\)^{k+1}+\(1-\frac{1}{D}\)^{k+1}\right]\\
        &\approx \(P_{\Delta\to\infty}\)^{2k}D^2\cosh(\alpha)\,.
    \end{aligned}
    \label{eq:TNSQden}
\end{equation}
We can finally compute the semi-quenched entropy,
\begin{equation}
    S_{SQ}^{(2)}(\rho)=\log\[\coth(\alpha)\]\,,
    \label{eq:TNSQres}
\end{equation}
which agrees with the result \eqref{eq:TNannealed} for small $\alpha$ (in which only the configurations of permutations that contribute at leading order in $D$ matter), but never becomes negative and instead vanishes as we take $\alpha$ large. See Figure \ref{fig:TNmoneyplot}. We can also check that, in the double-scaling limit, the re-summation of configurations with more domain walls does not rescue positivity for the annealed entropy. In fact, using equations \eqref{eq:TNannealednorm} and \eqref{eq:TNtrrho2}, we obtain
\begin{equation}
    S_{A}^{(2)}(\rho)_{\text{res.}}=-\log\[\sinh(\alpha)\]\,,
    \label{eq:TNannealedres}
\end{equation}
which becomes negative as we increase $\alpha$. We conclude that, just like in the puzzle discussed in Sections \ref{sec:2}, \ref{sec:bulk-resolution}, and \ref{sec:matrixmodels}, it is the re-summation of the additional terms included in the denominator of the semi-quenched entropy (i.e., the terms with $W>0$ in Eq.~\eqref{eq:TNSQden}) that rescues positivity and solves the puzzle. As we will see shortly, from a bulk perspective, these configurations precisely correspond to cylinder geometries connecting the two boundaries, leading to the same bulk interpretation of the solution of the puzzle as in the previous sections and in \cite{Antonini_Iliesiu_Rath_Tran_2025}.

Before discussing the relationship between RTN configurations of permutations and bulk matter diagrams, let us make two final comments. First, the puzzle of this section can be generalized, still using RTN techniques, to the case in which the operators $\tilde{O}_{\Delta}^{(i)}$ are Hermitian. To do that, one needs to consider an RTN where the random tensors are generated using random orthogonal matrices, similar to that used in \cite{Akers:2025ahe}. We discuss this briefly in Appendix~\ref{app:RTNs}. Second, we can comment on the relation to the case where all the operators are taken to be the same, similar to the previous sections. We could also do that calculation from the RTN perspective. However, this is now a much more complicated calculation in the tensor network language, because it involves $k n$ copies of the same tensor and would thus involve an Ising-like model on the permutation group on $k n$ elements, and we will not pursue it. For orthogonal random tensors (corresponding to Hermitian operators $\tilde{O}_\Delta$), this would be completely equivalent to studying the setup of the previous sections. From an RTN perspective, the resolution would be quite similar, with the only different feature being that the paradox arises at $k=O\(D^{2/3}\)$. This occurs due to fluctuations in the length of the Cauchy slice coming from self-contractions of tensors within the same bra or ket. In the case of different matter operators, there are no such self-contractions, hence the length of the Cauchy slice is always $O(k)$.

\subsection{Mapping between RTN and bulk matter diagrams}
\label{sec:TNmapping}

In this section, we will show that the combinatorics of bulk matter diagrams for non-Hermitian operators $\tilde{O}_\Delta^{(i)}$ precisely matches the combinatorics of the RTN. In particular, each choice of configuration of permutations in the RTN corresponds precisely to one matter diagram in JT (super)gravity. For simplicity, let us focus again on the R\'enyi-2 entropy calculation. Related discussions of the connection between the replica trick in RTN and fixed-area geometries in the gravitational path integral can be found in e.g. \cite{Penington:2019kki,Akers:2022zxr,Penington:2022dhr}.

Let us start by introducing a different graphical notation for the computation of $\overline{\tr(\rho^2)}$, which explicitly shows the multiple copies of the tensor network preparing the state \eqref{eq:TNstate} and its conjugate. To make the relation with bulk diagrams clearer, we also include an asymptotic boundary and wiggly ``matter lines'' anchored at the boundary and ending at the location of the tensors. The boundary conditions imposed by the replica trick lead to
{\small
\begin{equation}\inlinefig[3]{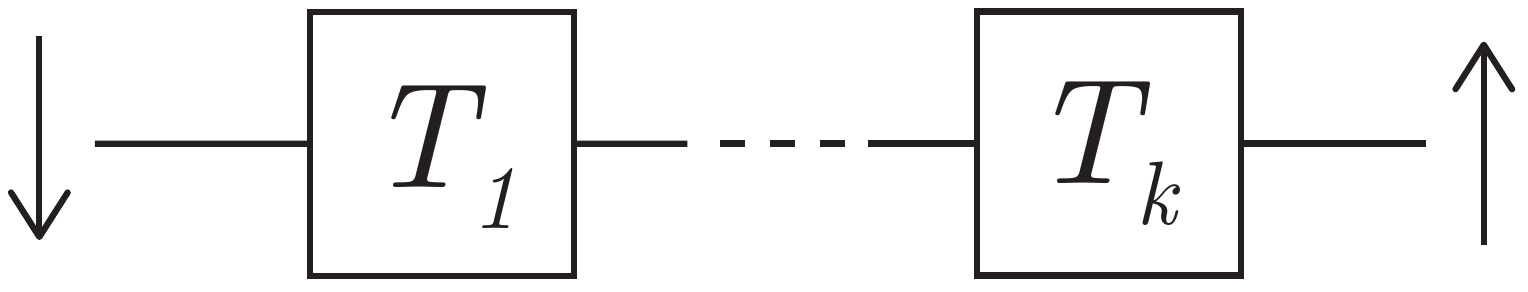} =  \inlinefig[12]{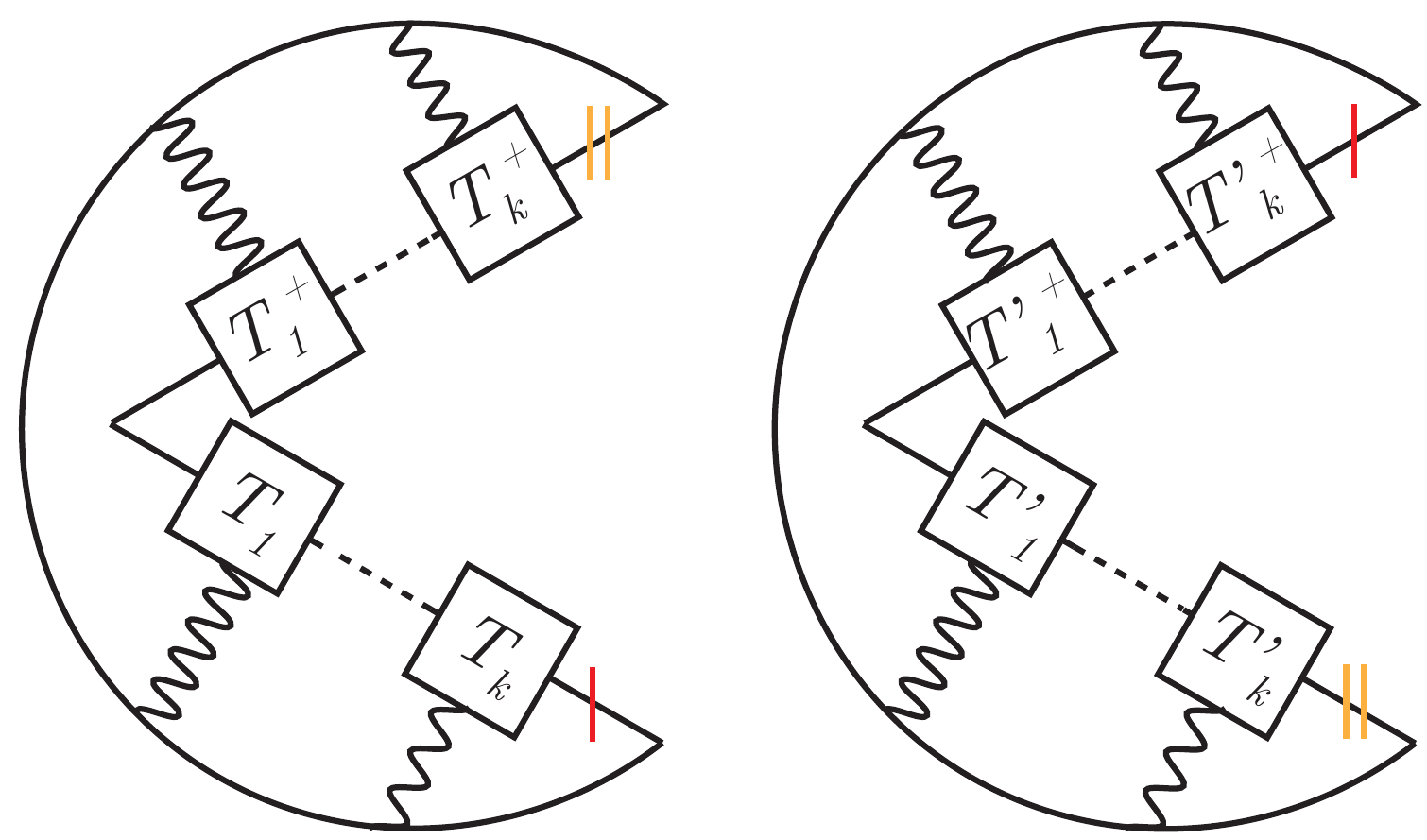} =\quad   \inlinefig[14]{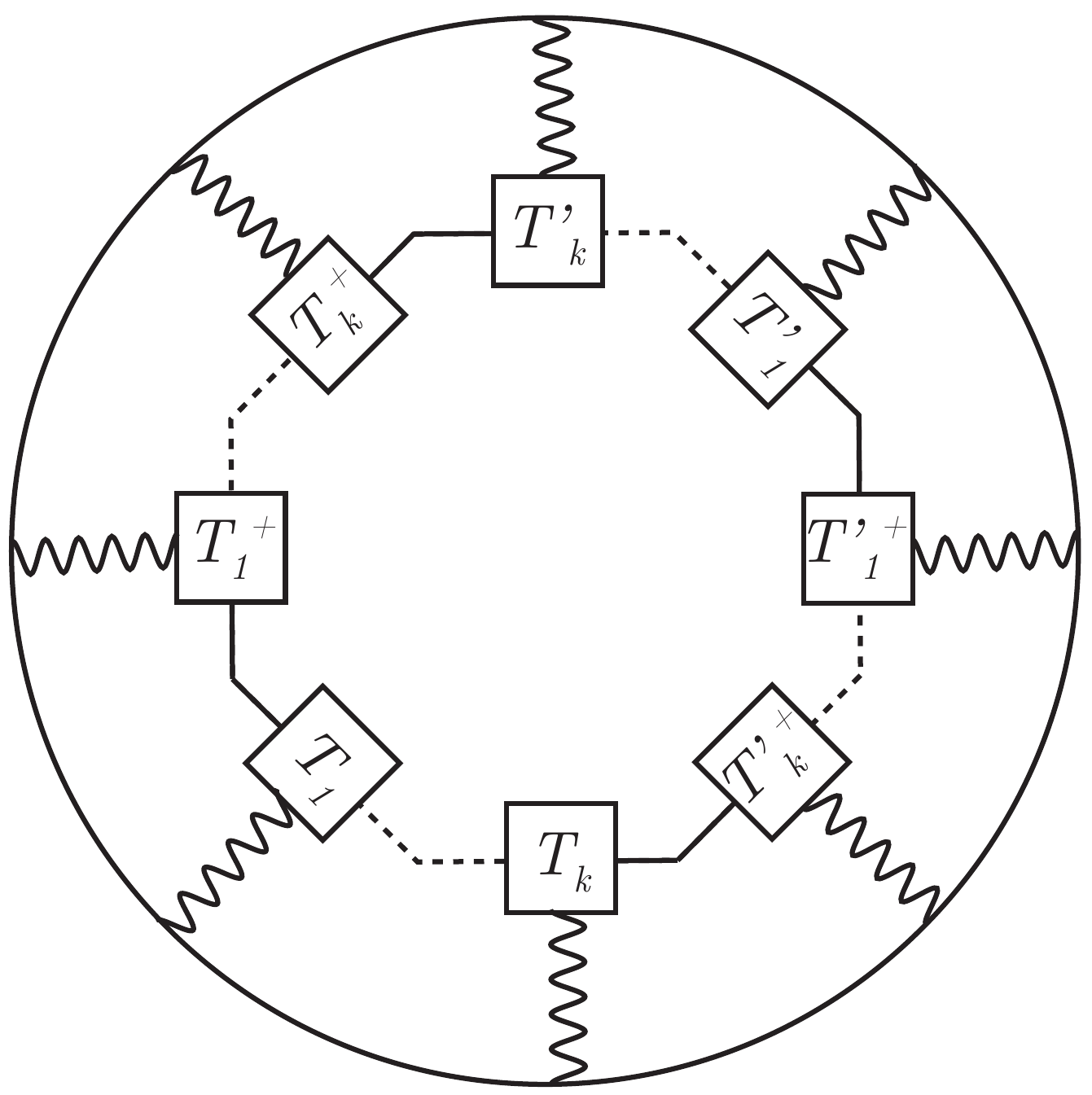}  
\end{equation}}where in the middle ``Pac-Man'' figure we show the gluing of the two replicas imposed by the replica trick, in the right figure we show the glued replicas, and we used primes to distinguish tensors in the second replica from tensors in the first replica.

In order to compute $\overline{\tr(\rho^2)}$, we must now sum over all possible contractions of the tensors. For a given tensor $T_i$, a choice of permutation $\downarrow$ corresponds to contracting the tensors labeled by $i$ with their counterparts in the same replica, while a choice of permutation $\uparrow$ corresponds to contracting the tensor between the two replicas. In the gravitational path integral, these correspond to different gluings of geometries, whose action matches the RTN calculation as long as there is a fixed-area projection or a microcanonical projection as in Section~\ref{sec:bulk-resolution}. For example, for $k=2$ this leads to the following diagrams:
\begin{equation}
\begin{aligned}
        &\inlinefig[3]{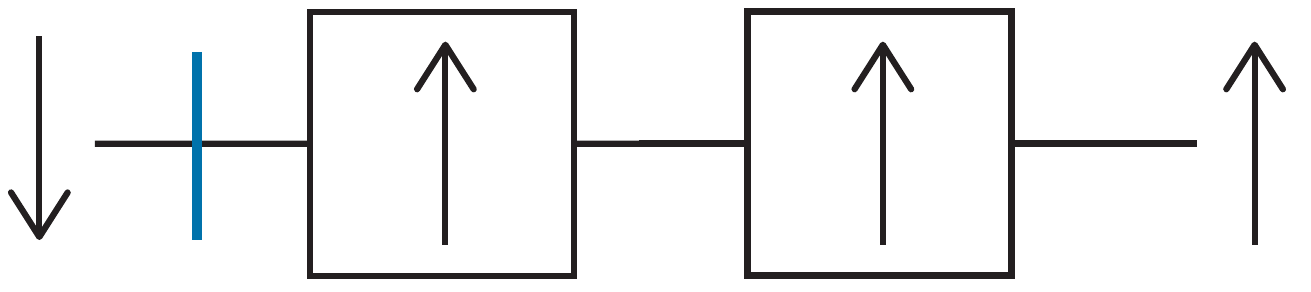} =\quad  \inlinefig[12]{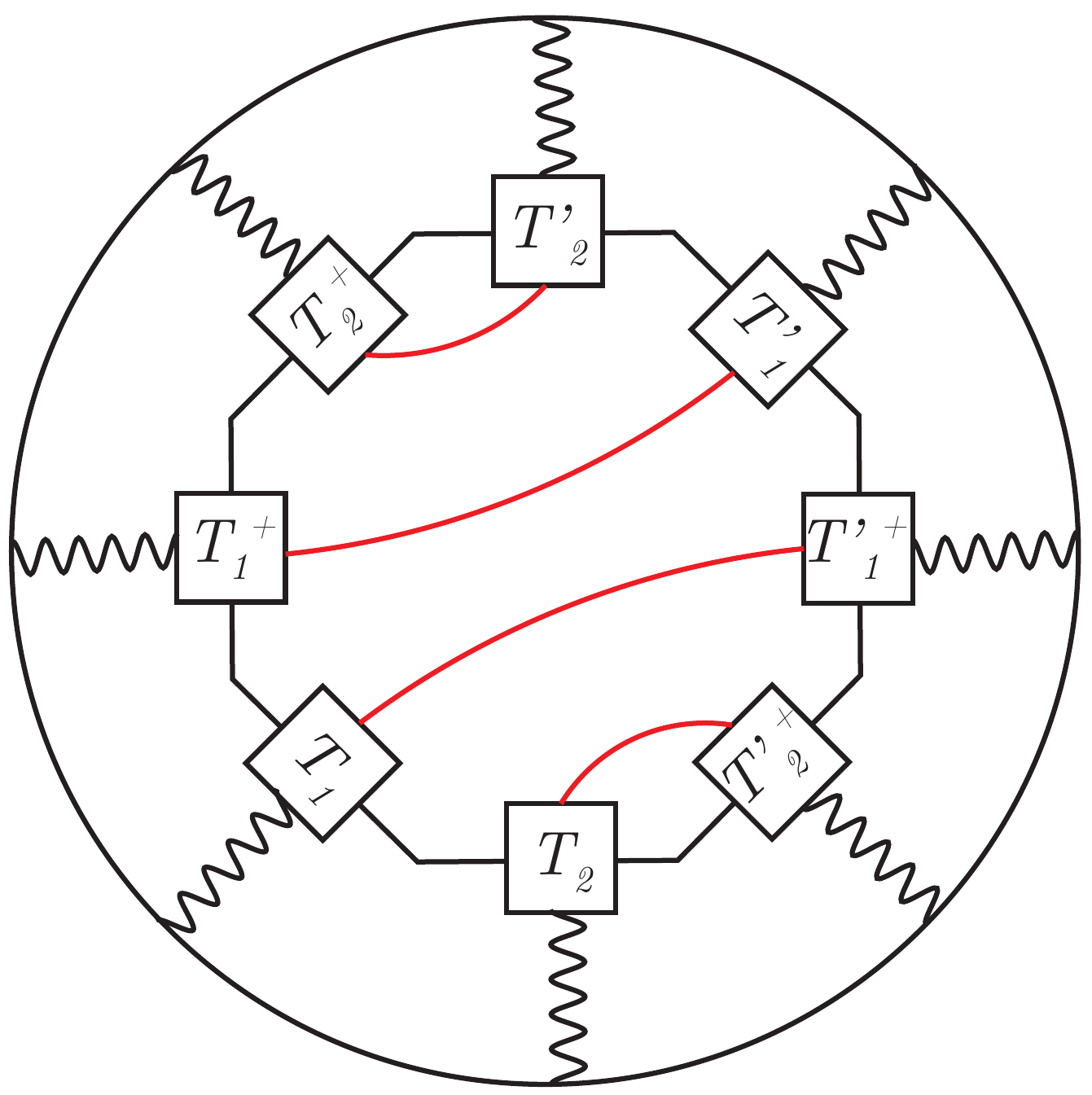}\quad=\quad  \inlinefig[12]{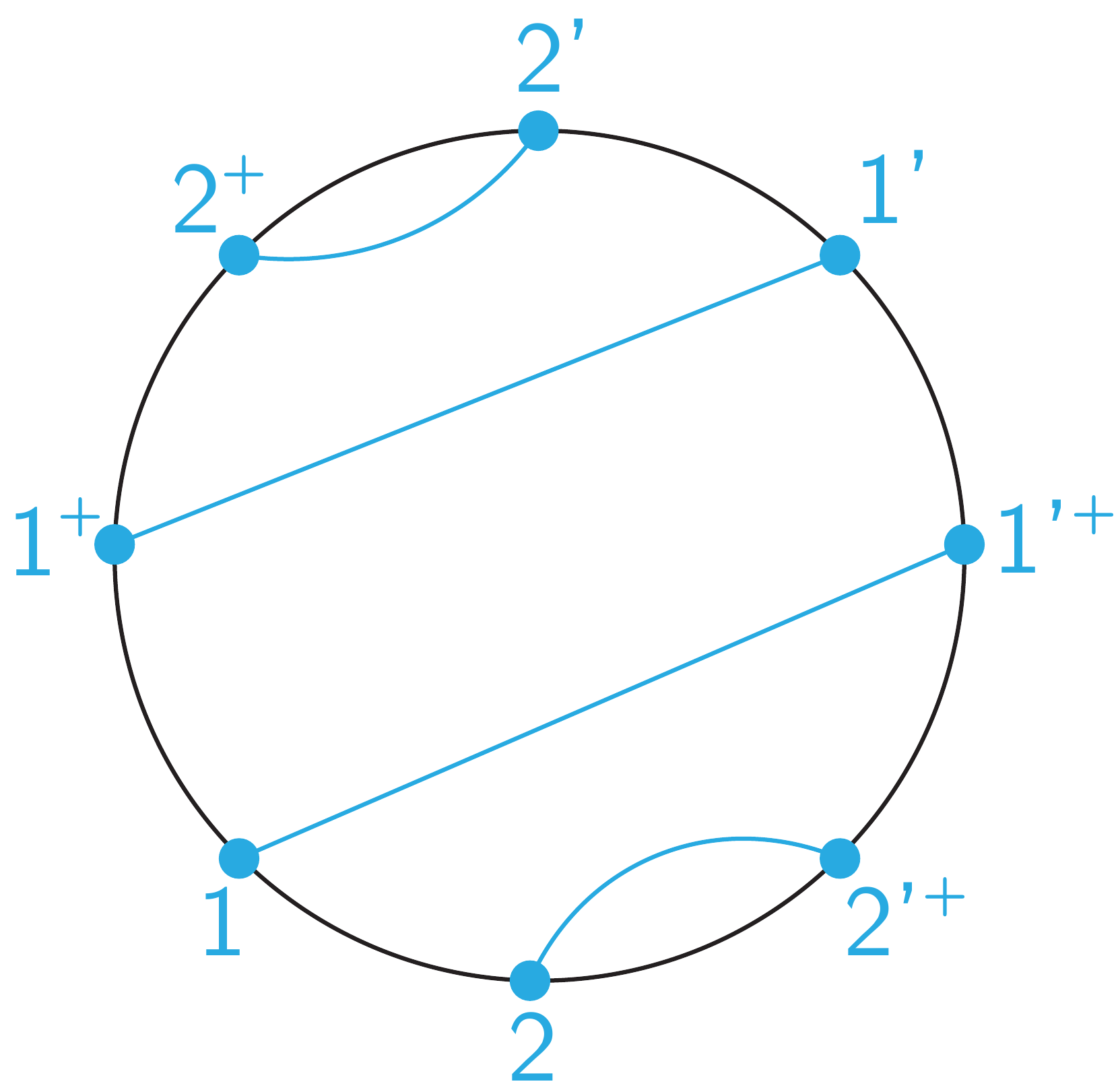},\\
        &\inlinefig[3]{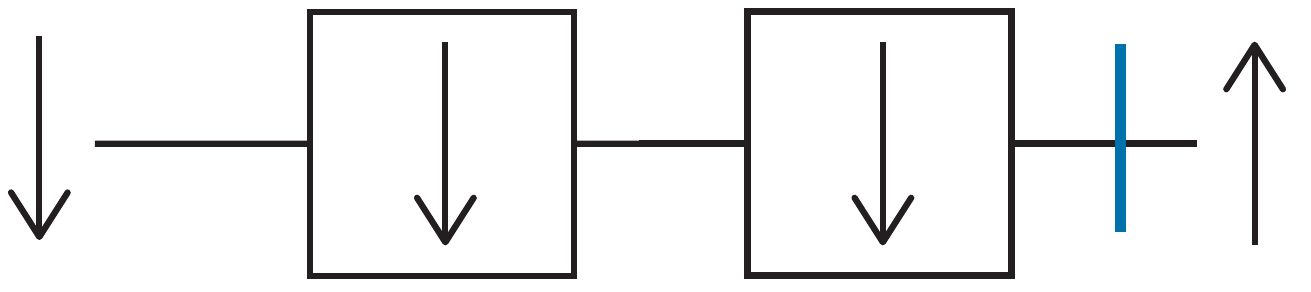} =\quad  \inlinefig[12]{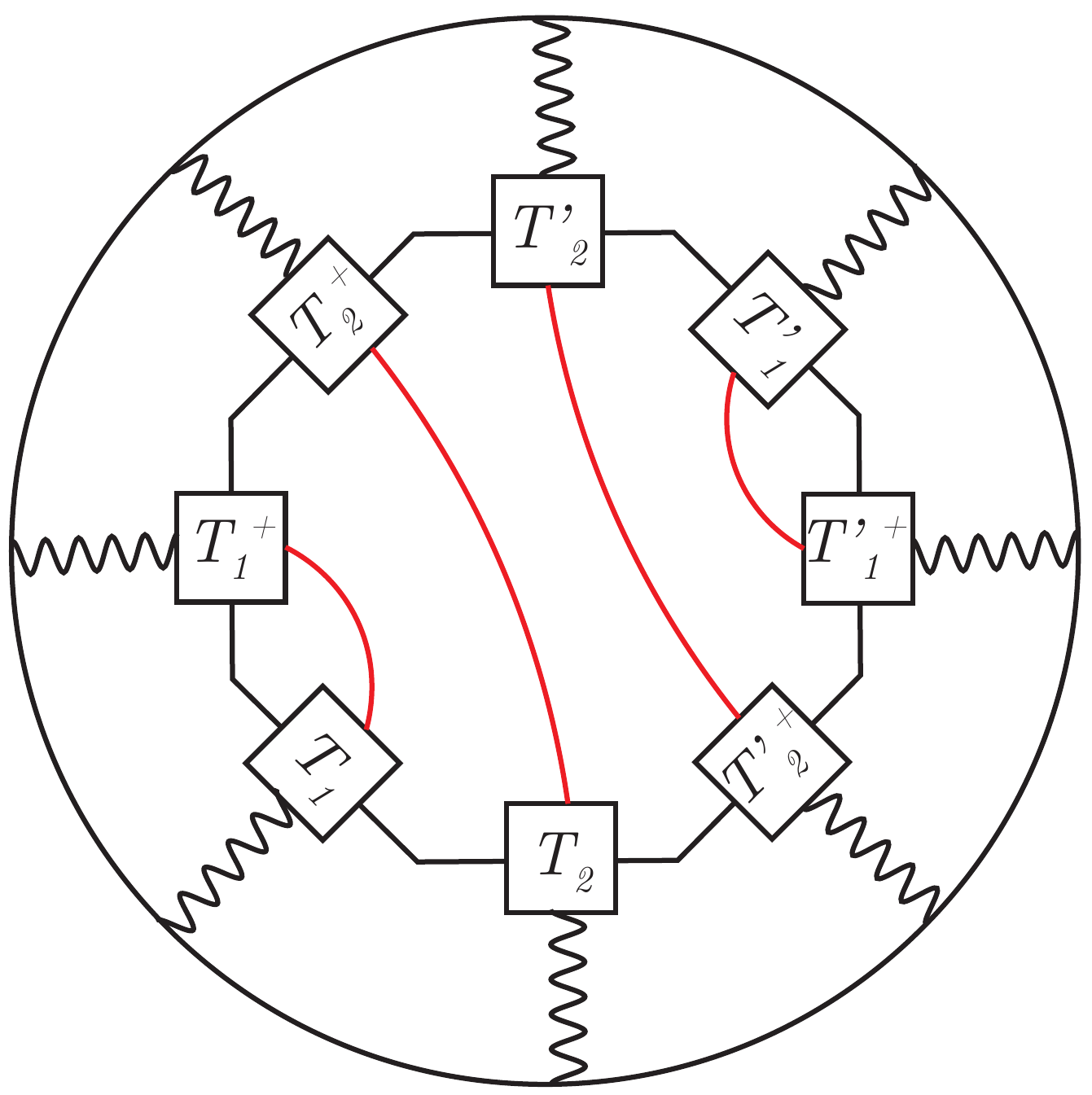}\quad=\quad  \inlinefig[12]{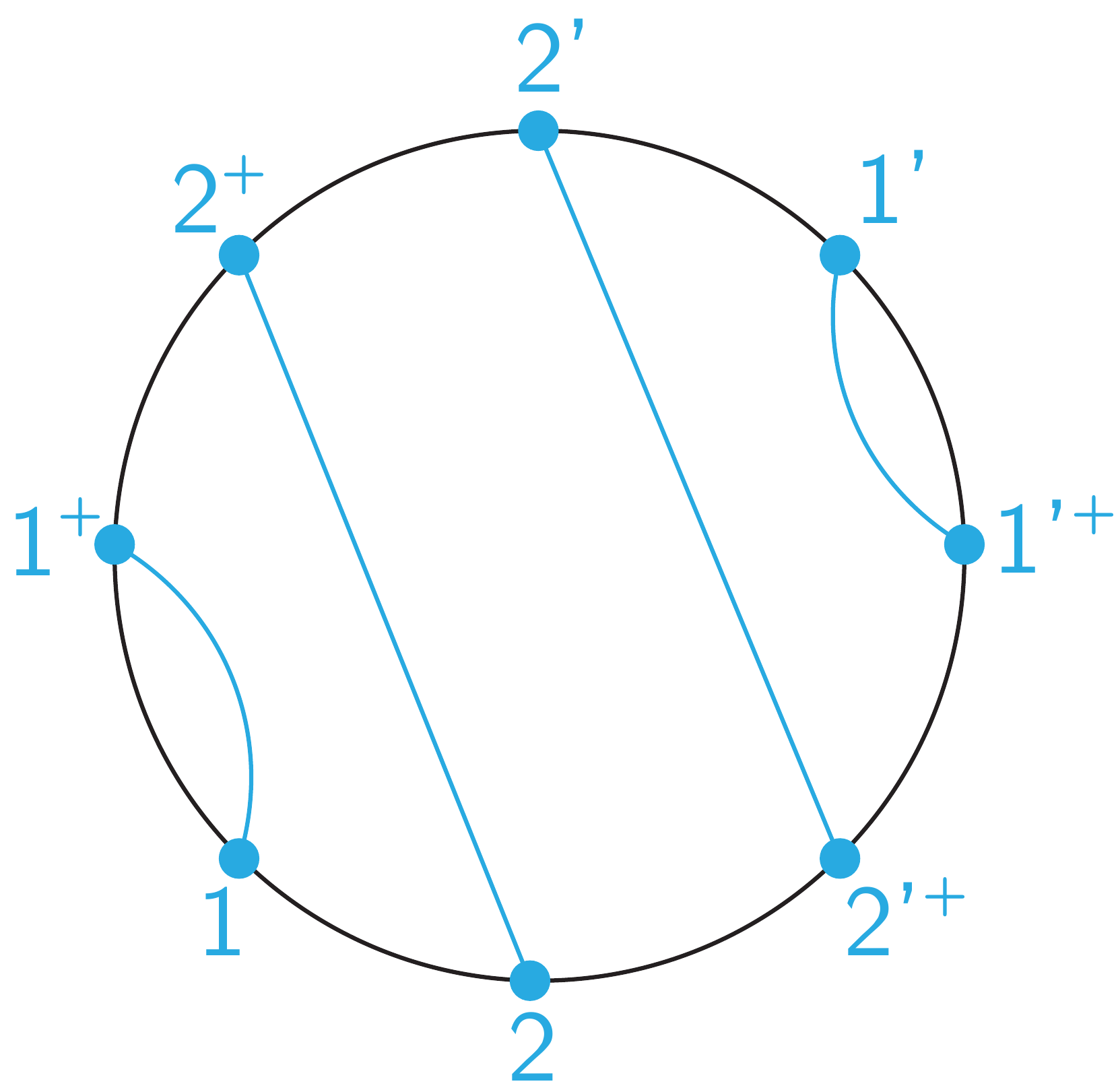},\\
        &\inlinefig[3]{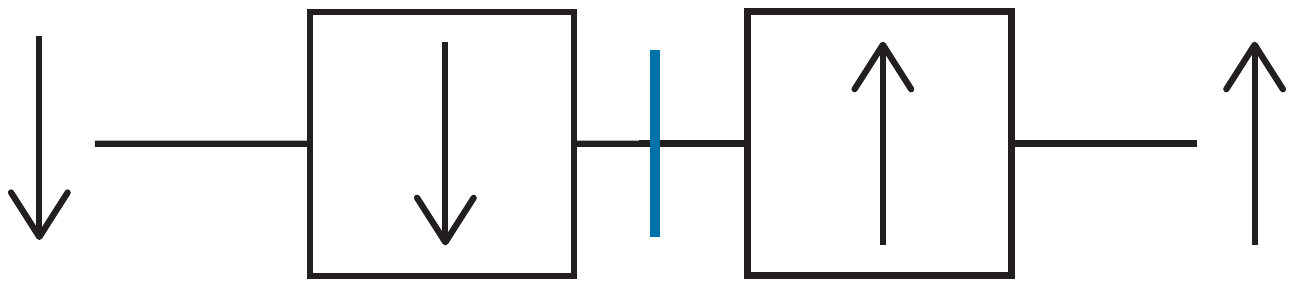} =\quad  \inlinefig[12]{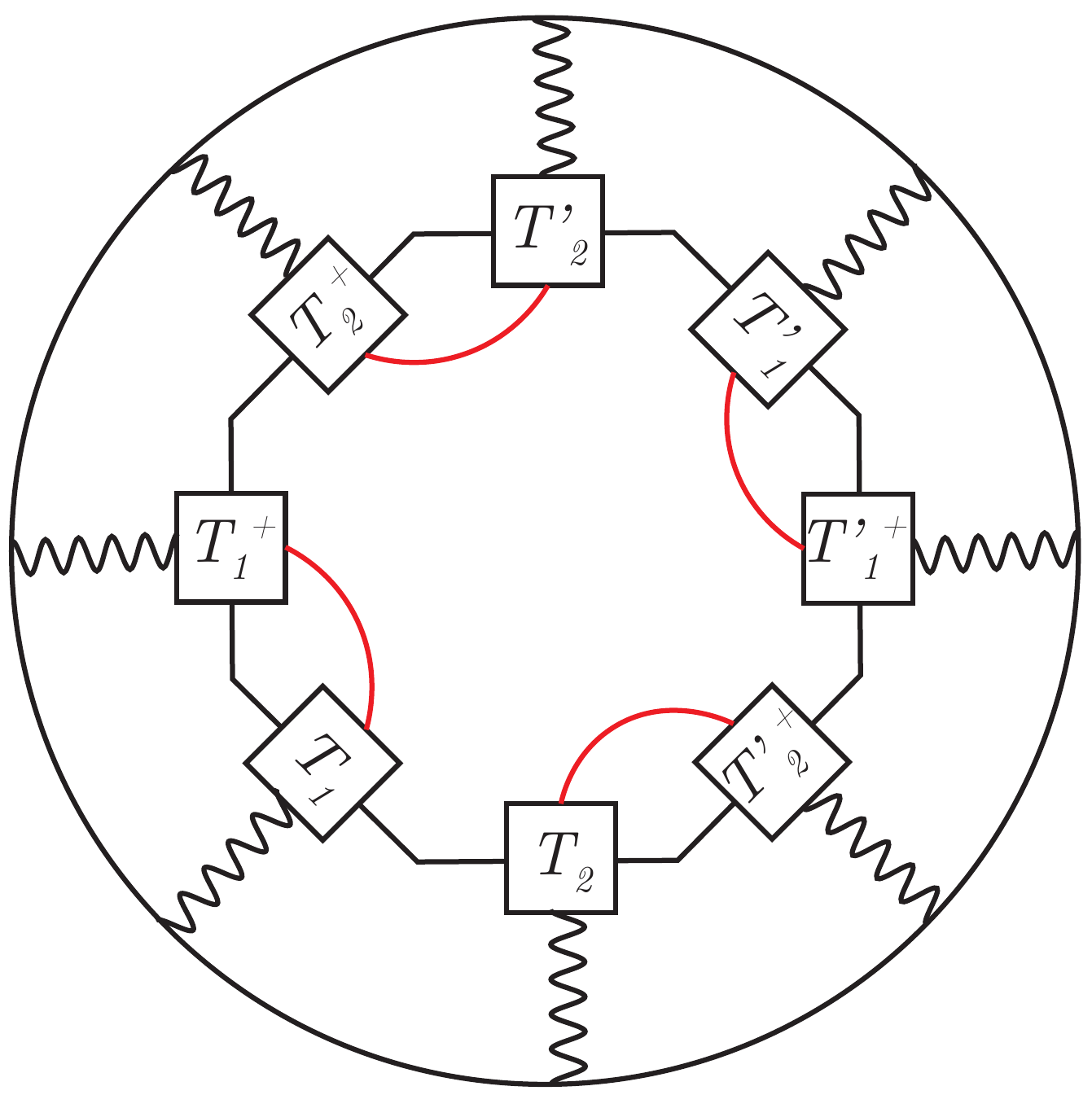}\quad=\quad  \inlinefig[12]{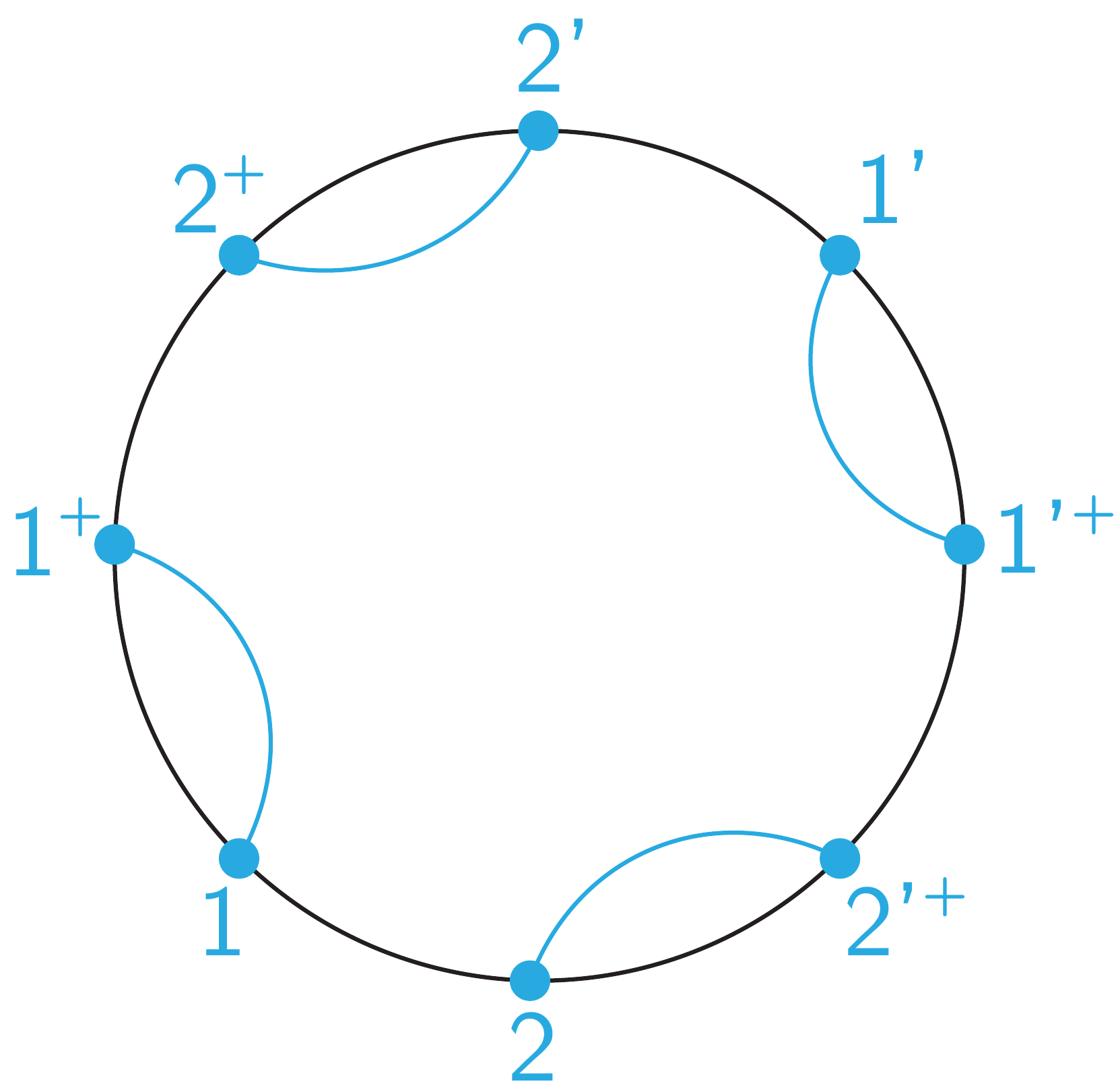},\\
        &\inlinefig[3]{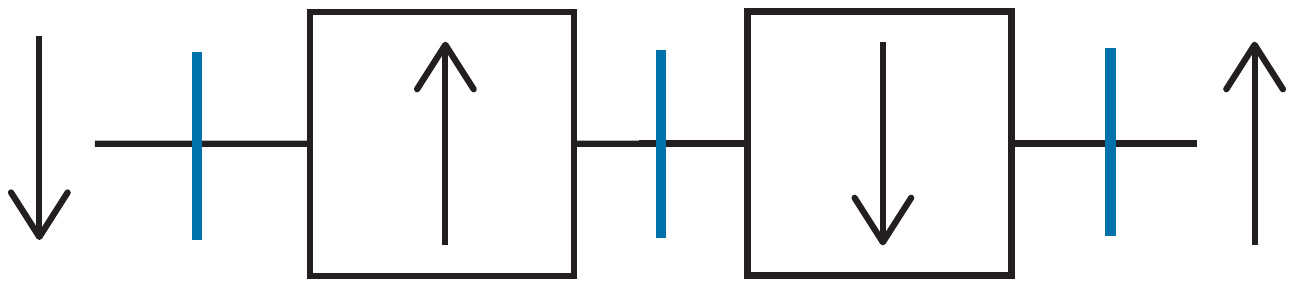} =\quad  \inlinefig[12]{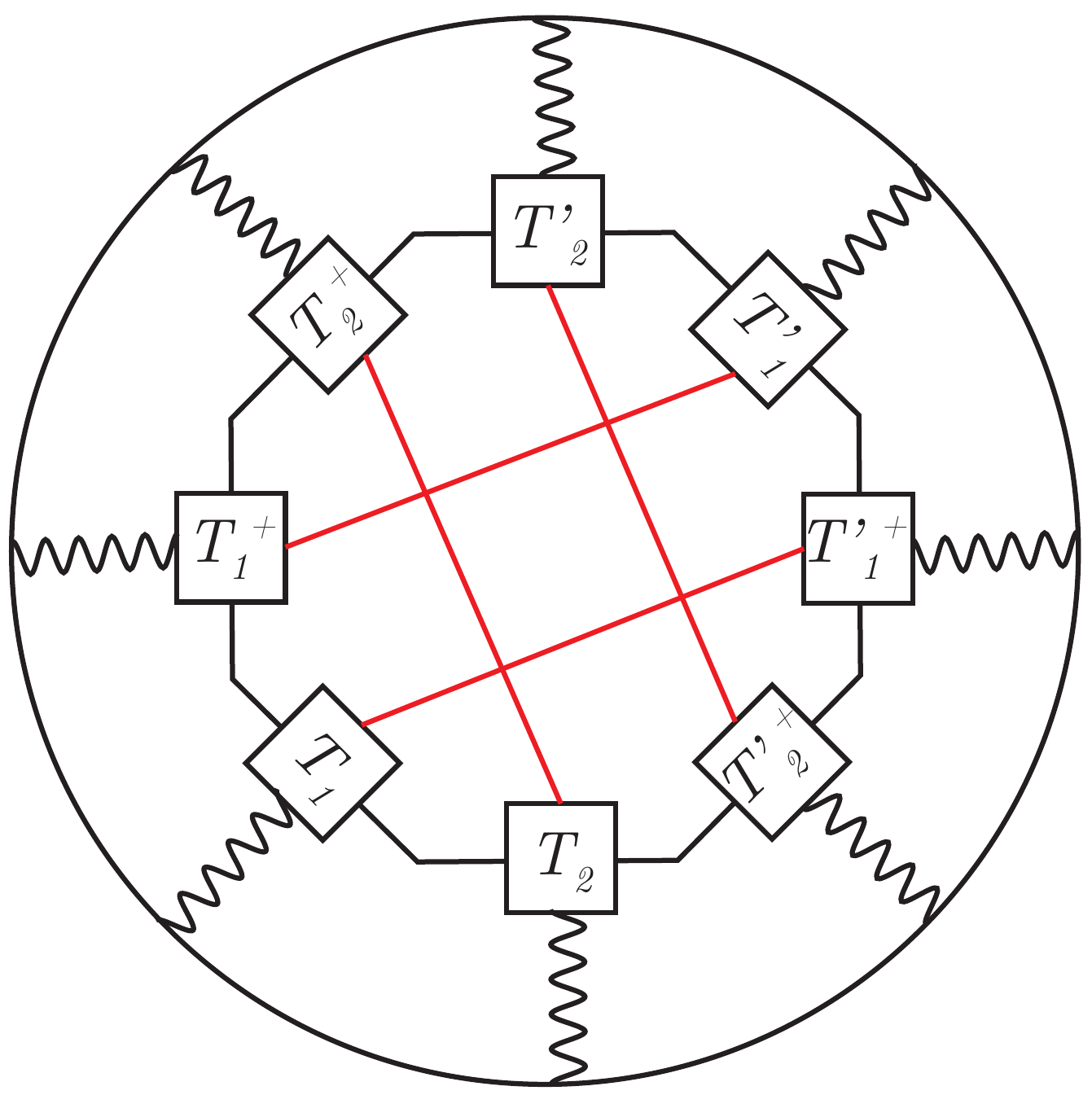}\quad=\quad  \inlinefig[10]{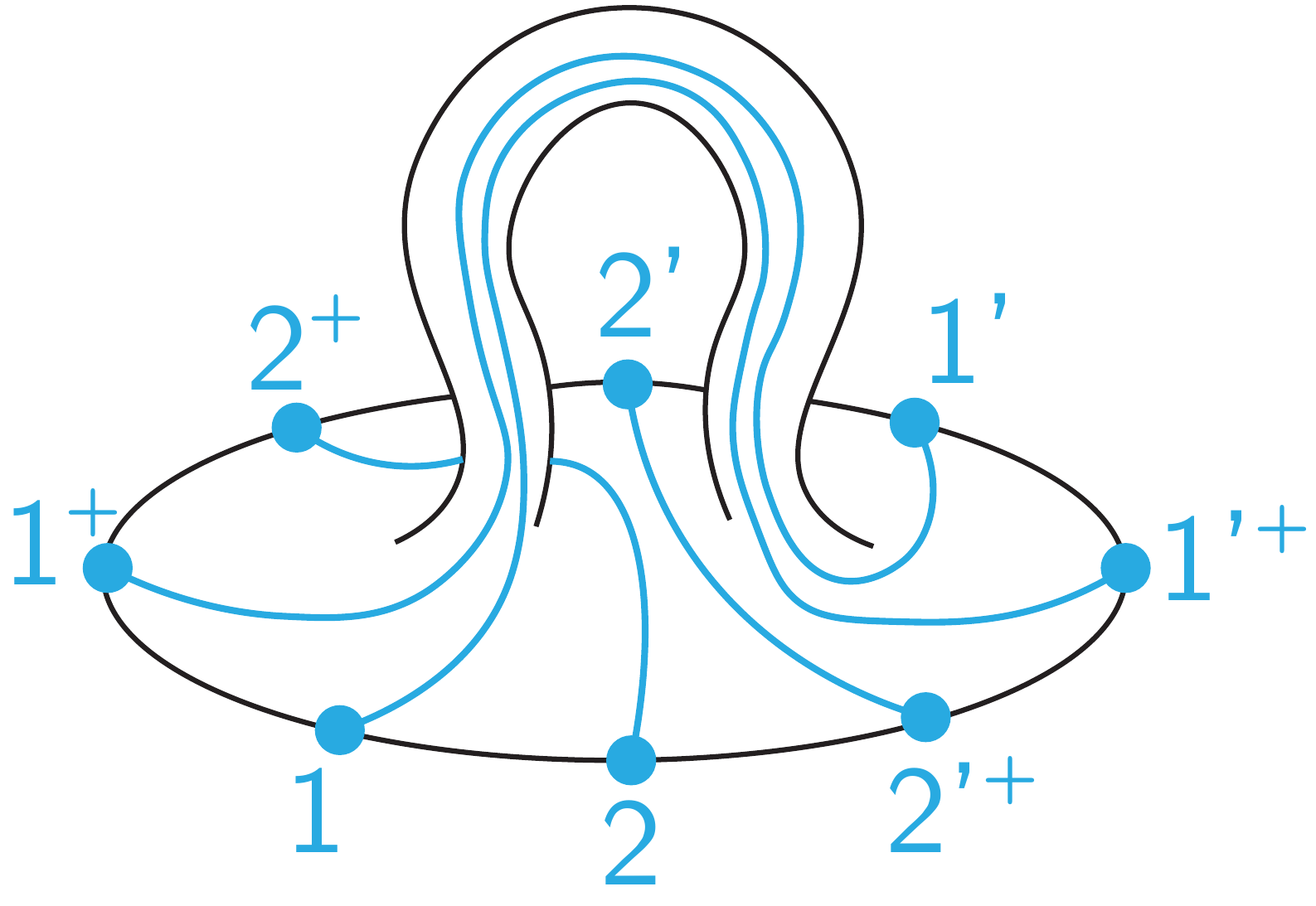}.
\end{aligned}
\end{equation}
where the rightmost diagrams are the corresponding bulk diagrams in JT (super)gravity coupled to matter. Notice that the tensor contractions, which correspond to choices of configurations of permutations in the Ising model, exhaust all possible matter diagrams for non-interacting non-Hermitian operators $\tilde{O}_{\Delta}^{(i)}$, including higher genus diagrams. The exact same procedure can be used to match matter diagrams to RTN configurations of permutations for any value of $k$ and for generic R\'enyi-$n$ entropies with $n>2$. For $n>2$, one must consider permutations on $n$ elements, corresponding to all possible contractions of a given tensor across the $n$ different replicas. Just like in Section \ref{sec:bulk-resolution}, diagrams with ``empty handles'' vanish in the BPS case and, in the double-scaling limit, only correct the answer for the semi-quenched entropy at subleading order in $e^{-S(\mathfrak{E})}$ (i.e., $1/D$) for the non-SUSY JT gravity case. Accordingly, they are not accounted for in the RTN calculation.

Finally, let us study, using the same approach, the match between RTN configurations of permutations and bulk matter diagrams in the computation of $\overline{\(\tr\rho\)^2}$. The only difference is in the choice of boundary condition $\downarrow$ at the right end of the network. This leads to the boundary conditions:
\begin{equation}
 \inlinefig[3]{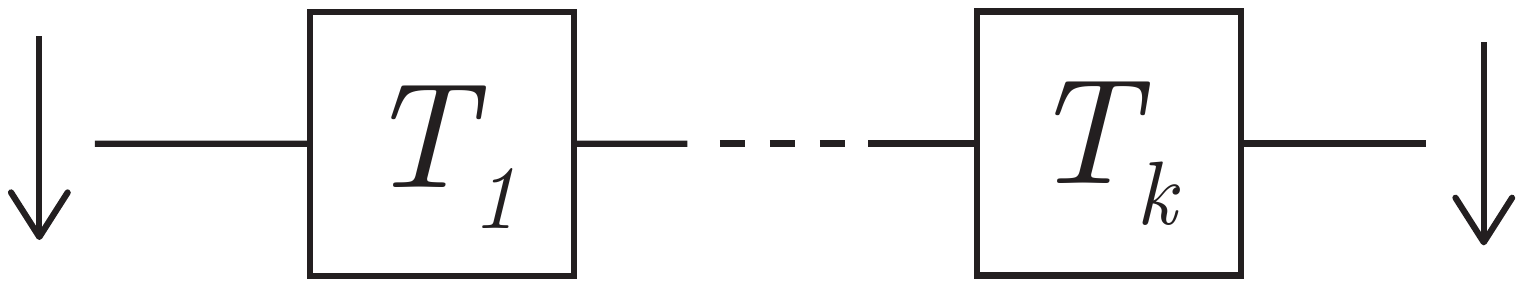} =\quad  \inlinefig[12]{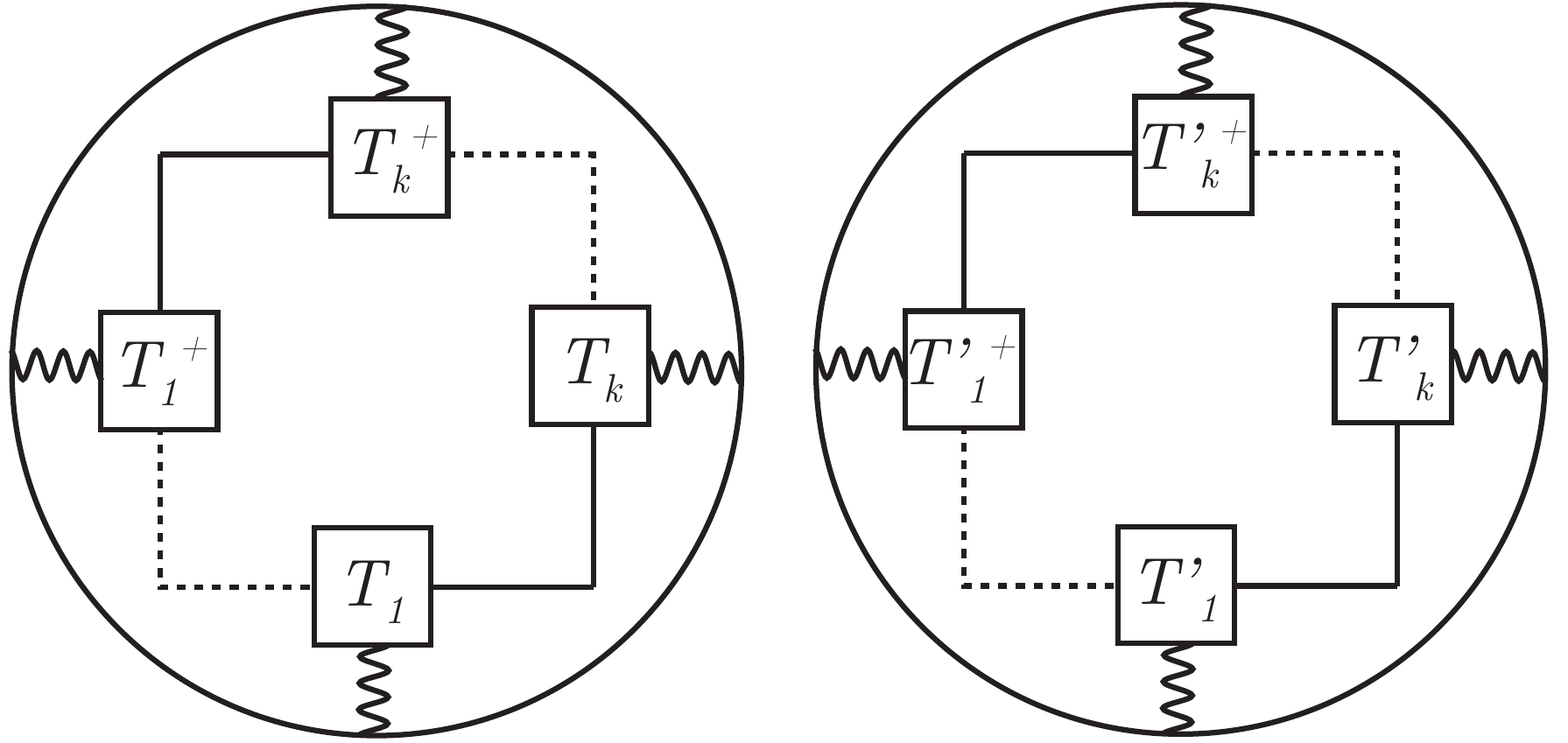}
\end{equation}
Choosing permutations $\downarrow$ or $\uparrow$ for each tensor again corresponds to contracting the tensor within the same replica or between the two replicas, respectively. For $k=2$, the different possible configurations of permutations thus lead to the following diagrams:
\begin{equation}
\begin{aligned}
        &\inlinefig[3]{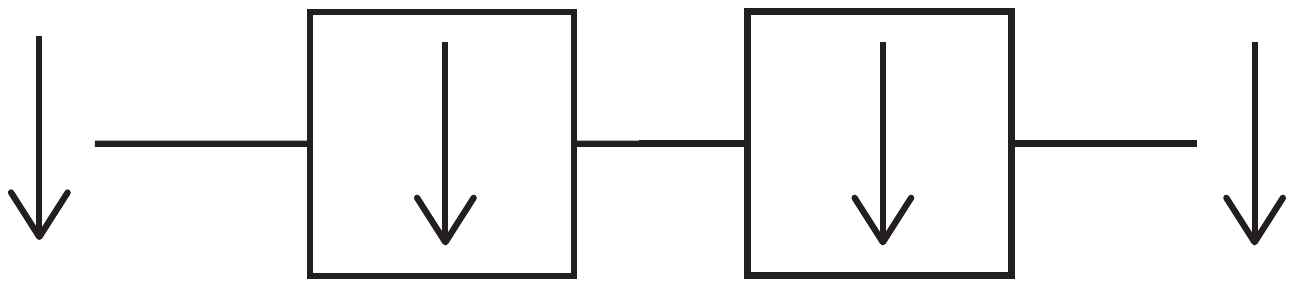} =\quad  \inlinefig[8]{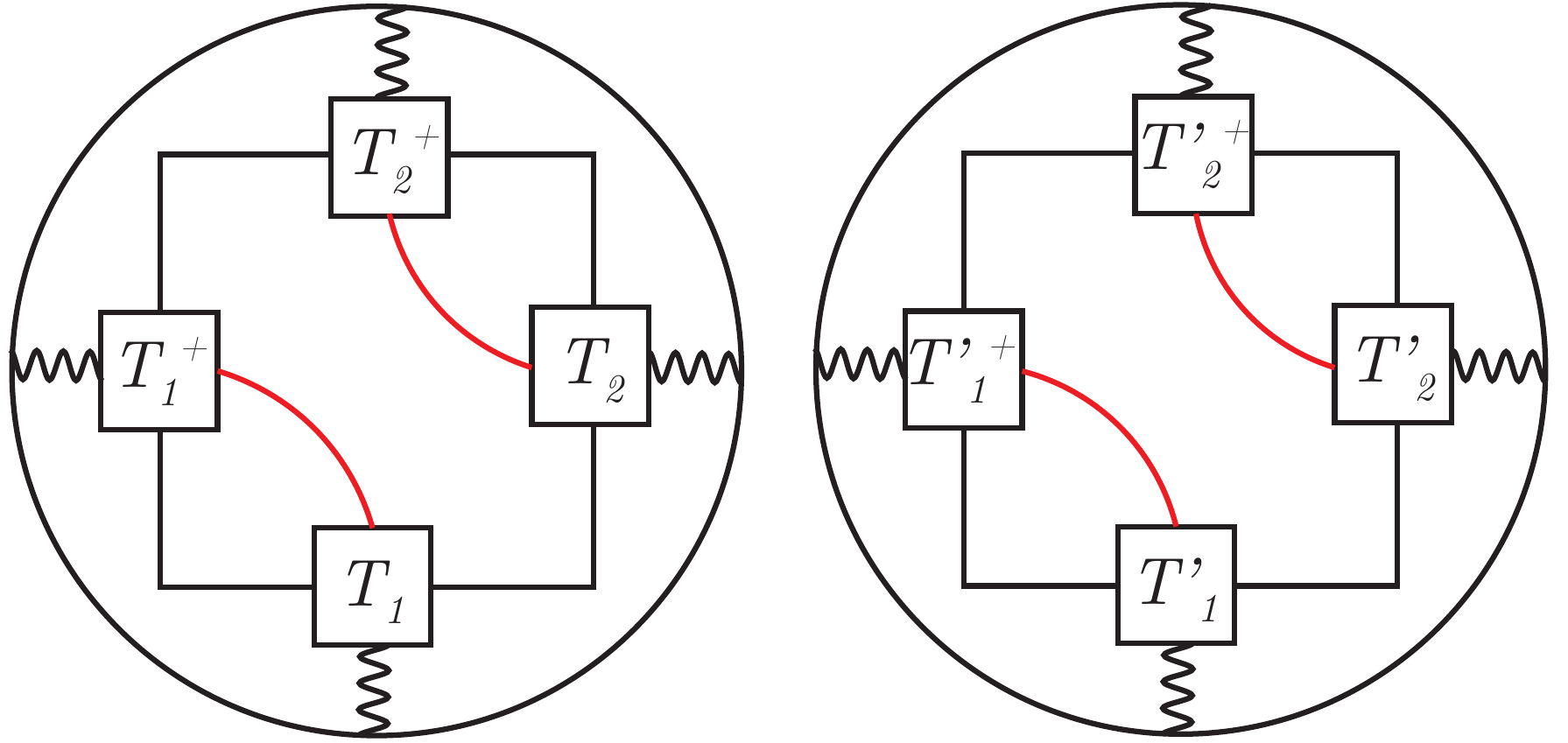}\quad=\quad\inlinefig[8]{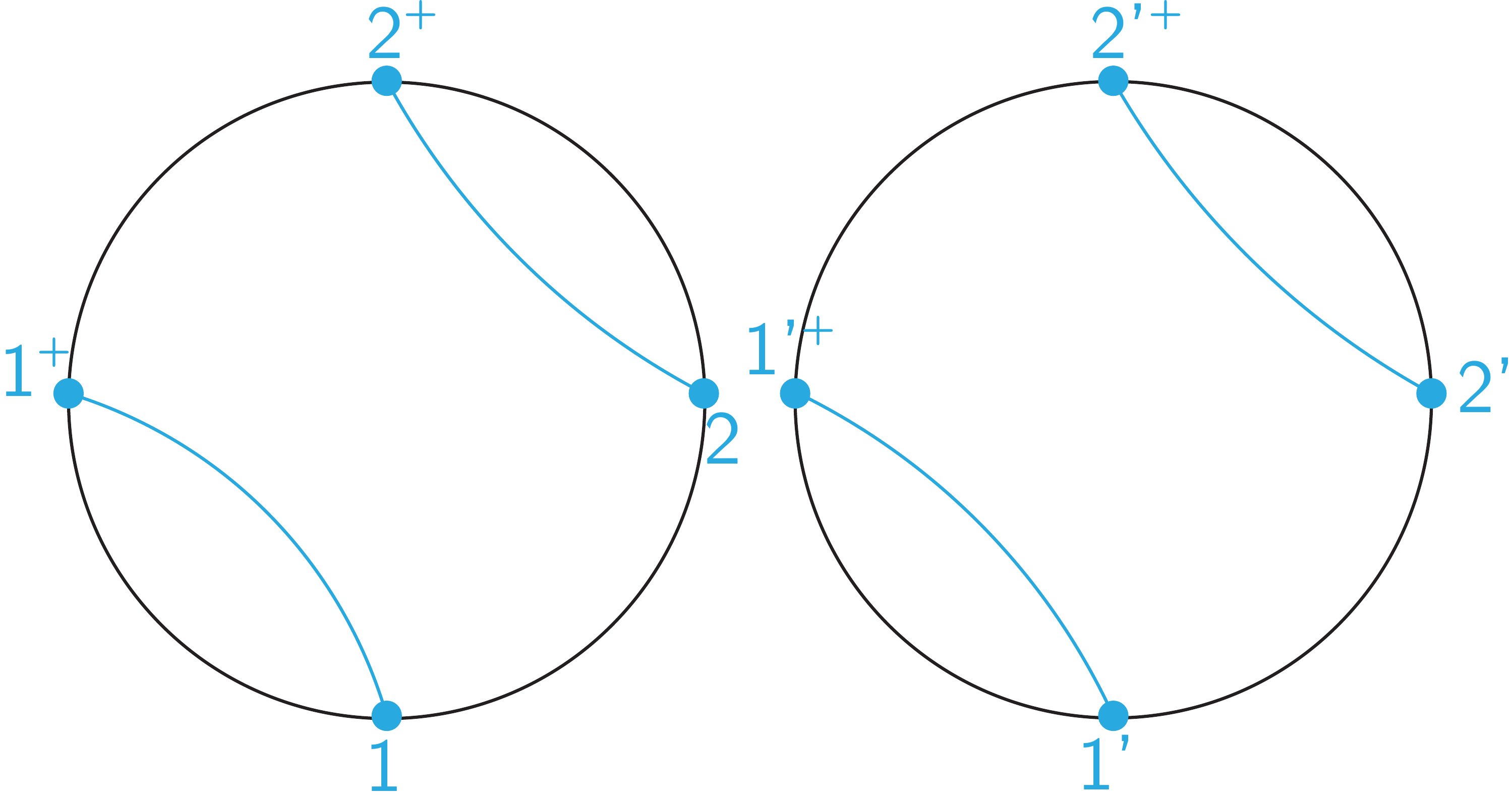},\\
        &\inlinefig[3]{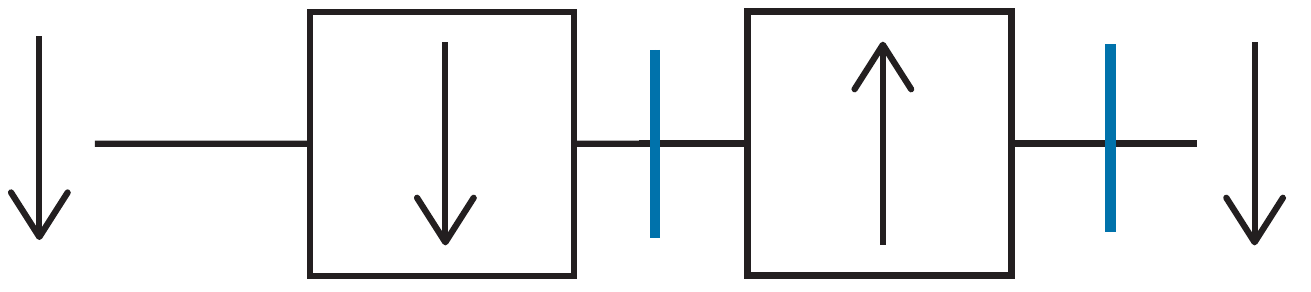} =\quad  \inlinefig[8]{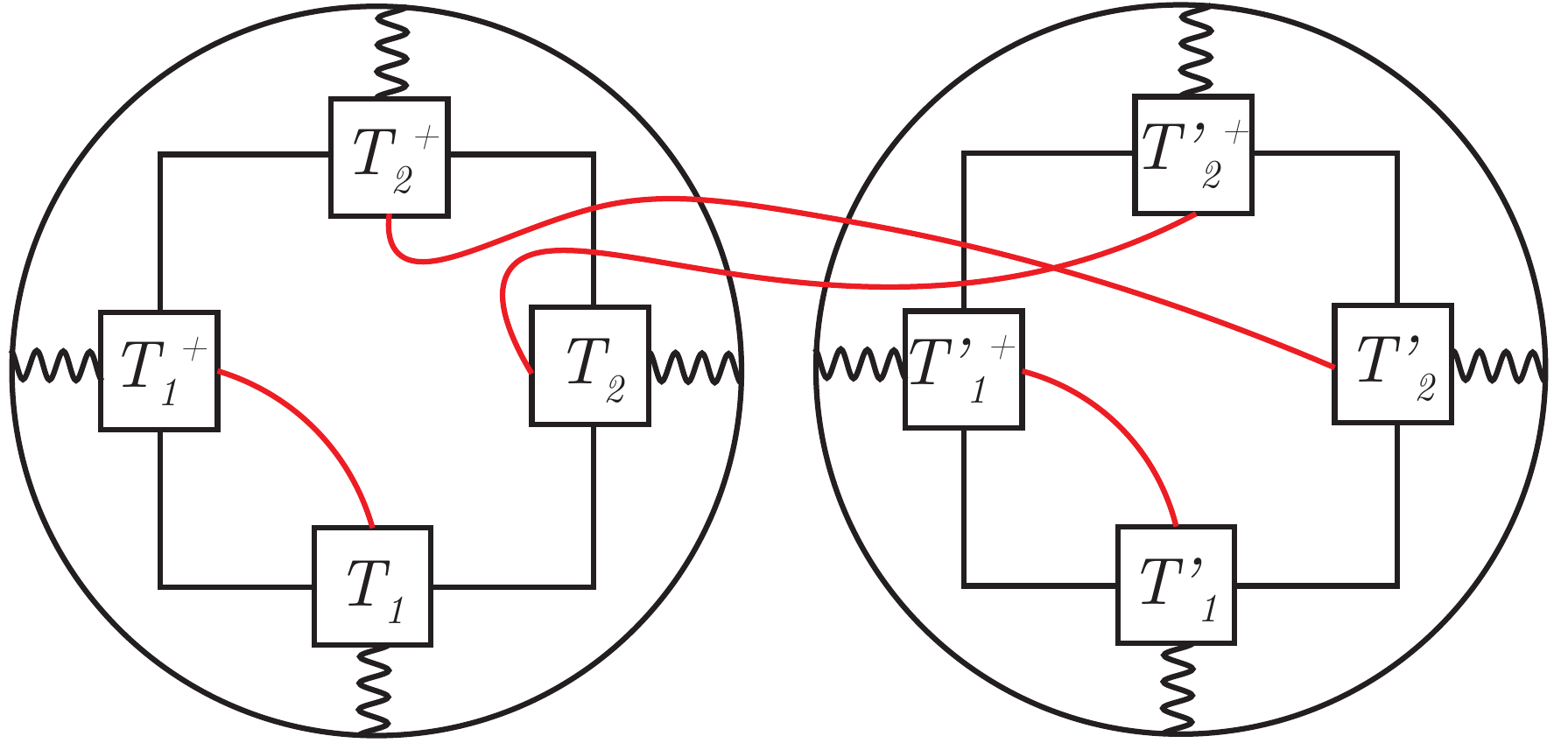}\quad=\quad\inlinefig[8]{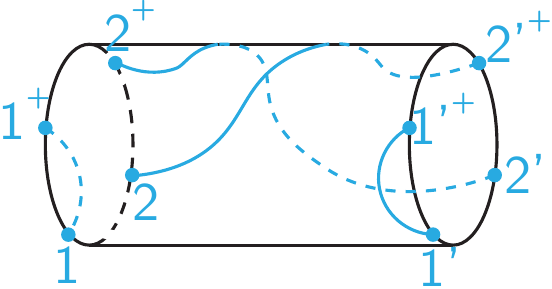},\\
        &\inlinefig[3]{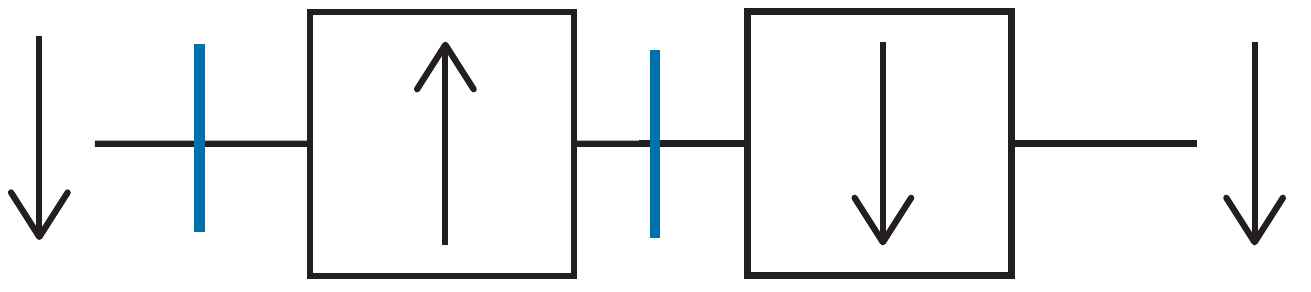} =\quad  \inlinefig[8]{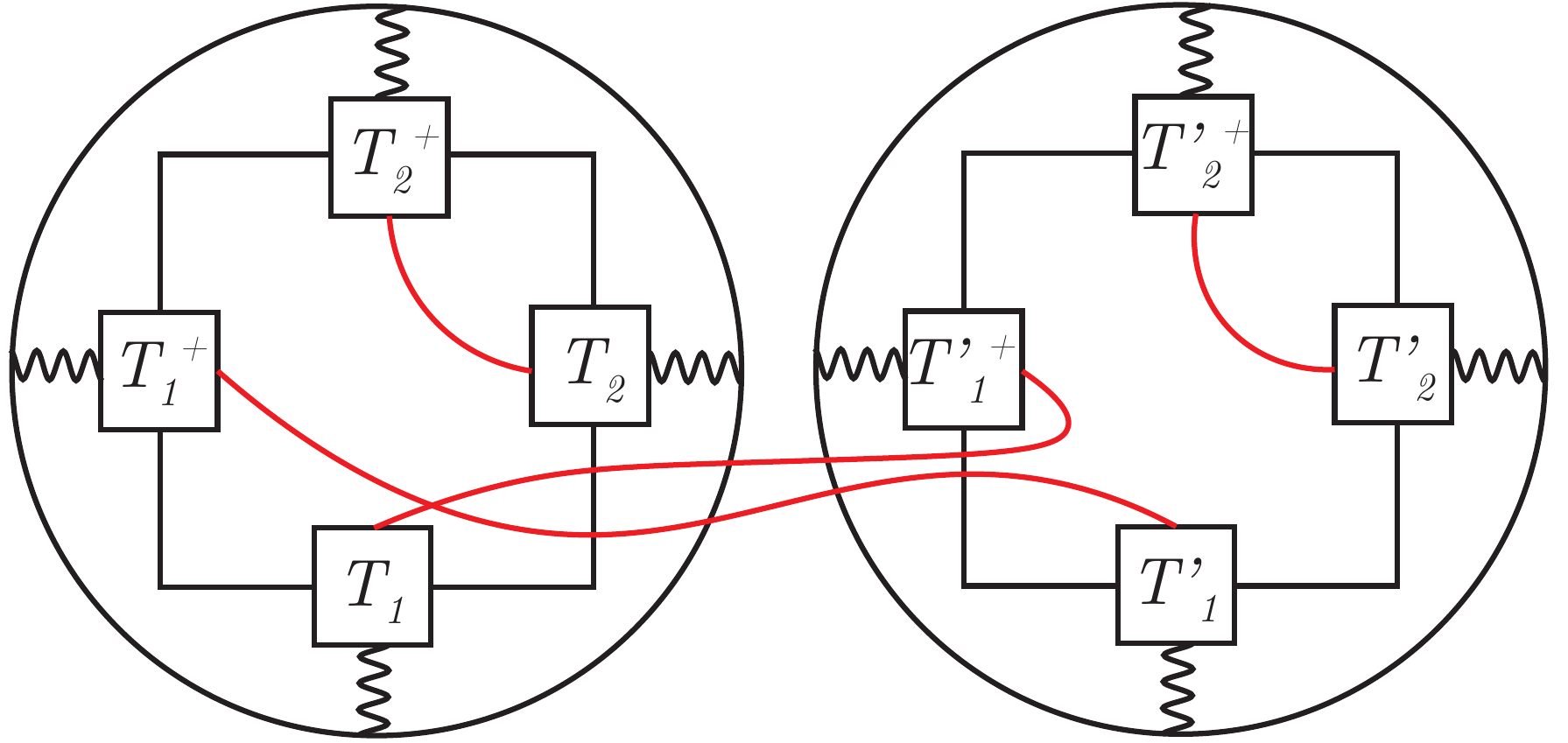}\quad=\quad\inlinefig[8]{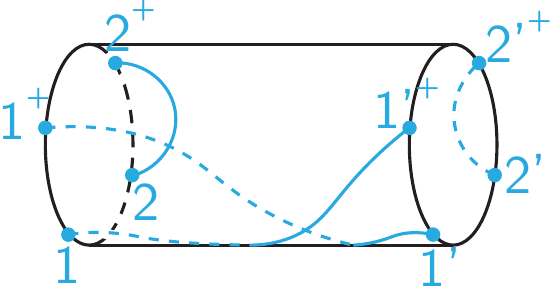},\\
        &\inlinefig[3]{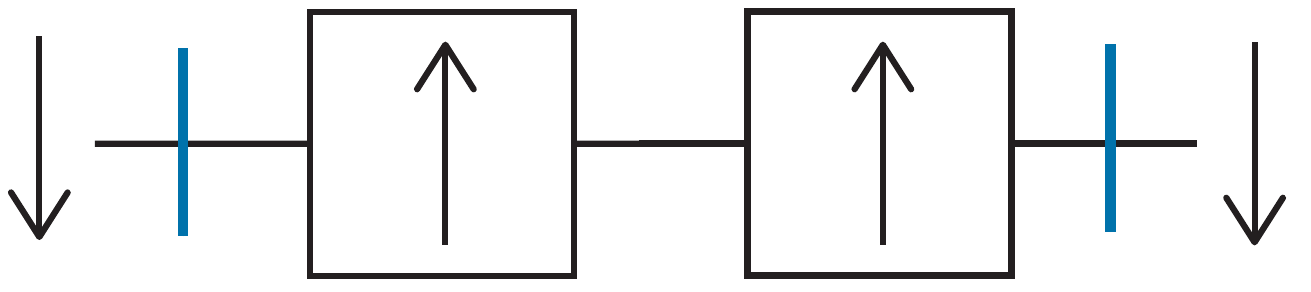} =\quad  \inlinefig[8]{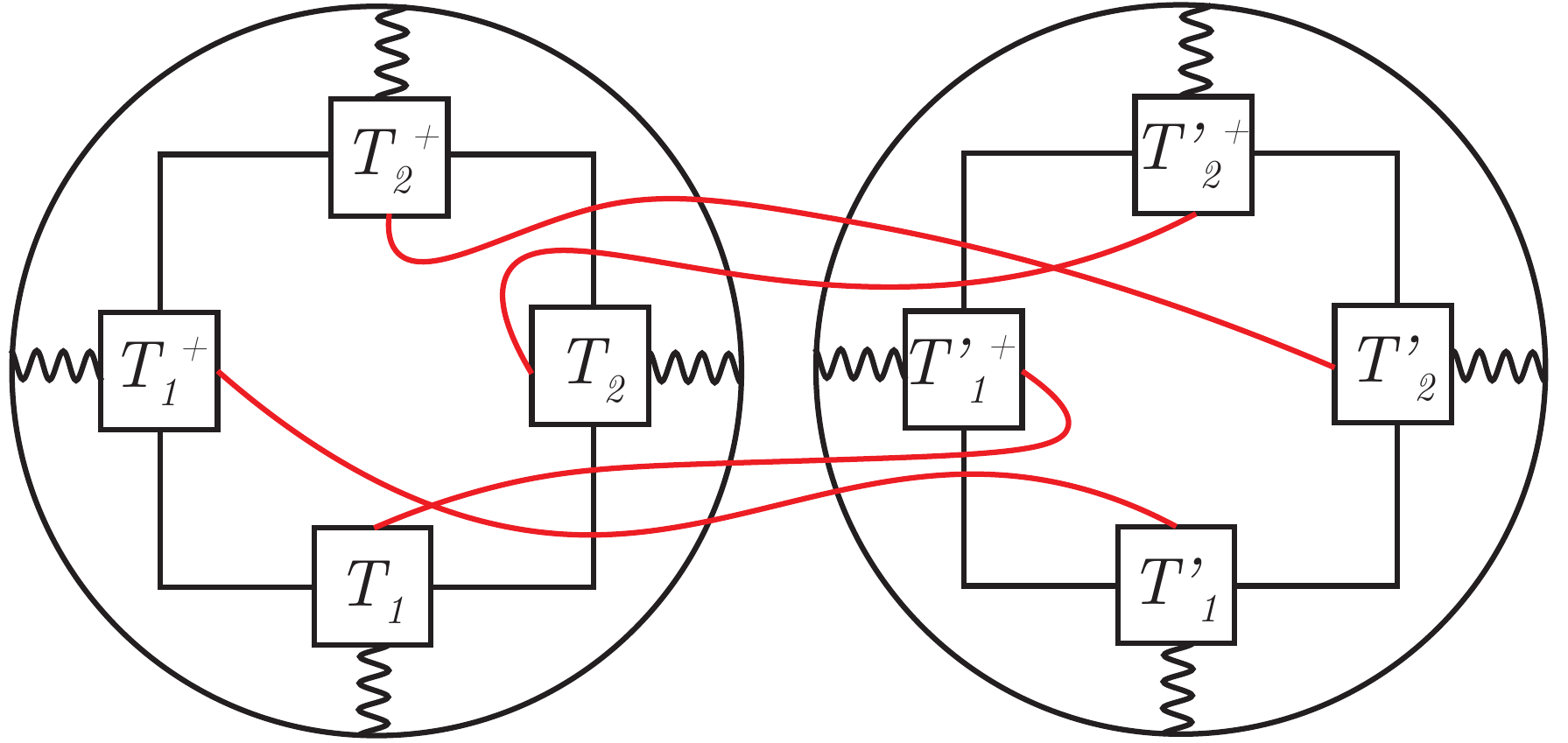}\quad=\quad\inlinefig[8]{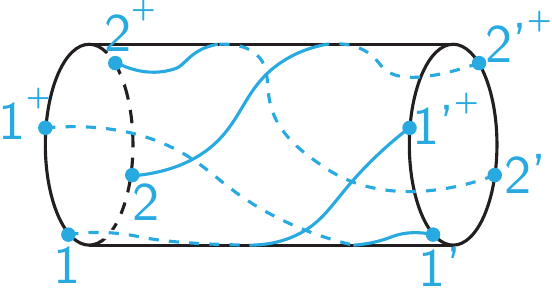}.
\end{aligned}
\end{equation}
Notice that, at leading order in $D=e^{\SBPS}$ (first configuration), we obtain the answer \eqref{eq:TNannealednorm}, which remains true for any value of $k$ because there is only one fully disconnected diagram. Therefore, just like in Section \ref{sec:2}, if one computes the semi-quenched entropy only keeping the leading order in $D$, it becomes negative just like the annealed entropy. What guarantees the positivity of the semi-quenched entropy is again the connected cylinder contributions between the two replicas (second, third, and fourth configuration), which are not present when computing the annealed entropy. 

Finally, we remark that one advantage of the setup of the present section (involving non-interacting, non-Hermitian operators) over that studied in the previous sections is that the simpler combinatorics, together with tensor network techniques, allow us to explicitly compute R\'enyi-$n$ semi-quenched entropies for $n>2$ (we provide the expressions for $n=3,4$ in Appendix \ref{app:RTNs}). A second advantage is that tensor networks are good toy-models for holography even in higher dimensions, see e.g \cite{Chandra:2023dgq} for 3D gravity. We can thus hope that our results extend beyond the realm of two-dimensional quantum gravity.

\section{Discussion}
\label{sec:discussion}

In this paper, we studied the entropic puzzle introduced by LMRS in the BPS sector of JT supergravity \cite{Lin_Maldacena_Rozenberg_Shan_2023}, in which the entropy for one side of a two-sided black hole becomes negative when inserting a sufficiently large number of operators in the Euclidean preparation of the two-sided state. We extended the puzzle to the case of non-SUSY JT gravity within a microcanonical band, and showed how positivity can be rescued by summing over topologies and including connected multi-boundary contributions when computing R\'enyi entropies. The crux of the argument is that the naive gravitational entropy that becomes negative is an annealed entropy, resulting from an incorrect averaging procedure in the matrix model describing matter operators. We then studied a novel quantity, the semi-quenched entropy (introduced in \cite{Antonini_Iliesiu_Rath_Tran_2025}), and showed that it stays positive using a purely bulk argument in the regime where the number of insertions $k$ scales with $e^{2\SBPS/3}$. Along the way, we analyzed in detail a surprising match between JT (super)gravity with matter and an effective pure JT gravity description, which emerges in the limit $k=O(e^{2\SBPS/3})$ of interest. To study the regime $k=\omega(e^{2\SBPS/3})$, we resorted to the matrix integral description of matter operators and studied how new saddles, the one-eigenvalue instanton (dominant) and the two-eigenvalue instanton (subdominant), start to dominate the matrix integral, again yielding a positive entropy. The matrix model description also offered us the tools to understand in detail the emergence of the effective pure JT gravity description, which is due to the universal statistics of eigenvalues near the edge of matrix models. Finally, we formulated an alternative form of the puzzle involving different non-interacting, non-Hermitian matter operators and solved it using random tensor network techniques. The RTN combinatorics precisely matches the bulk combinatorics of Witten diagrams, making the RTN an exact reformulation of the bulk setup and the resolution easily interpretable from a bulk perspective. Once again, connected multi-boundary wormhole contributions are the necessary ingredient to rescue positivity.
We will now conclude with some comments and open questions.

\subsubsection*{Self-averaging properties and higher-dimensional generalization}

One important fact about all entropies that we have discussed is that they are not self-averaging in the low-temperature regime. In fact, in the regime of our interest where $k=O(e^{2\SBPS/3})$ (Airy limit) or $k=\omega(e^{2\SBPS/3})$ (one- and two-eigenvalue instantons), the (powers of) partition functions we have been computing are dominated by instances of the matrix ensemble with very small lowest eigenvalue and/or very large highest eigenvalue, outside the ``classically allowed region'' bounded by the square root edge. In other words, our quantities are dominated by outliers in the ensemble rather than typical members, and they are thus not self-averaging. For the thermal entropy puzzle studied in \cite{Antonini_Iliesiu_Rath_Tran_2025}, in which the ensemble under examination is that of microscopic Hamiltonians, this implies that the generalization of the entropy calculations to higher dimensions is challenging. In fact, in higher dimensions, the bulk quantum gravity theory is expected to be dual to a single, microscopic theory (e.g., a single CFT in AdS/CFT) rather than an ensemble of theories.\footnote{See \cite{Marolf:2024jze} for a different perspective on this matter.} In that case, the ensemble the GPI averages over is parametrizing ignorance: namely, it arises from coarse-graining over microscopic data inaccessible from the semiclassical bulk theory \cite{Belin:2020hea,Chandra:2022bqq,Chandra:2022fwi,Sasieta:2022ksu,Balasubramanian:2022gmo,deBoer:2023vsm}. The single, microscopic Hamiltonian describing the fundamental bulk theory should then be regarded as a typical realization of such an ensemble. Therefore, the only quantities averaged over the ensemble that reliably capture features of the single microscopic theory are the self-averaging ones. A property that is indeed self-averaging in the results of \cite{Antonini_Iliesiu_Rath_Tran_2025} is the positivity itself of the entropy.\footnote{To be more precise, we can, for example, consider the observable $\Theta\left((\tr \rho)^n-\tr \rho^n \right)$ where $\Theta$ is the Heavyside step function. Our analysis suggests that this observable has an average equal to $1$ and a zero standard deviation, making it self-averaging.} In fact, this is a consequence of the existence of an isolated ground state in every member of the Hamiltonian ensemble. Thus, this property is meaningful also in higher-dimensional gravity theories.

The positivity of entropy can be probed in higher dimensions by a slightly different calculation: that of the thermal entropy for a low-temperature PETS obtained by inserting a heavy shell in the preparation of the state, which was carried out in \cite{Antonini:2023hdh}. In fact, \cite{Antonini:2023hdh} computed precisely the semi-quenched R\'enyi-2 entropy using only saddle-point geometries (see equations (3.4)--(3.10) in \cite{Antonini:2023hdh}).\footnote{The averaging in the setup of \cite{Antonini:2023hdh} arises from an ensemble of possible heavy shell operators, as explained in more detail in \cite{Sasieta:2022ksu,Antonini:2025ioh}.} It is interesting to note that, in that setup, the annealed entropy---obtained by dropping the contribution from Eq.~(3.8) to the denominator of (3.10)---becomes indeed negative in the $\beta\to\infty$ limit. On the other hand, the semi-quenched entropy (3.10) remains positive for any $\beta$ and vanishes in the infinite $\beta$ limit. We remark that this thermal entropy computation probes the actual isolated ground state of the dual CFT's Hamiltonian,\footnote{This is also different from the entropy studied in \cite{Antonini_Iliesiu_Rath_Tran_2025}. In \cite{Antonini:2023hdh}, one is probing properties of the ground state that is separated by an $O(1)$ gap in scaling dimension from the first excited state; in \cite{Antonini_Iliesiu_Rath_Tran_2025}, one is instead probing properties of black hole ground states in sectors of fixed charge which are separated by exponentially small gaps from the first excited state. } rather than the operator's matter ``ground state'' as in the present paper.

For the setup of this paper, it is plausible that our results for the entropies can be directly generalized to higher dimensions. A first reason is that all the matter diagrams considered in this paper are on-shell, i.e., they are saddle points of the gravitational path integral for large operator scaling dimensions (in contrast with many of the contributions to the genus expansion for the thermal entropy in JT gravity). It is thus possible that a similar calculation can be carried out directly in higher dimensions; in fact, whenever a saddle-point exists in lower dimensions, this signals the existence of a saddle-point with spherically-symmetric shell matter insertions in higher dimensions.
A second reason is that the ensemble we are considering is not an ensemble of Hamiltonians, but rather an ensemble describing matter operators. As an example, consider matter operators in higher dimensions with a large scaling dimension (possibly scaling with $1/G$). Suppose that our precision in measuring the energy is limited, so that we can only determine the scaling dimension of operators up to some $O(1)$ error. We then have a large number of operators with roughly equal scaling dimensions, which are indistinguishable to us. We can thus model them as an ensemble, and not all operators in the ensemble will be typical: we can have outliers. Now, operationally, one could pick one operator at random from the ensemble, consider the state obtained by a large number of insertions of this operator, and compute its moments. Then we can repeat this experiment an exponential number of times and compute the semi-quenched entropy. With a similar experiment, one can compute the quenched entropy. Entropies originating from outliers will dominate the answer, and the theoretical analysis of this experiment would be precisely that studied in this paper. It would be interesting to make these ideas more precise by generalizing the puzzle to higher dimensions in the presence of insertions of heavy operators (for example, dust shells as in \cite{Balasubramanian:2022gmo}), exactly quantifying the number of operators in a given window of scaling dimension, and the probability of obtaining outliers. 

\subsubsection*{Extension of the puzzle resolution to generic $\Delta$}

Another interesting future direction would be to prove that the eigenvalue spectrum for the projected operators $\tilde{O}_\Delta$ is bounded for any value of the scaling dimension $\Delta>0$ and that the edge is a (universal) square-root edge. If the operators can always be modeled by random matrices, regardless of their scaling dimensions, then the existence and properties of the edge follow from edge universality in RMT. However, showing the existence and deriving the properties of such an edge 
beyond the large $\Delta$ limit is a very non-trivial task from the bulk point of view. We discuss this topic further in Appendix \ref{sec:ETH}. In particular, we explain the difficulties in mapping operators to random matrices (and therefore in identifying an edge), discuss hints to the existence of a finite edge, and propose a random matrix toy model for generic $\Delta$. Once the existence of the square-root edge is established, the resolution of the puzzle we presented in this paper, which solely relies on such an edge, can immediately be extended to generic $\Delta$.

\subsubsection*{Quenched entropy from the bulk}

One more important open question that remains is how to compute the quenched entropy directly from the bulk theory. One option, which has been explored in the past \cite{Engelhardt_Fischetti_Maloney_2021}, is to use the so-called no-replica trick, which involves computing the partition function in the presence of $m$ boundaries and then taking the limit $m\to 0$. In this regime, replica symmetry breaking effects \cite{Kondor_1983,PARISI1979203,Parisi:1979ad,Parisi:1979mn,Parisi:1983dx} are expected to become important, and the analytic continuation of the bulk partition function is ambiguous \cite{Engelhardt_Fischetti_Maloney_2021}. A possible alternative route, which is currently under investigation, is to first realize that, in the $k=O(e^{2\SBPS/3})$ regime, the Airy kernel can be computed from a purely gravitational calculation. The quenched entropy can then be computed from the Airy kernel \cite{tracy1994level}, which implies the Tracy-Widom distribution \cite{Janssen:2021mek}.\footnote{We thank Gabriele Di Ubaldo for discussions on these points.}

\vspace{0.5cm}

Finally, it would also be important to obtain a bulk dual description of the one-eigenvalue instanton matrix integral saddles studied in Section \ref{sec:resolution}. A conjectured semiclassical bulk dual involving UV objects such as end-of-the-world eigenbranes was suggested in \cite{Hernandez-Cuenca_2024}, building on the work of \cite{Zamolodchikov:2001ah,Fateev:2000ik,Teschner:2000md,Blommaert:2019wfy,Saad_Shenker_Stanford_2019,Goel:2020yxl,Gao:2021uro}.\footnote{See also \cite{Hernandez-Cuenca:2024xlg} for a proposal of a single-geometry description capturing the genus re-summation we considered in the Airy limit.} It would be interesting to explore these ideas further.

\section*{Acknowledgments}

We would like to thank Sarthak Chandra, Yiming Chen, Gabriele Di Ubaldo, Abhijit Gadde, Sergio Hernandez-Cuenca, Matt Headrick, Henry Lin, Shiraz Minwalla, Thomas Mertens, Onkar Parrikar, Geoff Penington, Martin Sasieta, Steve Shenker, Douglas Stanford, Brian Swingle and Wayne Weng for valuable discussions. This work was supported in part by the Leinweber Institute for Theoretical Physics, by the Department of Energy, Office of Science, Office of High Energy Physics through DE-SC0025522, DE-SC0019380, DE-SC0025293, and  DE-FOA-0002563, by AFOSR award FA9550-22-1-0098, and by a Sloan Fellowship.
This material is based upon work supported by the National Science Foundation Graduate Research Fellowship under Grant No. DGE-2146752. Any opinions, findings, and conclusions or recommendations expressed in this material are those of the authors and do not necessarily reflect the views of the National Science Foundation.

\appendix

\section{Properties of the $6j$-symbol in the $\Delta\to\infty$ and $\Delta\to 0$ limits}\label{6j}

\subsection{$\Delta\to\infty$ Limit}

Here we show that the $6j$-symbol of non-SUSY JT with scalar matter is exponentially suppressed in the $\Delta\to\infty$ limit. A similar result for the BPS case was obtained in \cite{Lin_Maldacena_Rozenberg_Shan_2023}.
To start, we utilize the following formula for the $6j$-symbol of $\text{SL}(2,\mathbb{R})$ \cite{Jafferis_Kolchmeyer_Mukhametzhanov_Sonner_2023,Yang_2019,Mertens_Turiaci_Verlinde_2017,Mertens_Turiaci_2023}:
{\small \begin{equation}
	\begin{split}
\begin{Bmatrix}
	\Delta & s_{1} & s_{2}\\
	\Delta & s_{3} & s_{4}
\end{Bmatrix}&=\left(\Gamma(\Delta\pm is_{1}\pm is_{2})\Gamma(\Delta\pm is_{2}\pm is_{3})\Gamma(\Delta\pm is_{3}\pm is_{4})\Gamma(\Delta\pm is_{4}\pm is_{1})\right)^{1/2}\\
&\times \left[\frac{2is_{1}\Gamma(\pm 2is_{1})}{\Gamma(\Delta-is_{1}\pm is_{2})\Gamma(\Delta-is_{1}\pm is_{4})}{}_{4}\tilde{F}_{3}\left(
\begin{matrix}
	\Delta+is_{1}\pm is_{2} & \Delta + is_{1}\pm is_{4}\\
	2\Delta+is_{1}\pm is_{3} & 2is_{1}+1
\end{matrix};1\right)+\textrm{c.c.}\right]\\
&=\left(\frac{\Gamma(\Delta\pm is_{2}\pm is_{3})\Gamma(\Delta\pm is_{3}\pm is_{4})}{\Gamma(\Delta\pm is_{1}\pm is_{2})\Gamma(\Delta\pm is_{4}\pm is_{1})}\right)^{1/2}2is_{1}\Gamma(\pm 2is_{1})\\
&\quad\times\left[\sum_{n=0}^{\infty}\frac{1}{n!}\frac{\Gamma(\Delta+n+is_{1}\pm is_{2})\Gamma(\Delta+n+is_{1}\pm is_{4})}{\Gamma(2\Delta+n+is_{1}\pm is_{3})\Gamma(2is_{1}+n+1)}-\textrm{c.c.}\right],
	\end{split}
    \label{eq:6j-symbol}
\end{equation}}
where in the second line we use the summation representation of the regularized hypergeometric function ${}_4\tilde{F}_3$. Taking the $\Delta\to\infty$ limit (with $s_i$ small compared to $\Delta$), the product of $\Gamma$'s in the prefactor gives unity. A nontrivial contribution comes from the sum:
{\small \begin{equation}
\begin{split}
	\sum_{n=0}^{\infty}\frac{1}{n!}\frac{\Gamma(\Delta+n+is_{1}\pm is_{2})\Gamma(\Delta+n+is_{1}\pm is_{4})}{\Gamma(2\Delta+n+is_{1}\pm is_{3})\Gamma(2is_{1}+n+1)}&\sim\frac{4\pi}{2^{4\Delta}\Delta}\sum_{n=0}^\infty\frac{1}{n!}\frac{1}{\Gamma(2is_{1}+n+1)}\left(\frac{\Delta}{2}\right)^{2n+2is_{1}}\\
			&=\frac{4\pi}{2^{4\Delta}\Delta}I_{2is_{1}}(\Delta).
\end{split}
\end{equation}}
where we first apply Stirling's approximation on the $\Gamma$'s to take the leading terms in the series.
Thus, including the complex conjugate, we obtain
\begin{equation}
	\begin{split}
\begin{Bmatrix}
	\Delta & s_{1} & s_{2}\\
	\Delta & s_{3} & s_{4}
\end{Bmatrix}&\sim\frac{4\pi}{2^{4\Delta}\Delta}2is_{1}\Gamma(\pm 2is_{1})(I_{2is_{1}}(\Delta)-I_{-2is_{1}}(\Delta))\\
&=\frac{8\pi}{2^{4\Delta}\Delta}K_{2is_{1}}(\Delta)\\
&\sim 2\left(\frac{2\pi}{\Delta}\right)^{3/2}e^{
-(1+4\log 2)\Delta}.
	\end{split}
\end{equation}
Therefore, $6j$-symbols are exponentially suppressed by $e^{-(1+4\log 2)\Delta}\approx e^{-3.77\Delta}$, so we are justified in dropping Witten diagrams that involve intersecting propagators.

\subsection{$\Delta\to0$ Limit}

Another limit that we are interested in is the $\Delta \to 0$ limit. Here we will show that, in non-SUSY JT gravity, an intersecting diagram is equal to a nonintersecting diagram. A similar result for the BPS case was obtained in \cite{Lin_Maldacena_Rozenberg_Shan_2023}. We will show this in the case of the 4-point function, but the result can also be generalized for an arbitrary $2k$-point function.

Following the Feynman rules of \cite{Jafferis_Kolchmeyer_Mukhametzhanov_Sonner_2023}, we will consider the crossing diagram
\begin{equation}
\mathcal{A}_{\text{cross}}=\inlinefig[10]{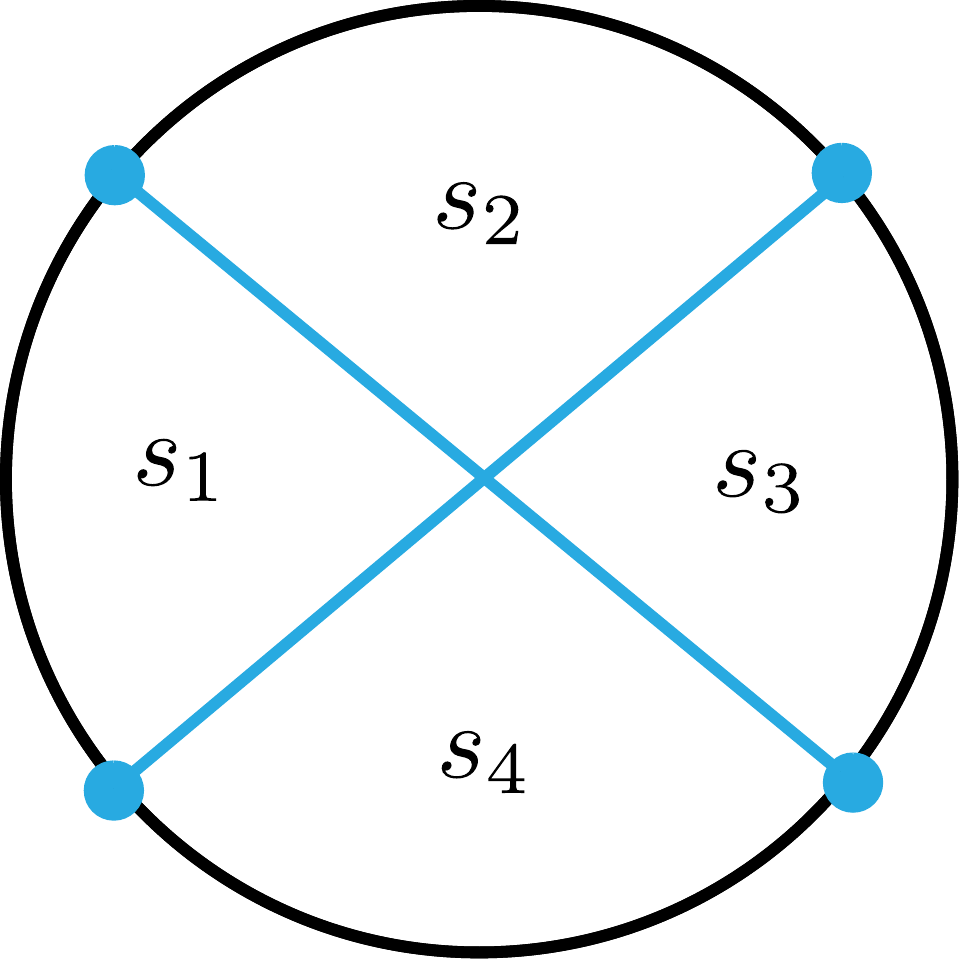}=\sqrt{\Gamma^{\Delta}_{12}\Gamma^{\Delta}_{23}\Gamma^{\Delta}_{34}\Gamma^{\Delta}_{41}}\begin{Bmatrix}
	\Delta & s_{1} & s_{2}\\
	\Delta & s_{3} & s_{4}
\end{Bmatrix},
\end{equation}
and compare this to the noncrossing diagram
\begin{equation}
    \mathcal{A}_{\text{uncross}}=\inlinefig[10]{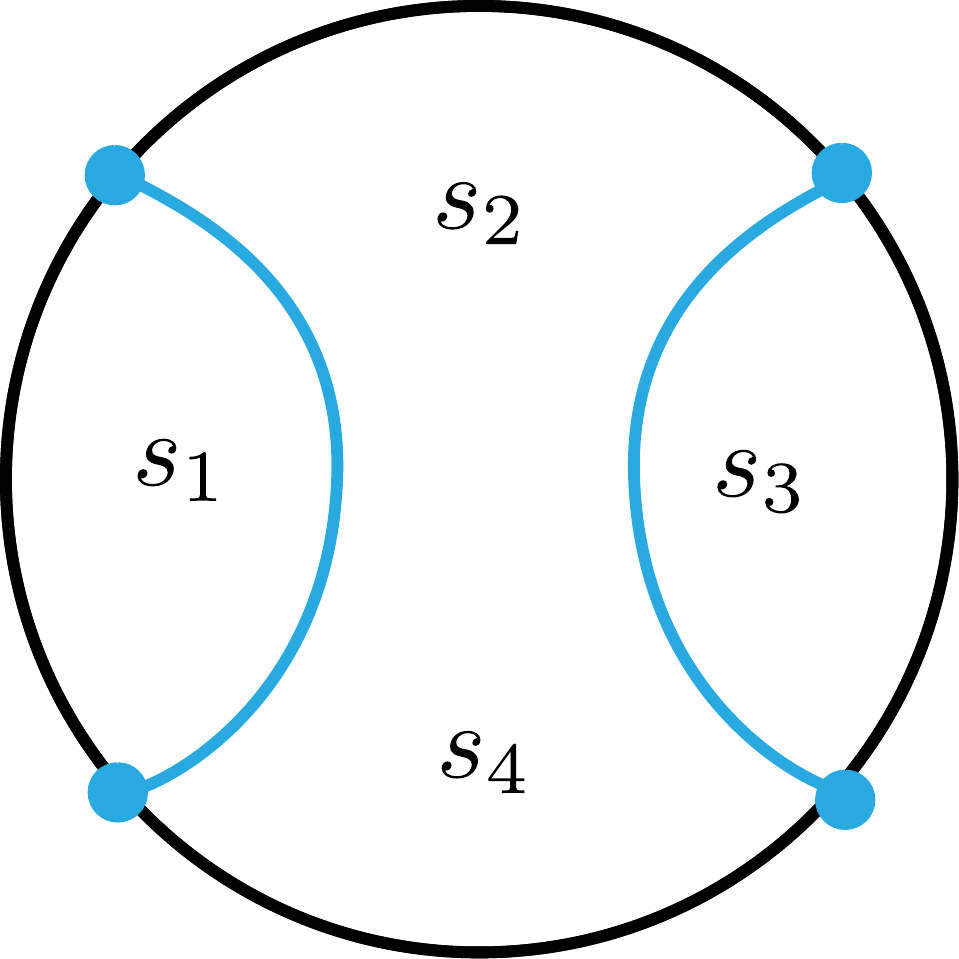}=\Gamma_{12}^{\Delta}\Gamma_{34}^{\Delta}\frac{\delta(s_2-s_4)}{\rho(s_2)}.
\end{equation}
where
\begin{equation}
    \Gamma^{\Delta}_{12}\equiv\frac{\Gamma(\Delta\pm is_1\pm is_2)}{\Gamma(2\Delta)}.
\end{equation}
We will show that in the limit $\Delta\to 0$, $\mathcal{A}_{\text{cross}}=\mathcal{A}_{\text{uncross}}$.

To this end, we first start with $\mathcal{A}_{\text{uncross}}$, for which we use the useful identity
\begin{equation}
    \lim_{\Delta\to 0}\Gamma_{12}^{\Delta}=2\pi\delta(s_1-s_2)\Gamma(\pm i(s_1+s_2))=\frac{\delta(s_1-s_2)}{\rho(s_1)},\label{eq:delta}
\end{equation}
where $\rho(s)=\frac{1}{2\pi\Gamma(\pm 2is)}$ is the density of states in the $s=\sqrt{2E/E_{\text{brk.}}}$ basis.
Therefore,
\begin{equation}
    \lim_{\Delta\to 0}\mathcal{A}_{\text{uncross}}=\frac{\delta(s_1-s_2)}{\rho(s_1)}\frac{\delta(s_3-s_4)}{\rho(s_3)}\frac{\delta(s_2-s_4)}{\rho(s_2)}
\end{equation}

For $\mathcal{A}_{\text{cross}}$, we will use the representation Eq. \eqref{eq:6j-symbol} of the $6j$-symbol. With that, we can rewrite $\mathcal{A}_{\text{cross}}=\Gamma_{23}^{\Delta}\Gamma_{34}^\Delta\mathcal{B}$ so that we can exploit the identity Eq. \eqref{eq:delta} and set $s_2=s_3=s_4$. Thus, the remaining factor $\mathcal{B}$ becomes
\begin{equation}
    \mathcal{B}\equiv 2is_1\Gamma(\pm 2is_1)\Gamma(\Delta+is_1\pm is_2)^2{}_{4}\tilde{F}_{3}\left(
		\begin{matrix}
			\Delta+is_{1}\pm is_{2},&\Delta+is_{1}\pm is_{2}\\
			2\Delta+is_{1}\pm is_{2},&2is_{1}+1
		\end{matrix};1
	\right)+\textrm{c.c}.
    \label{eq:B-def}
\end{equation}
Now taking the limit $\Delta\to 0$ on $\mathcal{B}$, we obtain
\begin{equation}
\begin{split}
    \lim_{\Delta\to 0}\mathcal B&=2is_1\Gamma(\pm 2is_1)\Gamma(is_1\pm is_2)^2{}_{4}\tilde{F}_{3}\left(
		\begin{matrix}
			is_{1}\pm is_{2},&is_{1}\pm is_{2}\\
			is_{1}\pm is_{2},&2is_{1}+1
		\end{matrix};1
	\right)+\textrm{c.c}.\\
    &=2is_1\Gamma(\pm 2is_1)\Gamma(is_1\pm is_2){}_2\tilde{F}_1\left(\begin{matrix}
			is_{1}\pm is_{2}\\
			2is_{1}+1
		\end{matrix};1
	\right)+\textrm{c.c}.\\
    &=2is_1\Gamma(\pm 2is_1)\frac{\Gamma(i(s_1\pm s_2))}{\Gamma(i(s_1\pm s_2)+1)}+\text{c.c.}\\
    &=2\Gamma(\pm 2is_1)\frac{2s_1}{s_1+s_2}\text{Im}\frac{1}{s_1-s_2}.
\end{split}
\end{equation}
Since these quantities are integrated over $s$'s, we add a regulating ``mass" $\epsilon$ to lift $s_1-s_2$ off the real axis: $i(s_{1}-s_{2})\to \epsilon+i(s_{1}-s_{2})$. This regulating ``mass'' is in fact the scaling dimension $\Delta \to 0$ appearing in \eqref{eq:B-def}. 
Then, using the Sokhotski–Plemelj theorem
\begin{equation}
	\Im\left[\frac{1}{s_{1}-s_{2}-i\epsilon}\right]=\pi\delta(s_{1}-s_{2}),
\end{equation}
we obtain
\begin{equation}
	\lim_{\Delta\to 0}\mathcal A_{\text{cross}}=\frac{\delta(s_2-s_3)}{\rho(s_2)}\frac{\delta(s_3-s_4)}{\rho(s_3)}\frac{\delta(s_1-s_2)}{\rho(s_1)}=\lim_{\Delta\to 0}\mathcal A_{\text{uncross}}.
\end{equation}
Hence, the diagrams are equally weighted in the $\Delta\to 0$ limit. Using similar manipulations, this can be generalized to show that diagrams contributing to $2k$-point correlators in the $\Delta\to 0$ limit are all equally weighted, regardless of crossings.

\section{Derivation of $2k$-point function in non-supersymmetric JT}
\label{sec:derivation-2n-pt-func-non-SUSY-JT}

In SUSY JT in the BPS sector, we showed that $2n$-point functions are given by summing over Wick contractions on all possible topologies. This calculation is simplified by the fact that all handles have to be ``threaded" by matter worldlines, i.e., every closed geodesic on a higher genus surface should have at least one matter worldline intersecting it. In fact, it was argued in Section \ref{sec:3.1} that diagrams 
 with empty handles vanish in the BPS sector. For each Wick contraction, only one diagram then contributes, the one with the lowest possible genus among those in which the Wick contraction can be realized without intersection. In Section \ref{sec:bulk-resolution}, we called this the ``minimal embedding'' diagram of the Wick contraction.

In non-SUSY JT, we cannot appeal to this argument since diagrams with empty handles do contribute. Nevertheless, we will show that the $2n$-point functions have a similar structure to the BPS setting, and that the additional diagrams will only affect the density of states within an energy window at subleading order without affecting the OPE coefficients of the inserted matter operators. Specifically, in this appendix, we will derive the correlation functions similar to the BPS calculation in Section \ref{sec:bulk-resolution}, starting with disk diagrams in Section \ref{subsec:disk-diagrams}, then generalizing to diagrams without empty handles in Section \ref{subsec:diagrams-without-empty-handles}, and then generalizing to diagrams with more complicated topologies in Section \ref{subsec:handles-between-patches}. 

\subsection{$2k$-point correlators at disk level}

\label{subsec:disk-diagrams}

Let us first briefly review the Feynman rules for non-SUSY JT at disk level, which can be computed using the rules derived in \cite{Mertens_Turiaci_Verlinde_2017,Blommaert:2018oro,Iliesiu:2019xuh,Yang_2019}. The disk $2k$-point functions $\tr\left(\widetilde{O}_{\Delta}^{2k}\right)$ are computed by Wick contractions on a circle. In the large $\Delta$ limit, which is the limit of interest that will be assumed throughout this appendix, the leading diagrams are diagrams without intersections since intersections are exponentially suppressed in $\Delta$.

We begin with a two-point function
\begin{equation}
    \tr_{\mathfrak E} \widetilde{O}_{\Delta}^2=\inlinefig[10]{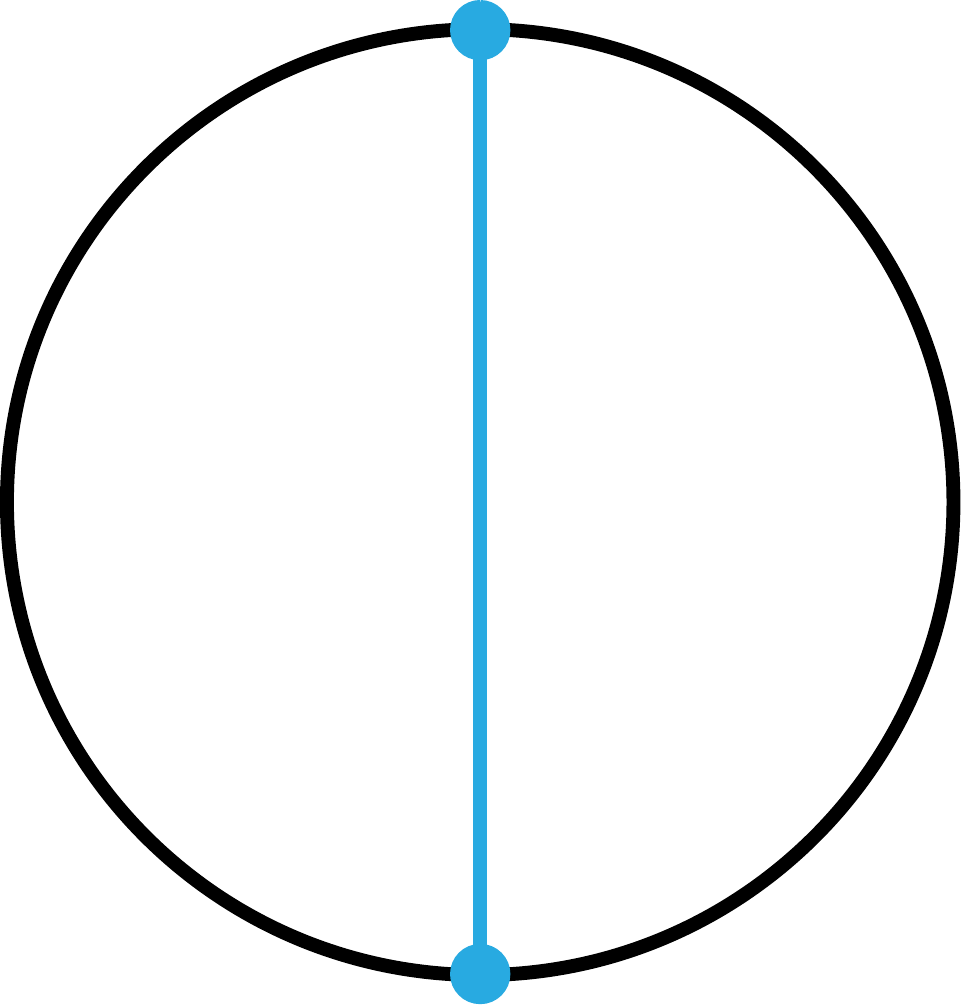}+\dots,
\end{equation}
where we neglect diagrams with higher genera for now. Slicing the picture along the propagator, we obtain 
\begin{equation}
\begin{split}
    \inlinefig[10]{Figures/jt_2pt.pdf}&=\int d\ell\, e^{-\Delta\ell}\,\inlinefig[10]{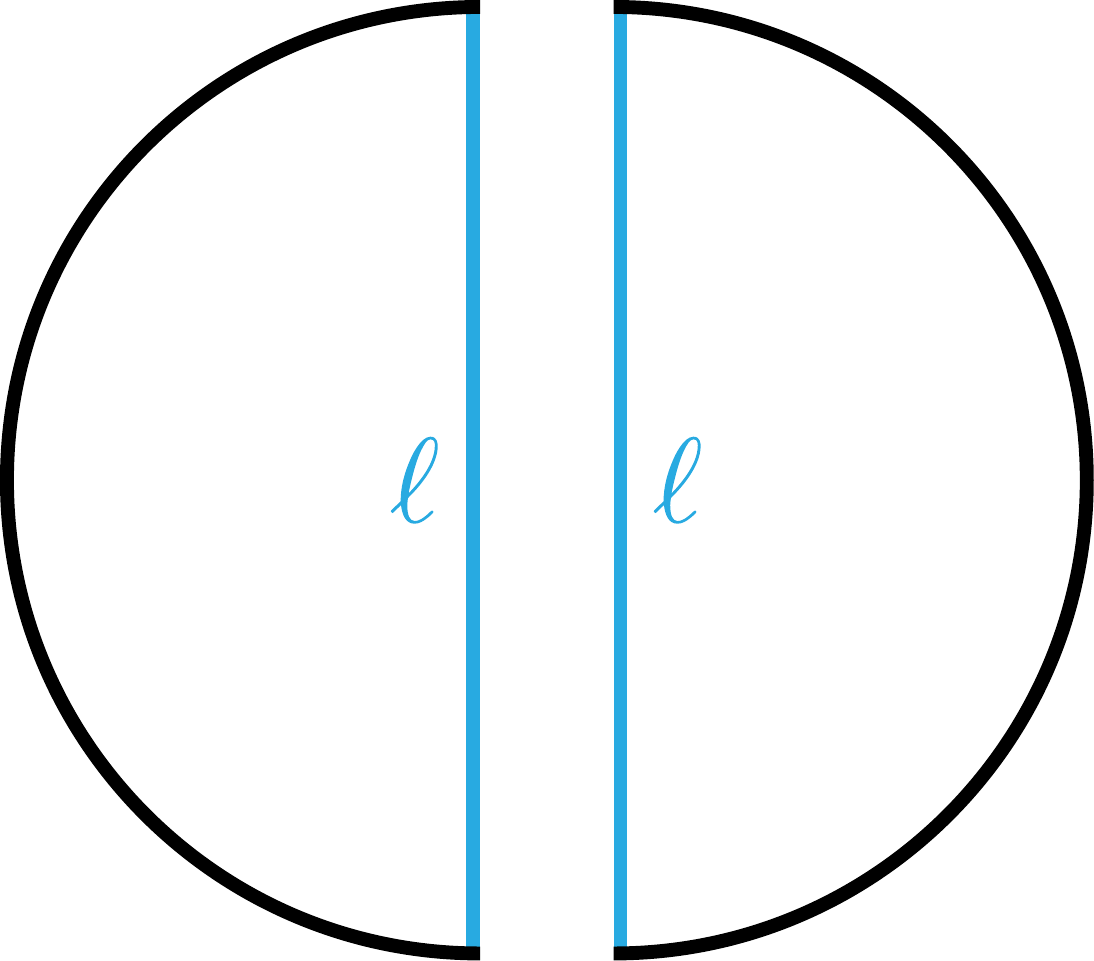}\\
    &=e^{S_0}\int d\ell\,e^{-\Delta\ell} |\Psi_{\mathfrak E}(\ell)|^2,
\end{split}
\end{equation}
where we weighed matter propagators with $e^{-\Delta\ell}$, glued along the matter geodesic, and introduced the microcanonical wavefunction in the length basis 
\begin{equation}
    \Psi_{\mathfrak{E}}(\ell)=\int_{\mathfrak{E}} dE\,\rho_0(E)\,\varphi_E(\ell),
\end{equation}
where
\begin{equation}
    \rho_0(E)=\frac{1}{2\pi^2 E_{\text{brk}}}\sinh(2\pi\sqrt{2\frac{E}{E_{\text{brk}}}}),
\end{equation}
is the JT density of states at the disk level and 
\begin{equation}
    \varphi_E(\ell)=4K_{2i\sqrt{2\frac{E}{E_\text{brk.}}}}(4e^{-\ell/2}),
\end{equation}
is the wavefunction of an energy eigenstate in the length basis. For simplicity, we will assume $E_{\text{brk}}=1$ in this appendix.
The energy eigenstates are normalized so that
\begin{equation}
    \int d\ell\,\varphi_{E}(\ell)\varphi_{E'}(\ell)=\frac{\delta(E-E')}{\rho_0(E)},
\end{equation}
which implies the microcanonical wavefunctions are normalized as
\begin{equation}
    \int d\ell\,|\Psi_{\mathfrak E}(\ell)|^2=\int_{\mathfrak E}dE\,\rho_0(E)\,.
\end{equation}
Integrating over $\ell$, we obtain
\begin{equation}
    \inlinefig[9]{Figures/jt_2pt.pdf}=e^{S_0}\int_{\mathfrak E}dE_1\,dE_2\,\rho_0(E_1)\rho_0(E_2)\, \gamma_{12}^{\Delta}
\end{equation}
with matter propagator
\begin{equation}
    \gamma_{12}^{\Delta}\equiv\frac{\Gamma(\Delta\pm i\sqrt{2E_1}\pm i\sqrt{2E_2})}{2^{2\Delta+1}\Gamma(2\Delta)}.
    \label{eq:matterprop}
\end{equation}
Taking the microcanonical window $\mathfrak E$ small and centered around $\bar{E}$, we can express this as
\begin{equation}
    \inlinefig[10]{Figures/jt_2pt.pdf}=e^{S_0}\int_{\mathfrak E}dE_1\,dE_2\,\rho_0(E_1)\rho_0(E_2)\, P_{\Delta\to\infty}(\bar{E})
\end{equation}
with 
\begin{equation}
   P_{\Delta\to\infty}(\bar{E})\equiv\frac{\Gamma(\Delta)^2\Gamma(\Delta\pm 2i\sqrt{2\bar{E}})}{2^{2\Delta+1}\Gamma(2\Delta)}.
\end{equation}

To compute higher point correlators (at disk level), we must introduce an asymptotic polygon with alternating $n$ geodesic boundaries and $n$ asympototic boundaries:
\begin{equation}
    \inlinefig[10]{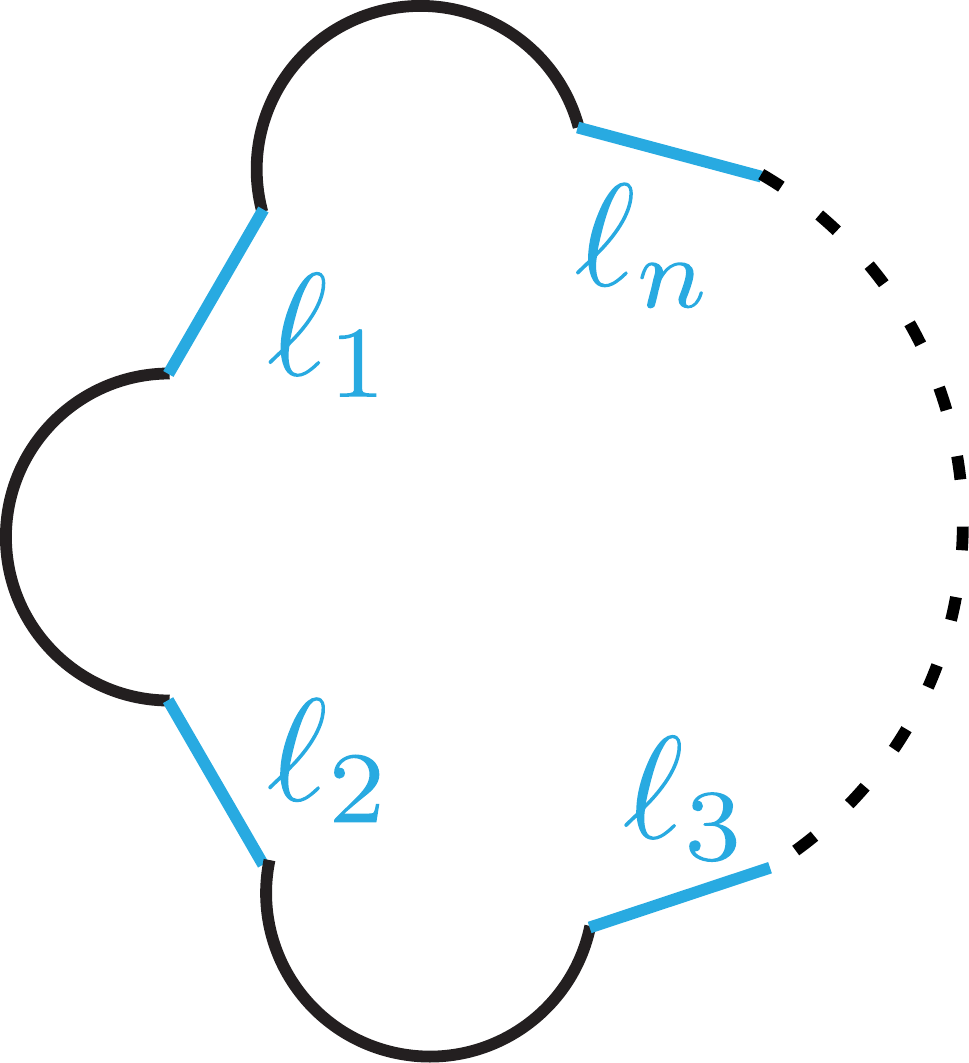}=I_n(\ell_1,\ell_2,\dots,\ell_n)=\int_{\mathfrak E} dE\,\rho_0(E) \varphi_E(\ell_1)\varphi_E(\ell_2)\dots\varphi_E(\ell_n).
\end{equation}
Clearly, the microcanonical wavefunction is a special case of this: $\Psi_{\mathfrak E}(\ell)=I_1(\ell)$.

With these asymptotic polygons in hand, we can compute correlators by cutting and gluing the resulting diagrams.
For example, the time-ordered four-point function diagram can be computed by
\begin{equation}
\begin{split}
    \inlinefig[10]{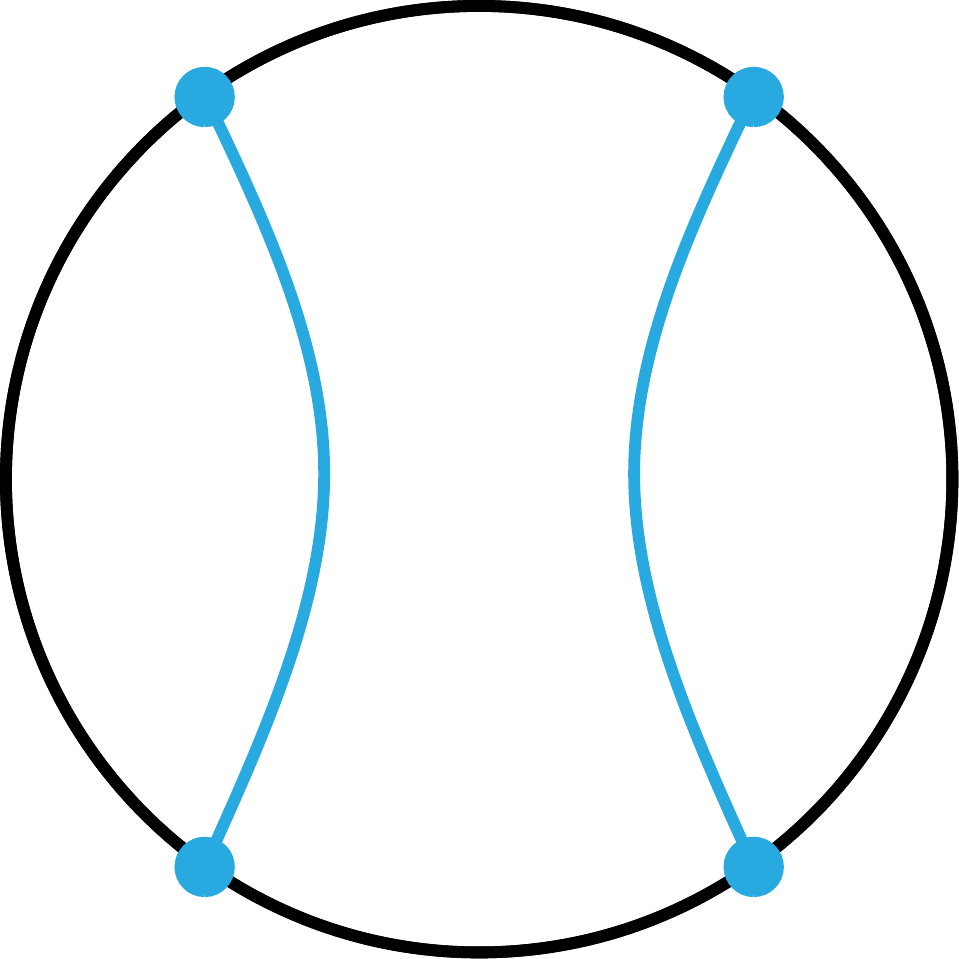}&=e^{S_0}\int d\ell_1\,d\ell_2\,e^{-\Delta\ell_1}e^{-\Delta\ell_2}\inlinefig[10]{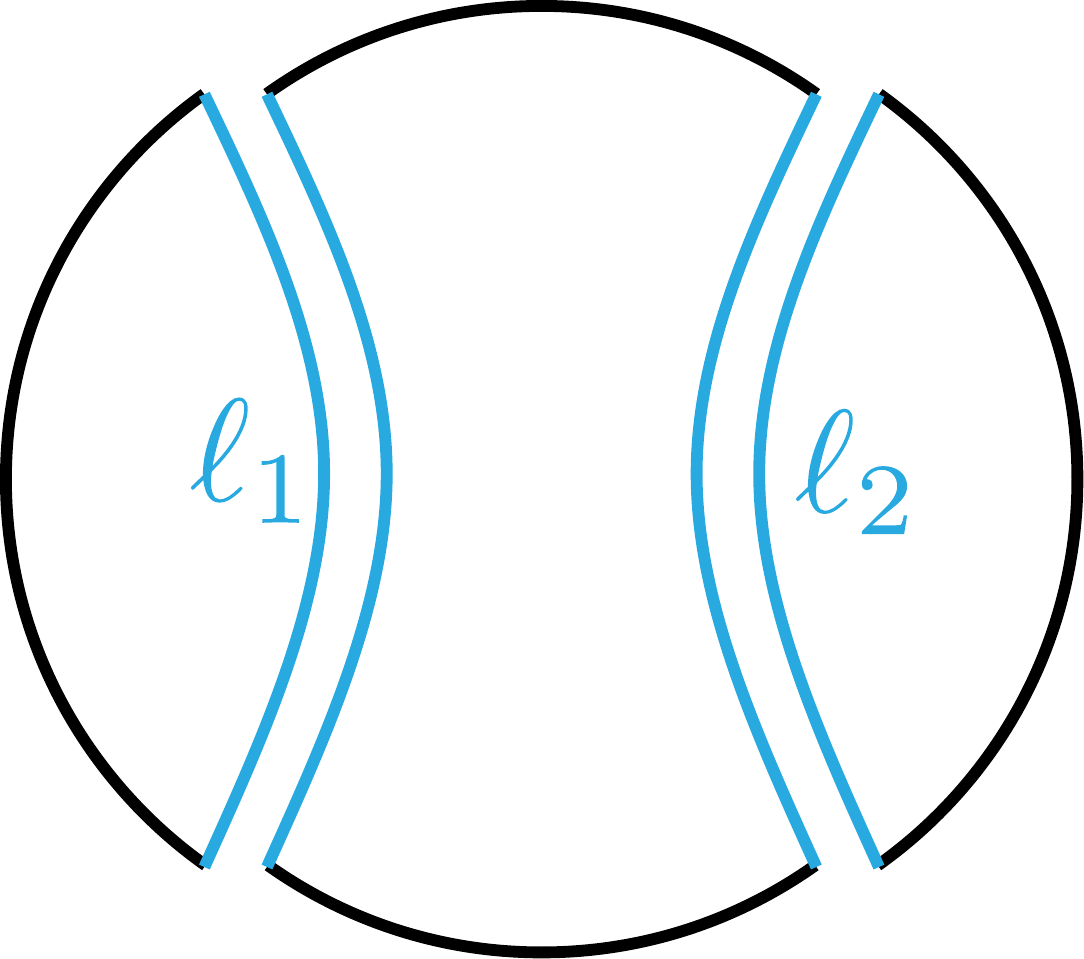}\\
    &=e^{S_0}\int_{\mathfrak{E}}dE_1\,dE_2\,dE_3\,\rho_0(E_1)\rho_0(E_2)\rho_0(E_3)\gamma_{12}^{\Delta}\gamma_{23}^{\Delta}\\
    &\to e^{S_0}\int_{\mathfrak{E}}dE_1\,dE_2\,dE_3\,\rho_0(E_1)\rho_0(E_2)\rho_0(E_3)(P_{\Delta\to\infty}(\bar{E}))^2,
\end{split}
\end{equation}
where, in the last line, we took the energy window small.

The rules should now be clear.
Given a $2k$-point diagram, there will be $k$ matter propagators, each contributing $P_{\Delta\to\infty}(\bar{E})$. Each matter propagator will cut the disk into $k+1$ energy patches, which we integrate over with $\int_{\mathfrak E}dE\,\rho_0(E)$. Since each diagram is equally weighted, we must sum over distinct ways of Wick contracting $2k$ points on a circle without intersections, which is the $k$th Catalan number $C_k$. Thus, we obtain the disk correlator
\begin{equation}
    \langle\tr_{\mathfrak E} O^{2k}\rangle_{\text{Disk}}=e^{S_0} C_k \int_{\mathfrak E}dE_1\,\dots dE_{k+1}\,\rho_0(E_1)\dots\rho_0(E_{k+1}) \,(P_{\Delta\to\infty}(\bar{E}))^{k}.\label{jt_disk}
\end{equation}

\subsection{Diagrams without empty handles}
\label{subsec:diagrams-without-empty-handles}

We can now turn our attention to contributions coming from higher topologies. As a starting point for our discussion of diagrams on higher topology surfaces, we focus on diagrams without ``empty handles". These diagrams, introduced in Section \ref{sec:3.1}, correspond to diagrams that are a product of asymptotic polygons after cutting along matter worldlines. In other words, after cutting along worldlines, all resulting diagrams have a disk topology.

As a single-boundary example, we can consider the disk with one handle contributing to the four-point function:
\begin{equation}
\begin{split}
    \langle\tr_{\mathfrak E}\widetilde{O}^4_{\Delta} \rangle\supset\inlinefig[10]{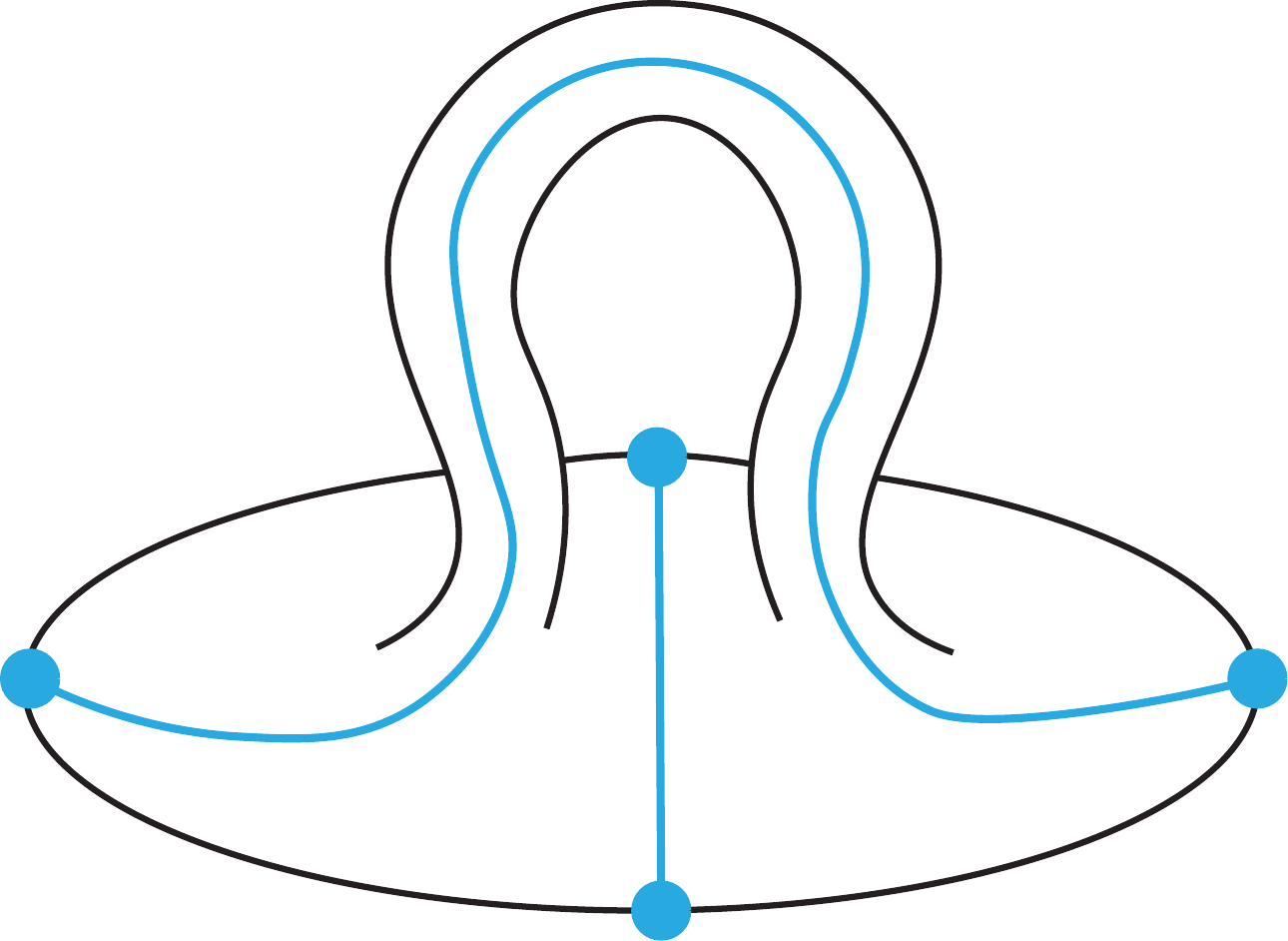}&=\int d\ell_1\,d\ell_2\,e^{-\Delta\ell_1}e^{-\Delta\ell_2}\,\inlinefig[10]{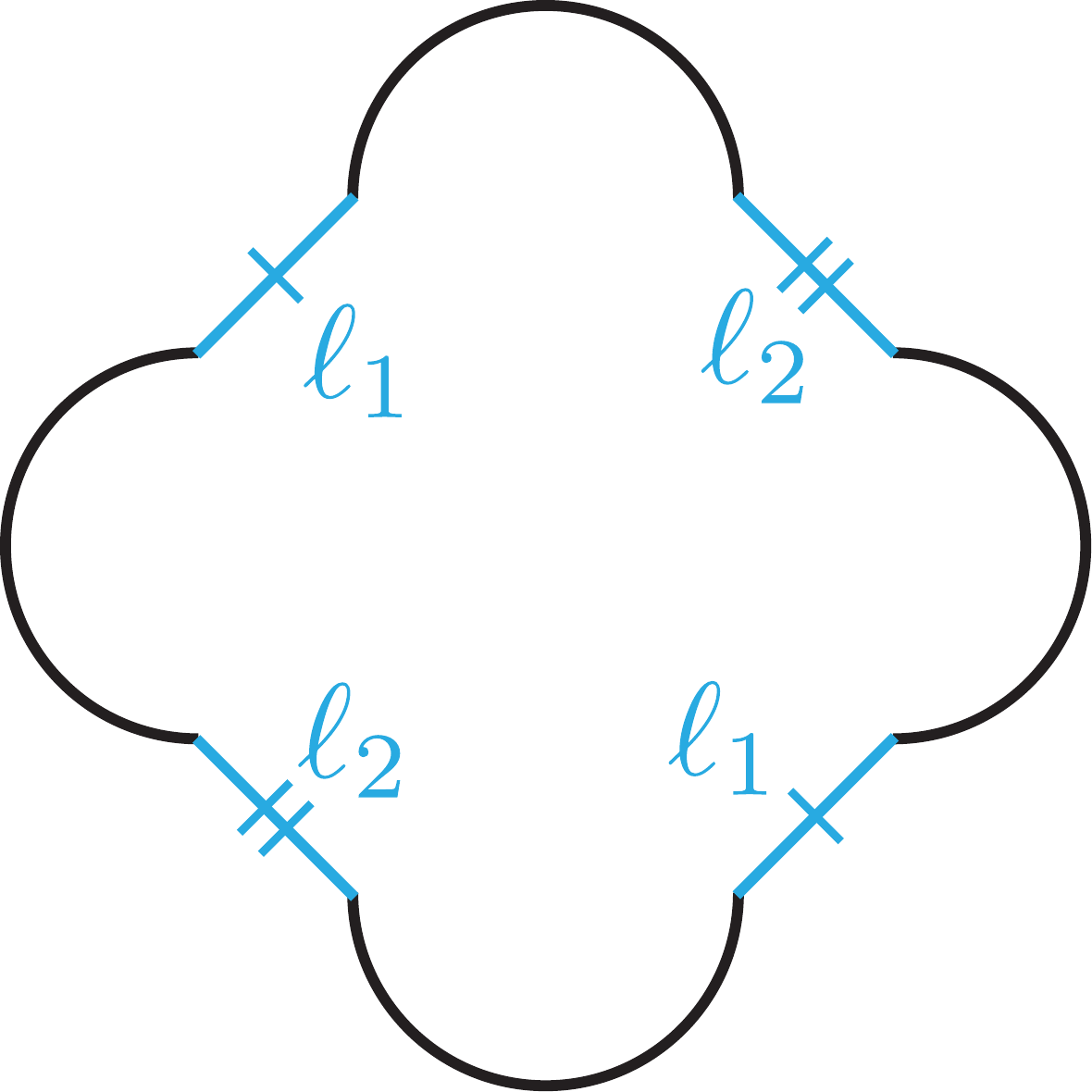}\\
    &=e^{-S_0}\int_{\mathfrak E} dE\,\rho_0(E)(P_{\Delta\to\infty}(\bar{E}))^2.
\end{split}
\end{equation}
We can also look at multi-boundary diagrams. We only have to consider topologies where the boundaries are connected, since general correlators can be constructed by a product of disconnected diagrams. One simple example is the genus-0 cylinder contributing to the four-point function on two boundaries:
\begin{equation}
\begin{split}
    \langle(\tr_{\mathfrak E}\widetilde{O}^2_{\Delta})^2\rangle_{\text{conn}}&\supset\inlinefig[8]{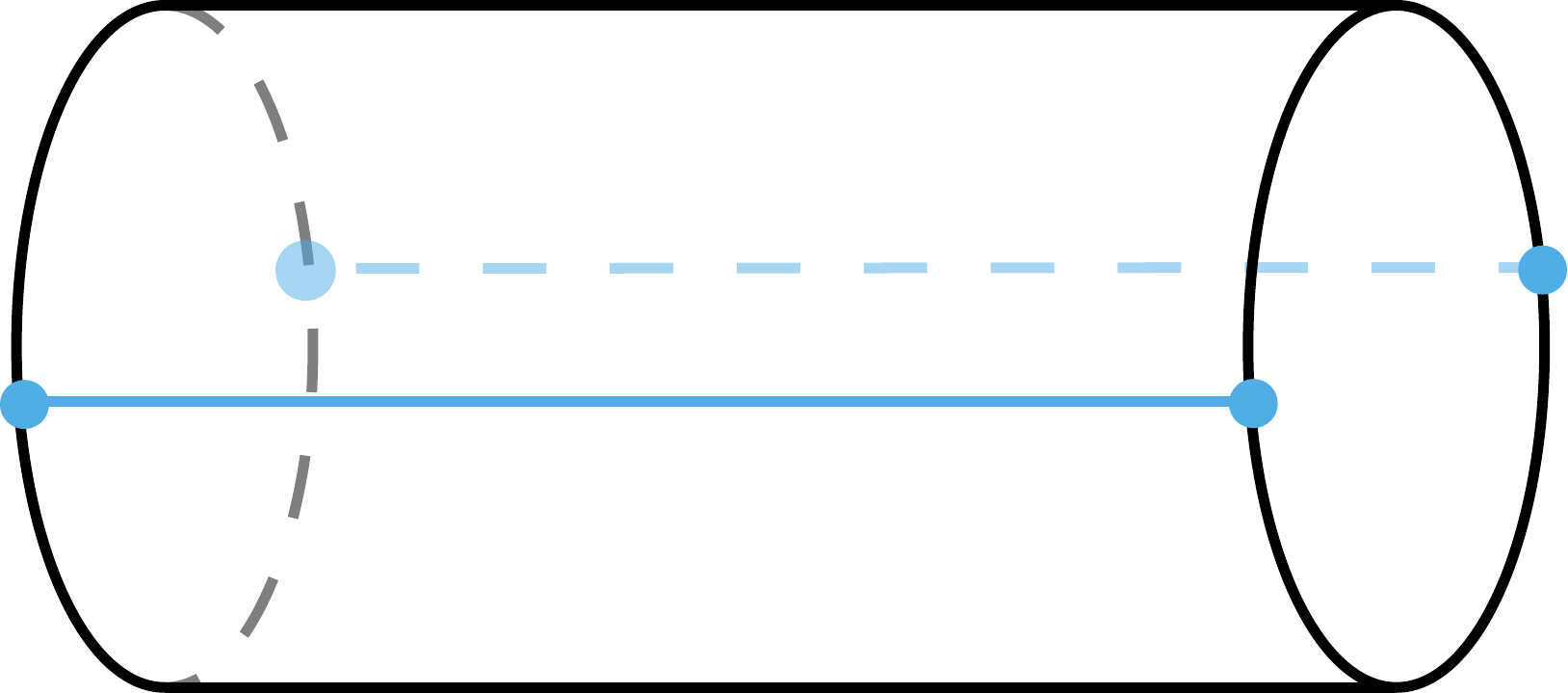}\\
    &=\int d\ell_1\,d\ell_2\,e^{-\Delta \ell_1}e^{-\Delta\ell_2}\,\left(\inlinefig[7]{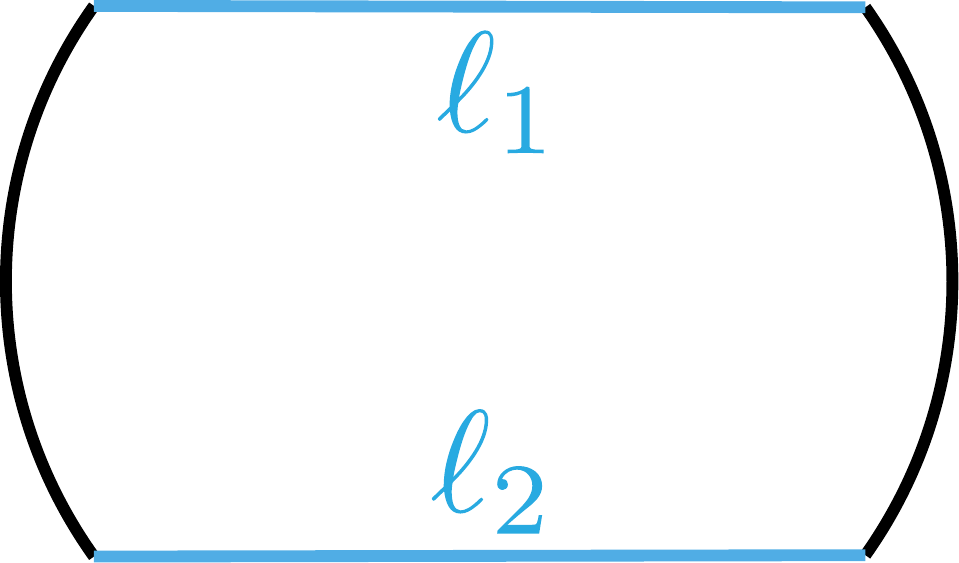}\right)^2\\
    &=\int_{\mathfrak{E}} dE_1\,dE_2\,\rho_0(E_1)\rho_0(E_2) (P_{\Delta\to\infty}(\bar{E}))^2.
\end{split}
\end{equation}

This procedure can easily be generalized to a higher number of boundaries and higher topologies.

It is straightforward to generalize Eq. \eqref{jt_disk} to diagrams without empty handles. Consider the fully-connected component of the $m$-boundary correlator $\langle(\tr O^{2k})^m\rangle_{\text{conn}}$ and consider a diagram with $g$ non-empty handles. This diagram will contribute at order $e^{(2-2g-m)S_0}$ and contains $(P_{\Delta\to\infty})^{km}$ from the $km$ propagators. 

The number of energy patches requires some thought and can be counted by considering the Euler characteristic. For a genus-$g$ and $m$-boundary surface, we have $\chi=2-2g-m$. Cutting along a geodesic increases the Euler characteristic by one. Since we require the diagram to become a product of $n$ asymptotic polygons with total Euler characteristic $n$, this implies $n=k+2-2g-m$. Thus, we obtain that each connected diagram of genus-$g$ with $m$ boundaries and without empty handles contributes
\begin{equation}
e^{(2-2g-m)S_0}\int \prod_{i=1}^{k+2-2g-m}dE_i\,\rho_0(E_i)(P_{\Delta\to\infty}(E))^{km}
\end{equation}
to the $m$-boundary correlator. We must now count how many such diagrams there are for a given $2knm$ point function with $2kn$ points on each of the $m$ boundaries. This counting has been discussed in Section \ref{sec:whichdiagrams} and Appendix \ref{appendix:f}, yielding the genus-dependent Fa\`a di Bruno coefficients $C^{(g)}_{2k,[2^k]}$ for single-boundary correlators and $C^{(m)}_g(2k,\dots ,2k)$ for multi-boundary correlators. Therefore, for single-boundary correlators, we can write
\begin{equation}
    \overline{\tr_{\mathfrak E}\widetilde{O}^{2k}}^{\text{min}}=(P_{\Delta\to\infty}(\bar{E}))^{k}\sum_{g=0}^{\lfloor k/2\rfloor}e^{(1-2g)S_0}C_{2k,[2^k]}^{(g)}\int_{\mathfrak E}\prod_{i=1}^{k+1-2g}dE_i\,\rho_0(E_i),\label{jt_nonempty}
\end{equation}
and, more generally,
\begin{equation}
\begin{split}
    \overline{\tr_{\mathfrak E}(\widetilde{O}^{2k})^m}_{\text{conn}}^{\text{min}}=&(P_{\Delta\to\infty}(\bar{E}))^{km}\\
    &\times\sum_{g=0}^{\lfloor (km+1-m)/2\rfloor}e^{(2-2g-m)S_0}C_g^{(m)}(2k,\dots,2k)\int_{\mathfrak E}\prod_{i=1}^{km+2-2g-m}dE_i\,\rho_0(E_i).\label{jt_nonempty_multi}
\end{split}
\end{equation}

\subsection{Handles between patches}
\label{subsec:handles-between-patches}

In the BPS puzzle, the diagrammatics end in the previous subsection because diagrams with empty handles vanish. However, in microcanonical non-SUSY JT, we must account for such diagrams since they account for fluctuations within the microcanonical window. Despite the vast number of diagrams, the results can be summarized by replacing the product of disk level densities of states $\rho_0(E_1)\dots\rho_0(E_{km+2-2g-m})$ by $\overline{\rho_0(E_1)\dots\rho_0(E_{km+2-2g-m})}$ in Eq. \eqref{jt_nonempty_multi}  \cite{Saad_2019, Iliesiu:2021ari}.

This can be argued as follows. Start with a connected non-empty handled diagram and slice along all matter propagators. By construction, this gives a product of asymptotic polygons, which all have disk topology. All diagrams, including those with empty handles, can then be obtained by adding handles on and between polygons. For example, for the diagrams contributing to the two-point function, we have
\begin{equation}
\inlinefig[3]{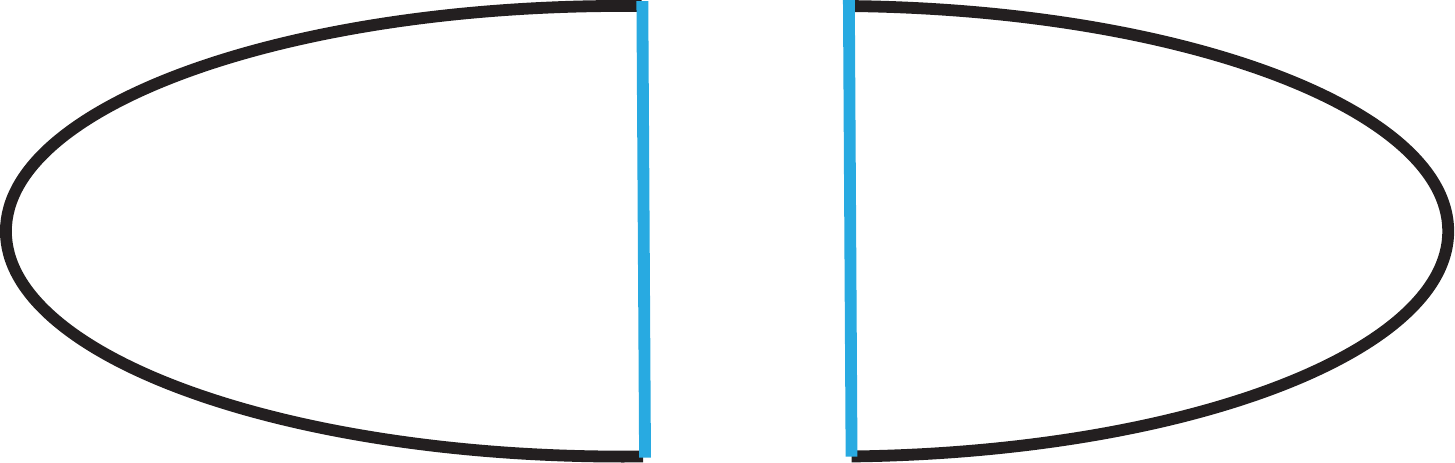}\to\inlinefig[3]{Figures/disk_side.pdf}+\inlinefig[6]{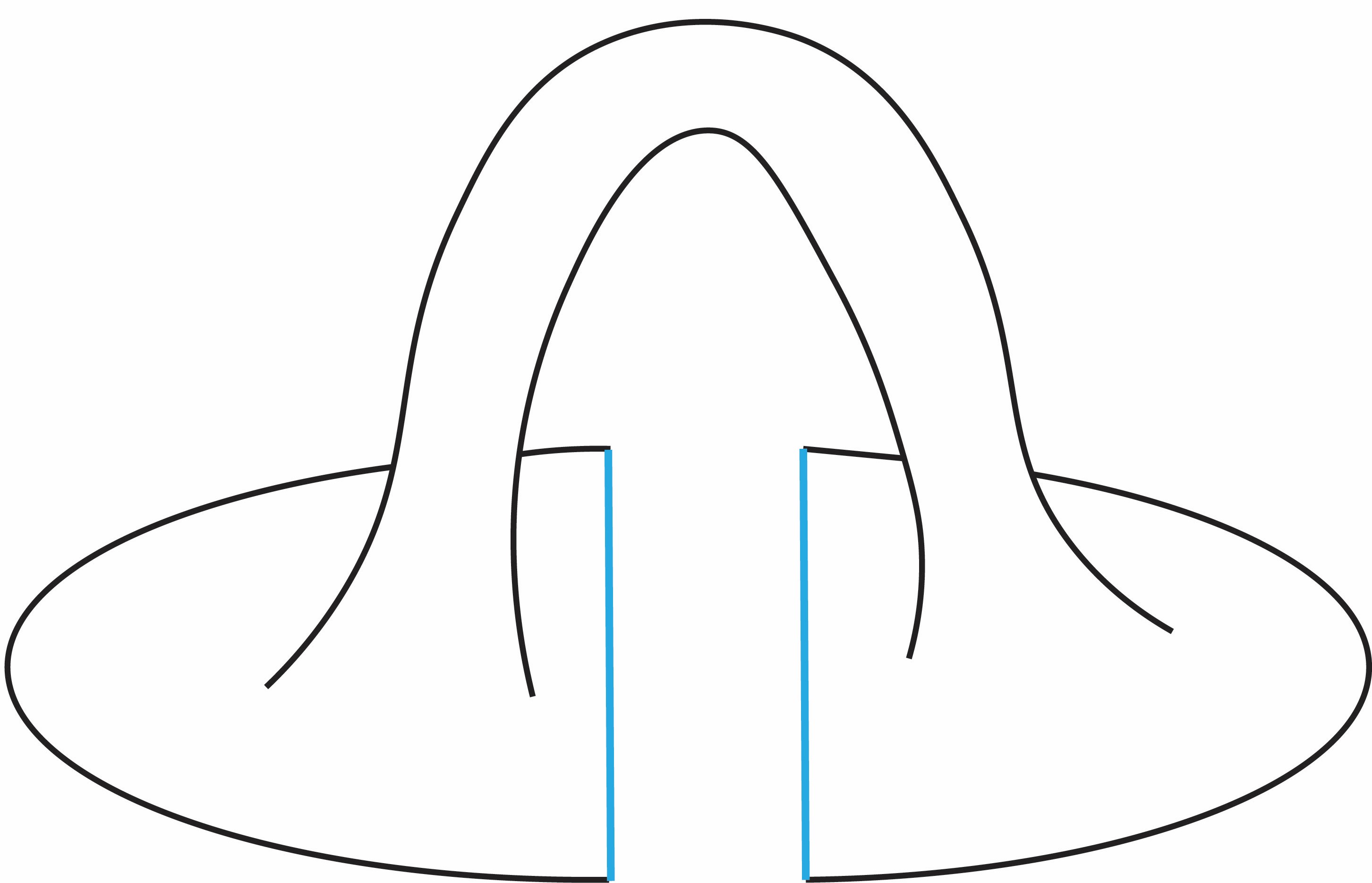}+\inlinefig[5]{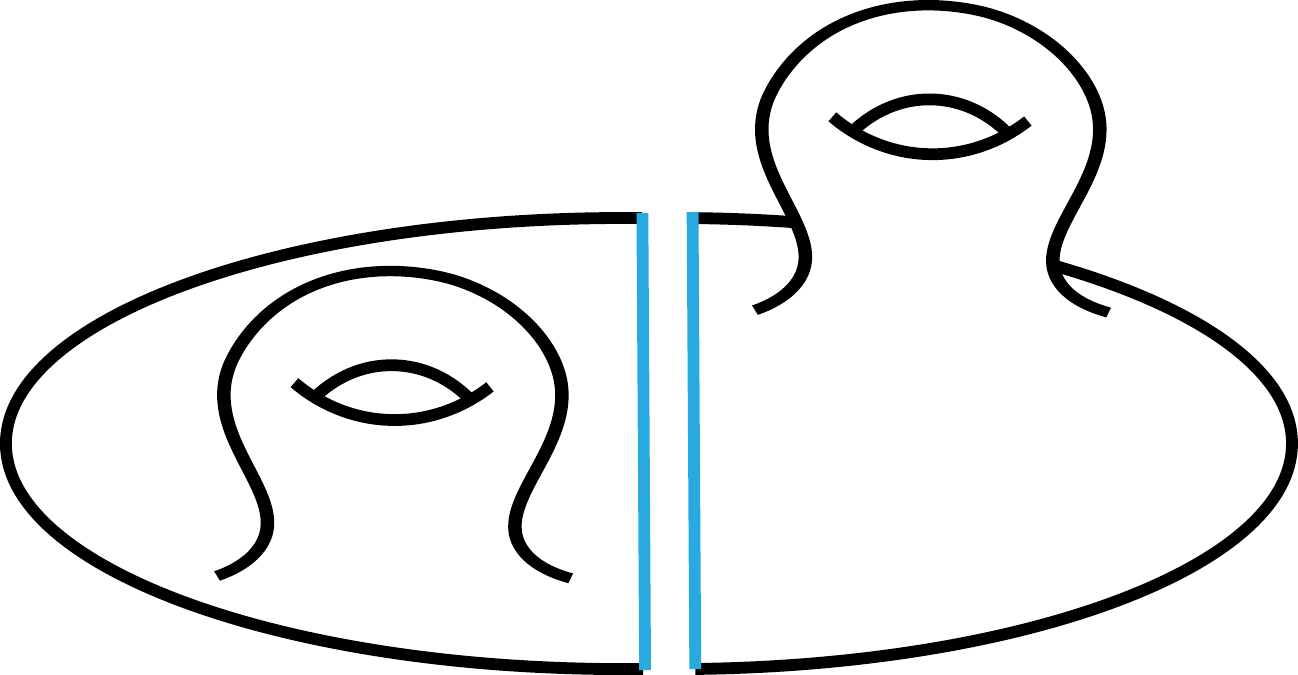}+\dots.
\label{eq:sum-over-genus}
\end{equation}
Notice that the right-hand side is equivalent to taking the gravitational path integral over two disconnected boundaries. Hence, accounting for fluctuations between asymptotic polygons is equivalent to the promotion of 
\begin{equation}
    \int_{\mathfrak E}dE_1\,dE_2\,\rho_0(E_1)\rho_0(E_2)\to\int_{\mathfrak{E}}dE_1\,dE_2\,\overline{\rho_0(E_1)\rho_0(E_2)},
\end{equation}
and, more generally,
\begin{equation}
    \int_{\mathfrak E}\prod_{i=1}^{n}dE_i\,\rho_0(E_i)\to\int_{\mathfrak E}\overline{\prod_{i=1}^ndE_i\,\rho_0(E_i)}.
\end{equation}
Note that in the two-point function, one has to sum over all possible paths between two boundary points, and in the path integral, one has to quotient by all large diffeomorphisms that can relate geodesics in different homotopy classes. Accounting for the diagrams in \eqref{eq:sum-over-genus} only once correctly accounts for the quotient by the mapping class group in the path integral \cite{Saad_2019, Iliesiu:2021ari, Iliesiu:2024cnh}.  

Thus, we can promote Eq.~\eqref{jt_nonempty} to the single-boundary correlator (this is Eq. \eqref{eq:1bdycorrelationJT})
\begin{equation}
\overline{\tr_{\mathfrak{E}}\widetilde{O}^{2k}}=(P_{\Delta\to\infty}(\bar{E}))^k\sum_{g=0}^{\lfloor k/2\rfloor}e^{(1-2g)S_0}C_{2k,[2^k]}^{(g)}\int_{\mathfrak{E}}\overline{\prod_{i=1}^{k+2-2g}dE_i\,\rho_0(E_i)}\,,\label{jt_single}
\end{equation}
and Eq. \eqref{jt_nonempty_multi} to the fully connected $m$-boundary correlator
\begin{equation}
\begin{split}
    \overline{\tr_{\mathfrak E}(\widetilde{O}^{2k})^m}_{\text{conn}}=&(P_{\Delta\to\infty}(\bar{E}))^{km}\\
    \times&\sum_{g=0}^{\lfloor (km+1-m)/2\rfloor}e^{(2-2g-m)S_0}C_g^{(m)}(2k,\dots,2k)\int_{\mathfrak E}\overline{\prod_{i=1}^{km+2-2g-m}dE_i\,\rho_0(E_i)}\,.\label{jt_multi}
\end{split}
\end{equation}
Finally, we can map these correlators back to the LMRS puzzle. By taking the microcanonical window small while keeping the overall number of states in the window $O(e^{S_0})$, in the $k=O(e^{2S(\mathfrak{E})/3})$ regime we can make the approximation\footnote{This approximation is in general valid for small $k$ if one only keeps diagrams with trivial topology. In Eqs. \eqref{jt_single} and \eqref{jt_multi}, however, we kept also higher genus contributions. For generic $k$, the fluctuations we are dropping here are not suppressed with respect to these higher genus contributions. However, for $k=O(e^{2S(\mathfrak{E})/3})$, diagrams with no empty handles are combinatorially enhanced in $k$ (as we have explained in Section \ref{sec:bulk-resolution}), whereas the fluctuations we are dropping here are not. This justifies our approximation.}
\begin{equation}
    \int_{\mathfrak E}\overline{\prod_{i=1}^ndE_i\,\rho_0(E_i)}\approx (\rho(\bar{E})\delta E)^n,
\end{equation}
Inserting this into Eq.~\eqref{jt_single} and \eqref{jt_multi}, we obtain
\begin{equation}
    \overline{\tr_{\mathfrak E}\widetilde{O}^{2k}}\approx (P_{\Delta\to\infty}\rho(\bar{E})\delta E)^{k}\sum_{g=0}^{\lfloor k/2\rfloor}e^{(1-2g)S(\mathfrak{E})}C_{2k,[2^k]}^{(g)}
\end{equation}
\begin{equation}
    \overline{(\tr_{\mathfrak E}\widetilde{O}^{2k})^m}_{\text{conn}}\approx (P_{\Delta\to\infty}(\bar{E})\rho(\bar{E})\delta E)^{km}\sum_{g=0}^{\lfloor (km+1-m)/2\rfloor}e^{(2-2g-m)S(\mathfrak{E})}C_g^{(m)}(2k,\dots 2k),
\end{equation}
where we used the $\approx$ sign to emphasize the formulae are valid in the double-scaling limit $k=O(e^{2S(\mathfrak{E})/3}$ and we defined the microcanonical entropy
\begin{equation}
    S(\mathfrak{E})\equiv S_0+\log(\rho(\bar{E})\delta E).
\end{equation}
Notice that the correlators have the same structure as the correlators in SUSY JT, only by a relabeling of the overall number of states and a rescaling of the propagator. Therefore, the puzzles as stated in SUSY JT and microcanonical non-SUSY JT are equivalent, and so are their resolutions.

\section{More correlators in the BPS sector from the effective JT description}\label{app:matter}

As a further comment to strengthen the correspondence to an effective JT gravity description as described in Section \ref{sec:mapping}, we can calculate correlators involving an additional matter probe $O_m$ with arbitrary scaling dimension $\Delta_m$ in $\mathcal N=2$ super-JT gravity.\footnote{We would like to thank the referee for pointing this out.} Explicitly, we claim the following correspondence between correlators in the microcanonical ensemble and those in the effective JT theory
\begin{equation}
    \tr(\tilde{O}_{\Delta}^{2k_1}O_m\tilde{O}_{\Delta}^{2k_2}O_m)\Leftrightarrow 2\tr(e^{-\beta_1 H_{\rm eff}} \tilde O_m e^{-\beta_2 H_{\rm eff}} \tilde O_m)\label{mattercorr}
\end{equation}
where the left-hand side is evaluated in the BPS sector of $\mathcal N=2$ JT supergravity or microcanonical non-SUSY JT in the Airy limit ($k=O(e^{2S_0/3})$). On the right-hand side, the correlator is evaluated for some operator $\tilde O_m$, whose properties we shall describe shortly, in the effective JT theory at finite temperature.\footnote{Higher-point correlators can be constructed using similar surgery rules, though they are more complicated due to the presence of 6j-symbols and their Yang-Baxter rules.}

We first begin at disk level, for which Eq. \eqref{mattercorr} is represented diagrammatically with
\begin{equation}
    \inlinefig[12]{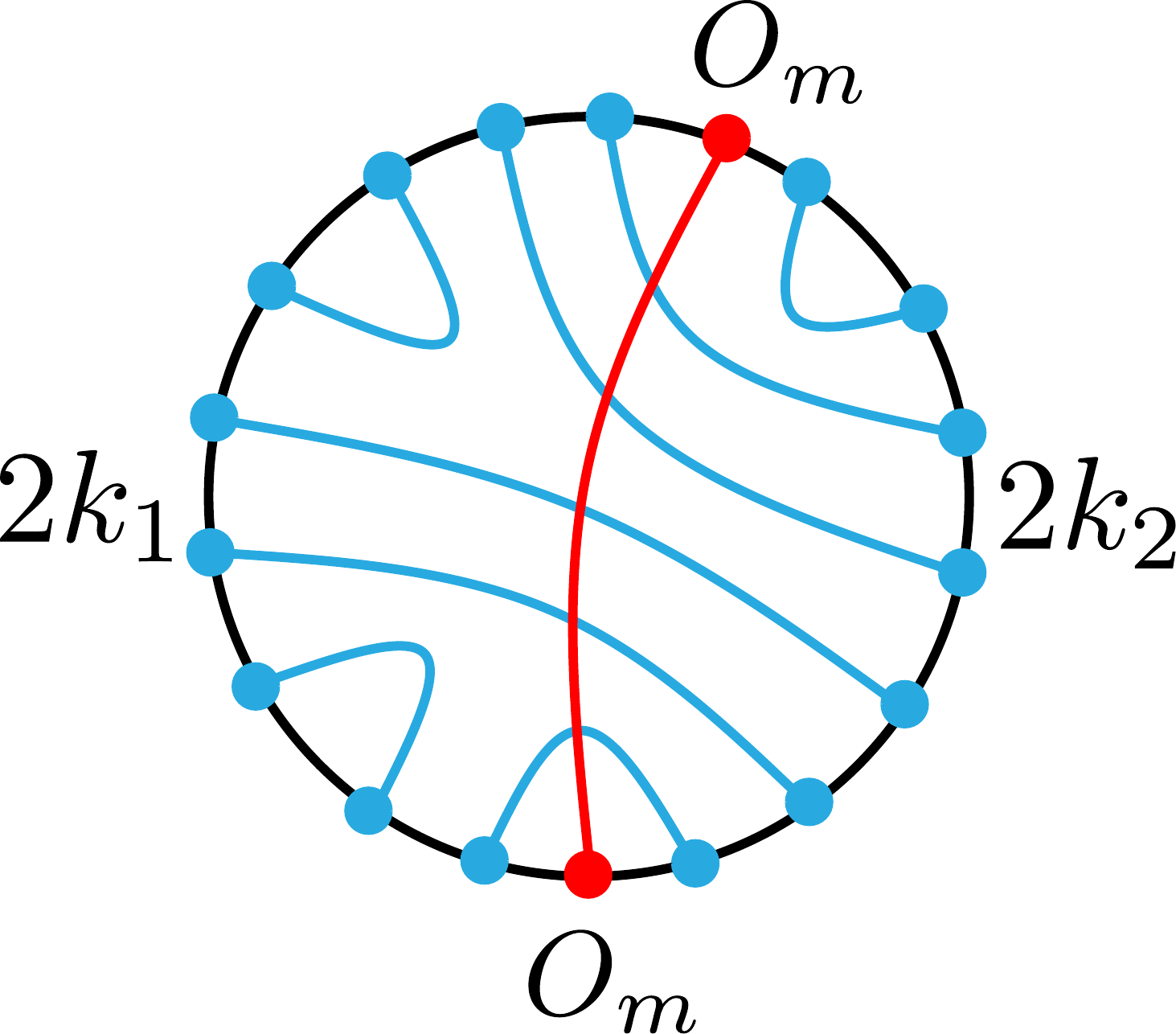}\quad+\quad\dots\qquad=\qquad 2\times\quad
    \inlinefig[12]{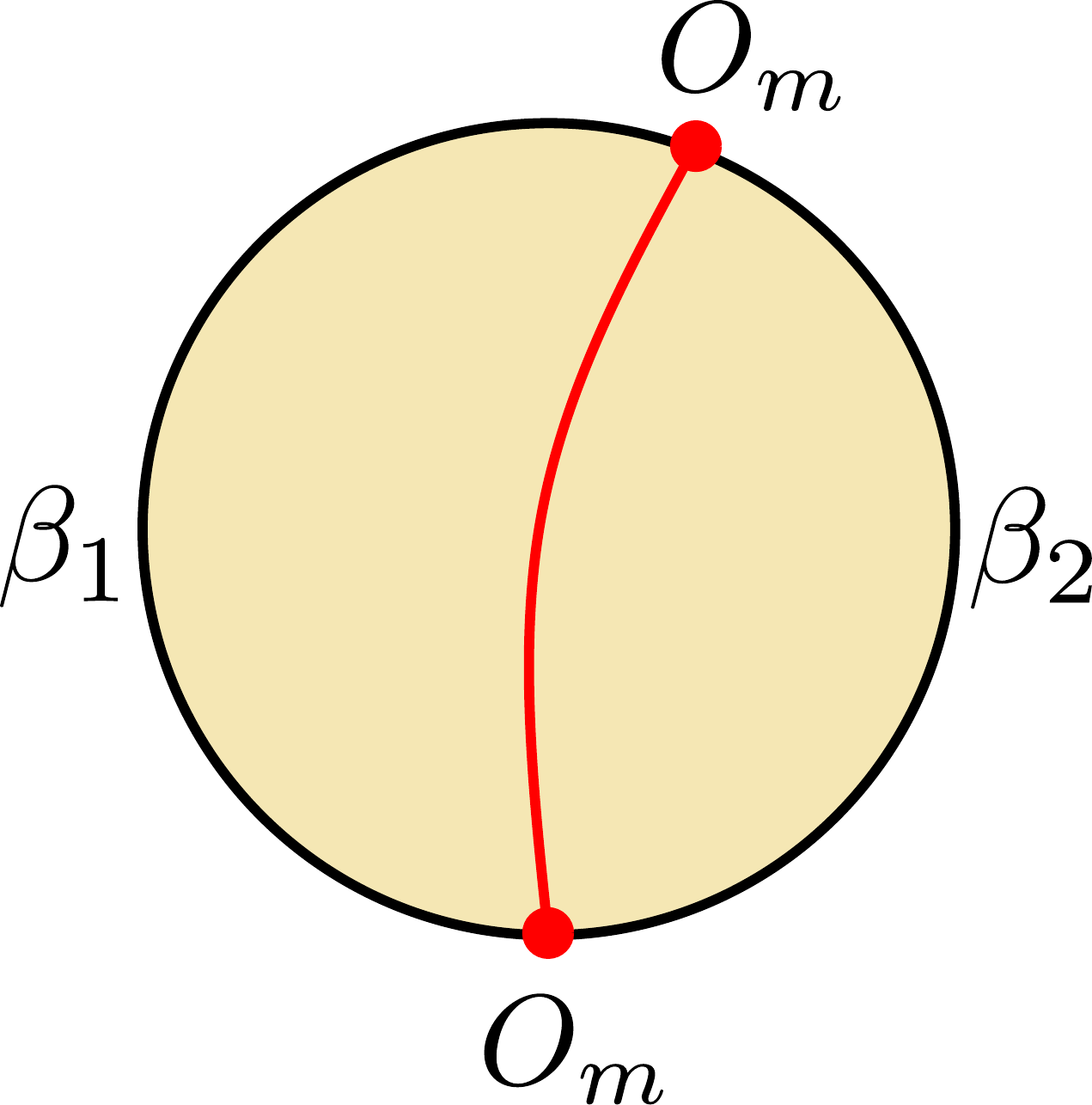},
\end{equation}
where $\dots$ represent all possible (non-crossing) Wick contractions between blue $O$ operators. On the effective JT side, this is given by
\begin{equation}
     \inlinefig[12]{Figures/matter_probe_RHS.pdf}=e^{S_0^{\rm eff}}\int d\ell\,\Psi_{\beta_1}(\ell)\Psi_{\beta_2}(\ell) e^{-\tilde \Delta_m\ell},
\end{equation}
where $\Psi_\beta(\ell)$ are Hartle-Hawking wavefunctions in the length basis for the effective JT theory and where $\tilde \Delta_m$ is the scaling dimension of the operator $\tilde O_m$. On the BPS side, we are interested in calculating correlators $\tr(\widetilde{O}_{\Delta}^{2k_1} O_m \widetilde{O}_\Delta^{2k_2}O_m)$. Given a diagram, we can slice along the red $O_m$ geodesic. If the blue geodesics cross the red geodesic $n$ times, we have
\begin{equation}
    \inlinefig[12]{Figures/matter_probe_LHS.pdf}\quad+\quad\dots\qquad=e^{S_{\rm BPS}}\sum_{n=0}^{\min(2k_1,2k_2)}(P_{\Delta_m})^n Z(k_1;n) Z(k_2;n),
\end{equation}
where $Z(k;n)$ denotes diagrams with $2k$ boundary insertions with $n$ boundary-to-slice geodesics:
\begin{equation}
    Z(k_1;n)= \inlinefig[8]{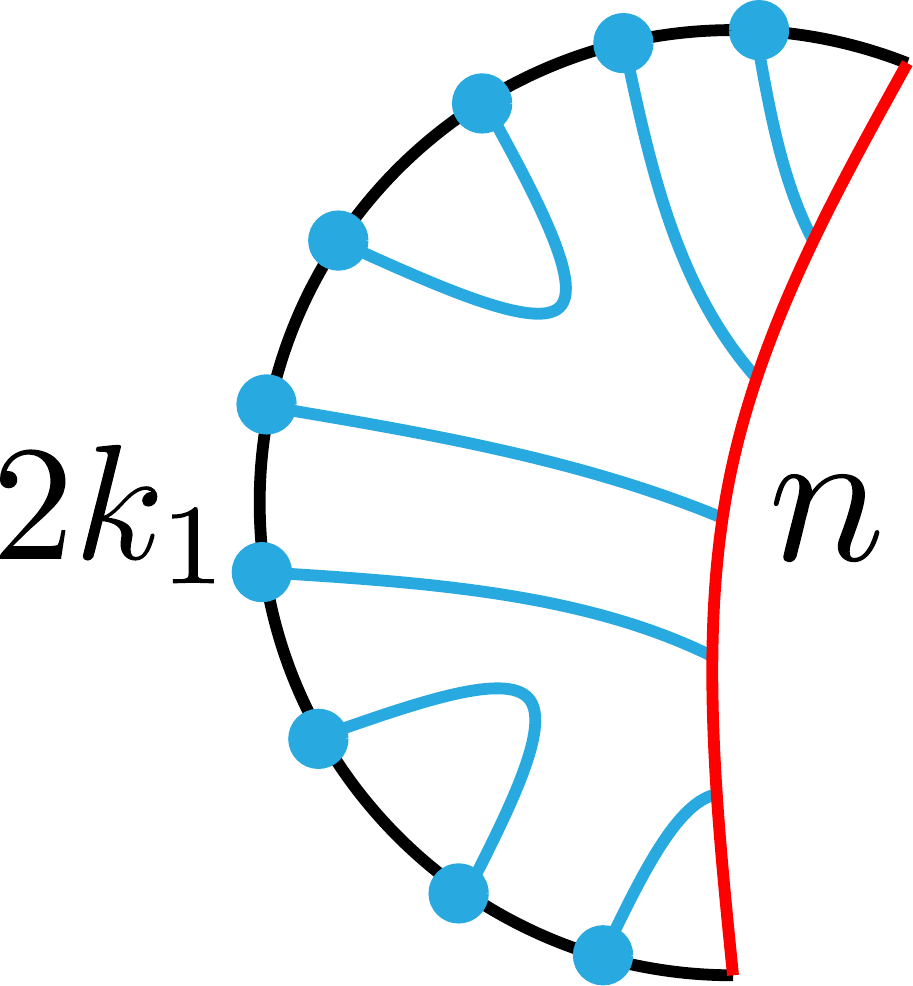}+\dots.
\end{equation}
Importantly, there can be no geodesics that start and end at the red slice.
$P_{\Delta_m}$ denotes the weight associated to intersections between the $O_m$ and $\widetilde{O}_\Delta$ worldlines, a complicated 6j-symbol whose exact form will not be of concern. Taking $k=O(e^{2S_{\rm BPS}/3})$, we can make the identifications $k_{1,2}=\beta_{1,2}$ and $S_{\rm BPS}=S_0^{\rm eff}$ similar to those found in Section \ref{sec:mapping}. In this limit, we can treat $n$ as continuous and identify the geodesic distance $\ell$ in the effective JT theory with $n$ as  $\sum_n\leftrightarrow \int d\ell$ and $n\leftrightarrow \ell$. This is precisely the identification of distances mentioned in Section \ref{sec:mapping}. This leads to the further simplification $Z(k;n)=\Psi_{\beta}(\ell)$ and $P_{\Delta_m}= e^{-\tilde\Delta_m}$, which defines what the scaling dimension $\tilde\Delta_m$ of the operator in the effective JT theory needs to be. 

Generalization to higher genera surfaces is straightforward. For example, at genus-1, we have the diagrams
\begin{equation}
    \inlinefig[10]{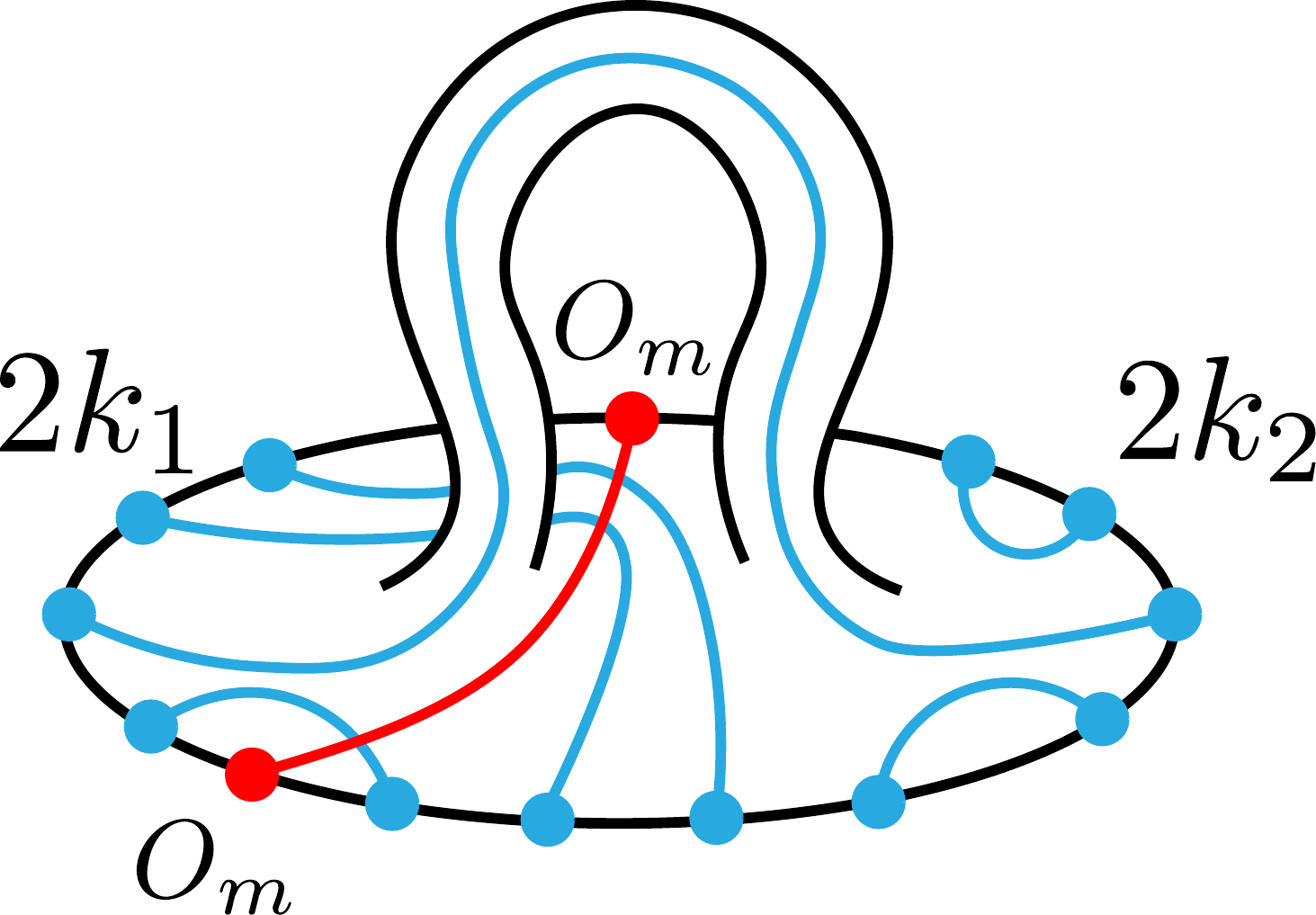}\quad+\quad\dots\qquad=\qquad 2\times\quad
    \inlinefig[10]{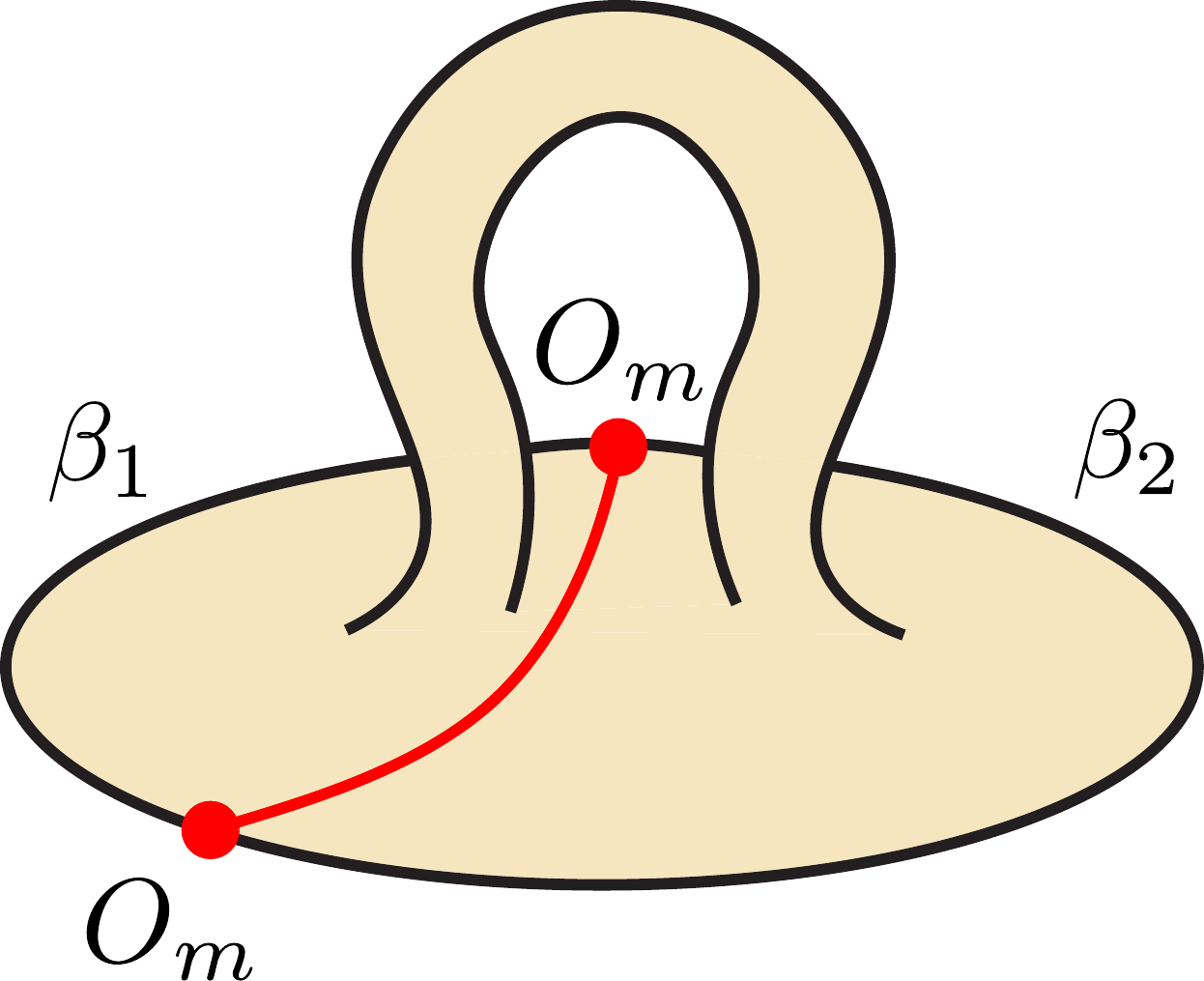}.
\end{equation}
Slicing along the propagator leads to a cylinder diagram. On the right-hand side, this amounts to 
\begin{equation}
     \inlinefig[10]{Figures/matter_probe_g1_RHS.pdf}=e^{-S_0^{\rm eff}}\int d\ell\,\overline{\Psi_{\beta_1}(\ell)\Psi_{\beta_2}(\ell)}_{\rm cyl} e^{-\Delta_m\ell},
\end{equation}
where $\overline{\Psi_{\beta_1}(\ell)\Psi_{\beta_2}(\ell)}_{\rm cyl}$ denotes the contribution from the cylinder. On the left-hand side, we have
\begin{equation}
    \inlinefig[10]{Figures/matter_probe_g1_LHS.pdf}\quad+\quad\dots\qquad=e^{-S_{\rm BPS}}\sum_n(P_{\Delta_m})^n\overline{Z(k_1;n)Z(k_2;n)}_{\rm cyl},
\end{equation}
where $\overline{Z(k_1;n)Z(k_2;n)}_{\rm cyl}$ denotes diagrams whose minimal embedding diagram is the cylinder. Such diagrams have $2k_1$ and $2k_2$ $\widetilde{O}_\Delta$ insertions, with $n$ boundary-to-slice geodesics. Naturally, we make the identification $\overline{Z(k_1;n)Z(k_2;n)}_{\rm cyl}\leftrightarrow \overline{\Psi_{\beta_1}(\ell)\Psi_{\beta_2}(\ell)}_{\rm cyl}$. This naturally leads to the generalization to include higher genera surfaces: $\overline{Z(k_1;n)Z(k_2;n)}\leftrightarrow \overline{\Psi_{\beta_1}(\ell)\Psi_{\beta_2}(\ell)}$.

\section{Matrix model for generic $\Delta$}
\label{sec:ETH}

In Section \ref{sec:matrixmodels} we explained how, within the BPS sector of $\mathcal{N}=2$ JT supergravity or a microcanonical band in non-SUSY JT gravity, correlation functions of the operator $\tilde{O}_{\Delta}$ at large $\Delta$ are exactly captured by a GUE matrix model. In this appendix, we will explain the difficulties in obtaining a similar result for generic $\Delta$, propose a matrix integral toy-model to capture the relevant feature of the generic $\Delta$ case, and finally explain a more generic perspective on matrix models for JT gravity coupled to matter. For simplicity, in most of this appendix, we will give formulas for non-SUSY JT gravity within a fixed microcanonical band, but the results also apply to the BPS case, as we will comment throughout.

\subsection{Difficulties for generic $\Delta$}
 \label{sec:genericd}

Let us consider a four-point function of the operator $\tilde{O}_{\Delta}$ (involving projectors on a microcanonical band $\mathfrak{E}$) in non-SUSY JT gravity. At the disk level, this can be computed by the following diagrams:
\begin{equation}
    \overline{\tr_{\mathfrak E}\(\tilde{O}_\Delta^{4}\)}=\inlinefig[10]{Figures/jt_disk1.pdf}+\inlinefig[10]{Figures/jt_disk1_rot.pdf}+\inlinefig[10]{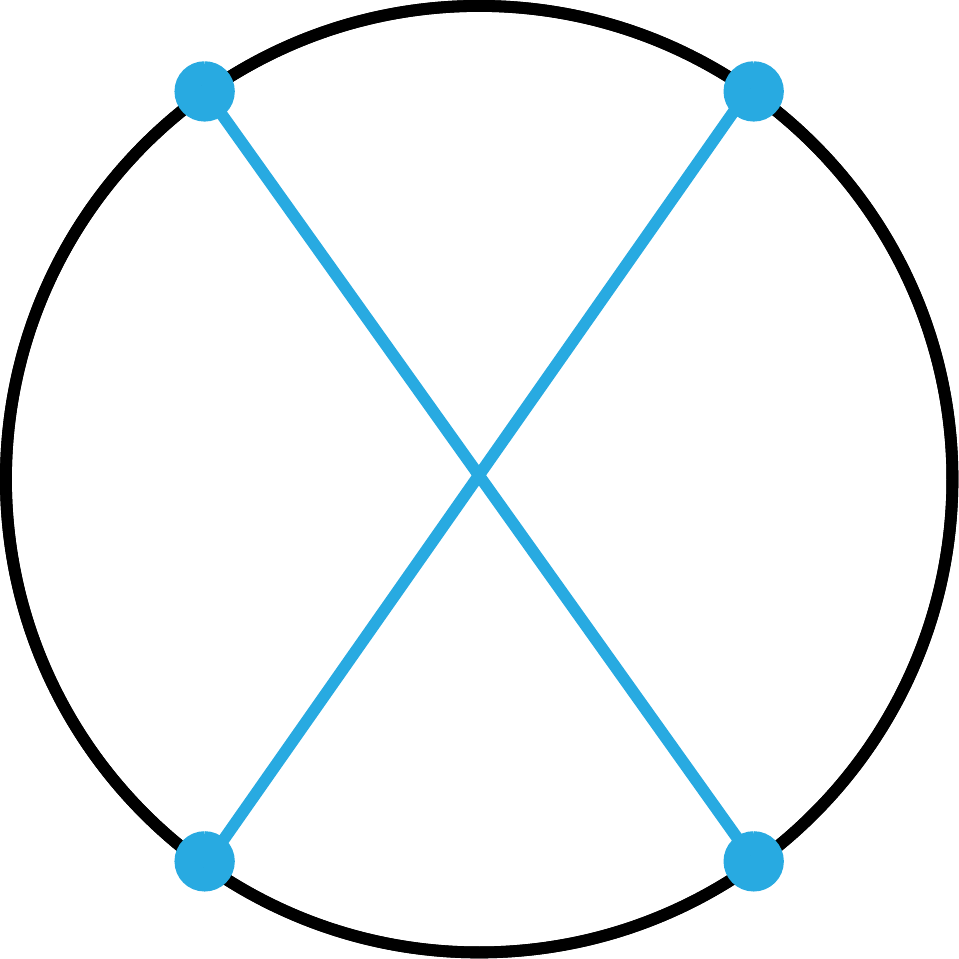}
    \label{eq:full4disk}
\end{equation}
The explicit expression for the four-point function in non-SUSY JT gravity is given by \cite{Mertens_Turiaci_Verlinde_2017,Blommaert:2018oro,Iliesiu:2019xuh,Jafferis_Kolchmeyer_Mukhametzhanov_Sonner_2023}
\begin{equation}
    \int_{\delta E}ds_1ds_2ds_3ds_4\rho_0(s_1)\rho_0(s_2)\rho_0(s_3)\rho_0(s_4)G_4(s_1,s_2,s_3,s_4)
    \label{eq:full4pf2}
\end{equation}
where $\delta E$ is the width of the microcanonical window of interest and\footnote{For a review of the Feynman rules leading to Eq.~\eqref{eq:4pf2} we refer the reader to \cite{Jafferis_Kolchmeyer_Mukhametzhanov_Sonner_2023}.}
\begin{equation}
    G_4(s_1,s_2,s_3,s_4)=e^{S_0}(\Gamma_{12}\Gamma_{23}\Gamma_{34}\Gamma_{41})^{\frac{1}{2}}\left(\frac{\delta(s_1-s_3)}{\rho_0(s_1)}+\frac{\delta(s_2-s_4)}{\rho_0(s_2)}+\begin{Bmatrix}
	\Delta & s_{1} & s_{2}\\
	\Delta & s_{3} & s_{4}
\end{Bmatrix}\right)
\label{eq:4pf2}
\end{equation}
with each term in the sum corresponding to each diagram in Eq.~\eqref{eq:full4disk}.

In the large $\Delta$ limit, the third diagram is exponentially suppressed in $\Delta$ with respect to the first two. In the non-supersymmetric case, this is because the 6j-symbol in the third term of Eq.~\eqref{eq:4pf2}---which weighs the intersection between two matter particles---satisfies (see Appendix \ref{6j}),
\begin{equation}
    \begin{Bmatrix}
	\Delta & s_{1} & s_{2}\\
	\Delta & s_{3} & s_{4}
\end{Bmatrix}\xrightarrow{\Delta\to\infty} Ce^{-(1+4\log 2)\Delta}\,.
\label{eq:larged6j2}
\end{equation}
 A similar suppression associated with intersections of matter lines in the $\Delta\to \infty$ limit, although with a slightly different coefficient in the exponent ($e^{-4\Delta\log(1+\sqrt{2})}$), is also present in JT supergravity \cite{Lin_Maldacena_Rozenberg_Shan_2023}. This suppression is the reason why, in the large $\Delta$ limit analyzed in the bulk of the paper, we only considered diagrams with non-intersecting matter worldlines. This approximation guarantees that the combinatorics of correlation functions in the large $\Delta$ limit matches the combinatorics of GUE correlation functions, as we have discussed in Section \ref{sec:GUE}.

What can we say about generic values of the scaling dimension $\Delta$? Certainly, the last term in Eq.~\eqref{eq:4pf2} cannot be neglected. More generically, when computing $2k$-point functions of the matter operator $\tilde{O}_\Delta$, Wick contractions involving intersections of matter geodesics---with each intersection weighted by a 6j-symbol---are comparable in size to non-intersecting diagrams, and the weighing of each diagram depends non-trivially on $\Delta$ through (products of) 6j-symbols. This immediately implies that the combinatorics associated with operators of generic scaling dimension $\Delta$ are much more involved.

A second, maybe even trickier issue manifests itself when computing higher-point functions. Let us consider a 6-point function and focus our attention on the diagram
\begin{equation}
    \inlinefig[12]{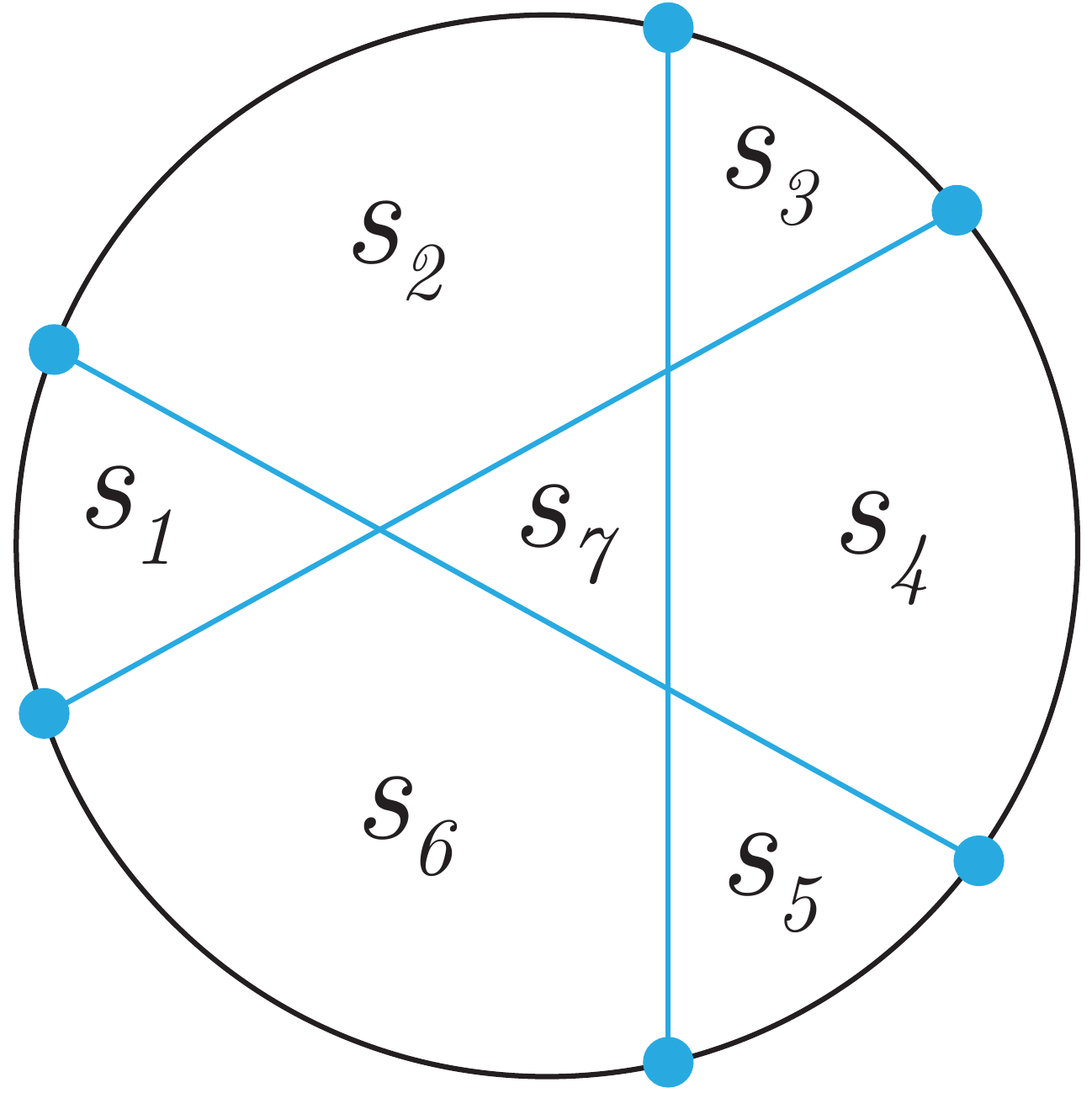}
\end{equation}
Using the Feynman rules reviewed in \cite{Jafferis_Kolchmeyer_Mukhametzhanov_Sonner_2023}, one finds the following expression for this diagram
\begin{equation}
    e^{S_0}\int \prod_{i=1}^7(ds_i \rho_0(s_i))\sqrt{\Gamma_{12}\Gamma_{23}\Gamma_{34}\Gamma_{45}\Gamma_{56}\Gamma_{61}}\begin{Bmatrix}
	\Delta & s_{1} & s_{2}\\
	\Delta & s_{7} & s_{6}
\end{Bmatrix}\begin{Bmatrix}
	\Delta & s_{2} & s_{3}\\
	\Delta & s_{4} & s_{7}
\end{Bmatrix}\begin{Bmatrix}
	\Delta & s_{4} & s_{5}\\
	\Delta & s_{6} & s_{7}
\end{Bmatrix}\,.
\end{equation}
The integral for the external energies $s_1,...,s_6$ is again fixed by assumption to be within the microcanonical window of interest. However, the energy $s_7$ appearing in the ``loop'' is unconstrained, and the associated integral runs over all possible energies. Naturally, this is true for all higher-point functions, for which diagrams involving more and more loops appear. A similar reasoning can be carried out for $\mathcal{N}=2$ JT supergravity. The energy spectrum in that case is determined by a different Hamiltonian ensemble \cite{Turiaci:2023jfa}, but the analysis is otherwise unchanged. In particular, even if one fixes all external energies to be equal to the BPS ground state energy as in the LMRS setup, energies in loops are arbitrary and need to be integrated over.

The presence of these loops, which are absent by construction in the large $\Delta$ limit, implies that the integral over the JT gravity energy spectrum contributes non-trivially to matter correlation functions at generic $\Delta$. In other words, we cannot hope to describe correlation functions simply in terms of a single matrix integral for the matter operator $\tilde{O}_\Delta$ as we did in the large $\Delta$ limit in Section \ref{sec:GUE}. In general, this will require a much more involved two-matrix integral \cite{Jafferis_Kolchmeyer_Mukhametzhanov_Sonner_2023}, whose details and regime of applicability remain quite unclear. We will comment more on this generic model in Appendix \ref{sec:twomatrix}.

To make progress, here we will instead work with a simple but effective toy-model of the generic $\Delta$ case: the $q$-deformed Gaussian matrix model. Before introducing the $q$-deformed Gaussian in Section \ref{sec:qgaussian}, we will study another treatable limit of our setup, namely the $\Delta\to 0$ limit in which one version of the LMRS puzzle was formulated (see Section \ref{sec:2}). As we will see, the $q$-deformed Gaussian is able to capture both the $\Delta\to 0$ and $\Delta\to\infty$ limits for appropriate values of its single parameter $q$, and it smoothly interpolates between the two.

\subsection{$\Delta\to 0$: Gaussian spectral density}
\label{sec:smalld}

 In the $\Delta\to 0$ limit, all Witten diagrams of fixed genus that contribute to a $k$-point function are weighted the same, regardless of the number of intersections of matter geodesics. In the BPS case, this was discussed in \cite{Lin_Maldacena_Rozenberg_Shan_2023}. In the non-supersymmetric JT case, this can be seen by taking the $\Delta\to 0$ limit of the 6j-symbol (see Appendix \ref{6j}).

Moreover, the matter propagator \eqref{eq:matterprop} also reduces to a delta function of the two energies. This is readily understood by recalling that it is given by $\langle E_a|e^{-\Delta \ell}|E_b\rangle$, where $\ell$ is the length operator in JT gravity \cite{Iliesiu:2024cnh}: as we take $\Delta\to 0$, we simply obtain a delta function. As we have discussed in Footnote \ref{foot:delta0}, $\Delta\to 0$ is a sick limit. Our $\Delta\to 0$ limit should be understood as the leading order approximation to a small but finite $\Delta$ limit, in which the operator is close to the identity.
The simple form taken by the product between the matter propagators and the 6j-symbol (see Appendix \ref{6j}) in the $\Delta\to 0$ limit also immediately implies that all energies, both external and running in loops, are set equal to each other.

As a result, both issues raised in Section \ref{sec:genericd} disappear, and $2kn$-point functions at the disk level are simply computed by counting all possible pairwise Wick contractions of the $2kn$ operator insertions with equal weights. This is the same conclusion reached by \cite{Lin_Maldacena_Rozenberg_Shan_2023} for the BPS case. This also suggests which matter matrix integral correctly captures this result: one with a Gaussian equilibrium spectral density, for which momenta are also captured by all possible pairwise Wick contractions. The normalized spectral density is thus given by
\begin{equation}
    \bar{\sigma}_{\text{Gauss}}(x)=\frac{1}{\sqrt{2\pi P_{\Delta\to 0}}}e^{-\frac{x^2}{2P_{\Delta\to 0}}}.
\label{eq:gaussspectrum}
\end{equation}
For a single-boundary $2kn$-point function this yields 
\begin{equation}
    \overline{\tr(\tilde{O}_{\Delta\to 0}^{2k})} = e^{S_0} \int dx\bar{\sigma}_\text{Gauss}(x)x^{2kn}=e^{S_0}(P_{\Delta\to 0})^{kn}(2kn-1)!!~,
\end{equation}
which reproduces the result in Eq.~\eqref{eq:moments-of-rho-Delta-to-0}, obtained by simply counting all possible pairwise Wick contractions. Given the equilibrium spectral density \eqref{eq:gaussspectrum} and using Eq.~\eqref{eq:extremization}, we can deduce the matrix potential:
\begin{equation}
	V_\text{Gauss}(\lambda)=\lambda^{2}{}_{2}F_{2}\left(1,1;\frac{3}{2},2;-\frac{1}{2}\lambda^{2}\right)\,,
\label{eq:gausspot}
\end{equation}
where ${}_{2}F_{2}(a,b;c;d)$ is a generalized hypergeometric function.
One important feature of this matrix model is that the spectrum is unbounded. Because our resolution of the paradox, as understood from the matrix model side, is based on the existence of a (square root) edge (see Section \ref{sec:resolution}), this is clearly an undesirable feature. Fortunately, we expect the absence of an edge to be a consequence of taking a strict $\Delta\to 0$ limit.

In fact, recall that taking $\Delta=0$ exactly is a sick limit (see Footnote \ref{foot:delta0}). Therefore, our analysis should be seen as a leading order approximation to the small but finite $\Delta$ case. Correlation functions thus generically receive corrections at order $\Delta$, implying the spectrum is not exactly Gaussian. To understand why one can expect the resulting equilibrium spectral density at small but finite $\Delta$ to have an edge, let us consider the following heuristic argument. 

Assume that for a generic matrix model, most eigenvalues are concentrated near the center of the spectrum at $x=0$, and let us study the behavior of the largest (or smallest) eigenvalue as we separate it from the rest of the eigenvalues. This experiences two contrasting forces: an attractive force towards the center of the spectrum from the matrix potential, and a repulsive force due to the presence of the other eigenvalues and captured by the Vandermonde determinant. From Eq.~\eqref{eq:effdiscrete}, the Vandermonde term in the effective potential takes the form\footnote{For simplicity, here we use the discrete form of the effective potential, but the exact same analysis can be carried out in the continuum limit.} 
\begin{equation}
    V_{V}(\lambda_{\max})=-\frac{\upbeta}{N}\sum_{i=1}^{N-1} \log|\lambda_i-\lambda_{\max}|\approx -\upbeta\log\lambda_{\max},
    \label{eq:vanderpot}
\end{equation}
where the last approximation is valid when we look at the potential for $\lambda_{\max}$ well separated from the rest of the eigenvalues, such that $\lambda_i/\lambda_{\max}\ll 1$, and we only kept the leading order in $1/N$. We recall that $\upbeta=2$ for the GUE. This leads to a repulsive force
\begin{equation}
    F_V=-V'(\lambda_{\max})\approx \frac{\upbeta}{\lambda_{\max}}.
\end{equation}
If in this regime the attractive force due to the matrix potential is larger in magnitude, the largest eigenvalue tends to move towards the rest of the eigenvalues: the matrix potential ``confines'' the spectrum to a finite support and an edge is present. This will be the case as long as the potential grows faster than the Vandermonde potential \eqref{eq:vanderpot} at large $\lambda$, i.e., $V(\lambda)=\omega( \upbeta \log\lambda)$.

The asymptotic behavior of the potential \eqref{eq:gausspot} leading to the Gaussian spectrum is given by
\begin{equation}
    V_\text{Gauss}(\lambda)\sim 2\log|\lambda|, \quad\quad\quad \lambda\to\infty
\end{equation}
which for the GUE $\upbeta=2$ is exactly the critical behavior for which the attractive and repulsive forces perfectly balance each other. Therefore, the largest eigenvalue does not experience any force as its value increases and behaves as a free particle allowed to ``escape'' to infinity. The fact that the potential leading to the Gaussian spectrum is marginally confining suggests that any modification to the spectrum due to finite $\Delta$ corrections will lead to the existence of an edge, with the location of the edge scaling as $\Delta^{-\gamma}$ for some $\gamma>0$. Edge universality arguments also suggest that the resulting edge should be a square root edge, with fluctuations captured by the Airy distribution \eqref{eq:airy}. We leave a thorough study of the edge at small but finite $\Delta$ for future work. We will instead focus on a toy model, the $q$-deformed Gaussian, which precisely reproduces this expectation, smoothly interpolating between the $\Delta\to 0$ and $\Delta\to\infty$ limits.

\subsection{A toy-model for generic $\Delta$: $q$-deformed Gaussian matrix model}
\label{sec:qgaussian}

In order to make progress in the generic $\Delta$ case, we introduce a simple toy-model. This can be summarized in the following three rules for computing a $2k$-point function of $\tilde{O}_\Delta$ at the disk level:\footnote{We emphasize that the rules specified here are essentially the ``chord rules'' of \cite{Berkooz:2018jqr,Lin:2022rbf}, which exactly reproduce Hamiltonian correlation functions in the double-scaled SYK model.}
\begin{enumerate}
    \item We sum over all possible pairwise Wick contractions of the $2k$ insertions;
    \item When the Witten diagram for a given Wick contraction can be drawn in different ways, leading to different numbers of intersections between matter worldlines, we consider the diagram with the smallest number of intersections;
    \item A diagram with $j$ intersections is weighted by a power $q^j$, where $0\leq q\leq 1$ is a fixed parameter of the model. 
\end{enumerate}
The first two rules are very natural from the point of view of JT (super)gravity. In particular, the second rule is exactly true due to mathematical properties of the 6j-symbols, namely orthogonality and the Yang-Baxter equations \cite{Jafferis_Kolchmeyer_Mukhametzhanov_Sonner_2023}. The third rule is the reason why this procedure can only be regarded as a toy-model. In fact, as we have seen in Section \ref{sec:genericd}, in JT gravity, intersections are not weighted by a constant but by a 6j-symbol, which is a complicated function of both the scaling dimension $\Delta$ of the operator and the four energies surrounding a vertex. Additionally, rule 3 cannot account for the existence of the ``loops'' seen in Section \ref{sec:genericd}, whose energy is not fixed by the external (boundary) energies being within the microcanonical band $\mathfrak{E}$ (or the ground state energy in the BPS case).

This simplification clearly eliminates both of the issues pointed out in Section \ref{sec:genericd} and allows us to study the resolution of the LMRS paradox within this toy-model. Before analyzing the properties of the resulting matrix integral---namely the $q$-deformed Gaussian matrix model \cite{Frisch:1970fa,greenberg,bozejko,Bozejko:1996yv,Effros_2003,Leeuwen,Berkooz:2018jqr}---we would like to remark that the $q$-deformed Gaussian exactly reproduces the JT (super)gravity results in the well-understood limits $\Delta\to\infty$ and $\Delta\to 0$. For $q=0$, all diagrams with intersections are excluded, and we clearly obtain the Gaussian matrix model considered in Section \ref{sec:GUE}. The $q$-deformed Gaussian is actually also able to reproduce JT (super)gravity results at disk level for $\Delta\gg 1$ but finite, as long as we only want to study correlation functions up to second order in $e^{-(1+4\log 2)\Delta}$ (or $e^{-4\Delta\log(1+\sqrt{2})}$ in the BPS case). In fact, in the large $\Delta$ limit, the 6j-symbol takes the form \eqref{eq:larged6j2}. If we only keep diagrams (at disk level) with up to two intersections---which are the only diagrams contributing at $O(e^{-2(1+4\log 2)\Delta})$ (or $O(e^{-8\Delta\log(1+\sqrt{2})})$ in the BPS case)---loops cannot be present. Therefore, all the JT correlation functions at this order are correctly captured by the $q$-deformed Gaussian at second order in $q$, with $q=e^{-(1+4\log 2)\Delta}$ (or $q=e^{-4\Delta\log(1+\sqrt{2})}$ in the BPS case).
The $\Delta\to 0$ limit is instead obtained by setting $q=1$, namely by weighting equally all diagrams with any number of intersections.

The rules explained above lead to the q-deformed Gaussian equilibrium spectral density \cite{Frisch:1970fa,greenberg,bozejko,Bozejko:1996yv,Effros_2003,Leeuwen,Berkooz:2018jqr,Cappelli_Colomo_1998,Nandy_2025} 
\begin{equation}
    \bar{\sigma}_q(x)=
    \frac{\sqrt{1-q}}{2\pi q^{1/8}}\vartheta_1\left(\arccos\left(\frac{\sqrt{1-q}}{2}x\right);q^{1/2}\right),
    \label{eq:qgaussspectrum}
\end{equation}
where $\vartheta_1(z;q)$ is the Jacobi $\vartheta$-function. In this appendix and in Appendix \ref{sec:genericdresolution}, we are omitting factors of the propagator $P_{\Delta}$ for simplicity of notation, but they can be easily reintroduced. Just like for the GUE, their effect is simply to shift the edge and modify the normalization of the spectral density. The correlation function will have a factor $P_{\Delta}$ for each pair of insertions, and the entropies remain independent of $P_\Delta$. 

\begin{figure}
    \centering
    \includegraphics[width=0.45\linewidth]{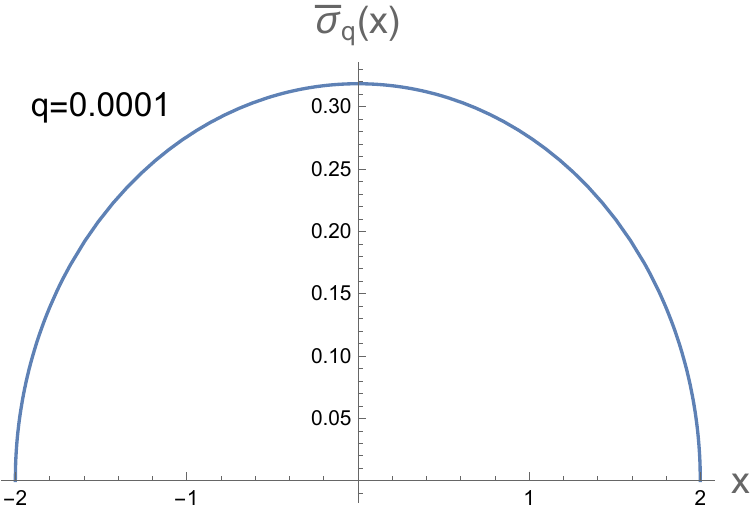}\quad \quad
    \includegraphics[width=0.45\linewidth]{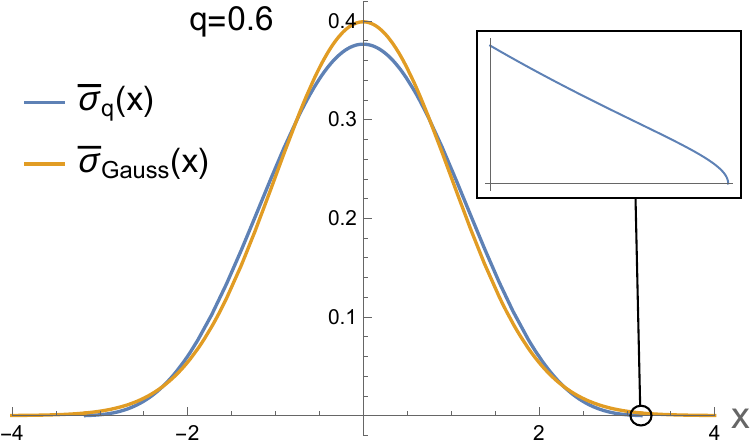}\\
    \vspace{0.5cm}
    \includegraphics[width=0.45\linewidth]{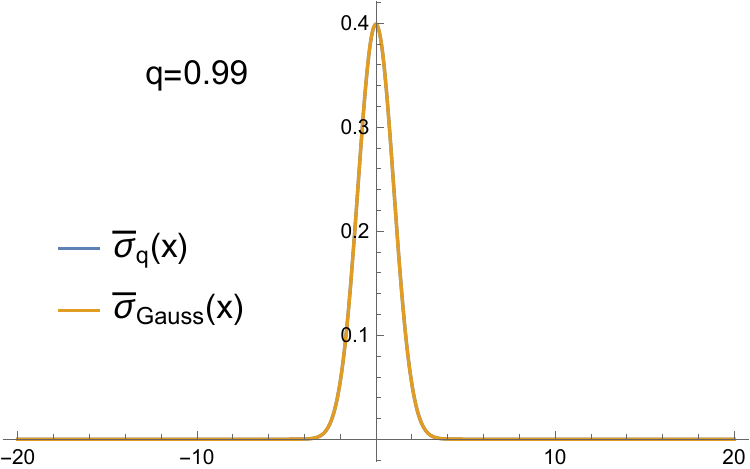}

    \caption{Top left: For $q\to 0$ (we plot here $q=0.0001$), the $q$-deformed Gaussian spectral density reduces to the Wigner semicircle. Top right: Increasing $q$ (we plot here $q=0.6$), the $q$-deformed Gaussian (in blue) is roughly a Gaussian for small $x$, but it retains a square-root edge at $x_{\text{edge}}=2/\sqrt{1-q}$, shown in the inset. $\bar{\sigma}_\text{Gauss}$ in Eq.~\eqref{eq:gaussspectrum} is plotted in yellow for reference. Bottom: As we take $q\to 1$ (we plot here $q=0.99)$, the spectrum becomes unbounded and the $q$-deformed Gaussian approaches the Gaussian spectrum \eqref{eq:gaussspectrum} (the $q$-deformed Gaussian is plotted in blue, indistinguishable from $\bar{\sigma}_\text{Gauss}$ in yellow).}
    \label{fig:qgaussianspectrum}
\end{figure}

The spectral density \eqref{eq:qgaussspectrum} has a square-root edge at $x_{\text{edge}}=2/\sqrt{1-q}$. Notice that for any value of $0\leq q<1$, there is a finite square-root edge.\footnote{In the bulk of the paper, we took for simplicity $\Delta$ to be the largest parameter in the theory. The fact that, away from the strict $q=0$ limit, the $q$-deformed Gaussian has a finite square root edge and that it exactly reproduces the disk results in JT gravity up to second order in $e^{-\#\Delta}$, is evidence that the consequence of taking $\Delta$ large but not the largest parameter in the theory is mainly a shift in the location of the edge of the eigenvalue spectrum for the operator $\tilde{O}_{\Delta}$.} At $q=0$, the $q$-deformed Gaussian spectral density reduces exactly to the GUE spectral density \eqref{eq:semicircle}, confirming our expectation that $q=0$ corresponds to the $\Delta\to\infty$ limit. As we increase $q$, the edge moves outward and $\bar{\sigma}_q(x)$ resembles more and more a Gaussian, although it retains a finite square-root edge. Finally, as we take $q\to 1$, the spectrum becomes unbounded and $\bar{\sigma}_q(x)$ approaches the Gaussian spectrum \eqref{eq:gaussspectrum}, recovering the $\Delta\to 0$ limit of JT gravity. See Figure \ref{fig:qgaussianspectrum}.

The corresponding matrix potential obtained from Eq. \eqref{eq:extremization} can be expressed as a sum of Chebyshev polynomials of the first kind $T_n(x)$ \cite{Cappelli_Colomo_1998, Nandy_2025}:
\begin{equation}
    V_q(x)=2\sum_{n=1}^{\infty}\frac{(-1)^{n-1}}{n}q^{n^2/2}(q^{n/2}+q^{-n/2})T_{2n}\left(\frac{\sqrt{1-q}}{2}x\right).
    \label{eq:qgausspot}
\end{equation}
By construction, the matrix integral defined by this potential leads to the equilibrium density of states \eqref{eq:qgaussspectrum}. In other words, in the large-$N$ limit and thus at leading order in the genus, it reproduces exactly the one-boundary correlation functions obtained with the Feynman rules described at the beginning of the present subsection. What are the exact higher-genus or multiple-boundary diagrams contributing to correlation functions in this matrix integral, and to what extent they match the corresponding bulk diagrams are interesting open questions that we leave for future work. Here we will simply take the potential \eqref{eq:qgausspot} as the definition of our simple toy-model for the operator $\tilde{O}_\Delta$ at generic $\Delta$ also beyond the disk level (both in non-SUSY JT in a microcanonical band and in the BPS case), and study the semi-quenched R\'enyi entropies from this matrix model's perspective.

Although the $q$-deformed Gaussian is only a toy-model for JT gravity, we expect its spectral properties to qualitatively match those of a more complete matrix model for JT gravity with matter. In fact, it not only reproduces the small and large $\Delta$ limits of JT gravity exactly, but it also smoothly interpolates between the two as we would expect. Moreover, it matches our key expectation that, for any finite $\Delta$, the matter matrix model presents a (square-root) edge, and only in the strict (and sick) $\Delta\to 0$ limit does the spectrum become unbounded. Because our resolution of the LMRS paradox only relied on the existence of such an edge, for our purposes, the $q$-deformed Gaussian toy-model captures all the relevant ingredients of JT (super)gravity. The solution to the paradox for generic $\Delta$ we will present in Appendix \ref{sec:genericdresolution} will immediately extend to a more complete matrix integral, as long as its spectrum is bounded and possesses a square-root edge.

\subsection{Generic matrix model for JT coupled to matter}
\label{sec:twomatrix}

We would like to conclude this appendix with some more generic comments about the dual description of JT (super)gravity coupled to matter. Our discussion will be mainly based on the model introduced in \cite{Jafferis_Kolchmeyer_Mukhametzhanov_Sonner_2023}. We will also comment on the relationship between this generic model and the specific cases we considered (large and small $\Delta$, as well as the $q$-deformed Gaussian toy-model).

The Eigenstate Thermalization Hypothesis reviewed at the beginning of Section \ref{sec:matrixmodels} suggests that, in order to describe JT (super)gravity coupled to matter, we can model each matter operator with a random matrix. For simplicity, let us focus on the case of a single matter operator. This leads to a two-matrix integral which, working in the energy eigenbasis, can be written as \cite{Jafferis_Kolchmeyer_Mukhametzhanov_Sonner_2023}
\begin{equation}
    Z=\int \mathcal{D}\mathcal{E}\,\mathcal{D}O\, e^{-\sum_a\left[V_{\text{JT}}(E_a)+V_{\text{c.t.}}(E_a)\right]-V_{\text{matter}}(\mathcal{E},O)}
    \label{eq:twomatrix}
\end{equation}
where $\mathcal{E}$ represents the set of energy eigenvalues, $V_{\text{JT}}$ is the pure JT (super)gravity matrix model potential (for example given by the Saad-Shenker-Stanford potential $V_{\text{SSS}}$ for non-SUSY JT \cite{Saad_Shenker_Stanford_2019}), and $V_{\text{c.t.}}$ is a counterterm needed to recover the correct pure JT (super)gravity density of states upon integrating out the matter operator $O$. The matter potential can be generically expressed as an infinite sum \cite{Jafferis_Kolchmeyer_Mukhametzhanov_Sonner_2023}
\begin{equation}
    V_{\text{matter}}(\mathcal{E},O)=\frac{e^{S_0}}{2}\sum_{a,b}F(E_a,E_b)O_{ab}O_{ba}+\frac{e^{S_0}}{4}\sum_{a,b,c,d}G(E_a,E_b,E_c,E_d)O_{ab}O_{bc}O_{cd}O_{da}+...
    \label{eq:matterpot}
\end{equation}
We explicitly extracted here the factors of $e^{S_0}$ for convenience, where $S_0$ should be interpreted as the ground state entropy $\SBPS$ in the JT supergravity case, or the background value of the dilaton in non-SUSY JT gravity.  $F,G,...$ are smooth functions of the energies that can be determined by matching correlation functions of $O$ computed with the matrix integral to disk correlation functions in JT (super)gravity coupled to matter. For example, $F$ is fixed by the inverse of the disk two-point function at fixed external energies, and $G$ is related to the connected disk four-point function \cite{Jafferis_Kolchmeyer_Mukhametzhanov_Sonner_2023}. 
As a caveat to the two-matrix model introduced in \cite{Jafferis_Kolchmeyer_Mukhametzhanov_Sonner_2023}, we emphasize that whether this match at the disk level is sufficient to correctly reproduce correlation functions in JT gravity coupled to matter in the presence of multiple boundaries and at higher genus is an important and not fully-understood open problem.

The matrix integral \eqref{eq:twomatrix} is valid for any value of the scaling dimension $\Delta$ of the matter operator. Determining the precise form of the potential \eqref{eq:matterpot} is possible but highly non-trivial \cite{Jafferis_Kolchmeyer_Mukhametzhanov_Sonner_2023}. In the rest of this subsection, we will connect this two-matrix model to the specific cases we have considered in this paper.

First, let us analyze the large $\Delta$ limit considered in the bulk of the paper. In the matrix model \eqref{eq:twomatrix}, the suppression of crossing diagrams corresponds to keeping only the first term in the potential \eqref{eq:matterpot}:\footnote{We remark that this matrix integral is exactly equivalent to the Gaussian ansatz for ETH \cite{Jafferis_Kolchmeyer_Mukhametzhanov_Sonner_2023}. Gaussian ETH is unsatisfactory in general \cite{Jafferis:2022uhu}. In fact, because it cannot capture the third diagram in Eq.~\eqref{eq:full4disk}, it cannot correctly describe OTOCs, nor reproduce crossing symmetry, which must hold in any (chaotic) conformal field theory we might be modeling using ETH. To correctly capture these features, we must include non-Gaussianities in the ETH ansatz. These are encoded in the higher order terms in the potential \eqref{eq:matterpot}, which we dropped in the $\Delta\to\infty$ limit. Nonetheless, as we have seen in Section \ref{sec:matrixmodels}, for $\Delta\to\infty$ the GUE matrix model correctly reproduces the $\Delta\to\infty$ limit of JT (super)gravity coupled to matter in the energy sector of interest.}
\begin{equation}
    V_{\text{matter}}^{\Delta\to\infty}(\mathcal{E},O)=\frac{e^{S_0}}{2}\sum_{a,b}F(E_a,E_b)O_{ab}O_{ba}.
    \label{eq:matterpotdeltainf}
\end{equation}
Moreover, in the BPS case, the energy of all states is obviously equal. In the JT microcanonical window for the non-supersymmetric puzzle, the energy eigenvalues lie in a small microcanonical window in which the spectrum is nearly flat. Recall that we are interested in computing correlation functions of operators $\tilde{O}_\Delta$ projected on the energy sector of interest $\mathcal{E}_0$ (the BPS sector or the microcanonical band $\mathfrak{E}$). In both cases, the two-point function is then (exactly for BPS, nearly for the microcanonical window) constant within the BPS sector/microcanonical window. Namely, $F(E_a,E_b)=F(\bar{E},\bar{E})\equiv F_{\bar{E}}$ for all $a,b$ indices running in the energy sector of interest, where $\bar{E}=0$ for the BPS case or $\bar{E}$ is the average energy of the microcanonical band for the non-SUSY JT case. $F_{\bar{E}}$ is the inverse of the propagator $P_{\Delta\to\infty}$ defined in Footnote \ref{foot:PBPS} for the BPS case, and the inverse of $P_{\Delta\to\infty}(\bar{E})$ defined in Footnote \ref{foot:non-susy-prop} for the non-SUSY case.

When computing correlation functions of $\tilde{O}_\Delta$ in the $\Delta\to\infty$ limit, the two-matrix integral \eqref{eq:twomatrix} then greatly simplifies. In fact, because we are fixing the energy on both sides of a matter operator insertion within the energy sector of interest, the only terms in the potential \eqref{eq:matterpotdeltainf} relevant for calculating these correlation functions are those for which both indices $a,b$ run within $\mathcal{E}_0$ (i.e., within the BPS sector/the microcanonical band $\mathfrak{E}$). In the BPS sector there are exactly $|\mathcal{E}_0|=e^{\SBPS}$ states, whereas in the JT microcanonical band the number of states is given by $|\mathcal{E}_0|=e^{S_0}\int_{\delta E}dE\,\rho_0(E)\equiv e^{S(\mathfrak{E})}$, where $\delta E$ is the width of the band and $\rho_0(E)$ is the JT density of states introduced in Section \ref{sec:2}.\footnote{This statement is exact for the BPS case, in which the number and energy of degenerate ground states is fixed in all realizations of the ensemble to all orders in $e^{-\SBPS}$ \cite{Iliesiu:2021are}. In the non-SUSY JT case, this is an approximate statement because the number of states in a given microcanonical band of fixed energy and width can vary across members of the ensemble. As we have explained in Appendix \ref{sec:derivation-2n-pt-func-non-SUSY-JT}, these fluctuations can be accounted for by appropriately replacing powers of the JT density of states by their corresponding moments when computing correlation functions of the matter operator. This, in principle, spoils the connection with the GUE. However, as we have explained in Section \ref{sec:mapping} (see Footnote \ref{footnote:suppression}), these corrections are exponentially suppressed in $S_0$ in the large-$k$ regime of our interest, and therefore are not important for our discussion. We can thus neglect them.} In either case, we can then split the potential \eqref{eq:matterpotdeltainf} as (in the BPS case $S_0$ should be replaced by $\SBPS$)
\begin{equation}
    V_{\text{matter}}^{\Delta\to\infty}=e^{S_0}F_{\bar{E}}\frac{1}{2}\tr_{\mathcal{E}_0}(\tilde{O}_{\Delta}^2)+\frac{e^{S_0}}{2}\sum_{\substack{a,b\\ E_a,E_b\notin \mathcal{E}_0}}F(E_a,E_b)\(\tilde{O}_{\Delta}\)_{ab}\(\tilde{O}_\Delta\)_{ba}
    \label{eq:matterinter}
\end{equation}
where the first term accounts for terms only involving energy eigenvalues in $\mathcal{E}_0$, whose (average) energy and number are known and fixed a priori in our setup. Note that the first term is independent of the Hamiltonian spectrum, and it only depends on the size and average energy of $\mathcal{E}_0$.

Using Eq.~\eqref{eq:matterinter}, we can then integrate out the Hamiltonian and all $(\tilde{O}_\Delta)_{ab}$ matrix elements involving energy eigenvalues $E_a,E_b\notin \mathcal{E}_0$ in Eq.~\eqref{eq:twomatrix}. Thanks to the Gaussian nature of the matter potential, this procedure does not affect the potential for the $\tilde{O}_\Delta$ matrix elements for $E_a,E_b\in \mathcal{E}_0$. As a result, we are left with a single-trace, single matrix integral able to capture $\tilde{O}_\Delta$ correlation functions in the energy sector $\mathcal{E}_0$ of interest. This is simply given by the GUE discussed in Section \ref{sec:GUE}:
\begin{equation}
    Z=\tilde{Z}\int\mathcal{D}\tilde{O}_\Delta \,e^{-e^{S_0}F_{\bar{E}}\frac{1}{2}\tr_{\mathcal{E}_0}(\tilde{O}_{\Delta}^2)},
\end{equation}
where we should identify $F_{\bar{E}}=1/P_{\Delta\to\infty}$ and $\tilde{Z}$ contains all data we integrated out and is canceled out by the normalization in all correlation functions. We thus see that the result obtained in Section \ref{sec:GUE} by simply matching bulk diagramatic combinatorics to the GUE can also be obtained starting from a more general two-matrix model for JT (super)gravity coupled to matter and taking the appropriate large $\Delta$ limit for the operator $\tilde{O}_\Delta$ of interest.

Finally, let us comment on the issues arising in the generic $\Delta$ case. The first issue we pointed out in Appendix \ref{sec:genericd}, namely that intersection diagrams such as the last one in Eq.~\eqref{eq:full4disk} are non-vanishing and weighted by the 6j-symbols (which are complicated functions of the scaling dimension and the energies), implies that one must keep non-Gaussianities in the potential \eqref{eq:matterpot}. The second issue, namely the presence of loops whose internal energy is not fixed within the energy sector $\mathcal{E}_0$ of interest and must be integrated over, is a manifestation of the fact that all energy scales mix non-trivially at generic $\Delta$. As a result, unlike the $\Delta\to\infty$ limit, we cannot simply integrate out the Hamiltonian and all matrix elements of $(\tilde{O})_{ab}$ involving energy eigenvalues $E_a,E_b\notin \mathcal{E}_0$ in our two-matrix integral, because this would affect very non-trivially the potential for matrix elements $(\tilde{O})_{ab}$ involving energies within $\mathcal{E}_0$. In other words, we cannot obtain a simple single-matrix integral describing correlation functions of $\tilde{O}_\Delta$. Therefore, without further simplifications, one truly needs to treat JT coupled to matter as a full two-matrix integral with a non-Gaussian potential \eqref{eq:matterpot} in order to solve the entropic puzzle of our interest. We will not attempt this, leaving such an interesting study to future work. In the next appendix, we will instead focus on the resolution of the LMRS puzzle at generic $\Delta$ using the simple $q$-deformed Gaussian toy-model introduced in Appendix \ref{sec:qgaussian}.

\section{Resolution of the puzzle at generic $\Delta$} 
\label{sec:genericdresolution}

As explained in Appendix \ref{sec:genericd}, computing correlators and, consequently, entropies at generic $\Delta$ is a very difficult task. However, motivated by the fact that it correctly captures the main features we expect at generic $\Delta$ and it interpolates between the large $\Delta$ and small $\Delta$ behaviors of JT gravity, we can use the $q$-deformed Gaussian toy-model introduced in Appendix \ref{sec:qgaussian} to formulate and resolve the LMRS puzzle at generic $\Delta$. We remark that the results of this appendix, although formulated in the specific $q$-deformed Gaussian toy model, continue to hold for any more general model with a compact spectrum and a square-root edge. 

The generic behavior we find is that, at generic $\Delta$, entropies fall off linearly when $k\ll 1/\Delta$, exactly as it happens in the $\Delta\to 0$ LMRS paradox reviewed in Section \ref{sec:2}. As $k$ becomes much larger than $1/\Delta$, the correlators become sensitive to the existence of the edge and the behavior changes: the entropy starts to fall off logarithmically like in the $\Delta\to\infty$ puzzle. The negativity then arises again for $k=O(e^{2\SBPS/3})\equiv O(N^{2/3})$. In that regime, provided there is a finite square-root edge, the Airy analysis studied in Section \ref{sec:resolution} applies and rescues the positivity of the semi-quenched entropy. Similarly, for $k=\omega(e^{2\SBPS/3})$, one- and two-eigenvalue instantons dominate the matrix integral and rescue positivity.

\subsection{Crossover at small $\Delta$}

\begin{figure}
    \centering
    \includegraphics[width=0.8\linewidth]{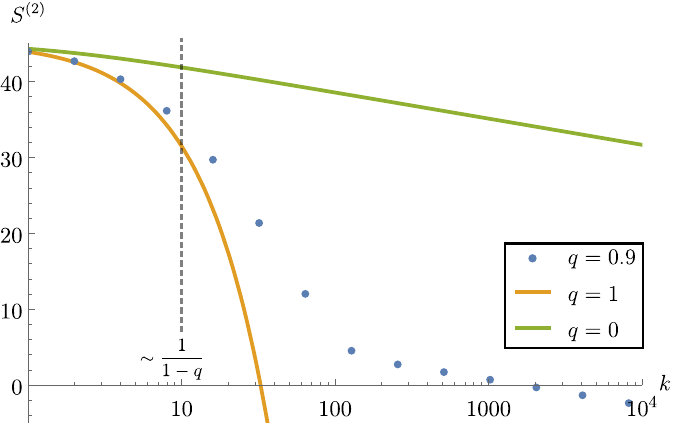}
    \caption{The (annealed) second R\'enyi entropy of the $q$-deformed Gaussian using $\log N=45$. This was obtained numerically for $q=0.9$ and analytically for $q=0$ and $q=1$. We mark the estimated location when the entropy crosses over from linear to logarithmic decay at $k\sim\frac{1}{1-q}$. Therefore, the annealed entropy for generic $q$ goes negative when $k=O(N^{2/3})$ rather than $k=O(\log N)$ as in the $q=1$ case.}
    \label{fig:crossover}
\end{figure}

For $q\sim 1$ (that corresponds to the regime of small $\Delta$), the spectrum of the $q$-deformed Gaussian is close to a Gaussian distribution, demonstrated in Figure \ref{fig:qgaussianspectrum}. This means that for sufficiently small $k$, the correlators like $\langle\tr M^{2k}\rangle$ will be determined by the bulk of the spectrum. This implies that the entropies for small $k$ will behave according to a Gaussian spectrum:
\begin{equation}
    S^{(n)}\sim \log N-\frac{n\log n}{n-1}k,\qquad S\sim \log N - k\qquad \text{when }k\lesssim\frac{1}{1-q}.
\end{equation}
We will explain shortly why this happens for $k\lesssim 1/(1-q)$. Importantly, the entropy falls off linearly in $k$ for small $k$.\footnote{Since $k$ is small and much less than the Airy limit $O(N^{2/3})$, we need not make the distinction between annealed, semi-quenched, or quenched because all entropies are self-averaging and agree with each other.}
Once $k$ becomes sufficiently large, these correlators will start to be dominated by eigenvalues near the edge. In the presence of a square-root edge (as is the case for the $q$-deformed Gaussian), the statistics of these eigenvalues are governed by the Airy model. Therefore, for $k\ll N^{2/3}$, entropies decay logarithmically, analogous to the $\Delta\to\infty$ behavior:
\begin{equation}
S^{(n)}\sim\log N-\frac{3}{2}\log k,\qquad S\sim\log N-\frac{3}{2}\log k,\qquad \text{when }\frac{1}{1-q}\lesssim k\ll O(N^{2/3}).
\end{equation}

To estimate where this crossover in behavior occurs, consider the correlator $\langle\tr M^{2k}\rangle$. For $k$ small, we expect this to be similar to the Gaussian answer, which gives $\langle\tr M^{2k}\rangle\sim (2k-1)!!$.\footnote{For simplicity, in this Appendix we have rescaled $M/\sqrt{P_{\Delta}}\to M$ so that correlators are independent of $P_{\Delta}$. Factors of $P_\Delta$ can be reintroduced in the correlators by simply multiplying a factor of $P_\Delta$ per pair of operator insertions. Regardless, entropies are independent of the value of $P_{\Delta}$.} On the other hand, the crossover occurs when the correlator is governed by near-edge statistics, so $\langle\tr M^{2k}\rangle\approx \lambda_e^{2k}=(2/\sqrt{1-q})^{2k}$. Comparing the two regimes, we find that the crossover occurs at $k=O(1/(1-q))$. This is confirmed numerically in Figure \ref{fig:crossover}.

Therefore, provided that the eigenvalue spectrum has a finite square-root edge, the entropy negativity problem occurs when $k=O(N^{2/3})$ rather than $k=O(\log N)$. In more generic models, the location of the edge (and therefore of the crossover between the two behaviors) may vary. However, we expect it to be controlled by $\Delta^{-\gamma}$ for some positive value of $\gamma$. The change in behavior should thus occur for $k\approx \Delta^{-2\gamma}$. For the specific case of the $q$-deformed Gaussian, $\Delta\approx (1-q)^{\eta}$ for some positive $\eta$, which confirms this expectation.

The annealed entropy thus becomes negative for $k=O(N^{2/3})$. In the presence of a square-root edge, the positivity of the semi-quenched R\'enyi entropies in the $k=O(N^{2/3})$ regime simply follows from exactly the same Airy limit argument carried out in Section \ref{sec:GUEairy}, which we will not repeat here. For $k=\omega(N^{2/3})$ we can again study one- and two-eigenvalue instantons in the matrix model. 

As we have explained in Sections \ref{sec:oneeigen} and \ref{sec:twoeigen}, unlike the Airy limit, the study of eigenvalue instantons is not universal and it depends on the specific matrix potential. The resulting entropies are also dependent on the matrix model. In the remainder of this appendix, we will explain how to find eigenvalue instantons for the $q$-deformed Gaussian matrix model introduced in Section \ref{sec:qgaussian}.

\subsection{Eigenvalue instantons for the $q$-deformed Gaussian matrix model}

\begin{figure}[h]
    \centering
    \includegraphics[width=0.65\linewidth]{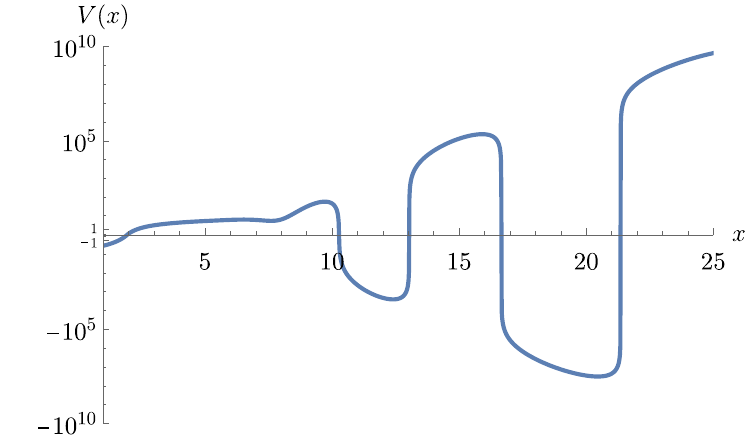}
    \caption{The matrix potential of the $q$-deformed Gaussian for $q=0.6$ on a symmetric log plot. Notice that the potential oscillates, implying the $q$-deformed Gaussian spectrum saddle is metastable. Therefore, to provide a nonperturbative definition to the matrix model, we must deform the contour of integration over eigenvalues.}
    \label{fig:qgaussianpotential}
\end{figure}

Beyond the Airy regime with $k=\omega(N^{2/3})$, the eigenvalue instanton keeps the semi-quenched entropy from going negative, just as before. However, as we will now discuss, finding the one-eigenvalue instanton for the $q$-deformed Gaussian is more nuanced than the GUE case studied in Section \ref{sec:resolution}. 

\begin{figure}
    \centering
    \includegraphics[width=0.7\linewidth]{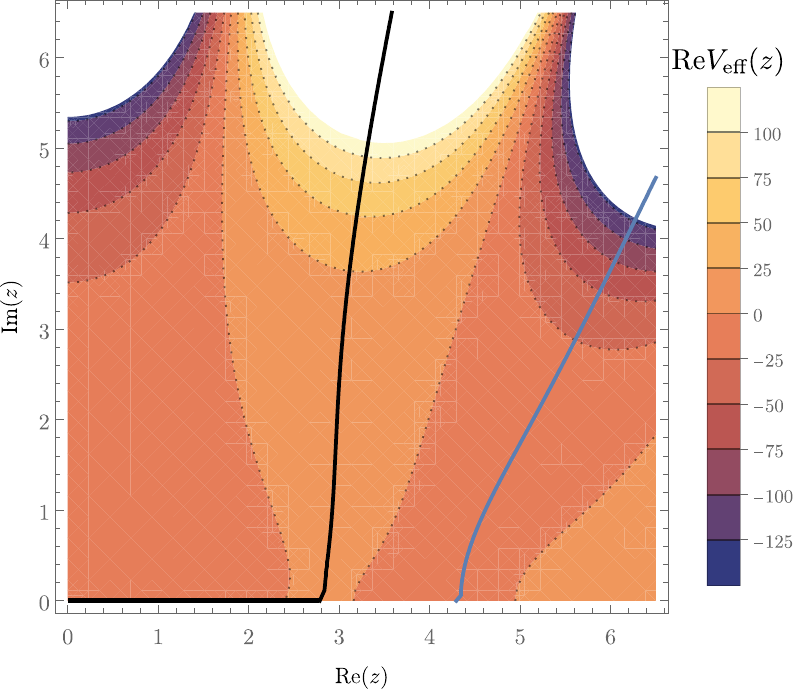}
    \caption{The deformed contour (thick black line) giving the non-perturbative stable definition of our $q$-deformed Gaussian matrix integral, defined by deforming the contour into the complex plane at the first maximum of $V_{\text{eff}}(z)$ along the line of steepest descent (steepest ascent for $V_{\text{eff}}(z)$). We plot here only the portion of the integration contour in the upper right quadrant of the complex plane. This specific contour is symmetric around the imaginary axis. Our full integration contour is the union of this contour and its reflection along the real axis in the lower complex plane. We also depict the line of steepest ascent for the second minimum of $V_{\text{eff}}(z)$ (light blue), which does not intersect the integration contour. The same applies to all other minima. This guarantees that we only pick up the first minimum of $V_{\text{eff}}(z)$ and not the following minima leading to the non-perturbative instability.}
    \label{fig:steepestascent}
\end{figure}

The matrix potential for the $q$-deformed Gaussian was given in Eq. \eqref{eq:qgausspot}.
For generic $q\in (0,1)$, this matrix potential grows and rapidly oscillates for large $|x|$, see Figure \ref{fig:qgaussianpotential}.
This means that the saddle-point yielding the $q$-deformed Gaussian density of states \eqref{eq:qgaussspectrum} is metastable and the matrix model is nonperturbatively unstable. To give a nonperturbative definition of the matrix model, we can make a contour deformation of the integral over eigenvalues off the real axis \cite{Saad_Shenker_Stanford_2019} (see also \cite{Marino:2008ya,Marino2022-sf, Marino2007-qe, Marino2008-sa, Johnson_2020, Johnson2022-yy}). This new contour allows the matrix integral to converge, and perturbation theory around the $q$-deformed Gaussian saddle is asymptotic to this convergent value. To define our new integration contour, following \cite{Saad_Shenker_Stanford_2019} we deform the contour along the line of steepest descent (i.e., steepest ascent for $V_{\text{eff}}(z)$), starting from the first maximum in $V_{\text{eff}}(z)$ on the real line.\footnote{This is a line of constant $\text{Im} V_{\text{eff}}(z)$. Since $\text{Im}V_{\text{eff}}(z)=0$ on the real line, the contour follows $\text{Im}V_{\text{eff}}(z)=0$.} This guarantees convergence of the integral. This deformation is shown in Figure \ref{fig:steepestascent}, where the solid black line is our deformed contour. Because the $q$-deformed Gaussian matrix potential is symmetric, we must also perform a similar contour deformation at the left edge. This contour deformation guarantees that all minima of $V_{\text{eff}}(z)$ on the real line except the first one are not picked up by our integration contour. In fact, saddle points are picked up by an integration contour only if their lines of steepest ascent intersect the integration contour, which is not the case for any of these minima. We show an example for the second minimum in Figure \ref{fig:steepestascent}. These were the minima leading to the non-perturbative instability, which is thus cured by our contour deformation.

This choice of contour deformation is not unique \cite{Saad_Shenker_Stanford_2019}. We could deform the contour either upward (as depicted in Figure \ref{fig:steepestascent}) or downward in the complex plane along the line of steepest descent, or take a linear combination of these two choices. These are all valid non-perturbative definitions of the matrix integral \cite{Saad_Shenker_Stanford_2019}. We choose the contour to be an equally weighted linear combination of these two contours, for a reason that will become clear shortly.

\begin{figure}
\centering
\includegraphics[width = 0.7\linewidth]{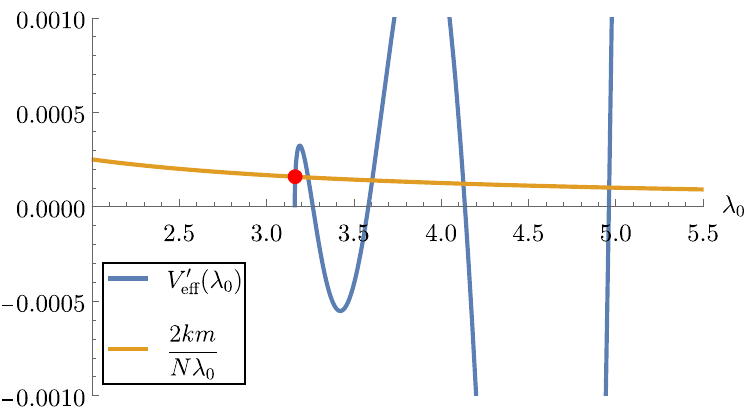}
\caption{A plot of $V_{\text{eff}}'(\lambda_0)$ with $q=0.6$ along with $2km/N\lambda_0$ with $2km/N=0.0005$. The location of the one-eigenvalue instanton is given by the intersection between these two functions (see Eq. \eqref{eq:1eigensaddle}). For sufficiently small $k$, the one-eigenvalue instanton is given by the first intersection of the two curves (red dot). However, as we increase $k$, there is no real intersection between the first peak of $V_{\text{eff}}'(\lambda_0)$and $2km/N\lambda_0$. Intersections with the following peaks at larger real $\lambda_0$ are also excluded by our contour deformation. However, there still exists (complex conjugate pairs of) complex solutions to Eq. \eqref{eq:1eigensaddle}, which are in fact picked up by our complex contour: the one-eigenvalue instanton becomes complex.}
\label{fig:qgaussianinst}
\end{figure}

Let us now discuss the one-eigenvalue instanton. Due to the oscillatory nature of $V_q(x)$ and, consequently, $V_{\text{eff}}(x)$, one-eigenvalue instantons become complex for $k$ sufficiently large. This is depicted in Figure \ref{fig:qgaussianinst}. To see this, recall that the location of the one-eigenvalue instanton is determined by the saddle-point equation \eqref{eq:1eigensaddle}. For small enough $k$, the one-eigenvalue instanton is real and given by the first intersection between $V_{\text{eff}}'(\lambda_0^*)$ and $2km/(N\lambda_0^*)$. However, by increasing $k$, $2km/(N\lambda_0^*)$ can become larger than the first maximum of $V_{\text{eff}}'(\lambda_0^*)$. When this happens, there is no real solution for Eq. \eqref{eq:1eigensaddle} with $\lambda_0^*$ smaller than the location of the first maximum of $V_{\text{eff}}(\lambda_0^*)$. Intersections with other peaks---i.e., for $\lambda_0^*$ real and larger than the location of the first maximum of $V_{\text{eff}}(\lambda_0^*)$---are also excluded by our contour deformation. However, there are still complex solutions to Eq.~\eqref{eq:1eigensaddle}. The locations of these one-eigenvalue instantons for the $q$-deformed Gaussian (both the real ones for small $k$ and the complex ones for larger $k$) cannot be obtained analytically, but they can be obtained by numerically solving Eq. \eqref{eq:1eigensaddle} for the $q$-deformed Gaussian effective potential. 

\begin{figure}
    \centering
    \includegraphics[width=0.7\linewidth]{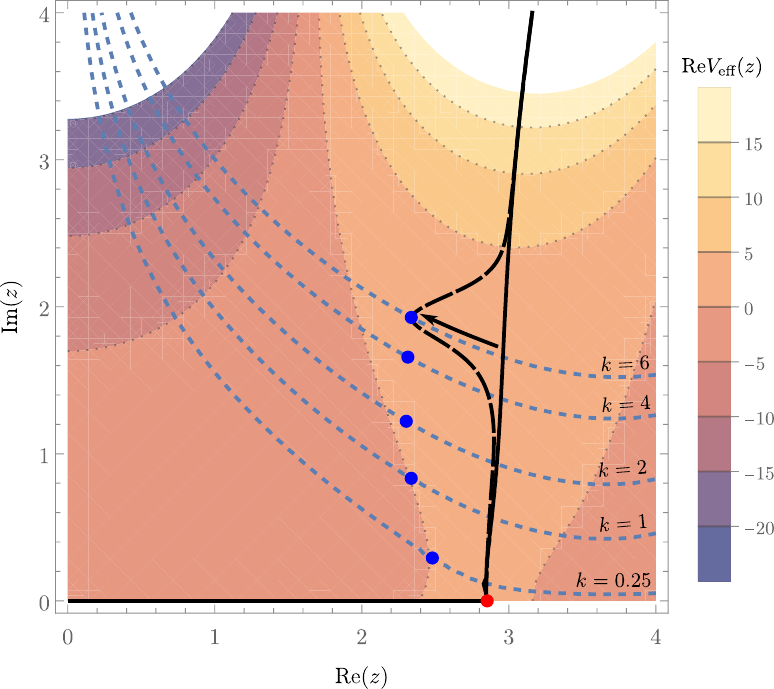}
    \caption{Our integration contour (thick black line) picks up complex saddles for the one-eigenvalue instanton. Complex one-eigenvalue instantons at various $k=2n/N$ are shown by blue dots, along with their steepest ascent lines (steepest descent for the instanton action $V_{\text{eff}}(z)-\frac{2km}{N}\log z$), depicted by dotted blue lines. These lines intersect our contour, so the contour can be locally deformed towards the lines of steepest descent to pick up these saddles. As an example, the deformed contour picking up the one-eigenvalue instanton for $k=6$ is sketched in dashed black.
    }
    \label{fig:contour}
\end{figure}
To pick up these complex one-eigenvalue instantons, we need to further locally deform the integration contour. These saddles are picked up by our contour provided that the line of steepest ascent (steepest descent in the instanton action $\text{Re}(V_{\text{eff}}(z)-\frac{2km}{N}\log z)$) passing through the one-eigenvalue instanton location intersects our contour. It is not hard to show that this is indeed the case. Thus, starting with our modified contour along the line of steepest descent, we can (locally) deform the contour to pick up the complex one-eigenvalue instantons. This deformation does not affect the result of the integral. One example of this secondary deformation is shown in Figure \ref{fig:contour} (dashed black line).

Notice that the complex one-eigenvalue instanton for each $k$ comes in complex conjugate pairs. Recall that we defined our modified integration contour to be an equally-weighted linear combination between a contour extending into the upper complex plane and one extending into the lower complex plane (both along lines of steepest descent). Then, for each $k$, both one-eigenvalue instantons can be picked up by an appropriate secondary deformation of the integration contours along lines of steepest ascent. Notice that each one of these eigenvalue instantons gives a complex value for the correlation functions we are computing, and the two results are complex conjugates of each other. The reason why we chose the integration contour to be the equally-weighted linear combination is precisely because this way we can pick up both eigenvalue instantons and obtain a real result for the correlators.

This procedure allows us to give a non-perturbatively stable version of the $q$-deformed Gaussian matrix integral, find one-eigenvalue instanton saddles by simply finding complex solutions to Eq. \eqref{eq:1eigensaddle}, and obtain real correlators by evaluating the matrix integral effective action on-shell for the two complex conjugate instantons and averaging the two results to take into account the linear combination of the two contours. The semi-quenched R\'enyi-$n$ entropy can then be computed, and it vanishes exactly like in Section \ref{sec:oneeigen}. A completely analogous procedure can be carried out to find two-eigenvalue instantons as in Section \ref{sec:twoeigen}. This yields the exponentially decaying behavior of the semi-quenched entropy for $k=\omega(N^{2/3})$ in the $q$-deformed Gaussian toy model. Although we did not numerically compute the semi-quenched entropy, it is easy to do so using the procedure outlined in this section.

\section{Two-boundary correlators from the GUE}
\label{appendix:f}

In this Appendix we explain how to obtain the genus expansion of the generic two-boundary correlation function $\overline{\tr\(\tilde{O}_{\Delta}^{2k_1}\)\tr\(\tilde{O}_{\Delta}^{2k_2}\)}_{\text{conn.}}$ and its re-summation at large $k_1$, $k_2$, leading to Eq.~\eqref{eq:cyl-resummed}. 

We will use the connection between our setup and the GUE discussed in Section \ref{sec:GUE}. In the GUE, the correlation function $\left\langle\tr\(M^{a+1}\)\tr\(M^{b+1}\)\right\rangle_{\text{conn.}}$ is known explicitly in closed form \cite{dubrovin}:\footnote{Here we include the same variance $\sigma=P_{\Delta\to\infty}$ considered in Section \ref{sec:GUE} to match the GUE to the bulk results.}
\small{
\begin{align}
&(P_{\Delta\to\infty})^{-\frac{a+b+2}{2}}\left\langle\tr\(M^{a+1}\)\tr\(M^{b+1}\)\right\rangle_{\text{conn.}}
\\
&=\frac{1}{N^{\frac{a+b}{2}}} \,(a+b+1)!! \,(1+b) \;
{}_2F_1\!\left(-\frac{a+b}{2},\, 1-N;\, 2;\, 2\right)
\nonumber \\[0.5em]
&\quad
+ 2 N^{1-\frac{a+b}{2}} \sum_{\substack{j=0 \\ j \equiv b \ (\mathrm{mod}\ 2)}}^{b-2}
(a+j+1)!! \,(b-j-1)!! \,(1+j)
\;
{}_2F_1\!\left(-\frac{a+j}{2},\, 1-N;\, 2;\, 2\right)
\nonumber \\[-0.3em]
&\hspace{8em} \times
{}_2F_1\!\left(-\frac{b-j-2}{2},\, 1-N;\, 2;\, 2\right)
\nonumber \\[0.5em]
&\quad
- N^{-\frac{a+b}{2}} \sum_{\substack{j=0 \\ j \equiv b-1 \ (\mathrm{mod}\ 2)}}^{b-1}
(a+j)!! \,(b-j-2)!! \,(1+j)
\nonumber \\[0.3em]
&\qquad \times
\Bigg[
{}_2F_1\!\left(-\frac{a+j+1}{2},\, -N;\, 1;\, 2\right) \;
{}_2F_1\!\left(-\frac{b-j-1}{2},\, 1-N;\, 1;\, 2\right)
\nonumber \\[-0.3em]
&\hspace{6em} +
{}_2F_1\!\left(-\frac{b-j-1}{2},\, -N;\, 1;\, 2\right) \;
{}_2F_1\!\left(-\frac{a+j+1}{2},\, 1-N;\, 1;\, 2\right)
\Bigg] .
\label{eq:2f1}
\end{align}}
This is valid for $a+b$ even. If $a+b$ is odd, the two-point function obviously vanishes. To make contact with the correlation function of our interest, we simply need to identify $M=\tilde{O}_{\Delta}$, $N=e^{\SBPS}$, $a=2k_1-1$, and $b=2k_2-1$. Using properties of the ${}_2F_1$, Eq.~\eqref{eq:2f1} can be written as a finite sum over genus
\begin{equation}
    \left\langle\tr\(M^{a+1}\)\tr\(M^{b+1}\)\right\rangle_{\text{conn.}}=\(P_{\Delta\to\infty}\)^{\frac{a+b+2}{2}}\sum_{g=0}^{\lfloor\frac{a+b}{4}\rfloor}C^{(2)}_g(a+1,b+1)N^{-2g}
    \label{eq:f2}
\end{equation}
where, for $a+b$ even and $m\equiv (a+b)/2$,\footnote{We acknowledge ChatGPT 5.0 for help in the derivation of the coefficients \eqref{eq:cgs} from the known result \eqref{eq:2f1}.}
{\small
\begin{align}
&C^{(2)}_g(a+1,b+1) = 
\ff{a+b+1} \, (1+b) \;
\sum_{i = \max(0,\,m-2g)}^{m}
A^{(2)}(-m, i) \, (-1)^i \, s(i+1, m+1-2g) \notag\\[2mm]
&\quad + 2
\sum_{\substack{j = 0 \\ j \equiv b \ (\mathrm{mod}\ 2)}}^{b-2}
\ff{a+j+1} \; \ff{b-j-1} \; (1+j)
\sum_{i_1=0}^{\frac{a+j}{2}} \sum_{i_2=0}^{\frac{b-2-j}{2}}
A^{(2)}\!\left(-\frac{a+j}{2}, i_1\right)
\notag\\
&\qquad\qquad\qquad\qquad \times
A^{(2)}\!\left(-\frac{b-2-j}{2}, i_2\right) (-1)^{i_1+i_2} \;
T_2(g; i_1, i_2) \notag\\[2mm]
&\quad - \sum_{\substack{j = 0 \\ j \equiv b-1 \ (\mathrm{mod}\ 2)}}^{b-1}
\ff{a+j} \; \ff{b-j-2} \; (1+j)\[\Sigma_{j,g}^{(1)}+\Sigma_{j,g}^{(2)}\]
 \,,
 \label{eq:cgs}
\end{align}}
where $s(n,k)$ are the Stirling numbers of the first kind, we defined
\begin{align}
&A^{(1)}(\alpha,i) = \frac{2^{i} \, (\alpha)_i}{(i!)^2},\\
&A^{(2)}(\alpha,i) = \frac{2^{i} \, (\alpha)_i}{(2)_i \, i!}, \\
&T_2(g;i_1,i_2) =
\sum_{t=0}^{\min(i_1+1,i_2+1)}
\binom{i_1+1}{t} \binom{i_2+1}{t} \, t! \;
s\(i_1+i_2+2-t,\, m+1-2g\), \\[2mm]
&\Sigma_{j,g}^{(1)}= \sum_{i_1=0}^{\frac{a+j+1}{2}} \sum_{i_2=0}^{\frac{b-1-j}{2}}
A^{(1)}\!\left(-\frac{a+j+1}{2}, i_1\right)
A^{(1)}\!\left(-\frac{b-1-j}{2}, i_2\right)
(-1)^{i_1+i_2} K(g; i_1, i_2) ,\\
&\Sigma_{j,g}^{(2)}= \sum_{i_1=0}^{\frac{b-j-1}{2}} \sum_{i_2=0}^{\frac{a+1+j}{2}}
A^{(1)}\!\left(-\frac{b-j-1}{2}, i_1\right)
A^{(1)}\!\left(-\frac{a+1+j}{2}, i_2\right)
(-1)^{i_1+i_2} \;
K(g; i_2, i_1),\\
&K(g;i_1,i_2) =
\sum_{t=0}^{\min(i_1,i_2+1)}
\binom{i_1}{t} \binom{i_2+1}{t} \, t! \;
s\big(i_1+i_2+1-t,\, m+1-2g\big)\,,
\end{align}
and $(x)_i$ is the Pochhammer symbol. Notice that by setting $a=b=2kn-1$, these are exactly the coefficients $C^{(2)}_g(2kn,2kn)$ introduced in Eq.~\eqref{eq:mbdycorrelation} for $m=2$ boundaries. For specific values of the genus $g$, it is possible to obtain much simpler expressions for these coefficients. For instance\footnote{The form taken by these coefficients suggests that the full expression \eqref{eq:cgs} at arbitrary $g$ also simplifies non-trivially after setting $a=b=2kn-1$. However, we could not find an explicit form for the polynomials and some of the prefactors for generic $g$.}
\begin{equation}
\begin{aligned}
    C^{(2)}_0(2kn,2kn)&=2^{4kn-3}\frac{\[\(\frac{3}{2}\)_{kn-1}\]^2}{\Gamma\(kn\)\(1\)_{kn-1}}P_{-1}(kn)=\frac{kn}{2}\frac{\[\(2kn\)!\]^2}{\[(kn)!\]^4}\,,\\
    C^{(2)}_1(2kn,2kn)&=3\times 2^{4kn-6}\frac{\[\(\frac{5}{2}\)_{kn-2}\]^2}{\Gamma\(kn-1\)\(2\)_{kn-2}}P_{1}(kn)\,,\\
    C^{(2)}_2(2kn,2kn)&=5\times 2^{4kn-12}\frac{\[\(\frac{7}{2}\)_{kn-3}\]^2}{\Gamma\(kn-2\)\(3\)_{kn-3}}P_{3}(kn)\,,\\
    C^{(2)}_3(2kn,2kn)&=\frac{35}{27}\times 2^{4kn-15}\frac{\[\(\frac{9}{2}\)_{kn-4}\]^2}{\Gamma\(kn-3\)\(4\)_{kn-4}}P_{5}(kn)\,,
    \end{aligned}
    \label{eq:specialcgs}
\end{equation}
where $P_l(x)$ are polynomials of degree $l$ in $x$, in particular
\begin{equation}
    \begin{aligned}
        &P_{-1}(x)=x^{-1}\,,\\
        &P_{1}(x)=3x-1\,,\\
        &P_{3}(x)=(2x-3)(49x^2-43x+6)\,,\\
        &P_{5}(x)=(2x-5)(1181x^4-4282 x^3+4969 x^2-1868x+120).
    \end{aligned}
\end{equation}
Starting from the explicit expression \eqref{eq:cgs}, we can also find the expansion of the coefficients when the number of matter insertions on each boundary is large. Let us set $a=2k_1-1$, $b=2k_2-1$. Then at large $k_1$, $k_2$ (with $k_1/k_2=O(1)$) we find
\begin{equation}
    C_g^{(2)}(2k_1,2k_2)\approx\frac{2^{2k_1+2k_2+1}}{2\pi}\frac{\sqrt{k_1k_2}}{k_1+k_2}\sum_{p=0}^{g}\frac{(-1)^{g-p}(k_1k_2)^{g-p}(k_1+k_2)^{g+2p}}{(g-p)!(2g-2p+1)2^{2(g-p)}12^pp!}.
\end{equation}
Summing over the genus $g$, defining $l=g-p$, and using the Taylor series representation for the error function, we obtain\footnote{Notice that the upper bound of the sum in Eq.~\eqref{eq:f2} diverges because $a+b=2k_1+2k_2-2\to\infty$ in the large $k$ limit.}
\begin{equation}
\begin{aligned}
    \left\langle\tr\(M^{2k_1}\)\right.&\left.\tr\(M^{2k_2}\)\right\rangle_{\text{conn.}}\approx\(P_{\Delta\to\infty}\)^{k_1+k_2}\sum_{g=0}^\infty C_g^{(2)}(2k_1,2k_2)N^{-2g}\\[10pt]
    &=\(P_{\Delta\to\infty}\)^{k_1+k_2}\frac{2^{2k_1+2k_2}}{\pi}\frac{\sqrt{k_{1}k_{2}}}{k_{1}+k_{2}}\,
\sum_{p=0}^\infty \frac{1}{p!}\[\frac{(k_1+k_2)^3}{12N^2}\]\,
\sum_{l=0}^{\infty}
\frac{\left[-\,k_{1}k_{2}(k_{1}+k_{2})\right]^{l}}
{\,l!\,(2l+1)\,(4N^{2})^{l}}\\
&=\(P_{\Delta\to\infty}\)^{k_1+k_2}\frac{2^{2k_1+2k_2}}{\sqrt{\pi}}\(\frac{N^{2/3}}{k_{1}+k_{2}}\)^{\frac{3}{2}}e^{\frac{(k_1+k_2)^3}{12N^2}}\erf\left(\sqrt{\frac{k_1k_2(k_1+k_2)}{4N^2}}\right).
    \end{aligned}
\end{equation}

This equation matches the analogous result obtained in \cite{Brezin_Hikami_2007} (reported in Eq. \eqref{eq:2pfairy} for $k_1=k_2$), up to a factor of $(-1)^{k_1+k_2+1}/(2\pi)$, which was unaccounted for in \cite{Brezin_Hikami_2007}.\footnote{One way to quickly check the presence of this factor is to consider the genus $g=0$ coefficient and compare it to the known result obtained by the same authors in the large-$N$ limit (with $k_1$, $k_2$ fixed) \cite{Brezin:2016eax}. The result of \cite{Brezin:2016eax} (which can be easily checked to correctly reproduce the combinatorics for small $k_1$, $k_2$) exactly matches our result \eqref{eq:specialcgs} and, when expanded at large $k_1$, $k_2$, is off by the same factor of $(-1)^{k_1+k_2+1}/(2\pi)$ with respect to the result in \cite{Brezin_Hikami_2007}. We also identified the missing factor of $(-1)^{k_1+k_2+1}/(2\pi)$ from the evaluation of prefactors in the saddle point approximation after Eq.~(54) of \cite{Brezin_Hikami_2007}.}

By setting $k_1=k_2=k$ and defining $\alpha=k/N^{2/3}$, we obtain
\begin{equation}
    \left\langle\tr\(M^{2k}\)\tr\(M^{2k}\)\right\rangle_{\text{conn.}}\approx \(P_{\Delta\to\infty}\)^{2k}\frac{2^{4k}e^{2\alpha^3/3}}{2\sqrt{2\pi}\alpha^{3/2}}\erf\left(\sqrt{\frac{\alpha^3}{2}}\right)\,.
\end{equation}
Finally, taking into account the match between the GUE and the diagrams in $\mathcal{N}=2$ JT supergravity coupled to matter and identifying $N=e^{\SBPS}$, $k=\beta_{\text{eff}}$, and $E_0=-\log\(4P_{\Delta\to\infty}\)$, we finally obtain
\begin{equation}
    \left\langle\tr\(M^{2k}\)\tr\(M^{2k}\)\right\rangle_{\text{conn.}}=\overline{\tr\(\tilde{O}_{\Delta}^{2k}\)\tr\(\tilde{O}_{\Delta}^{2k}\)}_{\text{conn.}}\equiv \overline{(\tr\rho)^2}\approx 2 \overline{Z^2(\beta_{\text{eff}})}\,,
\end{equation}
which is Eq.~\eqref{eq:cyl-resummed}.

\section{Quenched entropy from the Airy model}
\label{app:quenched}

For the states of our interest that are prepared by multiple insertions of a matter operator, we are interested in computing the asymptotics of the quenched entropy analogous to the results of \cite{Janssen:2021mek} for the thermal entropy. Representing the matter operator by a random matrix $M$, the quenched entropy is given by
\begin{equation}\label{eq:quench}
     S_Q^{(n)}=\frac{1}{1-n}\left\langle\log\frac{\sum_i|\lambda_i|^{2kn}}{\(\sum_i|\lambda_i|^{2k}\)^n}\right\rangle,
\end{equation}
where $\lambda_i$ are the eigenvalues of $M$. The eigenvalues of $M$ are distributed in a Wigner semicircle between $[-2\sqrt{P_{\Delta\to\infty}},2\sqrt{P_{\Delta\to\infty}}]$. For large $k$, the leading contribution to the quenched entropy is governed by the largest and second largest eigenvalues \textit{in magnitude}. To see this, we can order the eigenvalues $\lambda_0,\lambda_1,\dots$ with $|\lambda_0|\geq|\lambda_1|\geq\dots$. Then,
\begin{equation}
    S_Q^{(n)}\approx\frac{1}{1-n}\left\langle\log\frac{1+R^{2kn}}{\(1+R^{2k}\)^n}\right\rangle,\label{eq:mm-quench}
\end{equation} 
where $R=|\lambda_1/\lambda_0|$ so that $0\leq R\leq 1$. We then see that $S_Q^{(n)}$ is dominated by $R\approx1$ for large $k$.

Now we can consider two cases: we could have $\lambda_0,\lambda_1$ coming from the same edge or opposite edges. If they come from the same edge, there is eigenvalue repulsion between them, which ensures that $R=1$ has vanishing probability density. If, on the other hand, they come from opposite edges, then there is a very weak correlation between them, and it allows for $R$ to be much closer to 1. For computing the asymptotics of the quenched entropy, we will thus be interested in this latter case.

Let us label the maximum eigenvalue $X$ and the minimum eigenvalue $Y$. using the fact that $X,Y$ are uncorrelated \cite{Bornemann_2010}, $X$ and $-Y$ are i.i.d. random variables that are both drawn from the Tracy-Widom distribution, which describes the distribution of the largest (or smallest) eigenvalue~\cite{Tracy_Widom_1993}, i.e.,
\begin{equation}
    p_{X}(x)=p_Y(-x)=\frac{N^{2/3}}{\sqrt{P_{\Delta\to\infty}}}p_{\text{TW}_2}\left(\frac{N^{2/3}}{\sqrt{P_{\Delta\to\infty}}}(x-2\sqrt{P_{\Delta\to\infty}})\right),
\end{equation}
where $p_{\text{TW}_2}(x)$ is the usual definition of the Tracy-Widom distribution for the GUE. 
From Eq. \eqref{eq:mm-quench}, we are interested in the distribution of the random variable $R\equiv\frac{\min(|X|,|Y|)}{\max(|X|,|Y|)}$, which is given by
\begin{equation}\label{eq:probR}
    p_R(r)=2\int_0^\infty d\lambda\,\lambda p_{|X|}(\lambda)p_{|X|}(r\lambda),
\end{equation}
where $p_{|X|}(x)\equiv p_{X}(x)+p_{X}(-x)$ is the distribution of the random variable $|X|$. Since $S_Q^{(n)}$ is dominated by $R\approx 1$, we focus on the leading behaviour of $p_R(r)$ in this neighbourhood, which in the large $N$ limit becomes
\begin{equation}
    p_R(1)= 4\chi N^{2/3},\qquad\qquad \chi\equiv\int_{-\infty}^{\infty}dz\,p_{\text{TW}_2}(z)^2=0.31470\dots.
\end{equation}

Thus, taking the leading behavior in $k$, we have
\begin{gather}
    S_Q^{(n)}\approx \frac{p_R(1)}{1-n}\int_0^1 dr\,\log\frac{1+r^{2kn}}{(1+r^{2k})^n}
    \approx \chi\frac{\pi^2}{6}\left(1+\frac{1}{n}\right)\frac{N^{2/3}}{k}\label{eq:quench-renyi-large},\\
    S_Q\approx \chi\frac{\pi^2}{3}\frac{N^{2/3}}{k}.\label{eq:quench-large}
\end{gather}
The quenched second R\'enyi entropy is plotted in Figure \ref{fig:moneyplot}, obtained numerically using 1000 GUE $1000\times1000$ matrices. The leading behavior Eq. \eqref{eq:quench-renyi-large} is shown in the inset and matches the $k^{-1}$ power-law decay. 

There are various corrections to this result that we can discuss. Firstly, $X,Y$ are the eigenvalues with the two largest magnitudes ($\l_1,\l_2$) only with a finite probability ($\approx0.66$) even in the large $N$ limit. Thus, one may consider corrections coming from the case where $\l_1,\l_2$ come from the same edge. However, due to eigenvalue repulsion, the probability density of $R$ in this case vanishes at $r=1$ as $(1-r)^2$. Thus, it leads to a correction to the quenched entropy that decays as $k^{-3}$ as in the case of the thermal entropy \cite{Janssen:2021mek, Johnson:2021rsh,Antonini_Iliesiu_Rath_Tran_2025}. Further, one may consider situations where $\l_3$ is also close to $\l_1,\l_2$, but in this case, eigenvalue repulsion exists between at least two of them, and thus, it also leads to a $k^{-3}$ correction. Each of these effects may be carefully estimated using the Airy kernel.

Finally, the above analysis can be repeated for the GOE and GSE ensembles. The only difference is the change in $\chi$ due to the different Tracy-Widom distributions. Numerically, we find $\chi\approx 0.225$ for GOE and $\chi\approx 0.393$ for GSE.

\section{Random tensor network chains}
\label{app:RTNs}

In this appendix, we provide some additional calculations related to those discussed in Section \ref{sec:TN}. We will first discuss the calculation of integer semi-quenched R\'enyi entropies in the standard random tensor network (RTN) built using random unitaries \cite{Hayden:2016cfa}. This relates to the puzzle discussed in Section \ref{sec:TN} where we consider many non-interacting and non-Hermitian operator insertions. We then move on to discussing the same puzzle for non-interacting, Hermitian operator insertions, in which case the relevant RTN is built using random orthogonal matrices \cite{Akers:2025ahe}.

\subsection{Unitary Matrices}

In the standard RTN, random tensors are generated by acting with random unitaries on a fixed reference state. As we discussed in Section \ref{sec:TN}, the calculation of $\overline{\tr\(\rho^n\)}$ in the standard RTN can be reformulated in terms of a sum over permutations $g_x$ on $n$ elements applied at each tensor (labeled by $x$) in one copy of the network, with boundary conditions at the endpoints of the network fixed by the replica trick. We discussed how the corresponding action is of an Ising-type with neighboring permutations $g_x$ and $g_y$ that do not align, costing $d(g_x,g_y)\log D$ in action, where $d(g,h)$ is the Cayley distance given by the number of swaps required in going from $g$ to $h$. We can reformulate the above action in a different manner that will allow us to generalize to RTNs built up using orthogonal matrices in Section \ref{sub:ortho}.

\begin{figure}
    \centering
    \includegraphics[width=\linewidth]{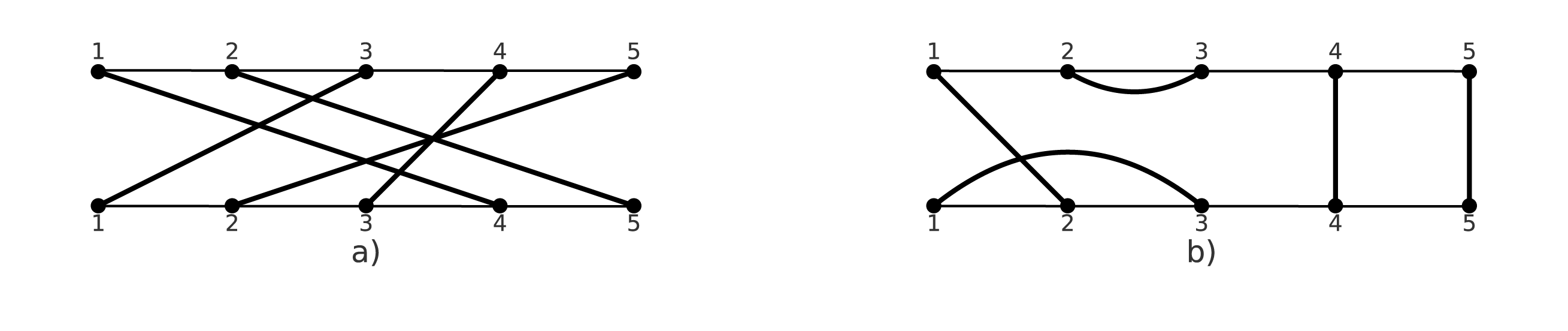}
    \caption{a) Permutation (143)(25) represented as a Brauer diagram, and b) an example of a Brauer diagram which is not a permutation.}
    \label{fig:perm_diag}
\end{figure}

We can represent a permutation as a diagram connecting $n$ points on the top line to $n$ points on the bottom line, see Figure \ref{fig:perm_diag}. Then, given neighboring permutations $g_x$ and $g_y$, the contribution to the partition function can be computed by reflecting the diagram for $g_y$ about the horizontal axis, gluing the strands together, closing the loops, and multiplying one power of $D$ for each closed loop. Since the permutations form a group, this is the same as computing $D^{f\(g_x, g_y\)}$, where $f(g,h)$ is the number of cycles in the permutation $gh^{-1}$.

We can now discuss R\'enyi-$n$ entropies with $n>2$ for the density matrix discussed in Section \ref{sec:TN}. The relevant partition functions can be computed using the standard transfer matrix method used for solving the 1D Ising model. We define the transfer matrix as (in the next two equations, we use matrix notation with matrix indices $[i,j]$)
\begin{equation}
    T\[g_x,g_y\] = D^{f\(g_x, g_y\)},
\end{equation}
in terms of which we have
\begin{equation}
    \overline{\tr\(\rho^{\otimes n} \Pi_{L}\Pi_{R}\)} =  T^{k+1}\[\Pi_L,\Pi_R\],
\end{equation}
where $\Pi_{L,R}$ are the permutations applied as boundary conditions at the left and right boundaries, respectively. For small $n$, we can explicitly compute the transfer matrix, e.g., at $n=2$, we have
\begin{equation}
    T=
\begin{pmatrix}
D^2 & D \\
D   & D^2
\end{pmatrix}.
\end{equation}
The $(k+1)$-th power can be taken by computing the eigenvalues and eigenvectors of this matrix. 

Using this method and taking the double-scaling limit $k,D\to \infty$ with $\a=\frac{k}{D}$ fixed, we have that for example,
\begin{equation}
\begin{aligned}
    S_{SQ}^{(3)}(\rho) &= \frac{1}{2}\log \[\frac{2+\cosh\(3\a\)}{\sinh \(3\a\)}\]\\
    S_{SQ}^{(4)}(\rho) &= \frac{1}{3}\log \[\frac{2+\cosh\(6\a\)+9\cosh\(2\a\)}{2+\cosh\(6\a\)-3\cosh\(2\a\)}\],
\end{aligned}
\label{eq:higherrenyi}
\end{equation}
where each of these can be checked to be positive and vanish as $\a\to\infty$.

\subsection{Orthogonal matrices}
\label{sub:ortho}

Here, we discuss how the puzzle of Section \ref{sec:TN} can be generalized, still using RTN techniques, to the case in which the operators $\tilde{O}_{\Delta}^{(i)}$ are Hermitian. In this case, one needs to create random states with an additional anti-linear $\mathbb{Z}_2$ symmetry so that the same tensor appears in both bra and ket. To do this, one can instead consider an RTN with random tensors generated by acting with random orthogonal matrices on a fixed reference state, as discussed in \cite{Akers:2025ahe}. Since the rules were not worked out in detail previously, we will briefly discuss how it works in analogy with the unitary case.

\begin{figure}
    \centering
    \includegraphics[width=\linewidth]{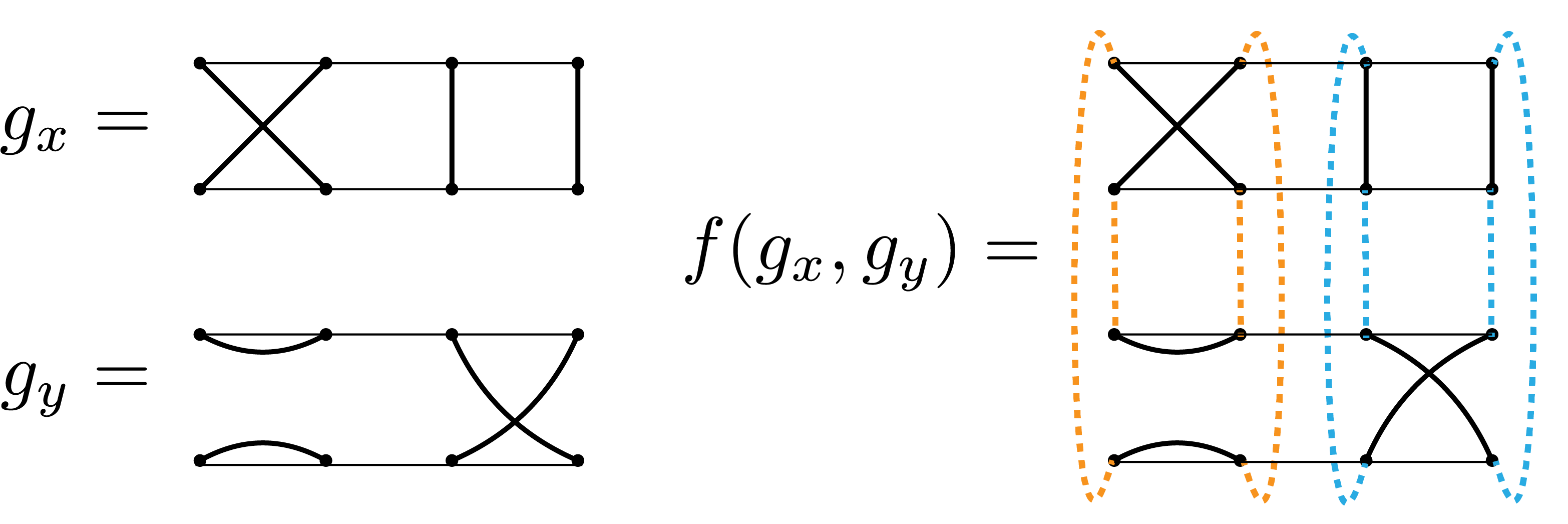}
    \caption{The contribution to the partition function from two neighbouring Brauer diagrams $g_x$ and $g_y$ is obtained by reflecting $g_y$, gluing the strands together as shown, and counting the number of loops. The two loops here are shown using orange and blue dashed lines, leading to a contribution of $D^2$.}
    \label{fig:trace_loops}
\end{figure}

Haar integrals over the orthogonal group result in a sum over Brauer diagrams, which are pairwise connections between $n$ points on the top line and $n$ points on the bottom line, see Figure \ref{fig:perm_diag}. Permutations are a subset of Brauer diagrams, where we do not allow connections within the top or bottom line. However, the existence of diagrams connecting within the top or bottom line implies that while the Brauer diagrams form an algebra, they do not form a group since the inverse of such elements does not exist. Nevertheless, there is still a natural bilinear form $f\(g_x,g_y\)$ given two Brauer diagrams $g_x$ and $g_y$, which is again defined by inverting the diagram for $g_y$, gluing the strands together, closing the loops and multiplying one power of $D$ for each closed loop, see Figure \ref{fig:trace_loops}.

The calculation of the R\'enyi entropies in this RTN can again be reformulated in terms of an Ising-like model where the degrees of freedom at each vertex are Brauer diagrams and the contribution to the action from a link connecting elements $g_x$ and $g_y$ is $D^{f(g_x,g_y)}$. We can then compute again the relevant partition functions using the transfer matrix method. For example, for $n=2$, the transfer matrix is
\begin{equation}
     T=
\begin{pmatrix}
D^2 & D & D \\
D   & D^2 &D\\
D   & D &D^2
\end{pmatrix},
\end{equation}
where the rows/columns label the three Brauer diagrams at $n=2$. Using this method and taking the double-scaling limit $k,D\to \infty$ with $\a=\frac{k}{D}$ fixed, we have, e.g.,
\begin{equation}
\begin{aligned}
    S_{SQ}^{(2)}(\rho) &= \log \[\frac{\exp\(2\a\)-\exp\(-\a\)}{\exp\(2\a\)+2\exp\(-\a\)}\],
\end{aligned}
\label{eq:higherrenyiTN}
\end{equation}
which is manifestly positive and vanishes as $\a\to\infty$.

Finally, we comment on the bulk interpretation of this setup. As we saw in Sections \ref{sec:2}, \ref{sec:bulk-resolution}, and \ref{sec:matrixmodels}, when we consider Hermitian operators that are uncharged, we are allowed to consider diagrams where we have bra-bra or ket-ket contractions. In a more general setting, we require an additional anti-linear $\mathbb{Z}$ symmetry like time-reversal to allow a bra-bra or ket-ket gluing for fixed-area geometries. These diagrams are necessarily non-planar and do not modify any previous results in the literature, such as the resolution of entanglement phase transitions \cite{Penington:2019kki,Dong:2020iod,Marolf:2020vsi,Akers:2020pmf}, but nevertheless should be accounted for when considering time-reflection symmetric setups. 

We would also like to mention the relation to the Temperley-Lieb algebra discussed in \cite{Akers:2022zxr}. In that setup, despite starting from a standard RTN, the canonical purification results in a $\mathbb{Z}_2$ symmetric state. The Temperley-Lieb algebra is the non-crossing subalgebra of the Brauer algebra. Thus, our discussion here is closely related to that of \cite{Akers:2022zxr} and agrees at leading order in the number of crossings.

\bibliographystyle{jhep}
\bibliography{biblio}
\end{document}